\documentclass[english,onecolumn,a4paper,sort&compress]{extarticle}
\usepackage[T1]{fontenc}
\usepackage[latin9]{inputenc}
\usepackage{geometry}
\usepackage{xcolor}
\usepackage{babel}
\usepackage{array}
\usepackage{float}
\usepackage{units}
\usepackage{textcomp}
\usepackage{mathrsfs}
\usepackage{mathtools}
\usepackage{multirow}
\usepackage{amsmath}
\usepackage{amsthm}
\usepackage{amssymb}
\usepackage{graphicx}
\usepackage[numbers]{natbib}
\usepackage[unicode=true,pdfusetitle,
 bookmarks=true,bookmarksnumbered=false,bookmarksopen=false,
 breaklinks=false,pdfborder={0 0 0},pdfborderstyle={},backref=false,colorlinks=true]
 {hyperref}

\makeatletter

\providecommand{\tabularnewline}{\\}

\numberwithin{equation}{section}
\numberwithin{figure}{section}
\theoremstyle{plain}
\newtheorem{thm}{\protect\theoremname}
  \theoremstyle{plain}
  \newtheorem{cor}[thm]{\protect\corollaryname}
  \theoremstyle{plain}
  \newtheorem{lem}[thm]{\protect\lemmaname}
  \theoremstyle{remark}
  \newtheorem{rem}[thm]{\protect\remarkname}

\usepackage{bbm}
\usepackage[auth-lg]{authblk}

\author[1]{Fl\'avio Eler De Melo\thanks{flavio.de-melo@liverpool.ac.uk}}
\author[1]{Simon Maskell\thanks{s.maskell@liverpool.ac.uk}}
\author[2]{Matteo Fasiolo\thanks{matteo.fasiolo@bristol.ac.uk}}
\author[3]{Fred Daum\thanks{frederick\_e\_daum@raytheon.com}}
\affil[1]{\small Department of Electrical Engineering and Electronics, University of Liverpool}
\affil[2]{\small Department of Mathematics, University of Bristol}
\affil[3]{\small Raytheon Company}

\usepackage{fancyhdr}
\usepackage[algo2e,linesnumbered,ruled,vlined]{algorithm2e}

\SetCommentSty{mycommfont}

\makeatother

  \providecommand{\corollaryname}{Corollary}
  \providecommand{\lemmaname}{Lemma}
  \providecommand{\remarkname}{Remark}
\providecommand{\theoremname}{Theorem}

\begin{document}

\title{Stochastic Particle Flow for Nonlinear High-Dimensional Filtering
Problems}
\maketitle
\begin{abstract}
A series of novel filters for probabilistic inference that propose
an alternative way of performing Bayesian updates, called particle
flow filters, have been attracting recent interest. These filters
provide approximate solutions to nonlinear filtering problems. They
do so by defining a continuum of densities between the prior probability
density and the posterior, i.e. the filtering density. Building on
these methods' successes, we propose a novel filter. The new filter
aims to address the shortcomings of sequential Monte Carlo methods
when applied to important nonlinear high-dimensional filtering problems.
The novel filter uses equally weighted samples, each of which is associated
with a local solution of the Fokker-Planck equation. This hybrid of
Monte Carlo and local parametric approximation gives rise to a global
approximation of the filtering density of interest. We show that,
when compared with state-of-the-art methods, the Gaussian-mixture
implementation of the new filtering technique, which we call Stochastic
Particle Flow, has utility in the context of benchmark nonlinear high-dimensional
filtering problems. In addition, we extend the original particle flow
filters for tackling multi-target multi-sensor tracking problems to
enable a comparison with the new filter.
\end{abstract}

\section{Introduction}

Stochastic filtering in high-dimensional spaces is a challenging estimation
task because of two fundamental issues:
\begin{itemize}
\item \emph{The curse of dimensionality}. In a statistical experiment, as
the sample space's dimensionality increases a finite number of realizations
can only populate the space to an increasingly sparse extent \citep{Novak2016}.
This issue makes it challenging to use approximation based on realizations
of the state. 
\item \emph{The infinite number of parameters required to describe a general
probability density on a continuous state space}. Such a density,
in common with any other real function, can always be exactly described
using a power series with infinitely many terms. In all but a very
few cases, where the density is known to have a specific parametric
form, using a finite set of parameters is necessarily an approximation
to this complete description. The fidelity of such approximations
falls rapidly as dimension increases. This issue makes it challenging
to define a parameterization that uses a number of parameters that
scales only gently with dimension.
\end{itemize}
The development of the vast majority of practical filters focuses
on how to accurately represent generic probability densities. However,
in the view of the authors, relatively few filters are systematically
developed with the explicit intent of efficiently expressing densities
in high-dimensional spaces. There does appear to be a consensus that
the statistical efficiency associated with expressing high-dimensional
filtering densities can be improved by simulating tempering distributions
\citep{Gelman1998,Deutscher2000,Neal2001,DelMoral2006}. Such approaches
involve introducing intermediate distributions such that it is easier
to migrate between these intermediate distributions than it is to
migrate directly from the prior to the posterior. The use of such
intermediate distributions stabilizes the sampling procedure and maintain
the variance of the Monte Carlo weights at an acceptable level. Bickel~\textit{et
al.} \citep{Bickel2008} considered, in the context of a bootstrap
particle filter, the number of intermediate distributions needed to
use such tempering successfully. They prove that, as the dimensionality
increases, the number of intermediate distributions needed to accurately
represent a high-dimensional density becomes practically infinite.
This implies that considering a continuum, i.e. infinite number, of
intermediating distributions \citep{Neal2001} might be the basis
of a successful approach. This implication is corroborated by the
reported success of Markov chain Monte Carlo (MCMC) algorithms that
populate high-dimensional state spaces efficiently using approximations
of problem-specific continuous-time processes \citep{Duane1987,Neal1996,Roberts2002,Girolami2011}\footnote{Note that while practical implementation of these techniques necessarily
involves finite time-horizons, the continuous-time processes are typically
designed such that, as time tends to infinity, the distribution of
the samples from the process tends to the distribution of interest.
This is in contrast to the use of tempering distributions, where samples
from the posterior are generated after a defined (finite) number of
steps between intermediate distributions.}. 

Techniques for continuous-time processes stem from the seminal work
by Stratonovich \citep{Stratonovich1959}, Kushner \citep{Kushner1964}
and Zakai \citep{Zakai1969} on filtering theory. The most popular
instance of such filters is the so-called Kalman-Bucy filter \citep{Bucy1968},
the continuous-time counterpart of the Kalman filter. More general
filters directly approximate solutions to the Kushner-Stratonovich
equation either by a finite-dimensional density parameterization \citep{BeneS1981,Daum1986}
or by Monte Carlo methods \citep{Crisan1999,Crisan1999a,Crisan2010}.
Other important finite-order filters that appeal to an unusual formalism
of multiple stochastic integrals \citep{Budhiraja1996,Budhiraja1997}
are worth mentioning as well.

Continuous-time filtering may be seen by some as an idealized problem
of limited practical utility. However, recent research \citep{Verge2015,Rebeschini2015}
has shown that continuous-time filtering can offer key insights into
the fundamental principles necessarily associated with successful
filtering in high-dimensions: the effects of local, continuous spatial
properties of the observation process need to be incorporated in the
solution. As identified by Bickel~\textit{et al.} \citep{Bickel2008},
the information in the data gives rise to the notion of the \emph{effective
dimension} of the space. It is this effective dimension, and not the
dimension of the state space itself, that actually affects the statistical
efficiency of inference algorithms. This implies that tempering only
addresses part of the problem, remaining the local observation properties
to be incorporated. Recently, based on a principled approach, Rebeschini
\& van Handel \citep{Rebeschini2015} proposed to decompose the state
space into separate blocks. The global solution to the inference problem
is then constructed by combining the local solutions for each of the
blocks. The paper goes on to demonstrate that, by using the \emph{decay
of correlations} property\footnote{A spatial counterpart of the stability property of nonlinear filters,
by which a probability mass is strongly correlated to masses within
its neighbourhood but has negligible correlation with respect to the
remaining areas of the state space.}, it is possible to develop particle filters based on local solutions
in such a way that the approximation error does not depend on the
dimension of the state space. 

Largely independently, the idea of filtering via a continuum of intermediate
distributions seems to have its appeal revigorated as several new
methods have been proposed for progressive Bayesian updates, whose
continuity is considered in the limit, aiming to gradually introduce
the effect of each observation. Those filters have emerged either
in a variational, ensemble-based, or sequential Monte Carlo framework.
In the variational framework, the new methods presented in \citep{Hanebeck2003,Hagmar2011,Hanebeck2012}
pose the filtering problem as a multi-step optimization problem for
which the cost function is an approximated distance between a parameterised
density and the actual filtering density. In the ensemble-based framework,
the methods \citep{Reich2011,Reich2013,Venugopal2015} are focused
on data assimilation problems and apply ideas of optimal transport
along with continuous-time filtering to generate multiple independent
solutions that are combined to obtain a single solution of an inference
problem. The methods in the sequential Monte Carlo framework explore
extensions or alternatives to particle filters (e.g., \citep{Oudjane2000,Crisan2006,Daum2007,Sarkar2014,Crisan2015}),
or simply capitalize on techniques for properly choosing a sequence
of \textit{bridging} importance densities (e.g., \citep{Deutscher2000,Neal2001,DelMoral2006,Bunch2013}),
carrying on the intent to overcome the widely known problem of particle
filters called \emph{degeneracy }or collapse of weights \citep{Arulampalam2002,Daum2003,Snyder2008,Doucet2009}.

Among those methods one in particular has recently attracted interest
in an Engineering context where it has been described as \emph{particle
flow}. The performance that has been reported is remarkable and the
literature is extensive with several variants having been developed
over recent years (see, for example, \citep{Daum2007,Daum2008,Daum2009,Daum2010b,Daum2013a}).
The development of particle flow draws on analogies to problems that
arise in Fluid Dynamics and Electromagnetism. These filters \emph{flow}
probability masses (particles) from a prior probability space to one
that is updated according to a set of measurements without the need
to perform a Bayesian update explicitly. All particle flow algorithms
explore the concept of a homotopy between the prior and posterior
probability spaces, implicitly describing a joint measure that couples
the prior and posterior probability measures. This idea is in the
heart of the Kantorovich's optimal tranportation problem \citep{Villani2003}
that, by evoking deterministic transport maps for very simple cost
functions and dynamic constraints, yields an essential explanation
on why original particle flow methods work well based on deterministic
rules to flow the particles.

When the sequential filtering problem involves non-compactly supported
densities, solving it via deterministic (optimal) transport is not
straightforward. A solution would require either a non-trivial approximation
of the highly nonlinear Monge-Amp\`ere equation \citep{Gutierrez2001}
or adapting classical solutions constructed for measures on bounded
sets \citep{Moser1965,Dacorogna1990}. In these approaches, severe
technical difficulties may arise and not all the effects on the estimation
errors are clearly known. A continuously evolving, exact, optimal
transport map would require a complete description, at all time instants,
of an embedding dynamic field that induces a transference plan to
correctly move particles. If the posterior density could be completely
characterized beforehand then the optimal transport problem could
be numerically solved by the multiple-step augmented-Lagrangian optimization
method as proposed by Benamou \& Brenier \citep{Benamou2000}. However,
detailed knowledge of the posterior would imply a direct answer to
the filtering problem. As an alternative, theoretically speaking,
a complete description of the optimal field could be achieved by solving
the Monge-Amp\`ere equation for any possible location of particles
on the state space. Notwithstanding, the Monge-Amp\`ere equation
admits exact solutions only for few particular cases \citep{Gutierrez2001}
and would also require a thorough description of the posterior density
in advance.

In this scenario, one feasible approach is that advocated by particle
flow methods, which take simplifying assumptions on the embedding
dynamic field in order to avoid both optimization over a parametric
class of transport maps and explicit solution of the associated elliptic
partial differential equation. However, in our experience, these symplifying
assumptions result in approximated filtering densities providing accurate
estimates for the first-order moment but estimates for second and
higher-order moments whose quality is highly dependent on the problem
and algorithm settings (e.g., \citep{Heng2015}). In practice, particle
flow methods address this latter issue by either relying on a companion
filter \citep{Choi2011,Ding2012} or using the sample covariance matrix
with shrinkage and Tikhonov regularization \citep{Khan2015} to be
able to estimate the second-order moment. 

We conjectured if appealing to stochastic transport could provide
a new avenue for solving the filtering problem. Fortunately, a variational
formulation of the Fokker-Planck equation as a gradient flow, as exposed
by Jordan~\emph{et al.} \citep{Jordan1998}, enables the precise
interpretation that, if a transport operation is to be understood
as a diffusion, then it minimizes the free energy functional of the
process with respect to the Wasserstein metric over an admissible
class of probability measures. Relying on this formulation, it is
straightforward to obtain a transport rule, optimal in terms of minimizing
the free energy functional, as a Langevin stochastic process. This
rule is based simply on the assumptions of stationarity of the filtering
distribution (Gibb's distribution) and on potential conditions, for
which an embedding stationary field is exactly derived.

In this article we take into consideration the findings presented
by Jordan~\emph{et al.} \citep{Jordan1998}, incorporate the description
of statistically efficient processes in high-dimensional spaces as
proposed by Girolami \& Calderhead \citep{Girolami2011}, and incorporate
local properties of the observation process to formulate a stochastic
variant of particle flow\footnote{Existing particle flow algorithms (including, perhaps surprisingly,
that known as non-zero diffusion particle flow \citep{Daum2013c})
propagates particles deterministically.}. This new \emph{stochastic particle flow} (SPF) involves defining
a Langevin diffusion such that a posterior measure from a previous
step, under a known stationary potential field, is diffused onto the
current posterior measure, satisfying the Fokker-Planck equation to
produce an accurate approximation of the filtered density. This process
involves guiding local solutions of the Fokker-Planck equation in
such a way that we construct a mixture that approximates the posterior.
As we will discuss later on, the SPF method we propose is essentially
built as a Gaussian sum filter (SPF-GS), nevertheless, it is possible
to use a similar formulation to define an implementation strategy
based on a marginal particle filter (SPF-MPF). This variant demonstrates
versatility of the SPF to algorithm settings. 

It is worth mentioning that our resulting SPF technique is in the
same ethos as the method recently developed by Bunch \& Godsill \citep{Bunch2013,Bunch2015}.
However, in constrast to our approach, their method (i) is based on
the homotopy between the prior and posterior spaces, (ii) assumes
the particle flow is an Ornstein-Uhlenbeck process whose scaling parameter
determines the rate of diffusion of samples' paths, (iii) proposes
weights that must be updated iteratively by a partial differential
equation (PDE) describing how the unnormalized log-density evolves
with a pseudo-time variable; (iv) is articulated as a standard (not
marginal) particle filter.

The outline of the article is as follows. We begin by reviewing the
stochastic filtering problem in a sequential Monte Carlo framework
in Section~\ref{sec:Sequential-Monte-Carlo}. We abstract the solution
in terms of a general map that could adopt any valid method to perform
the filtering update. In Section~\ref{sec:Particle-flow}, we present
a brief overview of the original particle flow methods. We discuss
their principles in order to further clarify these methods and motivate
the natural step towards stochastic particle flow. In Section~\ref{sec:Stochastic-particle-flow}
we derive the generic SPF algorithm by describing the proposed dynamics
of probability masses, describing the associated stationary solution
to the Fokker-Planck equation, and constructing the stochastic flow.
Algorithmic details are given and relate to how to compute the diffusion
matrix, how to integrate the stochastic flow, and how to select the
simulation time horizon and integration step size. We present the
stochastic particle flow implementation using a Gaussian sum filter
(SPF-GS) in Section~\ref{sec:SPF-GS}. We achieve this by considering
the posterior to be well approximated as a mixture of local solutions
to the flow. Similarly, in Section~\ref{sec:SPF-MPF} we show the
SPF articulated as a marginal particle filter (SPF-MPF) by setting
the importance density as a mixture analogous to that generated by
the SPF-GS. Section~\ref{sec:Examples} then illustrates the SPF's
properties by a series of toy problems, and compares the performance
of the SPF and other state-of-the-art methods in the context of three
instructive multi-sensor or multi-target tracking problems: multi-sensor
bearing-only tracking, convoy tracking and inference on a large network
of sensors (as in \citep{Septier2016}). In the comparisons for the
multi-sensor bearing-only and convoy tracking problems, we included
extensions to two of the most effective (original) particle flows,
namely, the Gaussian particle flow (GPF) \citep{Daum2010a} and the
scaled-drift particle flow (SDPF) \citep{Daum2013c}. Finally, Section~\ref{sec:Conclusions}
concludes.

\section{Sequential Monte Carlo Filtering\label{sec:Sequential-Monte-Carlo}}

In this section we report the filtering framework within which the
particle flows may be formalized. Let $\{\mathrm{x}_{t}\in\mathcal{X}:t\in\mathbb{R}_{+}\}$
be a sequence of states generated through time by a known continuous-time
state process, modelled as a Markov process, and $\{\mathrm{y}_{t_{k}}\in\mathcal{Y}:t_{k}\in\mathbb{R}_{+},k\in\mathbb{N}\}$
be a sequence of discrete-time observations of the process generated
by an observation model. In the classical filtering problem, one is
required to compute the best estimate of a function of interest $\varphi$
of the state, given all observations realized up to the time instant
$t_{k}$, i.e., 
\begin{equation}
\hat{\varphi}_{k}=\mathbb{E}\left[\varphi(\mathrm{x}_{t_{k}})|\mathrm{y}_{t_{1}},\mathrm{y}_{t_{2}},\dots,\mathrm{y}_{t_{k}}\right].\label{eq:filtered-function}
\end{equation}

To simplify notation, we will denote all variables at discretized
time instants by the time indexes $k\in\mathbb{N}$, and write $\mathrm{y}_{1:k}^{\vphantom{(i)}}\triangleq\{\mathrm{y}_{1}^{\vphantom{(i)}},\mathrm{y}_{2}^{\vphantom{(i)}},\dots,\mathrm{y}_{k}^{\vphantom{(i)}}\}$.
Now consider a set of particles $\{\mathrm{x}_{k-1}^{(i)},w_{k-1}^{(i)}:\,i=1,\,\dots,\,N\}$
constituting samples that can be used to approximate a filtering probability
density $p(\mathrm{x}_{k-1}|\mathrm{y}_{1:k-1})$ by means of a Monte
Carlo measure satisfying 
\begin{equation}
\sum_{i=1}^{N}w_{k-1}^{(i)}\delta(\mathrm{x}_{k-1}^{\vphantom{(i)}}-\mathrm{x}_{k-1}^{(i)})\overset{\vphantom{\tfrac{}{}}{\scriptstyle N\rightarrow\infty}}{\longrightarrow}p(\mathrm{x}_{k-1}^{\vphantom{(i)}}|\mathrm{y}_{1:k-1}^{\vphantom{(i)}}),\label{eq:MC-measure-convergence-prior}
\end{equation}

\noindent to represent the convergence as follows for any test function
$\varphi:\mathcal{X}\rightarrow\mathbb{R}$: 
\begin{equation}
\sum_{i=1}^{N}w_{k-1}^{(i)}\varphi(\mathrm{x}_{k-1}^{(i)})\overset{\vphantom{\tfrac{}{}}{\scriptstyle N\rightarrow\infty}}{\longrightarrow}\int_{\mathcal{X}}\varphi(\mathrm{x}_{k-1}^{\vphantom{(i)}})p(\mathrm{x}_{k-1}^{\vphantom{(i)}}|\mathrm{y}_{1:k-1}^{\vphantom{(i)}})d\mathrm{x}_{k-1}^{\vphantom{(i)}}\quad\text{almost surely}.\label{eq:MC-measure-convergence-prior-1}
\end{equation}

Given a new observation obtained at instant $k$, one wishes to find
a procedure to transform the set of particles $\{\mathrm{x}_{k-1}^{(i)},w_{k-1}^{(i)}\}$
into a new set of particles $\{\mathrm{x}_{k}^{(i)},w_{k}^{(i)}:\,i=1,\,\dots,\,N\}$
that incorporates the effect of the latest observation in order to
estimate the filtered entity as 
\begin{equation}
\hat{\varphi}_{k}\approx\sum_{i=1}^{N}w_{k}^{(i)}\varphi(\mathrm{x}_{k}^{(i)}).\label{eq:estimated-filtered-function}
\end{equation}

In theory, the filtering problem in the sequential Monte Carlo form
can be solved by any map $\mathcal{T}:\mathcal{X}\times\mathcal{Y}\rightarrow\mathcal{X}'$,
$\mathcal{T}\in\mathcal{C}^{1}(\mathbb{R}^{n_{x}})\times\mathcal{C}^{0}(\mathbb{R}^{n_{y}})$,
where $\left|\mathcal{X}'\right|=\left|\mathcal{X}\hphantom{'}\right|$,
that implements 
\begin{eqnarray}
\mathrm{x}_{k}^{(i)} & \coloneqq & \mathcal{T}(\mathrm{x}_{k-1}^{(i)},\mathrm{y}_{k}^{\vphantom{(i)}});\,i=1,\,\dots,\,N;\label{eq:mapping}\\
w_{k}^{(i)} & \coloneqq & \det\mathcal{J}_{\mathrm{x}_{k-1}}\left[\mathcal{T}\right]^{-1}w_{k-1}^{(i)};\label{eq:mapping-weight}
\end{eqnarray}
\noindent where $\mathcal{J}_{\mathrm{x}_{k-1}}\left[.\right]$ is
the Jacobian matrix with respect to $\mathrm{x}_{k-1}$, and such
that 
\begin{equation}
\sum_{i=1}^{N}w_{k}^{(i)}\delta(\mathrm{x}_{k}^{\vphantom{(i)}}-\mathrm{x}_{k}^{(i)})\overset{\vphantom{\tfrac{}{}}{\scriptstyle N\rightarrow\infty}}{\longrightarrow}p(\mathrm{x}_{k}^{\vphantom{(i)}}|\mathrm{y}_{1:k}^{\vphantom{(i)}}).\label{eq:MC-measure-convergence-posterior}
\end{equation}

Although most practical filters implement the mapping (\ref{eq:mapping})
in terms of discrete Bayesian updates, there should be no objection
to the general idea of considering the map $\mathcal{T}$ as a transform
continuous in time within $t_{k-1}<t\le t_{k}$. This idea establishes
the basis for the particle flow filters.

\section{Particle Flow\label{sec:Particle-flow}}

This section aims to present a brief overview on the particle flow
methods, to discuss their principles, and to set the background for
the introduction of the stochastic particle flow. The key idea of
the particle flow is to transfer a set of probability masses by an
operation that transports the prior probability measure onto the posterior
measure. This operation realizes the measurement update smoothly in
order to express a filtering entity, usually an estimate. The mechanism
implied is, therefore, a filtering algorithm that avoids the need
to perform a Bayesian measurement update explicitly.

Given a set of particles $\left\{ \mathrm{x}^{(i)}\left(\lambda\right)\in\mathbb{R}^{n_{x}}:\,i=1,\,\dots,\,N\right\} $
dependent on a continuous pseudo-time variable $\lambda\in\left[0,1\right]$,
where $n_{x}$ is the number of dimensions of the state space, and
such that $\mathrm{x}_{\vphantom{k}}^{(i)}\left(0\right)=\mathrm{x}_{k-1}^{(i)}$
and $\mathrm{x}_{\vphantom{k}}^{(i)}\left(1\right)=\mathrm{x}_{k}^{(i)}$,
the transformation of the particles is accomplished by solving through
$0<\lambda\le1$ an ordinary differential equation (ODE) referred
to as the flow equation 
\begin{eqnarray}
\frac{d\mathrm{x}}{d\lambda} & = & \mu\left(\mathrm{x},\lambda\right),\quad\mathrm{x}^{(i)}\left(0\right)\sim p_{0}\left(\mathrm{x}\right).\label{eq:flow-equation}
\end{eqnarray}
The varieties of particle flow methods rely on how one defines the
flow drift $\mu\left(\mathrm{x},\lambda\right)$, which in turn depends
on the assumptions made to solve the associated continuity equation
\begin{align}
\frac{\partial p}{\partial\lambda} & =-\nabla_{\mathrm{x}}\cdot\left(\mu\cdot p\right),\quad p\left(\mathrm{x},0\right)=p_{0}\left(\mathrm{x}\right).\label{eq:continuity-equation}
\end{align}
The operator $\nabla_{\mathrm{x}}\cdot\left(.\right)$ is the divergence
operator and the drift can be understood as a vector field $\mu\left(\mathrm{x},\lambda\right)\in\mathbb{R}^{n_{x}}$
that is not uniquely determined for a given probability density $p\left(\mathrm{x},\lambda\right)$.
In the optimal transportation literature the vector field is usually
determined by the constraint that it minimizes the kinetic energy.
In that case, the flow equation (\ref{eq:flow-equation}) can be written
in terms of a dynamic potential field as $\mu\left(\mathrm{x},\lambda\right)=\mathcal{M}^{-1}\nabla_{\mathrm{x}}\psi\left(\mathrm{x},\lambda\right)$
\citep{Villani2003}, where $\mathcal{M}$ is a positive-definite
mass matrix, $\nabla_{\mathrm{x}}$ is the gradient operator, and
$\psi\left(\mathrm{x},\lambda\right)$ is a dynamic potential function
that satisfies the elliptic PDE 
\begin{equation}
\nabla_{\mathrm{x}}\cdot\left(p\left(\mathrm{x},\lambda\right)\mathcal{M}^{-1}\nabla_{\mathrm{x}}\psi\left(\mathrm{x},\lambda\right)\right)=-\partial_{\lambda}p\left(\mathrm{x},\lambda\right).\label{eq:elliptic-equation-a}
\end{equation}

An exact solution to equation (\ref{eq:elliptic-equation-a}) has
been derived by Reich \citep{Reich2011} considering Gaussian likelihood
functions. In more general settings, if the target posterior density
$\pi\left(\mathrm{x}\right)$ could be thoroughly characterized in
advance, the numerical solution to this problem could be achieved
by the multiple-step augmented-Lagrangian optimization method as proposed
by Benamou \& Brenier \citep{Benamou2000}. However, availability
of a detailed description of the posterior density would constitute
a direct answer to the filtering problem. Similarly, the well known
flow constructed by Dacorogna \& Moser \citep{Dacorogna1990}, appropriate
for mapping measures on bounded open sets, could be adapted for problems
involving non-compactly supported densities as the solution of the
$p$-Laplacian equation \citep{Evans1999} 
\begin{equation}
\nabla_{\mathrm{x}}\cdot\left(a\left(\mathrm{x},\lambda\right)\nabla_{\mathrm{x}}\vartheta\left(\mathrm{x},\lambda\right)\right)=\pi\left(\mathrm{x}\right)-p\left(\mathrm{x},\lambda\right),\label{eq:elliptic-equation-b}
\end{equation}
\noindent where $p\left(\mathrm{x},\lambda\right)$ and $\pi\left(\mathrm{x}\right)$
are the intermediate and target densities respectively. Function $a\left(\mathrm{x},\lambda\right)\ge0$,
${a\left(\mathrm{x},\lambda\right)\in\mathbb{L}^{\infty}}$ ($\mathbb{L}^{\infty}$-space\footnote{The $\mathbb{L}^{\infty}$-space generalises the $\mathbb{L}^{p}$-spaces
to $p=\infty$. An $\mathbb{L}^{p}$-space describes the set of all
functions $f$ for which the norm $\left\Vert f\right\Vert _{p}=\left(\int_{\mathcal{X}}\left|f\right|^{p}\right)^{1/p}$
converges. The concept is analogous for the $\mathbb{L}^{\infty}$-space
although its norm is defined by the essential supremum.}), is a Lagrange multiplier that scales the distance of optimal transportation,
whereas the term $\nabla_{\mathrm{x}}\vartheta\left(\mathrm{x},\lambda\right)$
gives the direction of optimal tranportation. As mentioned before,
these transport-based solutions are not straightforwardly applicable
to filtering problems as they would require anticipative approximations
of the target probability density, and the solution by Dacorogna \&
Moser \citep{Dacorogna1990} would require truncation of the involved
densities to bound their support.

Indeed original particle flows do not follow the classical transport-based
methodology but rather take simplifying assumptions on the dynamic
potential field, avoiding the complexity of solving the elliptic PDEs
(\ref{eq:elliptic-equation-a}) and (\ref{eq:elliptic-equation-b}).
Specifically, the particle flows are derived from a programmed sequence
of a dynamic potential field that roughly solves the equation (\ref{eq:continuity-equation}).
As examples we refer the reader to the incompressible particle flow
\citep{Daum2007}, the Gaussian or exact particle flow \citep{Daum2010a},
and the non-zero \textit{``diffusion''} particle flow \citep{Daum2013c},
which is not actually a diffusion, but simply takes into account a
diffusion term to scale and/or offset the drift term.

In a closely related problem, as an alternative to the solution of
elliptical equations or to original particle flows, it is possible
to demonstrate that if the drift solves the continuity equation (\ref{eq:continuity-equation}),
under a stationary potential field (conservative) related to an invariant,
locally\footnote{Log-concave in the vicinity of the density maxima.}
log-concave density of the form ${p\left(\mathrm{x},T\right)=\pi\left(\mathrm{x}\right)\propto\exp\left(-\psi\left(\mathrm{x}\right)\right)}$,
then the flow (\ref{eq:flow-equation}) produces the maximum-a-posteriori
(MAP) estimate, $\hat{\mathrm{x}}_{MAP}$, after an appropriate time
horizon $\lambda\ge T$ (see \emph{Theorem~\ref{thm:T6}} in the
Appendix\emph{~}\ref{subsec:Proofs-2}). A similar concept is used
in optimization algorithms based on gradient descent. An evident problem
with this approach is that, regardless of providing a MAP estimate,
it is unable to capture higher-order aspects of a target posterior
density. Thus, under the assumption of a stationary potential field,
a stochastic particle flow seems suitable to describe a filtering
density precisely up to an arbitrary moment order, by following the
dynamics of a diffusion that minimizes the free energy functional
(see \citep{Jordan1998} for details). Such stochastic flow would
propagate a probability density according to the Fokker-Planck equation.
This observation becomes fundamental when we note that, loosely speaking,
obtaining a precise approximation of a stationary potential field
requires less effort than obtaining a sequence of accurate approximations
of a dynamic potential field. In this context, approximating a dynamic
potential field forms the basis for the classical transport methodology
(e.g., \citep{Reich2011}).

\section{Stochastic Particle Flow\label{sec:Stochastic-particle-flow}}

This section derives stochastic particle flow based on a stationary
solution to the Fokker Planck equation. We capitalize on the fact
that, under certain conditions on the drift and diffusion terms of
a stochastic process, there is a stationary solution that satisfies
a variational principle, minimizing a certain convex free energy functional
over an admissible class of probability densities. The Fokker\textendash Planck
equation is shown to follow the direction of steepest descent of the
associated free energy functional \citep{Jordan1998} at each instant
of time, rendering a process where the entropy is maximized, i.e.,
a diffusion.

In Section~\ref{subsec:Dynamics-of-particles} we set dynamics for
stochastic particle flow. Section~\ref{subsec:Stationary-solution}
derives the stationary solution to the Fokker-Planck equation such
that the particles follow the Langevin dynamics. In Section~\ref{subsec:The-stochastic-flow}
we show how to specify the Langevin dynamics to solve the specific
problem of interest. In Section~\ref{subsec:The-diffusion-matrix}
we discuss the interpretation of and possible choices for the diffusion
matrix; in Section~\ref{subsec:Integration-method} we present the
integration methods used to sample from the Langevin dynamics; and
in Section~\ref{subsec:Selection-of-time} we discuss criteria for
choosing the algorithm's parameters (the step size and time horizon).

\subsection{Dynamics of Particles\label{subsec:Dynamics-of-particles}}

Assuming that a set of particles $\left\{ \mathrm{x}^{(i)}\left(\lambda\right):\,i=1,\,\dots,\,N\right\} $
follows a diffusion process $\left\{ X_{\lambda}\right\} _{\lambda\ge0}$
when subject to a Bayesian measurement update, the dynamics of the
particles can, in general, be described by the \^Ito stochastic differential
equation 
\begin{equation}
dX_{\lambda}=\mathrm{\mu}\left(X_{\lambda},\lambda\right)d\lambda+\mathrm{\sigma}\left(X_{\lambda},\lambda\right)dW_{\lambda},\quad X_{0}=X\left(0\right);\label{eq:SDE}
\end{equation}
\noindent such that the associated probability distribution, $p\left(\mathrm{x},\lambda\right)$,
is continuously evolving with respect to the pseudo-time variable
$\lambda\in\mathbb{R}_{+}$, where $\left\{ W_{\lambda}\right\} $
is a standard Brownian motion, $\mathrm{\mu}\left(X_{\lambda},\lambda\right)$
is the drift vector and $\mathrm{\sigma}\left(X_{\lambda},\lambda\right)$
is the diffusion coefficient. It is well known \citep{Jazwinski1970,Gardiner1985}
that the probability density $p\left(\mathrm{x},\lambda\right)$ of
an $n_{x}$-dimensional random state vector $\mathrm{x}$ under the
dynamics of (\ref{eq:SDE}) has a deterministic evolution according
to the Fokker-Planck equation 
\begin{align}
\frac{\partial}{\partial\lambda}p\left(\mathrm{x},\lambda\right)= & -\sum_{i=1}^{n_{x}}\frac{\partial}{\partial x_{i}}\left[\mu_{i}\left(\mathrm{x},\lambda\right)p\left(\mathrm{x},\lambda\right)\right]\nonumber \\
 & +\frac{1}{2}\sum_{i=1}^{n_{x}}\sum_{j=1}^{n_{x}}\frac{\partial}{\partial x_{i}}\frac{\partial}{\partial x_{j}}\left[D_{ij}\left(\mathrm{x},\lambda\right)p\left(\mathrm{x},\lambda\right)\right],\label{eq:Fokker-Planck-equation}\\
 & \hphantom{+\ \ }p\left(\mathrm{x},0\right)=p_{0}\left(\mathrm{x}\right),\,\lambda\ge0;\nonumber 
\end{align}
\noindent where $\mathrm{x}=[x_{1},\dots,x_{n_{x}}]^{T}$, $\mathrm{\mu}=[\mu_{1},\dots,\mu_{n_{x}}]^{T}$,
and 
\begin{equation}
D_{ij}\left(\mathrm{x},\lambda\right)=\sum_{k=1}^{n_{x}}\mathrm{\sigma}_{ik}\left(\mathrm{x},\lambda\right)\mathrm{\sigma}_{jk}\left(\mathrm{x},\lambda\right),\label{eq:diffusion-coefficient}
\end{equation}
\noindent for an $n_{x}$-dimensional Wiener process $\left\{ W_{\lambda}\right\} $.
In its usual form, as described in Physics, the equation reads 
\begin{equation}
\frac{\partial}{\partial\lambda}p=-\nabla_{\mathrm{x}}\cdot\left[\mathrm{\mu}p\right]+\frac{1}{2}\nabla_{\mathrm{x}}\cdot\left[D\nabla_{\mathrm{x}}p\right].\label{eq:Fokker-Planck-equation-concise}
\end{equation}

We assume that the diffusion coefficient $\mathrm{\sigma}$ is locally
independent of $\mathrm{x}$, giving rise to a local diffusion matrix
$D\left(\lambda\right)=\mathrm{\sigma}\left(\lambda\right)\mathrm{\sigma}\left(\lambda\right){}^{T}$
that is invariant to the divergence operator in the vicinity of each
particle. This means that, at a given time instant, the diffusion
term in (\ref{eq:Fokker-Planck-equation-concise}) evolves at a rate
proportional to the curvature of a (Riemann) manifold that is approximately
constant in the neighbourhood of each particle. This assumption does
not affect the generality of the concepts applied in our derivation
for two reasons: it results in a stochastic particle flow that is
missing a simple term, of the form ${\sigma\left(x,\lambda\right)\cdot\partial_{x}\left[\sigma\left(x,\lambda\right)\right]}$,
that could be incorporated if needed; in practice, any probability
density can be well approximated by a mixture of densities whose covariances
are locally constant with respect to the state \citep{Silverman1986}
(i.e., $\partial_{x}\left[\sigma\left(x,\lambda\right)\right]=0$
locally). Additionally, as evidenced in \citep{Girolami2011}, keeping
the diffusion coefficient fixed for each sampling step does not perturb
the target distribution.

\subsection{Stationary Solution of the Fokker-Planck Equation\label{subsec:Stationary-solution}}

A stationary solution to the equation (\ref{eq:Fokker-Planck-equation-concise})
should satisfy 
\begin{equation}
\frac{\partial}{\partial\lambda}p\left(\mathrm{x},\lambda\right)\overset{\vphantom{\tfrac{}{}}{\scriptstyle \lambda\rightarrow\infty}}{\longrightarrow}0.\label{eq:condition-1}
\end{equation}
By writting 
\begin{equation}
\nabla_{\mathrm{x}}\cdot S\triangleq\nabla_{\mathrm{x}}\cdot\left[\mathrm{\mu}\,p\right]-\frac{1}{2}\nabla_{\mathrm{x}}\cdot\left[D\nabla_{\mathrm{x}}p\right],\label{eq:derivative-of-probability-current}
\end{equation}
\noindent the definition of the probability current becomes clear:
\begin{align}
S\left(\mathrm{x},\lambda\right) & =\mu\left(\mathrm{x},\lambda\right)p\left(\mathrm{x},\lambda\right)-\frac{1}{2}D\left(\lambda\right)\cdot\nabla_{\mathrm{x}}p\left(\mathrm{x},\lambda\right)\nonumber \\
 & =p\left(\mathrm{x},\lambda\right)\left[\mathrm{\mathrm{\mu}\left(\mathrm{x},\lambda\right)}-\frac{1}{2}D\left(\lambda\right)\cdot\nabla_{\mathrm{x}}\log p\left(\mathrm{x},\lambda\right)\right].\label{eq:probability-current}
\end{align}
Since the stationary condition requires 
\begin{equation}
\frac{\partial}{\partial\lambda}p\left(\mathrm{x},\lambda\right)=-\nabla_{\mathrm{x}}\cdot S\left(\mathrm{x},\lambda\right)\overset{\vphantom{\tfrac{}{}}{\scriptstyle \lambda\rightarrow\infty}}{\longrightarrow}0,\label{eq:condition-1-1}
\end{equation}
\noindent the probability current is required to vanish as $\lambda\rightarrow\infty$.
The probability current can only vanish if the drift $\mathrm{\mu}\left(\mathrm{x},\lambda\right)$
can be expressed as the gradient of a potential function \citep{Risken1989},
cancelling out the terms within brackets in (\ref{eq:probability-current}).
We write the drift as the gradient of a stationary potential function
according to 
\begin{equation}
\mathrm{\mu}\left(\mathrm{x},\lambda\right)=-\frac{1}{2}D\left(\lambda\right)\cdot\nabla_{\mathrm{x}}\Phi\left(\mathrm{x}\right).\label{eq:drift}
\end{equation}
The necessary and sufficient conditions for the existence of $\Phi\left(\mathrm{x}\right)$
are the potential conditions \citep{Risken1989} 
\begin{equation}
\frac{\partial\mu_{i\vphantom{j}}}{\partial x_{j}}=\frac{\partial\mu_{j}}{\partial x_{i}},\quad\forall i\neq j.\label{eq:potential-conditions}
\end{equation}
Provided that the probability current vanishes as $\nabla_{\mathrm{x}}\log p\left(\mathrm{x},\lambda\right)\rightarrow-\nabla_{\mathrm{x}}\Phi\left(\mathrm{x}\right)$,
we obtain the stationary solution, $p_{st}\left(\mathrm{x}\right)$,
as 
\begin{equation}
p\left(\mathrm{x},\lambda\right)\overset{\vphantom{\tfrac{}{}}{\scriptstyle \lambda\rightarrow\infty}}{\longrightarrow}p_{st}\left(\mathrm{x}\right)=\frac{1}{Z}e^{-\Phi\left(\mathrm{x}\right)},\label{eq:stationary-solution}
\end{equation}
\noindent where 
\begin{equation}
Z=\int_{\mathbb{R}^{n_{x}}}e^{-\Phi\left(\mathrm{x}\right)}d\mathrm{x}\label{eq:Gibbs-distribution-normalisation-constant}
\end{equation}
\noindent must be positive and finite. We promptly recognise (\ref{eq:stationary-solution})
as analogous to the Gibbs distribution. It is verifiable that (see,
for example, \citep{Jordan1996}) the Gibbs distribution minimizes
the free energy functional over all probability densities on $\mathbb{R}^{n_{x}}$.
It can also be shown that the stationary solution is the first eigenfunction
of the Fokker-Planck equation, corresponding to the eigenvalue zero
\citep{Risken1989}.

\subsection{The Stochastic Flow\label{subsec:The-stochastic-flow}}

The general stochastic particle flow is derived by setting the stationary
solution, $p_{st}\left(\mathrm{x}\right)$, to be the target posterior
density, $\pi\left(\mathrm{x}\right)=p\left(\mathrm{x}|\mathrm{y}_{1:k}\right)$,
to give 
\begin{align}
p_{st}\left(\mathrm{x}\right) & \coloneqq p\left(\mathrm{x}|\mathrm{y}_{1:k}\right),\nonumber \\
\frac{e^{-\Phi\left(\mathrm{x}\right)}}{Z\vphantom{p\left(\mathrm{y}_{1:k}\right)}} & =\frac{p\left(\mathrm{y}_{k}|\mathrm{x}\right)p\left(\mathrm{x}|\mathrm{y}_{1:k-1}\right)}{p\left(\mathrm{y}_{k}|\mathrm{y}_{1:k-1}\right)},\nonumber \\
\Phi\left(\mathrm{x}\right) & =-\log p\left(\mathrm{y}_{k}|\mathrm{x}\right)-\log p\left(\mathrm{x}|\mathrm{y}_{1:k-1}\right).\label{eq:solution-1}
\end{align}

Given a valid potential function $\Phi\left(\mathrm{x}\right)$ provides
the stationary solution, all potential functions of the form $\Phi\left(\mathrm{x}\right)\pm K$
for any constant $K\in\mathbb{R}$ are also valid. We can therefore
choose a valid potential function in (\ref{eq:solution-1}) such that
$p\left(\mathrm{y}_{k}|\mathrm{y}_{1:k-1}\right)=Z$. By using equation
(\ref{eq:drift}), we obtain 
\begin{align}
\mu\left(\mathrm{x},\lambda\right) & =-\frac{1}{2}D\left(\lambda\right)\cdot\nabla_{\mathrm{x}}\Phi\left(\mathrm{x}\right)=\frac{1}{2}D\left(\lambda\right)\cdot\nabla_{\mathrm{x}}\log\pi\left(\mathrm{x}\right)\nonumber \\
 & =\hphantom{-}\frac{1}{2}D\left(\lambda\right)\cdot\left[\nabla_{\mathrm{x}}\log p\left(\mathrm{y}_{k}|\mathrm{x}\right)+\nabla_{\mathrm{x}}\log p\left(\mathrm{x}|\mathrm{y}_{1:k-1}\right)\right].\label{eq:drift-2}
\end{align}
Substituting (\ref{eq:drift-2}) into (\ref{eq:probability-current}),
it is easy to see that the probability current vanishes as $\lambda\rightarrow\infty$.
Additionally, it is important to note that continuous multivariate
probability densities commonly used in parametric statistics (e.g.,
Gaussian, Student's t, Mises-Fisher, Pareto of first kind, Cauchy
etc) satisfy the potential conditions (\ref{eq:potential-conditions})
that suffice for $\Phi\left(\mathrm{x}\right)$ to exist. 

Based on equation (\ref{eq:SDE}) and on the drift obtained from the
stationary solution (\ref{eq:drift-2}), the dynamics of a set of
particles $\left\{ \mathrm{x}^{(i)}\left(\lambda\right):\,i=1,\,\dots,\,N\right\} $
can be described by the stochastic differential equation 
\begin{flalign}
d\mathrm{x} & =\mathrm{\mu}\left(\mathrm{x},\lambda\right)d\lambda+\mathrm{\sigma}\left(\mathrm{x},\lambda\right)d\mathrm{w}_{\lambda},\quad\mathrm{x}_{0}^{(i)}=\mathrm{x}_{k-1}^{(i)};\nonumber \\
d\mathrm{x} & =\frac{1}{2}D\nabla_{\mathrm{x}}\log\pi\left(\mathrm{x}\right)d\lambda+D^{\nicefrac{1}{2}}d\mathrm{w}_{\lambda};\label{eq:state-flow-equation}
\end{flalign}
\noindent where $\pi\left(\mathrm{x}\right)$ is the target (posterior)
probability density, $\left\{ \mathrm{w}_{\lambda}\right\} $ is the
standard Wiener process (Brownian motion) and $D$ is the diffusion
matrix. The stochastic process described by (\ref{eq:state-flow-equation})
is known in the literature to follow the Langevin dynamics and, except
for few special cases, cannot be exactly simulated. Thus, the most
common way to ensure the simulation provides samples from the correct
target distribution, $\pi\left(\mathrm{x}\right)$, is to set the
discretized dynamics as a proposal within a Markov chain Monte Carlo
framework, which leads to the Metropolis-adjusted Langevin algorithm
(MALA) \citep{Roberts2002}. 

By defining the distribution at $\lambda$ and target distribution
as ${d\mathcal{P}=p_{\vphantom{1}}\left(\mathrm{x},\lambda\right)d\mathrm{x}}$
and ${d\mathcal{P}_{\pi}=\pi\left(\mathrm{x}\right)d\mathrm{x}}$
respectively, one can articulate the total-variation distance between
the probability measures $\mathcal{P}\left(d\mathrm{x}\right)$ and
$\mathcal{P}_{\pi}\left(d\mathrm{x}\right)$ defined on $\left(\mathbb{R}^{n_{x}},\mathscr{B}(\mathbb{R}^{n_{x}})\right)$\footnote{$\mathscr{B}(\mathbb{R}^{n_{x}})$ is the $\sigma$-field of Borel
sets of $\mathbb{R}^{n_{x}}$.} as $\left\Vert \mathcal{P}_{\vphantom{1}}-\mathcal{P}_{\pi}\right\Vert _{\text{TV}}$.
It can be shown that (e.g., \citep{Dalalyan2014}), if the SDE (\ref{eq:state-flow-equation})
is integrated over a sufficiently long (finite) time horizon $T\in\mathbb{R}_{+}$,
then\footnote{The total variation norm for probability measures have an equivalence
to the $\mathrm{\mathbb{L}}^{1}$-norm as presented in (\ref{eq:total-variation-norm-exact}).
A simple argument for this equivalence is given in \citep{Levin2009},
chapter 4, proposition 4.2.} 
\begin{align}
\left\Vert \mathcal{P}_{\vphantom{1}}-\mathcal{P}_{\pi}\right\Vert _{\text{TV}}= & \frac{1}{2}\int_{\mathbb{R}^{n_{x}}}\left|p_{\vphantom{1}}\left(\mathrm{x},\lambda\right)-\pi\left(\mathrm{x}\right)\right|d\mathrm{x}\le\varepsilon\label{eq:total-variation-norm-exact}
\end{align}
\noindent for any $\lambda>T$, under a desired precision $\varepsilon$.
The implication is that stochastic particle flow implements the filtering
mapping (\ref{eq:mapping}) with increasing accuracy as $\lambda$
progresses.

Stochastic particle flow can be interpreted as a continuous-time filtering
method in the classical sense. Under the abstraction of a continuously
interpolated observation process, the method has a direct correspondence
to the Kallianpur-Striebel formula and satisfies the Zakai equation
as we demonstrate by the \emph{Theorem~\ref{thm:T11}} and \emph{Corollary~\ref{cor:C13}}
in the Appendix~\ref{subsec:Proofs-2}. Practically speaking, the
major difference between stochastic particle flow and other Langevin-based
algorithms is the way that the target density is sequentially approximated
via local representations that compose a mixture. This will be discussed
later in Section \ref{sec:SPF-GS}.

\subsection{The Diffusion Matrix\label{subsec:The-diffusion-matrix}}

As explained by Girolami \& Calderhead \citep{Girolami2011}, the
space of parameterized probability density functions is endowed with
a natural Riemann geometry, where the diffusion matrix arises as the
inverse of a position-specific metric tensor, $G\left(\mathrm{x}\left(\lambda\right)\right)$.
This metric tensor maps the distances inscribed in a Riemann manifold
to distances in the Euclidean space and, therefore, constitutes a
means to constrain the dynamics of any stochastic process to the geometric
structure of the parametric probability space. Rao \citep{Rao1945}
showed the tensor $G\left(\mathrm{x}\left(\lambda\right)\right)$
to be the expected Fisher information matrix 
\begin{equation}
G\left(\mathrm{x}\left(\lambda\right)\right)=-\mathbb{E}_{\mathrm{y}|\mathrm{x}}\left[\mathcal{H}_{\mathrm{x}}\left[\log p\left(\mathrm{y}|\mathrm{x}\right)\right]\right],\label{eq:expected-FIM}
\end{equation}
\noindent where $\mathcal{H}_{\mathrm{x}}\left[.\right]$ is the
Hessian matrix with respect to $\mathrm{x}$. In a Bayesian context,
Girolami \& Calderhead \citep{Girolami2011} suggested a metric tensor
that includes the prior information as 
\begin{equation}
G\left(\mathrm{x}\left(\lambda\right)\right)=-\mathbb{E}_{\mathrm{y}|\mathrm{x}}\left[\mathcal{H}_{\mathrm{x}}\left[\log p\left(\mathrm{y}|\mathrm{x}\right)\right]\right]-\mathcal{H}_{\mathrm{x}}\left[\log p_{x}\left(\mathrm{x}\right)\right],\label{eq:expected-posterior-FIM}
\end{equation}
\noindent although many possible choices of metric for a specific
manifold could be advocated. Because we are interested in local (curvature)
properties of stochastic particle flow, a sensible choice for the
metric tensor $G\left(\mathrm{x}\left(\lambda\right)\right)$ is the
observed Fisher information matrix incorporating the prior information.
In this case, the diffusion matrix becomes 
\begin{equation}
D=G\left(\mathrm{x}\left(\lambda\right)\right)^{-1}=\left[-\mathcal{H}_{\mathrm{x}}\left[\log\pi\left(\mathrm{x}\right)\right]\right]_{\mathrm{x}=\mathrm{x}_{\lambda}}^{-1},\label{eq:Diffusion-covariance}
\end{equation}
\noindent where the Hessian matrix is locally evaluated at $\mathrm{x}=\mathrm{x}_{\lambda}^{(i)}$
for the $i$th sample, and the resulting diffusion matrix is kept
constant for each integration step to obtain the subsequent sample.
A problem with this choice is that the expression (\ref{eq:Diffusion-covariance})
may not be strictly positive definite at specific points of the state
space for some types of probability distributions (e.g., mixtures).
In order to solve that problem, one could appeal to methods for regularizing
the diffusion matrix such as the Tikhonov regularization, the technique
to find the nearest (in terms of minimum Fr\"obenius norm) positive
definite matrix \citep{Higham2002}, or \emph{SoftAbs} \citep{Betancourt2013},
a technique that implements a smooth absolute transformation of the
eigenvalues to map the negative-Hessian metric into a positive-definite
matrix. Another possibility is adopting an empirical estimate to (\ref{eq:expected-posterior-FIM}).

\subsection{Integration Method\label{subsec:Integration-method}}

Among the discretization methods that could be used to integrate the
SDE (\ref{eq:state-flow-equation}), we advocate the use of Ozaki's
discretization \citep{Ozaki1992} of the Langevin diffusion. This
is more accurate than methods based on the Euler discretization. Ozaki's
discretization is only possible for target densities that are continuously
differentiable and have a smooth Hessian matrix. These requirements
may be fulfilled by a solution that constitutes a superposition of
conveniently parameterized local approximations to a density.

The algorithm that enables simulation from the SDE (\ref{eq:state-flow-equation})
using Ozaki's discretization is generally called \emph{Langevin Monte
Carlo with Ozaki discretization} (LMCO) in the MCMC community (see
\citep{Dalalyan2014}). Provided an appropriate time horizon, $T$,
by discretizing the interval $0\le\lambda\le T$ into $L$ sub-intervals
$\{\lambda_{0}=0,\,\lambda_{1},\,\dots\lambda_{l},\,\dots,\,\lambda_{L}=T\}$,
the discretized particle flow equation using Ozaki's method is given
by 
\begin{alignat}{1}
\mathrm{x}(\lambda_{l+1}) & =\mathrm{x}(\lambda_{l})+\left(\mathrm{\mathbb{I}}_{n_{x}}-e^{-\frac{\Delta\lambda}{2}D(\lambda_{l})^{-1}}\right)D(\lambda_{l})^{2}\nabla_{\mathrm{x}}\log\pi\left(\mathrm{x}(\lambda_{l})\right)\nonumber \\
 & +\left[\left(\mathrm{\mathbb{I}}_{n_{x}}-e^{-\Delta\lambda D(\lambda_{l})^{-1}}\right)D(\lambda_{l})^{2}\right]^{\nicefrac{1}{2}}\mathrm{w}_{l+1},\label{eq:SDE-Ozaki-discretisation}
\end{alignat}
\noindent where $\{\mathrm{w}_{l}:l=1,\dots,L\}$ is a sequence of
independent random vectors distributed according to $\mathrm{w}_{l}\sim\mathcal{N}(\mathrm{w};0_{n_{x}},\mathrm{\mathbb{I}}_{n_{x}})$.
The need to compute the exponential matrices in (\ref{eq:SDE-Ozaki-discretisation})
implies an increment in complexity typically bounded by $\mathcal{O}\left(NLn_{x}^{3}\right)$
computations, which may not be justifiable for some applications.
A cheaper alternative is achieved by linearizing (\ref{eq:state-flow-equation})
in the neighbourhood of the current state, assuming $D(\lambda)$
piecewise constant in pseudo-time, transforming the linearized equation
by the Laplace transform, solving it in the Laplace domain, and transforming
it back. The result is 
\begin{alignat}{1}
\mathrm{x}(\lambda_{l+1}) & =\mathrm{x}(\lambda_{l})+\left(1-e^{-\frac{1}{2}\Delta\lambda}\right)D(\lambda_{l})^{\vphantom{1}}\nabla_{\mathrm{x}}\log\pi\left(\mathrm{x}(\lambda_{l})\right)\nonumber \\
 & +\left(1-e^{-\Delta\lambda}\right)^{\nicefrac{1}{2}}D(\lambda_{l})^{\nicefrac{1}{2}}\mathrm{w}_{l+1},\label{eq:SDE-Ozaki-discretisation-1}
\end{alignat}
\noindent where the exponential matrices are avoided but the exponential
effect on the integration variable (time step) is kept. See Appendix~\ref{sec:Integration-rule-derivation}
for the derivation of this latter integration rule.

It is important to note that, upon integration of the SDE (\ref{eq:state-flow-equation})
by a numerical method, convergence to the invariant distribution is
no longer guaranteed for any finite step size. This is due to the
first-order integration error that is introduced. When tackling difficult
nonlinear filtering problems where the integration error becomes significant,
a correction can be carried out by employing a Metropolis acceptance
step after each integration step to ensure convergence to the invariant
measure. 

\subsection{Selection of Time Horizon and Integration Step Size\label{subsec:Selection-of-time}}

To successfully implement stochastic particle flow, it is necessary
to select an adequate time horizon, $T$, and an integration step
size, $\Delta\lambda$. These parameters need to be chosen such that
stationarity is reached and convergence to the invariant measure is
achieved. There are several routes one could take to solve this problem
with each making different assumptions about the probability measures
involved and about the regularity properties of the stationary distribution.
One could also pose related questions in the context of specific implementations.
Answering such questions might, for example, involve selecting the
time horizon and integration step size that minimizes the variance
of the samples' weights.

The view adopted here is that, since computational effort is a fundamental
issue for implementing stochastic particle flow, we should choose
these parameters to minimize computational effort. More specifically,
we want to minimize the number of integration steps that need to be
performed to achieve 
\begin{equation}
\Vert\mathcal{P}_{\mathcal{\tilde{\mathcal{L}}}[\Delta\lambda],T}-\mathcal{P}_{\vphantom{\tilde{\mathcal{L}}}\pi}\Vert_{\text{TV}}\le\varepsilon\label{eq:desired-precision-criteria}
\end{equation}
\noindent for an acceptable precision level $\varepsilon$, where
$\mathcal{P}_{\mathcal{\tilde{\mathcal{L}}}[\Delta\lambda],T}(d\mathrm{x})$
is the approximating probability measure achieved by sampling from
the discretized Langevin stochastic process over $L=\left\lceil T/\Delta\lambda\right\rceil $
steps.

Defining near-optimal choices of these parameters for general target
measures, including mixtures and highly skewed distributions, would
require a more thorough study that extrapolates the scope of this
work. Instead, in this paper, we propose two pragmatic approaches
to choosing both the time horizon and integration step size.

\subsubsection*{Approach 1}

Our first approach builds on results concerning the scaling of Langevin-based
MCMC algorithms: the interested reader is referred to Roberts \& Rosenthal
\citep{Roberts1998} and a recent extension by Pillai~\textit{et
al.} \citep{Pillai2012} that treat high-dimensional target measures
that are not of the product form. In summary, these analyses show
that the number of steps required to sample the target measure by
the Metropolis-Adjusted Langevin algorithm (MALA) grows as $L\sim\mathcal{O}(n_{x}^{\nicefrac{1}{3}})$.
In addition, these papers work out the optimal step size by maximizing
the ``speed function'' or, equivalently, producing the optimal average
acceptance rate. Although this optimal criterion is only applicable
to algorithms that employ a Metropolis acceptance step, tuning the
step size used in stochastic particle flow for an ``emulated'' (hypothetical)
acceptance rate of interest can guide the rate of convergence (even
if the accept-reject step is suppressed in practice). Abusing the
methodology presented by Pillai~\textit{et al.} \citep{Pillai2012}
and using \emph{Proposition~2.4} from Roberts~\textit{et al.} \citep{Roberts1997},
let us denote the asymptotic acceptance probability $\alpha(l)$ as
a function of a scaling parameter $l\in\mathbb{R}$, such that the
speed function $h(l)$ for high-dimensional MALA can be approximated
as \citep{Pillai2012} 
\begin{align}
h\left(l\right) & =l\cdot\alpha\left(l\right)\approx l\cdot\mathbb{E}_{\pi}\left[1\wedge e^{\mathcal{N}\left(-l^{3}/4,l^{3}/2\right)}\right]\nonumber \\
 & =l\cdot\left[\mathcal{N}_{\text{cdf}}\left(\frac{-l^{3}/4}{\sqrt{l^{3}/2}}\right)+\exp\left(-\frac{l^{3}}{4}+\frac{\left(\sqrt{l^{3}/2}\right)^{2}}{2}\right)\mathcal{N}_{\text{cdf}}\left(-\sqrt{l^{3}/2}-\frac{-l^{3}/4}{\sqrt{l^{3}/2}}\right)\right]\nonumber \\
 & =l\cdot2\mathcal{N}_{\text{cdf}}\left(-\sqrt{l^{3}/8}\right),\label{eq:speed-function}
\end{align}
\noindent where $\mathcal{N}_{\text{cdf}}(.)$ is a standard normal
cumulative distribution function (cdf). One can observe the maximum
occurs at $l_{opt}\approx1.3620$ and corresponds to the theoretical
optimal acceptance rate, $\alpha(l_{opt})\approx0.5741$. 

Our approach is then based on the notion that, for a conveniently
selected acceptance rate, the step size should scale as $\Delta\lambda\propto l\cdot n_{x}^{-\xi}$
such that the total number of steps is scaled as $L\propto n_{x}^{\xi}$,
and the exponent $\xi$ depends on whether the Metropolis adjustment
is used or not. Theoretically, if the accept-reject step is not adopted
then\footnote{In view of (\ref{eq:time-horizon}) and (\ref{eq:step-size}), $T/\Delta\lambda\sim\mathcal{O}(n_{x})$.}
$\xi=1$, otherwise the optimal choice is $\xi=1/3$ \citep{Pillai2012}. 

To exemplify the use of this approach, to produce an emulated (asymptotic)
acceptance rate of $\alpha=0.80$, a stochastic particle flow should
be scaled as $l\approx2\left[-\mathcal{N}_{\text{cdf}}^{-1}\left(\alpha/2\right)\right]^{2/3}=0.8008$.
Using the Langevin dynamics considered herein, if $n_{x}=10$ and
$\xi=1$ then $\Delta\lambda=2l\cdot n_{x}^{-\xi}\approx0.1602$.

If the accept-reject step is present, a criterion to stop the simulation
could be established online. Various MCMC convergence diagnostic methods
are applicable to this task. However, in our experience with stochastic
particle flow, such approaches to online determination of convergence
may indicate more steps are needed than are actually necessary to
obtain good results, and so give a pessimistic view of the amount
of computation required. To set the time horizon when accept-reject
step is not used, we note that $T=T_{0}+L_{s}\cdot\Delta\lambda$,
where $T_{0}$ is the time required to take the chain to the region
of high acceptance probability (``warm-up'') and $L_{s}$ is the number
of steps to explore the invariant measure. We determine both $L_{s}$
and $T_{0}$ either by presetting reasonable values, or based on the
second approach to be explained next. 

\subsubsection*{Approach 2}

Our second approach is an extension of the method proposed by Dalalyan
\citep{Dalalyan2014}. It is useful due to its ease of application
and suitability to ``nicely'' measurable filtering quantities although,
strictly speaking, the method is only applicable to target densities
that are log-concave. The criteria are presented as follows.
\begin{thm}
\label{thm:T1} Let $\Phi:\mathbb{R}^{n_{x}}\rightarrow\mathbb{R}$
be a measurable convex function satisfying 
\begin{alignat}{1}
\int_{\mathbb{R}^{n_{x}}}\exp\{-\Phi\left(\mathrm{x}\right)\} & <\infty,\label{eq:Energy-condition-1}\\
\begin{gathered}\Phi\left(\mathrm{x}\right)-\Phi\left(\bar{\mathrm{x}}\right)-\nabla_{\mathrm{x}}\Phi\left(\bar{\mathrm{x}}\right)^{T}\left(\mathrm{x}-\bar{\mathrm{x}}\right)\end{gathered}
 & \ge\frac{1}{2}m\left\Vert \mathrm{x}-\bar{\mathrm{x}}\right\Vert _{2}^{2},\label{eq:Lipschitz-conditions-1}\\
\left\Vert \nabla_{\mathrm{x}}\Phi\left(\mathrm{x}\right)-\nabla_{\mathrm{x}}\Phi\left(\bar{\mathrm{x}}\right)\right\Vert _{2} & \le M\left\Vert \mathrm{x}-\bar{\mathrm{x}}\right\Vert _{2},\quad\forall\mathrm{x},\bar{\mathrm{x}}\in\mathbb{R}^{n_{x}},\label{eq:Lipschitz-conditions-2}
\end{alignat}
\noindent for two existing positive constants $m$ and $M$. Let
$\bar{\mathrm{x}}\in\mathbb{R}^{n_{x}}$ be the global minimum of
$\Phi(\mathrm{x})$. Suppose a discrete-time Langevin Monte Carlo
algorithm integrates (\ref{eq:state-flow-equation}), targeting the
invariant density $\pi(\mathrm{x})\propto\exp\{-\Phi(\mathrm{x})\}$
with measure $\mathcal{P}_{\pi}(d\mathrm{x})$, and with the initial
density $\nu\left(\mathrm{x}\right)=\delta(\mathrm{x}-\mathrm{x}_{\nu})$
(a probability mass initially located at $\mathrm{x}=\mathrm{x}_{\nu}$).
In addition, assume that for some $\gamma\ge1$ we have $\Delta\lambda\le\left(\gamma M\right)^{-1}$,
and $K=\sup_{\mathrm{x}}\Vert D(\mathrm{x})\Vert_{2}$ where $D_{\lambda}=D(\mathrm{x}_{\lambda})$
is the diffusion matrix. Then, for a time horizon, $T$, and step
size, $\Delta\lambda$, the total-variation distance between the target
measure $\mathcal{P}_{\pi}$ and the approximated measure $\mathcal{P}_{\mathcal{\tilde{\mathcal{L}}}(\Delta\lambda),T}$
furnished by the discrete-time Langevin Monte Carlo algorithm satisfies
\begin{multline}
\Vert\mathcal{P}_{\mathcal{\tilde{\mathcal{L}}}[\Delta\lambda],T}-\mathcal{P}_{\vphantom{\tilde{\mathcal{L}}}\pi}\Vert_{\text{TV}}\le\frac{1}{2}\exp\left\{ -\frac{1}{2}mT+\frac{n_{x}}{2}\log\left(\frac{M}{m}\right)-\log\left[\Gamma_{u}\left(\frac{n_{x}}{2},\frac{M\Vert\bar{\mathrm{x}}-\mathrm{x}_{\nu}\Vert_{2}^{2}}{2}\right)\right]\right\} \\
+\frac{1}{2}-\frac{1}{2}\exp\left\{ -\frac{n_{x}}{2}\frac{M^{3}K^{4}\gamma}{48(2\gamma-1)}\left(\frac{1}{n_{x}}\Vert\mathrm{\bar{x}}-\mathrm{x}_{\nu}\Vert_{2}^{2}+2T\right)\Delta\lambda^{2}-\frac{n_{x}M^{2}K^{3}T}{16}\Delta\lambda\right\} ,\label{eq:total-variance-bound}
\end{multline}
\noindent where $\Gamma_{u}(s,x)\triangleq\Gamma(s)^{-1}\int_{x}^{\infty}t^{s-1}e^{-t}dt$
is the upper incomplete gamma function.
\end{thm}
\begin{cor}
\label{cor:C2} Let $n_{x}\ge2$, $\Phi$ satisfy (\ref{eq:Energy-condition-1}),
(\ref{eq:Lipschitz-conditions-1}) and (\ref{eq:Lipschitz-conditions-2}),
and $\varepsilon\in(0,\nicefrac{1}{2})$ be a desired precision level.
Let the time horizon $T$ and the step size $\Delta\lambda$ be defined
by 
\begin{alignat}{1}
T & \ge\frac{2\log\left(1/\varepsilon\right)+n_{x}\log\left(\frac{M}{m}\right)-2\log\left[\Gamma_{u}\left(\frac{n_{x}}{2},\frac{M\Vert\bar{\mathrm{x}}-\mathrm{x}_{\nu}\Vert_{2}^{2}}{2}\right)\right]}{m},\label{eq:time-horizon}\\
\Delta\lambda & \le\frac{-\frac{T}{16}+\sqrt{\left(\frac{T}{16}\right)^{2}+\frac{\gamma}{48(2\gamma-1)}\left(\frac{1}{n_{x}}\Vert\mathrm{\bar{x}}-\mathrm{x}_{\nu}\Vert_{2}^{2}+2T\right)M^{-1}K^{-2}\left[\frac{2}{n_{x}}\log\left(\frac{1}{1-\varepsilon}\right)\right]}}{\frac{\gamma}{48(2\gamma-1)}\left(\frac{1}{n_{x}}\Vert\mathrm{\bar{x}}-\mathrm{x}_{\nu}\Vert_{2}^{2}+2T\right)MK},\label{eq:step-size}
\end{alignat}
\noindent where $\gamma\ge1$. Then the resulting probability distribution
of a Langevin Monte Carlo algorithm that integrates (\ref{eq:state-flow-equation})
after $L=\left\lceil T/\Delta\lambda\right\rceil $ steps, satisfies
$\Vert\mathcal{P}_{\mathcal{\tilde{\mathcal{L}}}[\Delta\lambda],T}-\mathcal{P}_{\vphantom{\tilde{\mathcal{L}}}\pi}\Vert_{\text{TV}}\le\varepsilon$.
\end{cor}
\emph{Theorem \ref{thm:T1}} is thoroughly underpinned by the findings
of Dalalyan \citep{Dalalyan2016}, with specific settings changed
to match the Langevin algorithm proposed in this paper, and both a
more general bound for the time horizon and a tightened bound for
the step size (to reduce computational effort). We recommend the reader
interested in the proof to first refer to \citep{Dalalyan2016} and
then follow the missing arguments for its proof in the Appendix \ref{subsec:Proofs-1}.

\emph{Corollary \ref{cor:C2}} is a direct criterion for selecting
the time horizon and step size, arising from the right-hand side of
the inequality (\ref{eq:total-variance-bound}) being set to be a
desired precision level. It is essential to clarify that some practical
issues arise here: in this form, the method holds for log-concave
densities only; the positive constants, $m$ and $M$, are assumed
known a priori; and the approximation of the filtering density is
not taken into account in the error budget. Rigorously speaking, the
method does not apply to more general cases. However, the method has
utility as the basis of an approximate (and pragmatic) mechanism for
obtaining the required parameters. In making this approximation, we
explicitly acknowledge that we are assuming that: 
\begin{enumerate}
\item The target density can be well characterized by a central tendency
statistic, $\bar{\mathrm{x}}_{c}$, that replaces and roughly represents
$\bar{\mathrm{x}}$ in all aspects of the analysis.
\item The initial measure is composed of a superposition of $N$ probability
masses described by 
\[
\ensuremath{N^{-1}\sum_{i=1}^{N}\delta(\mathrm{x}-\mathrm{x}^{(i)})}
\]
 or, equivalently, an empirical distribution with mean $\mu_{\nu}$
and covariance matrix $\mathrm{V}$, spatially encompassing all initial
samples (from the previous filtering iteration), which is assumed
to constrain the constants $M$ and $m$ by 
\begin{align}
M & <\Vert\mathrm{V}\Vert_{2}/2,\label{eq:Constant-m-new-constraint-1}\\
(\mathrm{x}^{(i)}-\bar{\mathrm{x}}_{c})^{T}\frac{m}{2}(\mathrm{x}^{(i)}-\bar{\mathrm{x}}_{c}) & \ge\chi_{\text{cdf}}^{-2}(0.99,n_{x}),\quad\forall\mathrm{x}^{(i)}:\,i=1,\dots,N,\label{eq:Constant-m-new-constraint-2}
\end{align}
\noindent where $\chi_{\text{cdf}}^{-2}(P,\kappa)$ is the inverse
of chi-square cdf for probability $P$ and $\kappa$ degrees of freedom. 
\item In accordance with \emph{Lemma 4} in \citep{Dalalyan2014}, the constant
$M$ is also constrained by 
\begin{align}
\Phi(\mathrm{x}^{(i)})-\Phi(\bar{\mathrm{x}}_{c})-\nabla_{\mathrm{x}}\Phi(\bar{\mathrm{x}}_{c})^{T}(\mathrm{x}^{(i)}-\bar{\mathrm{x}}_{c}) & \le\frac{M}{2}\left\Vert \mathrm{x}-\bar{\mathrm{x}}_{c}\right\Vert _{2}^{2},\quad\forall\mathrm{x}^{(i)}:i=1,\dots,N.\label{eq:Constant-M-new-constraint}
\end{align}
\item The positive constants $M$ and $m$ can be roughly estimated by 
\begin{enumerate}
\item approximating the statistic $\bar{\mathrm{x}}_{c}$ of the target
density (e.g., obtaining a maximum-a-posteriori estimate by optimization
or an approximated mean by the EKF),
\item inverting conditions (\ref{eq:Lipschitz-conditions-1}) and (\ref{eq:Lipschitz-conditions-2}),
and incorporating the constraint (\ref{eq:Constant-M-new-constraint}),
to give 
\begin{align}
\tilde{M}= & 2\sup_{i\in[1,N]}\max\left[\frac{\left\Vert \nabla_{\mathrm{x}}\Phi(\mathrm{x}^{(i)})-\nabla_{\mathrm{x}}\Phi(\bar{\mathrm{x}}_{c})\right\Vert _{2}}{2\left\Vert \mathrm{x}^{(i)}-\bar{\mathrm{x}}_{c}\right\Vert _{2}},\frac{\Phi(\mathrm{x}^{(i)})-\Phi(\bar{\mathrm{x}}_{c})-\nabla_{\mathrm{x}}\Phi(\bar{\mathrm{x}}_{c})^{T}(\mathrm{x}^{(i)}-\bar{\mathrm{x}}_{c})}{\left\Vert \mathrm{x}^{(i)}-\bar{\mathrm{x}}_{c}\right\Vert _{2}^{2}}\right],\label{eq:Constant-M}\\
\tilde{m}= & 2\inf_{i\in[1,N]}\frac{\Phi(\mathrm{x}^{(i)})-\Phi(\bar{\mathrm{x}}_{c})-\nabla_{\mathrm{x}}\Phi(\bar{\mathrm{x}}_{c})^{T}(\mathrm{x}^{(i)}-\bar{\mathrm{x}}_{c})}{\left\Vert \mathrm{x}^{(i)}-\bar{\mathrm{x}}_{c}\right\Vert _{2}^{2}},\label{eq:Constant-m}
\end{align}
\noindent where all quantities can be computed from definition $\Phi\left(\mathrm{x}\right)\triangleq-\log p\left(\mathrm{y}_{k}|\mathrm{x}\right)-\log p\left(\mathrm{x}|\mathrm{y}_{1:k-1}\right)$
given an approximation to the prior pdf; the resulting values of $\tilde{M}$
and $\tilde{m}$ must also satisfy (\ref{eq:Constant-m-new-constraint-1})
and (\ref{eq:Constant-m-new-constraint-2}).
\end{enumerate}
\end{enumerate}
Once the positive constants $M$ and $m$ have been estimated, obtaining
$T$ and $\Delta\lambda$ follows from (\ref{eq:time-horizon}) and
(\ref{eq:step-size}) respectively. In our experience, for very simple
problems, (\ref{eq:time-horizon}) may produce overestimated time
horizons and, as a consequence, may cause (\ref{eq:step-size}) to
produce underestimated step sizes for stochastic particle flow. This
happens because the bound for the time horizon becomes loose, in view
of \emph{Lemma \ref{lem:L6}}, for initial distributions that are
far from the target distribution. In simple cases, a closer approximation
can by achieved by assuming 1-uniform ergodicity of the Markov chain
to give 
\begin{equation}
T\ge\frac{2\log\left(1/\varepsilon\right)+n_{x}\log R}{\tilde{m}},\label{eq:step-size-1-uniform-ergodicity}
\end{equation}
\noindent for a finite $R\in\mathbb{R}_{+},$ at the cost of having
to determine $R$ empirically. Similarly, for simple cases, expression
(\ref{eq:step-size}) can be replaced with an empirical rule of the
form 
\begin{equation}
\Delta\lambda\le2\frac{\sqrt{\tilde{m}}}{\tilde{M}},\label{eq:step-size-3}
\end{equation}
\noindent which satisfies $\Delta\lambda\le\left(\gamma\tilde{M}\right)^{-1}$
for $\gamma\le\left(2\sqrt{\tilde{m}}\right)^{-1}$ as required by
\emph{Theorem \ref{thm:T1}}, but is not guaranteed to satisfy \emph{Corollary
\ref{cor:C2}.}

\section{Stochastic Particle Flow as a Gaussian Sum Filter\label{sec:SPF-GS}}

In this section we use stochastic particle flow to derive a filter
that approximates the posterior probability density as a Gaussian
mixture. We refer to the resulting filter as the stochastic particle
flow Gaussian sum filter (SPF-GS).

\subsection{The Mixture-Based Approximating Measure\label{subsec:The-mixture-measure}}

Given a set of samples drawn from an importance distribution, $\left\{ \mathrm{x}^{(i)}\in\mathcal{X}:\,i\in1,\,\dots,\,N\right\} $,
if one is required to solve the filtering problem by a standard Monte
Carlo method, then the stochastic filter adopts the following approximation
\begin{align}
\hat{\varphi} & =\int_{\mathrm{\mathcal{X}}}\varphi\left(\mathrm{x}\right)\pi\left(\mathrm{x}\right)dx\nonumber \\
 & \approx\int_{\mathrm{\mathcal{X}}}\varphi\left(\mathrm{x}\right)\sum_{i=1}^{N}w(\mathrm{x}^{(i)})\delta(\mathrm{x}-\mathrm{x}^{(i)})d\mathrm{x}\nonumber \\
 & =\sum_{i=1}^{N}w^{(i)}\varphi(\mathrm{x}^{(i)}),\label{eq:approximation-filtering-problem}
\end{align}
\noindent where $w(\mathrm{x}^{(i)})=w^{(i)}$ are the importance
weights. Now suppose that we have access to an approximating measure
$\tilde{\mathcal{P}}_{\pi}\left(d\mathrm{x}\right)$ on $(\mathcal{X},\mathscr{B}(\mathcal{X}))$
with an associated density such that $d\tilde{\mathcal{P}}_{\pi}=\tilde{\pi}d\mathrm{x}$.
If the density $\tilde{\pi}$ involves a mixture of $N$ Gaussians
according to 
\begin{align}
\tilde{\pi}\left(\mathrm{x}\right) & =\sum_{i=1}^{N}w_{m}^{(i)}\mathcal{N}(\mathrm{x};\mu_{m}(\mathrm{x}^{(i)}),\Sigma_{m}(\mathrm{x}^{(i)}))\nonumber \\
 & =\sum_{i=1}^{N}w_{m}^{(i)}\mathcal{N}(\mathrm{x};\mu_{m}^{(i)},\Sigma_{m}^{(i)}),\label{eq:Gaussian-mixture-measure}
\end{align}
\noindent where $\{w_{m}^{(i)},\mu_{m}^{(i)},\Sigma_{m}^{(i)}\}$
are computed based on the samples $\{\mathrm{x}^{(i)}\}$, then the
solution is given by 
\begin{align}
\hat{\varphi} & =\int_{\mathrm{\mathcal{X}}}\varphi\left(\mathrm{x}\right)\pi\left(\mathrm{x}\right)d\mathrm{x}\nonumber \\
 & \approx\int_{\mathrm{\mathcal{X}}}\varphi\left(\mathrm{x}\right)\sum_{i=1}^{N}w_{m}^{(i)}\mathcal{N}(\mathrm{x};\mu_{m}^{(i)},\Sigma_{m}^{(i)})d\mathrm{x}\nonumber \\
 & =\sum_{i=1}^{N}w_{m}^{(i)}\int_{\mathrm{\mathcal{X}}}\varphi\left(\mathrm{x}\right)\mathcal{N}(\mathrm{x};\mu_{m}^{(i)},\Sigma_{m}^{(i)})d\mathrm{x}\nonumber \\
 & =\sum_{i=1}^{N}w_{m}^{(i)}\mathbb{E}_{\mathcal{N}}\left[\varphi\left(\mathrm{x}\right)|\mu_{m}^{(i)},\Sigma_{m}^{(i)}\right].\label{eq:approximation-filtering-problem-1}
\end{align}
In this setting, it is possible to prove that $\tilde{\pi}\left(\mathrm{x}\right)\rightarrow\pi\left(\mathrm{x}\right)$
as $N\rightarrow\infty$ almost surely if $\mu_{m}^{(i)}\rightarrow\mathrm{x}^{(i)}$
and $\Sigma_{m}^{(i)}\rightarrow0$, by appealing to convergence proofs
for mixture-based estimators (see \citep{Anderson1979}, pages 197\textendash 199).
Also, it is worth noting that this procedure is quite general in the
sense that (\ref{eq:Gaussian-mixture-measure}) could be replaced
by a mixture of any convenient parametric distribution.

\subsection{The Stochastic-Particle-Flow Gaussian Sum Filter\label{subsec:The-SPF-GS}}

Building upon the results presented in the previous section, SPF-GS
uses samples to propagate local Gaussian components, which can together
provide an accurate approximation to the posterior probability density
of the form (\ref{eq:Gaussian-mixture-measure}). More specifically,
given a new measurement and a set of samples and moments $\{\mathrm{x}_{k-1}^{(i)},\mu_{m,k-1}^{(i)},\Sigma_{m,k-1}^{(i)}:i=1,\dots,N\}$,
the filtering procedure consists of integrating the SDE (\ref{eq:state-flow-equation})
for each particle $\mathrm{x}^{(i)}\left(\lambda\right)$ and propagating
its associated moments $(\mu_{m}^{(i)}\left(\lambda\right),\Sigma_{m}^{(i)}\left(\lambda\right))$
through the interval $0<\lambda\le T$, which corresponds to the interval
$t_{k-1}<t\le t_{k}$. The integration process is performed until
one achieves the posterior set of samples and parameters $\{\mathrm{x}_{k}^{(i)},\mu_{m,k}^{(i)},\Sigma_{m,k}^{(i)}\}\coloneqq\{\mathrm{x}^{(i)}\left(T\right),\mu_{m\vphantom{k}}^{(i)}\left(T\right),\Sigma_{m\vphantom{k}}^{(i)}\left(T\right)\}$,
where $\mathrm{x}_{\vphantom{k-1}}^{(i)}\left(0\right)=\mathrm{x}_{k-1}^{(i)}$,
so that 
\begin{align*}
\frac{1}{2}\int_{\mathrm{\mathcal{X}}}\left|\sum_{i=1}^{N}\frac{1}{N}\mathcal{N}(\mathrm{x};\mu_{m}^{(i)}(T),\Sigma_{m}^{(i)}(T))-\pi\left(\mathrm{x}\right)\right|d\mathrm{x} & \le\varepsilon^{\prime},
\end{align*}
\noindent under a desired precision level $\varepsilon^{\prime}$
as $N\rightarrow\infty$. In practice, the integration of stochastic
particle flow (\ref{eq:state-flow-equation}) over $0<\lambda\le T$
involves multiple intermediate sampling steps that evolve the samples
$\mathrm{x}^{(i)}\left(\lambda\right)$ to populate the local regions
of the state space where the target distribution will be described.
At the same time, the mixture components are propagated to define
the local approximations and thereby the global approximation to the
filtering density. A fundamental aspect of the procedure proposed
herein is that, for each filtering step, the intermediate moments
$(\mu_{m}^{(i)}\left(\lambda\right),\Sigma_{m}^{(i)}\left(\lambda\right))$
are initialized as departing from the corresponding samples obtained
from the previous step, i.e., $\mu_{m}^{(i)}\left(0\right)\coloneqq\mathrm{x}_{k-1}^{(i)}$
and $\Sigma_{m}^{(i)}\left(0\right)\coloneqq0_{n_{x}\times n_{x}}$,
and evolved onto the local posterior moments, $\mu_{m}^{(i)}(T)$
and $\Sigma_{m}^{(i)}(T)$. In this setting, each component of the
filtering mixture is associated, via Fokker-Planck equation (Langevin
dynamics), with the mapping 
\begin{equation}
\frac{1}{N}\delta(\mathrm{x}_{k-1}^{\vphantom{(i)}}-\mathrm{x}_{k-1}^{(i)})\mapsto\frac{1}{N}\mathcal{N}(\mathrm{x}_{k}^{\vphantom{(i)}};\mu_{m,k}^{(i)},\Sigma_{m,k}^{(i)}).\label{eq:Mapping-to-mixture}
\end{equation}

It is also important to highlight that we perceive the most informative
approximation of the posterior from the previous iteration as provided
by the set of moments from the previous iteration, $\{\mu_{m,k-1}^{(i)},\Sigma_{m,k-1}^{(i)}\}$,
and not by the samples. We therefore use the previous iteration's
mixture to define the invariant target measure but samples from this
mixture to integrate the SDE. In our experience, this setting is beneficial
because: it avoids the bias that would result from propagating a mixture-only
approximation from one filtering iteration to the next; it dismisses
the need to explicitly compute mixture weights since the ergodic Markov
chain that carries out (\ref{eq:Mapping-to-mixture}) is known to
converge to the invariant measure no matter where it starts \citep{Meyn2009}.

More specifically, in the Langevin diffusion setting presented in
Section \ref{sec:Stochastic-particle-flow}, the mixture measure forms
the basis of approximating the target log-density. Each mixture component
from the previous filtering step enables a local approximation of
the prior density as
\begin{equation}
\tilde{p}^{(i)}\left(\mathrm{x}_{k}^{\prime}|\mathrm{y}_{1:k-1}^{\vphantom{\prime}}\right)=\int_{\mathbb{R}^{n_{x}}}p_{t}(\mathrm{x}_{k}^{\prime\vphantom{(i)}}|\mathrm{x}_{k-1}^{\vphantom{(i)}})\mathcal{N}(\mathrm{x}_{k-1}^{\vphantom{(i)}};\mu_{m,k-1}^{(i)},\Sigma_{m,k-1}^{(i)})d\mathrm{x}_{k-1}^{\vphantom{\prime}},\label{eq:predicted-pdf-per-particle-0}
\end{equation}
\noindent where $p_{t}\left(\mathrm{x}_{k}^{\prime}|\mathrm{x}_{k-1}^{\vphantom{\prime}}\right)$
is the state-process transition kernel, and the resulting prior density
is locally approximated as a Gaussian (e.g., via the Unscented Transform).
Thus, provided a known likelihood function, $p\left(\mathrm{y}_{k}|\mathrm{x}\right)$,
one Langevin transition kernel is computed per sample based on $\nabla_{\mathrm{x}}\log\tilde{\pi}^{(i)}(\mathrm{x})=\nabla_{\mathrm{x}}\log p\left(\mathrm{y}_{k}|\mathrm{x}\right)+\nabla_{\mathrm{x}}\log\tilde{p}^{(i)}\left(\mathrm{x}|\mathrm{y}_{1:k-1}\right)$.

We propose an approximate method to propagate the moments, $\mu_{m}(\lambda)$
and $\Sigma_{m}(\lambda)$, by linearizing the flow locally, in the
neighbourhood of a probability mass located at $\mathrm{x}_{l}$.
In the Appendix\emph{~}\ref{sec:Justification-for-the-local-flow-linearization}
we provide an argument (which we regard as useful, if not rigorous)
to justify why this local flow approximation should produce acceptable
errors on the propagated moments. The procedure produces a negligible
error for a small state displacement given a small increment of pseudo-time,
$\Delta\lambda$, so that stochastic particle flow (\ref{eq:state-flow-equation})
can be approximated within the region $\left\Vert \mathrm{x}-\mathrm{x}_{l}\right\Vert <\zeta$,
for a sufficiently small $\zeta\in\mathbb{R}_{+}$, as 
\begin{align}
d\mathrm{x} & =\frac{1}{2}D(\lambda)\nabla_{\mathrm{x}}\log\pi\left(\mathrm{x}\right)d\lambda+D(\lambda)^{\nicefrac{1}{2}}d\mathrm{w}_{\lambda},\quad\lambda\in(\lambda_{l},\lambda_{l}+\Delta\lambda],\,\mathrm{x}(\lambda_{l})=\mathrm{x}_{l};\nonumber \\
d\mathrm{x} & \approx\left[C(\mathrm{x}_{l},\lambda)\cdot\mathrm{x}+c(\mathrm{x}_{l},\lambda)\right]d\lambda+D(\lambda)^{\nicefrac{1}{2}}d\mathrm{w}_{\lambda}.\label{eq:flow-local-approximation}
\end{align}

As a consequence of integrating the flow, the corresponding component
moments are evolved according to the locally approximated ordinary
differential equations (Appendix\emph{~}\ref{sec:Justification-for-the-local-flow-linearization})
\begin{alignat}{1}
\frac{d\mu_{m}^{(i)}(\lambda)}{d\lambda} & =C(\mathrm{x}_{l}^{(i)})\mu_{m}^{(i)}(\lambda)+c(\mathrm{x}_{l}^{(i)}),\label{eq:mean-propagation}\\
\frac{d\Sigma_{m}^{(i)}(\lambda)}{d\lambda} & =C(\mathrm{x}_{l}^{(i)})\Sigma_{m}^{(i)}(\lambda)+\Sigma_{m}^{(i)}(\lambda)C^{T}(\mathrm{x}_{l}^{(i)})+D^{(i)}.\label{eq:covariance-propagation}
\end{alignat}
For nonlinear Gaussian problems, the locally approximated flow implies
that
\begin{align}
C\left(\mathrm{x}_{l},\lambda\right) & =-\frac{1}{2}D(\lambda)P_{k|k-1}^{-1}\nonumber \\
 & \hphantom{\,=}-\frac{1}{2}D(\lambda)\mathcal{J}_{\mathrm{x}}\left[h(\mathrm{x}_{l})\right]^{T}R_{k}^{-1}\mathcal{J}_{\mathrm{x}}\left[h(\mathrm{x}_{l})\right],\label{eq:Jacobian-of-the-drift-term}\\
c\left(\mathrm{x}_{l},\lambda\right) & =\hphantom{+}\frac{1}{2}D(\lambda)P_{k|k-1}^{-1}f(\mu_{m,k-1}^{\vphantom{-1}})\nonumber \\
 & \hphantom{\,=}+\frac{1}{2}D(\lambda)\mathcal{J}_{\mathrm{x}}\left[h(\mathrm{x}_{l})\right]^{T}R_{k}^{-1}\mathcal{J}_{\mathrm{x}}\left[h(\mathrm{x}_{l})\right]\cdot\mathrm{x}_{l}\nonumber \\
 & \hphantom{\,=}+\frac{1}{2}D(\lambda)\mathcal{J}_{\mathrm{x}}\left[h(\mathrm{x}_{l})\right]^{T}R_{k}^{-1}\left(\mathrm{y}_{k}-h(\mathrm{x}_{l})\right),\label{eq:offset-of-the-drift-term}
\end{align}
\noindent where $\mathcal{J}_{\mathrm{x}}\left[\cdot\right]$ is
the Jacobian matrix with respect to the state, $f(\cdot)$ is the
state process function, $h(\cdot)$ is the observation function, and
where 
\begin{align}
P_{k|k-1} & =\mathbb{E}\left[(\mathrm{x}_{k|k-1}-f(\mu_{m,k-1}))(\mathrm{x}_{k|k-1}-f(\mu_{m,k-1}))^{T}\right],\label{eq:predicted-covariance}\\
R_{k\hphantom{|k-1}} & =\mathbb{E}\left[(\mathrm{y}_{k}-h(\mathrm{x}_{k}))(\mathrm{y}_{k}-h(\mathrm{x}_{k}))^{T}\right],\label{eq:observation-noise-covariance}
\end{align}
 \noindent are, respectively, the covariance matrix of the prior
probability density and the covariance matrix of the observation noise.

Notice that the resulting algorithm is somewhat similar to the Kalman-Bucy
filter, in that it avoids an explicit discrete-time measurement update.
One could also interpret the SPF-GS as a Monte-Carlo and continuous-time
version of the original Gaussian sum filter \citep{Sorenson1971,Alspach1972},
albeit modified to explore the Riemannian geometric structure of the
probability space. However, in contrast to the Gaussian sum filter
(and the particle filter), the structure of the SPF-GS removes the
need to explicitly compute mixture weights: it relies on multiple
independent (ergodic) Markov chains that start at the samples to produce
equally weighted mixture components, each describing a local version
of the underlying geometric structure of the posterior measure. 

The SPF-GS has features that are apparently similar to those of the
Gaussian sum particle filter \citep{Kotecha2003}, however, in reality,
these filters rely on distinct fundamental principles that make them
very different. The principle of the Gaussian sum particle filter
is using importance sampling to estimate the moments of a mixture's
components for approximating a target density. In contrast, the SPF-GS
evolves a mixture through multiple intermediate steps by exploring
the local properties of a stochastic flow in order to translate probability
masses from the previous iteration to a mixture on the posterior probability
space. The SPF-GS is also very different from that proposed by Terejanu~\textit{et
al.} \citep{Terejanu2008}, which is a Gaussian sum filter analogous
to an extended Kalman-Bucy filter, but providing an estimate of the
predicted mixture weights based on an optimization procedure.

The stochastic particle flow Gaussian-sum filter (SPF-GS) is summarized
in \textit{Algorithm~\ref{alg:Stochastic-particle-flow}}. The algorithm
is expressed in a quite general form, including an accept-reject step
that enforces theoretical convergence to the invariant target measure.
Our empirical experience indicates that having this step was often
not necessary (at least in the cases we have considered). As has already
been touched on, if computational efficiency is a primary goal, the
SPF-GS should avoid the accept-reject step: calculating the acceptance
probability requires double\footnote{Because of the need to construct both the forward and backward transition
kernels.} the number of computations of the gradients and inverted Hessian
matrices. In addition, using this step requires the explicit evaluation
of the kernels and target densities themselves. When the step is removed,
there may be the need to deal with outliers, which can significantly
affect the estimates because all components in the propagated mixture
are treated as equally important. In all numerical examples studied
in this paper, none adopted the Metropolis adjustment. For the high-dimensional
examples, the (occasional) outliers are removed online by an empirical
test to identify the samples that fall outside a credible inference
region.
\begin{center}
\begin{algorithm2e}
\SetAlgoLined
\DontPrintSemicolon
\SetKwInOut{Input}{Input}\SetKwInOut{Output}{Output}
\textbf{Initialization:}\\
\If{time $k = 0$}{
Sample $\mathrm{x}_{0}^{(i)}\sim p(\mathrm{x}_{0}^{\vphantom{(i)}}),\,\forall i=1,\dots,N$\;
Set $w_{m,0}^{(i)}\coloneqq N^{-1}$, $\mu_{m,0}^{(i)}\coloneqq \mathbb{E}_{p_{0}}\left[\mathrm{x}_{0}\right]$, $\Sigma_{m,0}^{(i)}\coloneqq \mathbb{E}_{p_{0}}\left[(\mathrm{x}_{0}-\bar{\mathrm{x}}_{0})(\mathrm{x}_{0}-\bar{\mathrm{x}}_{0})^{T}\right]$, $\forall i=1,\dots,N$\;
}
\BlankLine
\textbf{Steps:}\\
\For{time $k \geq 1$}{
Compute the time horizon $T$ and step size $\Delta \lambda$ (section 4.6)\;
Discretize the interval $0\le\lambda\le T$ into $L$ sub-intervals $\{\lambda_{0}=0,\,\dots\lambda_{l},\,\dots,\,\lambda_{L}=T\}$\;
Set $\mathrm{x}_{l=0}^{(i)}\coloneqq\mathrm{x}_{k-1}^{(i)}$, $\mu_{l=0}^{(i)}\coloneqq\mathrm{x}_{k-1}^{(i)}$, $\Sigma_{l=0}^{(i)}\coloneqq0_{n_{x}\times n_{x}}$, $\forall i=1,\dots,N$\;

\For{$l = 1$ \KwTo $L$}{
\For{$i = 1,\dots,N$}{
Simulate 
\begin{equation}
\mathrm{x}_{l}^{\star(i)} \leftarrow \mathrm{x}_{l-1}^{(i)}+\frac{1}{2}\int_{\lambda_{l-1}}^{\lambda_{l}}D(\mathrm{x}_{l-1}^{(i)})\nabla_{\mathrm{x}}\log\tilde{\pi}(\mathrm{x}_{l-1}^{(i)})d\lambda +\int_{\lambda_{l-1}}^{\lambda_{l}}D(\mathrm{x}_{l-1}^{(i)})^{\nicefrac{1}{2}}d\mathrm{w}_{\lambda}
\end{equation}\label{eq:propagate-sample-eq}\;

Compute the MH acceptance probability $\rho^{(i)}=\min\left[1,\frac{\tilde{\pi}(\mathrm{x}_{l}^{\star(i)})}{q(\mathrm{x}_{l}^{\star(i)}|\mathrm{x}_{l-1}^{(i)})}\frac{q(\mathrm{x}_{l-1}^{(i)}|\mathrm{x}_{l}^{\star(i)})}{\tilde{\pi}(\mathrm{x}_{l-1}^{(i)})}\right]$\;
\tcc{We advocate using the Metropolis-adjustment step as practically optional, i.e., only for difficult problems. For many Engineering problems the approximation achieved by suppressing the MH step may be enough and will be more computationally efficient.}
Simulate $z^{(i)} \sim \mathcal{U}(0,1)$\;
\eIf{$z^{(i)} \leq \rho^{(i)}$}{
Set $\mathrm{x}_{l}^{(i)} \leftarrow \mathrm{x}_{l}^{\star(i)}$\;
Propagate
\begin{equation}
\mu_{l}^{(i)} \leftarrow \mu_{l-1}^{(i)}+\int_{\lambda_{l-1}}^{\lambda_{l}}\left[C(\mathrm{x}_{l-1}^{(i)})\mu_{l-1}^{(i)}+c(\mathrm{x}_{l-1}^{(i)})\right]d\lambda,
\end{equation}\label{eq:propagate-mean-eq}
\begin{equation}
\Sigma_{l}^{(i)} \leftarrow \Sigma_{l-1}^{(i)}+\int_{\lambda_{l-1}}^{\lambda_{l}}\left[C(\mathrm{x}_{l-1}^{(i)})\Sigma_{l-1}^{(i)}+\Sigma_{l-1}^{(i)}C^{T}(\mathrm{x}_{l-1}^{(i)})+D(\mathrm{x}_{l-1}^{(i)})\right]d\lambda
\end{equation}\label{eq:propagate-covariance-eq}\;
}{
Set $\mathrm{x}_{l}^{(i)} \leftarrow \mathrm{x}_{l-1}^{(i)}$, $\mu_{l}^{(i)} \leftarrow \mu_{l-1}^{(i)}$, $\Sigma_{l}^{(i)} \leftarrow \Sigma_{l-1}^{(i)}$\;
}
}
}
Set $\mathrm{x}_{k}^{(i)}\coloneqq\mathrm{x}_{l=L}^{(i)}$, $\mu_{m,k}^{(i)}\coloneqq\mu_{l=L}^{(i)}$, $\Sigma_{m,k}^{(i)}\coloneqq\Sigma_{l=L}^{(i)}$, $\forall i=1,\dots,N$\;
\BlankLine
\Output{Approximation of the filtering density as}
\begin{equation*}
\tilde{p}(\mathrm{x}_{k}^{\vphantom{(i)}}|\mathrm{y}_{1:k}^{\vphantom{(i)}})=\sum_{i=1}^{N}\frac{1}{N}\mathcal{N}(\mathrm{x}_{k}^{\vphantom{(i)}};\mu_{m,k}^{(i)},\Sigma_{m,k}^{(i)})
\end{equation*}
}
\caption{Stochastic particle flow - Gaussian sum filter\label{alg:Stochastic-particle-flow}}
\end{algorithm2e}
\par\end{center}

\section{Stochastic Particle Flow as a Marginal Particle Filter\label{sec:SPF-MPF}}

In this section we derive a marginal particle filter whose proposal
density is built upon a Gaussian mixture obtained via stochastic particle
flow. The resulting filter is referred to as the stochastic-particle-flow
marginal particle filter (SPF-MPF).

\subsection{Marginal Particle Filtering\label{subsec:Marginal-particle-filtering}}

In the standard setting, particle filters don't target the marginal
filtering distribution $p\left(\mathrm{x}_{k}|\mathrm{y}_{1:k}\right)$,
a characteristic inherited from the first particle filters, which
were designed to be relatively simple to implement. The main problem
with the standard particle filters arises because they construct importance
densities that target a joint filtering density $p\left(\mathrm{x}_{0:k}|\mathrm{y}_{1:k}\right)$.
A typical particle filter incrementally draws \emph{path samples,}
$\{\mathrm{x}_{0:k}^{(i)}\in\mathcal{X}^{k+1}:\,i=1,\,\dots,\,N\}$,
from a joint importance density $q\left(\mathrm{x}_{0:k}|\mathrm{y}_{1:k}\right)$,
and ignores the past of the sampled paths ($\{\mathrm{x}_{0:k-1}^{(i)}\in\mathcal{X}^{k}\}$)
when computing (filtered) expectations of interest. Thus, although
these algorithms provide a simple way to perform measurement update,
they perform importance sampling in the joint space along all time
steps, i.e., in $\mathcal{X}^{k+1}=\mathcal{X}\left(0\right)\times\mathcal{X}\left(1\right)\times\dots\times\mathcal{X}\left(k\right)$.
The result is precipitation of the degeneracy phenomenon: the set
of paths become increasingly sparse on the joint space $\mathcal{X}^{k+1}$,
leading to a quick increase in the weights' variance while most paths
have vanishingly small probability. In high-dimensional applications
this problem becomes even more pronounced, rendering the standard
particle filters to be practically infeasible.

With the mindset of improving this shortcoming in particle filters,
Klaas~\textit{et al.} \citep{Klaas2012} proposed the marginal particle
filter. The marginal particle filter targets the marginal posterior
distribution $p\left(\mathrm{x}_{k}|\mathrm{y}_{1:k}\right)$, performing
importance sampling on the marginal state space, $\mathcal{X}\left(k\right)$,
to produce samples with commensurate sparsity over time. The samples
are drawn from an importance density of the form 
\begin{equation}
q\left(\mathrm{x}_{k}|\mathrm{y}_{1:k}\right)\propto\int_{\mathcal{X}}q\left(\mathrm{x}_{k}|\mathrm{x}_{k-1},\mathrm{y}_{k}\right)q\left(\mathrm{x}_{k-1}|\mathrm{y}_{1:k-1}\right)d\mathrm{x}_{k-1},\label{eq:marginal-proposal-density}
\end{equation}
\noindent to target the posterior density 
\begin{equation}
p\left(\mathrm{x}_{k}|\mathrm{y}_{1:k}\right)\propto p\left(\mathrm{y}_{k}|\mathrm{x}_{k}\right)\int_{\mathcal{X}}p_{t}\left(\mathrm{x}_{k}|\mathrm{x}_{k-1}\right)p\left(\mathrm{x}_{k-1}|\mathrm{y}_{1:k-1}\right)d\mathrm{x}_{k-1},\label{eq:marginal-target}
\end{equation}
\noindent with the importance weights 
\begin{equation}
w\left(\mathrm{x}_{k}\right)\propto\frac{p\left(\mathrm{x}_{k}|\mathrm{y}_{1:k}\right)}{q\left(\mathrm{x}_{k}|\mathrm{y}_{1:k}\right)}.\label{eq:marginal-importance-weight}
\end{equation}

In practical terms, particles and weights from the previous iteration
are used to compose both an approximation of the target density (\ref{eq:marginal-target})
and the importance density (\ref{eq:marginal-proposal-density}),
in order to obtain particles and weights for the current iteration.
Even though the marginal particle filter is more robust than the standard
particle filter against degeneracy, and thereby more suitable to high-dimensional
problems in principle, its success is highly dependent on the validity
of sequential representations of the target density. Problems may
arise in situations where the usual approximation 
\begin{equation}
\tilde{p}\left(\mathrm{x}_{k}|\mathrm{y}_{1:k}\right)\propto p(\mathrm{y}_{k}^{\vphantom{(i)}}|\mathrm{x}_{k}^{\vphantom{(i)}})\sum_{i=1}^{N}w_{k-1}^{(i)}p_{t}(\mathrm{x}_{k}^{\vphantom{(i)}}|\mathrm{x}_{k-1}^{(i)}),\label{eq:marginal-target-approximation}
\end{equation}
\noindent is prone to relevant statistical or numerical errors, e.g.,
when the transition density $p_{t}\left(\mathrm{x}_{k}|\mathrm{x}_{k-1}\right)$
describes a Markov process with small variance and the observation
$\mathrm{y}_{k}$ lies relatively far from the current set of particles
$\{\mathrm{x}_{k-1}^{(i)},\mathrm{w}_{k-1}^{(i)}\}$ on the state
space (see the linear, univariate example in Section~\ref{sec:Examples}).
Moreover, owing to the curse of dimensionality, the usual approximation
(\ref{eq:marginal-target-approximation}) is corrupted by a Monte
Carlo error that increases geometrically with the number of state
dimensions. This may cripple the marginal particle filter in very
high-dimensional problems.  Because of this limitation, marginal particle
filters are likely to perform well only in moderately high-dimensional
problems. We illustrate this limitation of marginal particle filters
by numerical examples in Section~\ref{sec:Examples}.

As well covered in \citep{Klaas2012}, there exist several possibilities
to choose the marginal importance density (\ref{eq:marginal-proposal-density}),
among which the auxiliary marginal proposal density is particularly
interesting because it emulates an optimal importance density in the
sense of minimizing the weights' variance. The marginal optimal (auxiliary)
proposal density is usually approximated as 
\begin{align}
\tilde{q}\left(\mathrm{x}_{k}|\mathrm{y}_{1:k}\right) & =\sum_{i=1}^{N}w_{q,k-1}^{(i)}p(\mathrm{x}_{k}^{\vphantom{(i)}}|\mathrm{x}_{k-1}^{(i)},\mathrm{y}_{k}^{\vphantom{(i)}}),\label{eq:optimal-marginal-proposal}\\
w_{q,k-1}^{(i)} & \propto w_{k-1}^{(i)}p(\mathrm{y}_{k}^{\vphantom{(i)}}|\mathrm{x}_{k-1}^{(i)}).\nonumber 
\end{align}
It is straightforward to verify that, in the usual setting, the marginal
optimal proposal implies that weights never change: 
\begin{align*}
w_{k} & \propto\frac{\tilde{p}\left(\mathrm{x}_{k}|\mathrm{y}_{1:k}\right)}{\tilde{q}\left(\mathrm{x}_{k}|\mathrm{y}_{1:k}\right)}\propto\frac{p(\mathrm{y}_{k}^{\vphantom{(i)}}|\mathrm{x}_{k}^{\vphantom{(i)}})\sum_{i=1}^{N}w_{k-1}^{(i)}p_{t}(\mathrm{x}_{k}^{\vphantom{(i)}}|\mathrm{x}_{k-1}^{(i)})}{\sum_{i=1}^{N}w_{q,k-1}^{(i)}p(\mathrm{x}_{k}^{\vphantom{(i)}}|\mathrm{x}_{k-1}^{(i)},\mathrm{y}_{k}^{\vphantom{(i)}})}\\
 & \propto\frac{p(\mathrm{y}_{k}^{\vphantom{(i)}}|\mathrm{x}_{k}^{\vphantom{(i)}})\sum_{i=1}^{N}w_{k-1}^{(i)}p_{t}(\mathrm{x}_{k}^{\vphantom{(i)}}|\mathrm{x}_{k-1}^{(i)})}{\sum_{i=1}^{N}w_{k-1}^{(i)}p(\mathrm{y}_{k}^{\vphantom{(i)}}|\mathrm{x}_{k-1}^{(i)})\cdot\frac{p(\mathrm{y}_{k}^{\vphantom{(i)}}|\mathrm{x}_{k}^{\vphantom{(i)}})p_{t}(\mathrm{x}_{k}^{\vphantom{(i)}}|\mathrm{x}_{k-1}^{(i)})}{p(\mathrm{y}_{k}^{\vphantom{(i)}}|\mathrm{x}_{k-1}^{(i)})}}\\
 & =\text{constant.}
\end{align*}

This feature is crucial because it endows a particle filter with low
variance of weights, which essentially turns into statistical efficiency.
This finding motivates the marginal optimal proposal density as the
foundation for a marginal particle filter based on the stochastic
particle flow. The resulting filter is expected to work well for moderately
high-dimensional problems.

\subsection{Difficulties from a Usual Marginal Importance Density\label{subsec:Difficulties}}

This section discusses the problems that naturally arise when considering
a standard Monte Carlo setting as (\ref{eq:approximation-filtering-problem})
to build a marginal importance density based on the stochastic particle
flow. If one regards the proposal distribution as the result of a
sequence of $L$ Markov transitions through a discretization of the
interval $0<\lambda\le T$ onto the sub-intervals $\{\lambda_{0}=0,\,\lambda_{1},\,\dots\lambda_{l},\,\dots,\,\lambda_{L}=T\}$,
where $\mathrm{x}_{k}\triangleq\mathrm{x}_{L}$ and $\mathrm{x}_{k-1}\triangleq\mathrm{x}_{0}$,
then the sequence of transitions would provide the importance density
\begin{align}
q\left(\mathrm{x}_{k}|\mathrm{y}_{1:k}\right) & =\int_{\mathcal{X}}\int_{\mathcal{X}}\dots\int_{\mathcal{X}}q(\mathrm{x}_{L}|\mathrm{x}_{L-1},\mathrm{y}_{k})q(\mathrm{x}_{L-1}|\mathrm{x}_{L-2},\mathrm{y}_{k})\dots\nonumber \\
 & \hphantom{=\ \ }q(\mathrm{x}_{1}|\mathrm{x}_{0},\mathrm{y}_{k})q(\mathrm{x}_{0}|\mathrm{y}_{1:k-1})d\mathrm{x}_{L-1}d\mathrm{x}_{L-2}\dots d\mathrm{x}_{0}\cdot\label{eq:proposal-sequence-of-transitions}
\end{align}

In order to evaluate this importance density over a set of $N$ particles,
incorporating the previous set of samples and importance weights,
one would be required to compute 
\begin{align}
\tilde{q}(\mathrm{x}_{k}^{(i)}|\mathrm{y}_{1:k}^{\vphantom{(i)}}) & =\sum_{j=1}^{N}w_{k-1}^{(j)}\tilde{q}(\mathrm{x}_{k}^{(i)}|\mathrm{x}_{k-1}^{(j)},\mathrm{y}_{k}^{\vphantom{(i)}}),\label{eq:marginal-proposal-density-1}\\
 & \hphantom{=}i=1,\dots,N.\nonumber 
\end{align}

This implementation depends on the set of conditional kernels $\tilde{q}(\mathrm{x}_{k}|\mathrm{x}_{k-1},\mathrm{y}_{k})=\tilde{q}(\mathrm{x}_{L}|\mathrm{x}_{0},\mathrm{y}_{k})$
that could be achieved in terms of a recursion of the form 
\begin{align}
w_{1}^{(i|j)} & \triangleq q(\mathrm{x}_{1}^{(i)}|\mathrm{x}_{0}^{(j)},\mathrm{y}_{k}^{\vphantom{(i)}}),\quad i,\,j=1,\dots,N;\nonumber \\
w_{2}^{(i|j)} & \triangleq\tilde{q}(\mathrm{x}_{2}^{(i)}|\mathrm{x}_{0}^{(j)},\mathrm{y}_{k}^{\vphantom{(i)}})=\sum_{n=1}^{N}w_{1}^{(n|j)}q(\mathrm{x}_{2}^{(i)}|\mathrm{x}_{1}^{(n)},\mathrm{x}_{0}^{(j)},\mathrm{y}_{k}^{\vphantom{(i)}});\nonumber \\
 & \hphantom{\triangleq q(\mathrm{x}_{2}^{(i)}|)}\vdots\nonumber \\
w_{l}^{(i|j)} & \triangleq\tilde{q}(\mathrm{x}_{l}^{(i)}|\mathrm{x}_{0}^{(j)},\mathrm{y}_{k}^{\vphantom{(i)}})=\sum_{n=1}^{N}w_{l-1}^{(n|j)}q(\mathrm{x}_{l}^{(i)}|\mathrm{x}_{l-1}^{(n)},\mathrm{x}_{0}^{(j)},\mathrm{y}_{k}^{\vphantom{(i)}});\nonumber \\
 & \hphantom{\triangleq q(\mathrm{x}_{2}^{(i)}|)}\vdots\nonumber \\
 & \hphantom{\triangleq\:\:\,}\tilde{q}(\mathrm{x}_{L}^{(i)}|\mathrm{x}_{0}^{(j)},\mathrm{y}_{k}^{\vphantom{(i)}})=\sum_{n=1}^{N}w_{L-1}^{(n|j)}q(\mathrm{x}_{L}^{(i)}|\mathrm{x}_{L-1}^{(n)},\mathrm{x}_{0}^{(j)},\mathrm{y}_{k}^{\vphantom{(i)}});\label{eq:recursion-transition-pdfs}
\end{align}
\noindent where $q(\mathrm{x}_{l}|\mathrm{x}_{l-1},\mathrm{x}_{0},\mathrm{y}_{k})$
are the one-step proposal kernels conditioned on the initial state
(prior samples), which are directly available from the discretized
version of (\ref{eq:state-flow-equation}). Computing the conditional
proposal components (\ref{eq:recursion-transition-pdfs}) and the
marginal proposal (\ref{eq:marginal-proposal-density-1}) involves
high computational effort, bounded by $\mathcal{O}\left((L-1)N^{3}+N^{2}\right)$
evaluations. In addition, the main complication of this realization
is due to the mixing properties of (\ref{eq:marginal-proposal-density-1}),
leading to significant errors built up through the sequence of finite-sample
approximations in (\ref{eq:recursion-transition-pdfs}) along with
the prohibitively high variance of the resulting importance weights
(\ref{eq:marginal-importance-weight}).

While these problems could be tentatively worked around by a judicious
choice of a variance reduction method, it is worth looking how the
implementation difficulties would turn out to be by evoking a hypothetical
``continuity'' between sampling steps. It is well known that in the
limit $\Delta\lambda\rightarrow0$, the proposal density (\ref{eq:proposal-sequence-of-transitions})
defines a path integral. Based on the concept of path probability
density \citep{Graham1973} of a Markov process 
\begin{equation}
W_{\infty}\left[\mathrm{x}\left(\lambda\right)\right]\left[d\mathrm{x}\right]\propto e^{-\int_{0}^{T}\left[\frac{1}{2}\left(\mathrm{\dot{x}-\mathrm{\mu\left(\mathrm{x}\right)}}\right)^{T}D^{-1}\left(\mathrm{\dot{x}-\mathrm{\mu\left(\mathrm{x}\right)}}\right)+\frac{1}{2}\nabla_{\mathrm{x}}\cdot\mu\left(\mathrm{x}\right)\right]d\lambda},\label{eq:path-probability-density}
\end{equation}
\noindent for samples describing continuous paths, the proposal could
be written as a functional integral \citep{Janssen1976} of the form
\begin{equation}
q_{c}\left(\mathrm{x}_{k}|\mathrm{y}_{1:k}\right)\propto\int e^{-\int_{0}^{T}\left[\frac{1}{2}\left(\mathrm{\dot{x}-\mathrm{\mu\left(\mathrm{x}\right)}}\right)^{T}D^{-1}\left(\mathrm{\dot{x}-\mathrm{\mu\left(\mathrm{x}\right)}}\right)+\frac{1}{2}\nabla_{\mathrm{x}}\cdot\mu\left(\mathrm{x}\right)\right]d\lambda}\left[d\mathrm{x}\right],\label{eq:proposal-as-a-path-pdf}
\end{equation}
\noindent where $\left[d\mathrm{x}\right]=d\mathrm{x}_{L-1}\dots d\mathrm{x}_{0}$
as $\Delta\lambda\rightarrow0$. Solving path integrals in general
is a daunting task, nevertheless, a density of interest could be approximately
obtained in terms of a Gaussian mixture, under the assumption of local
Gaussianity of probability paths. Within this framework, an ensemble
of independently selected Gaussian densities can be analytically integrated
to achieve local solutions to (\ref{eq:proposal-as-a-path-pdf}).
This fundamental idea is equivalent to what stochastic particle flow
proposes when the filtering solution is formulated as the mixture
(\ref{eq:Gaussian-mixture-measure}).

\subsection{The Stochastic-Particle-Flow Marginal Particle Filter\label{subsec:The-SPF-MPF}}

In marginal particle filtering, the best importance density one could
achieve is the proposal density $q\left(\mathrm{x}_{k}|\mathrm{y}_{1:k}\right)$
when computed exactly. This density enables inference of the actual
posterior pdf, $p\left(\mathrm{x}_{k}|\mathrm{y}_{1:k}\right)$. Composing
the marginal optimal proposal requires computing $p(\mathrm{y}_{k}^{\vphantom{(i)}}|\mathrm{x}_{k-1}^{(i)})$
exactly, which is not possible in general. In addition, the same scenarios
that produce considerable errors in computing the empirical target,
$\tilde{\pi}\left(\mathrm{x}_{k}\right)=\tilde{p}\left(\mathrm{x}_{k}|\mathrm{y}_{1:k}\right)$,
according to (\ref{eq:marginal-target-approximation}), will also
affect evaluation of the proposal $\tilde{q}\left(\mathrm{x}_{k}|\mathrm{y}_{1:k}\right)$,
computed by (\ref{eq:optimal-marginal-proposal}), as illustrated
by the first example in Section~\ref{sec:Examples}. In these cases,
one can benefit from the inherent characteristics of stochastic particle
flow to construct a proposal density with better regularity properties
by doing 
\begin{align}
\tilde{q}\left(\mathrm{x}_{k}|\mathrm{y}_{1:k}\right) & =\sum_{i=1}^{N}w_{k-1}^{(i)}p(\mathrm{y}_{k}^{\vphantom{(i)}}|\mathrm{x}_{k}^{\vphantom{(i)}})p_{t}(\mathrm{x}_{k}^{\vphantom{(i)}}|\mathrm{x}_{k-1}^{(i)})\nonumber \\
 & =\sum_{i=1}^{N}w_{k-1}^{(i)}\frac{p_{t}(\mathrm{x}_{k}^{\vphantom{(i)}}|\mathrm{x}_{k-1}^{(i)})}{\tilde{p}^{(i)}(\mathrm{x}_{k}^{\vphantom{(i)}}|\mathrm{y}_{1:k-1}^{\vphantom{(i)}})}p(\mathrm{y}_{k}^{\vphantom{(i)}}|\mathrm{x}_{k}^{\vphantom{(i)}})\tilde{p}^{(i)}(\mathrm{x}_{k}^{\vphantom{(i)}}|\mathrm{y}_{1:k-1}^{\vphantom{(i)}})\nonumber \\
 & \propto\sum_{i=1}^{N}w_{k-1}^{(i)}\frac{p_{t}(\mathrm{x}_{k}^{\vphantom{(i)}}|\mathrm{x}_{k-1}^{(i)})}{\tilde{p}^{(i)}(\mathrm{x}_{k}^{\vphantom{(i)}}|\mathrm{y}_{1:k-1}^{\vphantom{(i)}})}w_{m,k}^{(i)}\mathcal{N}(\mathrm{x}_{k}^{\vphantom{(i)}};\mu_{m,k}^{(i)},\Sigma_{m,k}^{(i)}),\nonumber \\
\tilde{q}\left(\mathrm{x}_{k}|\mathrm{y}_{1:k}\right) & =\sum_{i=1}^{N}w_{q}^{(i)}(\mathrm{x}_{k}^{\vphantom{(i)}})\mathcal{N}(\mathrm{x}_{k}^{\vphantom{(i)}};\mu_{m,k}^{(i)},\Sigma_{m,k}^{(i)});\label{eq:SPF-marginal-proposal}
\end{align}
\noindent where 
\begin{align}
\bar{w}_{q}^{(i)}(\mathrm{x}_{k}^{\vphantom{(i)}}) & =w_{k-1}^{(i)}w_{m,k}^{(i)}\frac{p_{t}(\mathrm{x}_{k}^{\vphantom{(i)}}|\mathrm{x}_{k-1}^{(i)})}{\tilde{p}^{(i)}(\mathrm{x}_{k}^{\vphantom{(i)}}|\mathrm{y}_{1:k-1}^{\vphantom{(i)}})},\label{eq:SPF-marginal-weights}\\
w_{q}^{(i)}(\mathrm{x}_{k}^{\vphantom{(i)}}) & =\frac{\bar{w}_{q}^{(i)}(\mathrm{x}_{k}^{\vphantom{(i)}})}{\sum_{i=1}^{N}\bar{w}_{q}^{(i)}(\mathrm{x}_{k}^{\vphantom{(i)}})};\label{eq:SPF-marginal-weights-1}
\end{align}
\noindent and $\tilde{p}^{(i)}(\mathrm{x}_{k}|\mathrm{y}_{1:k-1})$
is a per-sample, local prior density. For a known Markov transition
density $p_{t}\left(\mathrm{x}_{k}^{\prime}|\mathrm{x}_{k-1}^{\vphantom{\prime}}\right)$,
we recall the local prior density as given by 
\begin{align}
\tilde{p}^{(i)}\left(\mathrm{x}_{k}^{\prime}|\mathrm{y}_{1:k-1}^{\vphantom{\prime}}\right) & =\int_{\mathcal{X}}p_{t}(\mathrm{x}_{k}^{\prime\vphantom{(i)}}|\mathrm{x}_{k-1}^{\vphantom{(i)}})\mathcal{N}(\mathrm{x}_{k-1}^{\vphantom{(i)}};\mu_{m,k-1}^{(i)},\Sigma_{m,k-1}^{(i)})d\mathrm{x}_{k-1}^{\vphantom{\prime}}.\label{eq:predicted-pdf-per-particle-1}
\end{align}

As in the classical Gaussian-sum setting, the mixture weights $\{w_{m,k}^{(i)}\}$
are given by (see \citep{Anderson1979}, pages 214 and 215) 
\begin{align}
w_{m,k}^{(i)} & \propto\frac{1}{N}\int_{\mathcal{X}}p\left(\mathrm{y}_{k}^{\vphantom{\prime}}|\mathrm{x}_{k}^{\prime}\right)\tilde{p}^{(i)}\left(\mathrm{x}_{k}^{\prime}|\mathrm{y}_{1:k-1}^{\vphantom{\prime}}\right)d\mathrm{x}_{k}^{\prime},\label{eq:mixture-weights-1}
\end{align}
\noindent where the proportionality to $N^{-1}$ holds because stochastic
particle flow generates equally weighted mixture components. The mixture
weights $\{w_{m,k}^{(i)}\}$ generated by this method are only applicable
in the context of the proposal (\ref{eq:SPF-marginal-proposal}),
and shall be re-evaluated in the same way whenever a new instance
of the marginal proposal is constructed. It is relevant to make clear
the distinction $\mathrm{x}_{k}^{\prime}\neq\mathrm{x}_{k}^{\vphantom{\prime}}$
in the expressions (\ref{eq:mixture-weights-1}) and (\ref{eq:predicted-pdf-per-particle-1}),
bearing in mind that $\mathrm{x}_{k}^{\prime}$ corresponds to the
state that the flow would reach when considering only the prior density
as the target ${\pi_{\text{prior}}(\mathrm{x}^{\prime})=p_{x}(\mathrm{x}^{\prime})\triangleq p(\mathrm{x}^{\prime}|\mathrm{y}_{1:k-1})}$.
We note that the involved integrals may not be tractable in general
and may require approximation either by a Gaussian representation
of the likelihood, or adequate quadrature rules (e.g., Gauss-Hermite).

This formulation evokes stochastic particle flow to promote an accurate
approximation to the marginal optimal proposal density. Given a set
of samples and parameters $\{\mathrm{x}_{k-1}^{(i)},w_{k-1}^{(i)},\mu_{m,k-1}^{(i)},\Sigma_{m,k-1}^{(i)}\}$
from a previous filtering iteration, where $w_{k-1}^{(i)}$ are importance
weights, the algorithm integrates the SDE (\ref{eq:state-flow-equation})
for each sample and propagates the associated parameters through the
interval $0<\lambda\le T$. As result, the procedure acquires posterior
samples and parameters, $\{\mathrm{x}_{k}^{(i)},w_{k}^{(i)},\mu_{m,k}^{(i)},\Sigma_{m,k}^{(i)}\}$,
which are used to evaluate the marginal proposal (\ref{eq:SPF-marginal-proposal})
and enable filtering as by a marginal particle filter. The moments
of the mixture's components are evolved in accordance with (\ref{eq:mean-propagation})
and (\ref{eq:covariance-propagation}), and the importance weights
are updated by 
\begin{equation}
w_{k}^{\vphantom{(i)}}(\mathrm{x}_{k}^{\vphantom{(i)}})\propto\frac{\sum_{j=1}^{N}w_{k-1}^{(j)}p(\mathrm{y}_{k}^{\vphantom{(i)}}|\mathrm{x}_{k}^{\vphantom{(i)}})p_{t}(\mathrm{x}_{k}^{\vphantom{(i)}}|\mathrm{x}_{k-1}^{(j)})}{\sum_{j=1}^{N}w_{q}^{(j)}(\mathrm{x}_{k}^{\vphantom{(i)}})\mathcal{N}(\mathrm{x}_{k}^{\vphantom{(i)}};\mu_{m,k}^{(j)},\Sigma_{m,k}^{(j)})}.\label{eq:importance-weights-update}
\end{equation}

The resulting filter, called stochastic-particle-flow marginal particle
filter (SPF-MPF), is summarized in \textit{Algorithm~\ref{alg:Stochastic-particle-flow-2}}.
It is worth noting that a simpler alternative to (\ref{eq:SPF-marginal-proposal})
could be chosen by considering 
\begin{equation}
\tilde{q}\left(\mathrm{x}_{k}|\mathrm{y}_{1:k}\right)=\sum_{i=1}^{N}\frac{1}{N}\mathcal{N}(\mathrm{x}_{k}^{\vphantom{(i)}};\mu_{m,k}^{(i)},\Sigma_{m,k}^{(i)}),\label{eq:SPF-marginal-proposal-alternative}
\end{equation}
\noindent however, in that case, the importance density would not
be affected by the same errors as the empirical target, $\tilde{\pi}\left(\mathrm{x}_{k}\right)=\tilde{p}\left(\mathrm{x}_{k}|\mathrm{y}_{1:k}\right)$,
as computed by (\ref{eq:marginal-target-approximation}), because
each component in (\ref{eq:SPF-marginal-proposal-alternative}) targets
a local instance of the posterior density itself. As a result, even
though the importance density could approximate the true posterior
density accurately, it would not directly approach the target density.
In situations where the empirical target density cannot represent
the true posterior density as well as a mixture of the form (\ref{eq:SPF-marginal-proposal-alternative}),
the SPF-MPF with such a proposal would fail because of the mismatch
originated from distinctions in the approximation methods. As a consequence,
the importance weights would have infeasibly high variance. The described
issue is equivalent to treat errors in the standard Monte Carlo measure
(\ref{eq:approximation-filtering-problem}) as comparable to errors
in the mixture measure (\ref{eq:approximation-filtering-problem-1}),
which is not true except for rare cases. This scenario is well illustrated
by two examples in Section~\ref{sec:Examples}.
\begin{center}
\begin{algorithm2e}
\SetAlgoLined
\DontPrintSemicolon
\SetKwInOut{Input}{Input}\SetKwInOut{Output}{Output}
\textbf{Initialization:}\\
\If{time $k = 0$}{
Sample $\mathrm{x}_{0}^{(i)}\sim p(\mathrm{x}_{0}^{\vphantom{(i)}})$ and set $w_{0}^{(i)}\coloneqq N^{-1}$, $\forall i=1,\dots,N$\;
Set $w_{m,0}^{(i)}\coloneqq N^{-1}$, $\mu_{m,0}^{(i)}\coloneqq \mathbb{E}_{p_{0}}\left[\mathrm{x}_{0}\right]$, $\Sigma_{m,0}^{(i)}\coloneqq \mathbb{E}_{p_{0}}\left[(\mathrm{x}_{0}-\bar{\mathrm{x}}_{0})(\mathrm{x}_{0}-\bar{\mathrm{x}}_{0})^{T}\right]$, $\forall i=1,\dots,N$\;
}
\BlankLine
\textbf{Steps:}\\
\For{time $k \geq 1$}{
Compute the time horizon $T$ and step size $\Delta \lambda$ (section 4.6)\;
Discretize the interval $0\le\lambda\le T$ into $L$ sub-intervals $\{\lambda_{0}=0,\,\dots\lambda_{l},\,\dots,\,\lambda_{L}=T\}$\;
Set $\mathrm{x}_{l=0}^{(i)}\coloneqq\mathrm{x}_{k-1}^{(i)}$, $\mu_{l=0}^{(i)}\coloneqq\mathrm{x}_{k-1}^{(i)}$, $\Sigma_{l=0}^{(i)}\coloneqq0_{n_{x}\times n_{x}}$, $\forall i=1,\dots,N$\;

\For{$l = 1$ \KwTo $L$}{
\For{$i = 1,\dots,N$}{
Simulate 
\begin{equation*}
\mathrm{x}_{l}^{(i)} \leftarrow \mathrm{x}_{l-1}^{(i)}+\frac{1}{2}\int_{\lambda_{l-1}}^{\lambda_{l}}D(\mathrm{x}_{l-1}^{(i)})\nabla_{\mathrm{x}}\log\tilde{\pi}(\mathrm{x}_{l-1}^{(i)})d\lambda +\int_{\lambda_{l-1}}^{\lambda_{l}}D(\mathrm{x}_{l-1}^{(i)})^{\nicefrac{1}{2}}d\mathrm{w}_{\lambda}
\end{equation*}\;

Propagate
\begin{equation*}
\mu_{l}^{(i)} \leftarrow \mu_{l-1}^{(i)}+\int_{\lambda_{l-1}}^{\lambda_{l}}\left[C(\mathrm{x}_{l-1}^{(i)})\mu_{l-1}^{(i)}+c(\mathrm{x}_{l-1}^{(i)})\right]d\lambda,
\end{equation*}
\begin{equation*}
\Sigma_{l}^{(i)} \leftarrow \Sigma_{l-1}^{(i)}+\int_{\lambda_{l-1}}^{\lambda_{l}}\left[C(\mathrm{x}_{l-1}^{(i)})\Sigma_{l-1}^{(i)}+\Sigma_{l-1}^{(i)}C^{T}(\mathrm{x}_{l-1}^{(i)})+D(\mathrm{x}_{l-1}^{(i)})\right]d\lambda
\end{equation*}\;
}
}
Set $\mathrm{x}_{k}^{(i)}\coloneqq\mathrm{x}_{l=L}^{(i)}$, $\mu_{m,k}^{(i)}\coloneqq\mu_{l=L}^{(i)}$, $\Sigma_{m,k}^{(i)}\coloneqq\Sigma_{l=L}^{(i)}$, $\forall i=1,\dots,N$\;
Compute the normalized proposal weights, $\forall i=1,\dots,N$, by
\begin{equation*}
w_{m,k}^{(i)}\propto\frac{1}{N}\int_{\mathcal{X}}p\left(\mathrm{y}_{k}^{\vphantom{\prime}}|\mathrm{x}_{k}^{\prime}\right)\tilde{p}^{(i)}\left(\mathrm{x}_{k}^{\prime}|\mathrm{y}_{1:k-1}^{\vphantom{\prime}}\right)d\mathrm{x}_{k}^{\prime},
\end{equation*}
\begin{equation*}
w_{q}^{(i)} \propto w_{k-1}^{(i)}w_{m,k}^{(i)}\frac{p_{t}(\mathrm{x}_{k}^{(i)}|\mathrm{x}_{k-1}^{(i)})}{\tilde{p}^{(i)}(\mathrm{x}_{k}^{(i)}|\mathrm{y}_{1:k-1}^{\vphantom{(i)}})}
\end{equation*}\;
Compute the normalized importance weights, $\forall i=1,\dots,N$, by
\begin{equation*}
w_{k}^{(i)}\propto\frac{\sum_{j=1}^{N}w_{k-1}^{(j)}p(\mathrm{y}_{k}^{\vphantom{(i)}}|\mathrm{x}_{k}^{(i)})p_{t}(\mathrm{x}_{k}^{(i)}|\mathrm{x}_{k-1}^{(j)})}{\sum_{j=1}^{N}w_{q}^{(j)}\mathcal{N}(\mathrm{x}_{k}^{(i)};\mu_{m,k}^{(j)},\Sigma_{m,k}^{(j)})}
\end{equation*}\;
\lIf{$ESS_{k} < 0.5 N$}{ resample: $\{\mathrm{x}_{k}^{(i)},N^{-1}\} \leftarrow \{\mathrm{x}_{k}^{(i)},w_{k}^{(i)}\}$ }
\BlankLine
\Output{Approximation of the filtering distribution by the empirical measure}
\begin{equation*}
\tilde{p}(\mathrm{x}_{k}^{\vphantom{(i)}}|\mathrm{y}_{1:k}^{\vphantom{(i)}})=\sum_{i=1}^{N}w_{k}^{(i)}\delta(\mathrm{x}_{k}^{\vphantom{(i)}}-\mathrm{x}_{k}^{(i)})
\end{equation*}
}
\caption{Stochastic particle flow - marginal particle filter\label{alg:Stochastic-particle-flow-2}}
\end{algorithm2e}
\par\end{center}

\section{Examples\label{sec:Examples}}

In this section we present some illustrative toy examples and experimental
results for three instructive applications in the multi-sensor multi-target
tracking context: a multi-sensor bearing-only problem, a convoy tracking
problem, and inference on a large spatial sensor network as presented
by Septier \& Peters \citep{Septier2016}.

In the experimental results for the bearing-only and convoy tracking
examples, we compared the SPF-GS against standard target trackers
and extensions of two of the most effective particle flows, namely,
the Gaussian particle flow (GPF) and the scaled-drift particle flow
(SDPF). The GPF was first called \emph{exact particle flow} in \citep{Daum2010a}
and the SDPF was first called \emph{non-zero diffusion particle flow}
in \citep{Daum2013c}. Actually, this latter is a particle flow with
the drift scaled by a diffusion coefficient, but the filter itself
is not a diffusion. 

It is important to mention that, in order to work properly, both the
Gaussian particle flow and the scaled-drift particle flow are implemented
with the aid of a companion filter such that the state covariance
matrix can be correctly estimated. Implementation details have been
presented by Choi~\textit{et al.} \citep{Choi2011} and Ding \& Coates
\citep{Ding2012}, who advocate using the EKF (or UKF) as a companion
filter to estimate the associated covariance matrices. Another option
is to shrink the empirical covariance and apply Tikhonov regularization
\citep{Khan2015}. In contrast, the stochastic particle flow does
not require any auxiliary technique to estimate the second order moment,
relying solely on its mixture measure. In the toy examples a companion
filter was not necessary for the original particle flows since a single
filtering cycle has been analyzed. In the multi-sensor and multi-target
examples we adopted baseline filters, which are the most structurally
similar to the EKF, as companion filters for the particle flows (GPF,
SDPF).

In the example of the large spatial sensor network, we compared the
SPF-GS, a particle filter (Sequential Importance Resampling - SIR),
a block particle filter (block SIR), and two of the best sequential
MCMC filters \citep{Girolami2011,Septier2016}: the Sequential manifold
Metropolis-Adjusted Algorithm (SmMALA) and the Sequential manifold
Hamiltonian Monte Carlo (SmHMC). The block particle filter partitions
the state space into separate subspaces of smaller dimensions and
run a particle filter on each subspace \citep{Rebeschini2015}.

\subsection{Toy Examples}

The toy examples are Gaussian processes chosen to demonstrate the
properties of the stochastic particle flow, summarized as
\begin{itemize}
\item Univariate
\begin{itemize}
\item linear,
\item quadratic,
\item cubic;
\end{itemize}
\item Bivariate
\begin{itemize}
\item multimodal, linear,
\item nonlinear (banana-shaped pdf).
\end{itemize}
\end{itemize}
In all cases, we analyze the filters for a single filtering cycle.
Generally, we describe the state process, the observation process
and the initial distribution for these examples as
\begin{align}
\mathrm{x}_{k}^{\vphantom{(i)}} & =f(\mathrm{x}_{k-1}^{\vphantom{(i)}})+\mathrm{u}_{k},\,\mathrm{u}_{k}\sim\mathcal{N}(\mathrm{u}_{k}^{\vphantom{(i)}};0,Q_{k}),\label{eq:state-process}\\
\mathrm{y}_{k}^{\vphantom{(i)}} & =h(\mathrm{x}_{k\hphantom{-1}}^{\vphantom{(i)}})+\mathrm{v}_{k},\,\mathrm{v}_{k}\sim\mathcal{N}(\mathrm{v}_{k}^{\vphantom{(i)}};0,R_{k}),\label{eq:observation-process}\\
p_{0}(\mathrm{x}_{k-1}^{\vphantom{(i)}}) & =\mathcal{N}(\mathrm{x}_{k-1}^{\vphantom{(i)}};\bar{\mathrm{x}}_{k-1}^{\vphantom{(i)}},P_{k-1}).\label{eq:initial-distribution}
\end{align}

We consider four different types of particle filters based on the
marginal importance density
\[
\tilde{q}\left(\mathrm{x}_{k}|\mathrm{y}_{1:k}\right)=\sum_{i=1}^{N}w_{k-1}^{(i)}q(\mathrm{x}_{k}^{\vphantom{(i)}}|\mathrm{x}_{k-1}^{(i)},\mathrm{y}_{k}^{\vphantom{(i)}}),
\]
\noindent where
\begin{itemize}
\item for the marginal bootstrap particle filter (MBPF), the proposal's
components are set as the Markov transition kernel: $q(\mathrm{x}_{k}^{\vphantom{(i)}}|\mathrm{x}_{k-1}^{(i)},\mathrm{y}_{k}^{\vphantom{(i)}})=p_{t}(\mathrm{x}_{k}^{\vphantom{(i)}}|\mathrm{x}_{k-1}^{(i)})$;
\item for the marginal EKF-based particle filter (MEPF), the proposal's
components are computed by the EKF: $q(\mathrm{x}_{k}^{\vphantom{(i)}}|\mathrm{x}_{k-1}^{(i)},\mathrm{y}_{k}^{\vphantom{(i)}})=p_{\text{EKF}}(\mathrm{x}_{k}^{\vphantom{(i)}}|\mathrm{x}_{k-1}^{(i)},\mathrm{y}_{k}^{\vphantom{(i)}})$;
\item for the marginal UKF-based particle filter (MUPF), the proposal's
components are computed by the UKF: $q(\mathrm{x}_{k}^{\vphantom{(i)}}|\mathrm{x}_{k-1}^{(i)},\mathrm{y}_{k}^{\vphantom{(i)}})=p_{\text{UKF}}(\mathrm{x}_{k}^{\vphantom{(i)}}|\mathrm{x}_{k-1}^{(i)},\mathrm{y}_{k}^{\vphantom{(i)}})$;
and
\item for the marginal auxiliary particle filter (MAPF) \citep{Klaas2012},
the importance density is given by (\ref{eq:optimal-marginal-proposal}).
\end{itemize}
When comparing probability densities furnished by different filters,
we include the empirical marginal target, $\tilde{\pi}\left(\mathrm{x}_{k}\right)=\tilde{p}\left(\mathrm{x}_{k}|\mathrm{y}_{1:k}\right)$,
evaluated according to (\ref{eq:marginal-target-approximation}) for
samples obtained by stochastic particle flow. For all filters, when
applicable, we calculate the average of the effective sample size
\begin{equation}
\text{ESS}=\left(\sum_{i=1}^{N}w_{k}^{(i)\,2}\right)^{-1}\label{eq:effective-sample-size}
\end{equation}
\noindent over 100 Monte Carlo runs, for 1000 particles. For all
marginal proposal densities, we analyze their similiarity to the true
posterior probability density by averaging their empirical Jensen-Shannon
divergence (JSD) with respect to the true posterior, which is obtained
to high numerical precision. The Jensen-Shannon divergence is defined
as
\begin{align}
\text{JSD}\left(P\parallel Q\right) & =\frac{1}{2}\text{D}_{\text{KL}}(P\parallel\left(P+Q\right)/2)\nonumber \\
 & +\frac{1}{2}\text{D}_{\text{KL}}(Q\parallel\left(P+Q\right)/2),\label{eq:Jensen-Shannon-divergence}
\end{align}
\noindent where the Kullback\textendash Leibler divergence, $\text{D}_{\text{KL}}(\cdot\parallel\cdot)$,
is computed using the base-2 logarithm such that the Jensen-Shannon
divergence is bounded as $0\le\text{JSD}\left(P\parallel Q\right)\le1$.
The Jensen-Shannon divergence is symmetric and equals zero when the
compared densities are equal. In the bivariate examples we also consider
the original particle flow methods, the Gaussian particle flow (GPF)
and scaled-drift particle flow (SDPF), for which the Jensen-Shannon
divergence with respect to the true posterior is evaluated based on
empirical densities constructed by (bidimensional) histograms of samples.\medskip{}

\subsubsection{Linear, Univariate Model}

The simplest example is a linear, univariate model, with parameters
set as in the table below.

\medskip{}

\begin{center}
\begin{tabular}{|l|l|}
\hline 
\multicolumn{2}{|l|}{\emph{Parameters for the linear, univariate model}}\tabularnewline
\hline 
\hline 
Initial distribution & $\bar{\mathrm{x}}_{k-1}^{\vphantom{(i)}}=0$, $P_{k-1}=20$\tabularnewline
\hline 
Markov transition pdf & $f(\mathrm{x}_{k-1}^{\vphantom{(i)}})=\mathrm{x}_{k-1}^{\vphantom{(i)}}$,
$Q_{k}=5$\tabularnewline
\hline 
Likelihood function & $h(\mathrm{x}_{k}^{\vphantom{(i)}})=\mathrm{x}_{k}^{\vphantom{(i)}}$,
$R_{k}=10$\tabularnewline
\hline 
Observation & $\mathrm{y}_{k}=30$\tabularnewline
\hline 
\end{tabular}\smallskip{}
\par\end{center}

Although very simple, this example was proposed to demonstrate a scenario
where the empirical marginal target, $\tilde{\pi}\left(\mathrm{x}_{k}\right)=\tilde{p}\left(\mathrm{x}_{k}|\mathrm{y}_{1:k}\right)$,
is prone to relevant statistical and numerical errors. This is done
by setting a situation where the transition kernel describes a Markov
process with small variance and the observation lies relatively far
from the initial distribution. In this scenario, statistical inefficiency
emerges because the observation provides little information in the
space region where probability masses are more densely distributed
by the state process. Not incidentally, this is also the main source
of degeneracy in standard particle filters. Additionally, there may
exist round-off errors when evaluating the empirical marginal target
owing to samples being located relatively far from the posterior mean,
several standard deviations apart, in the tail of each proposal component.

\begin{figure}[H]
\begin{centering}
\includegraphics[width=1\textwidth]{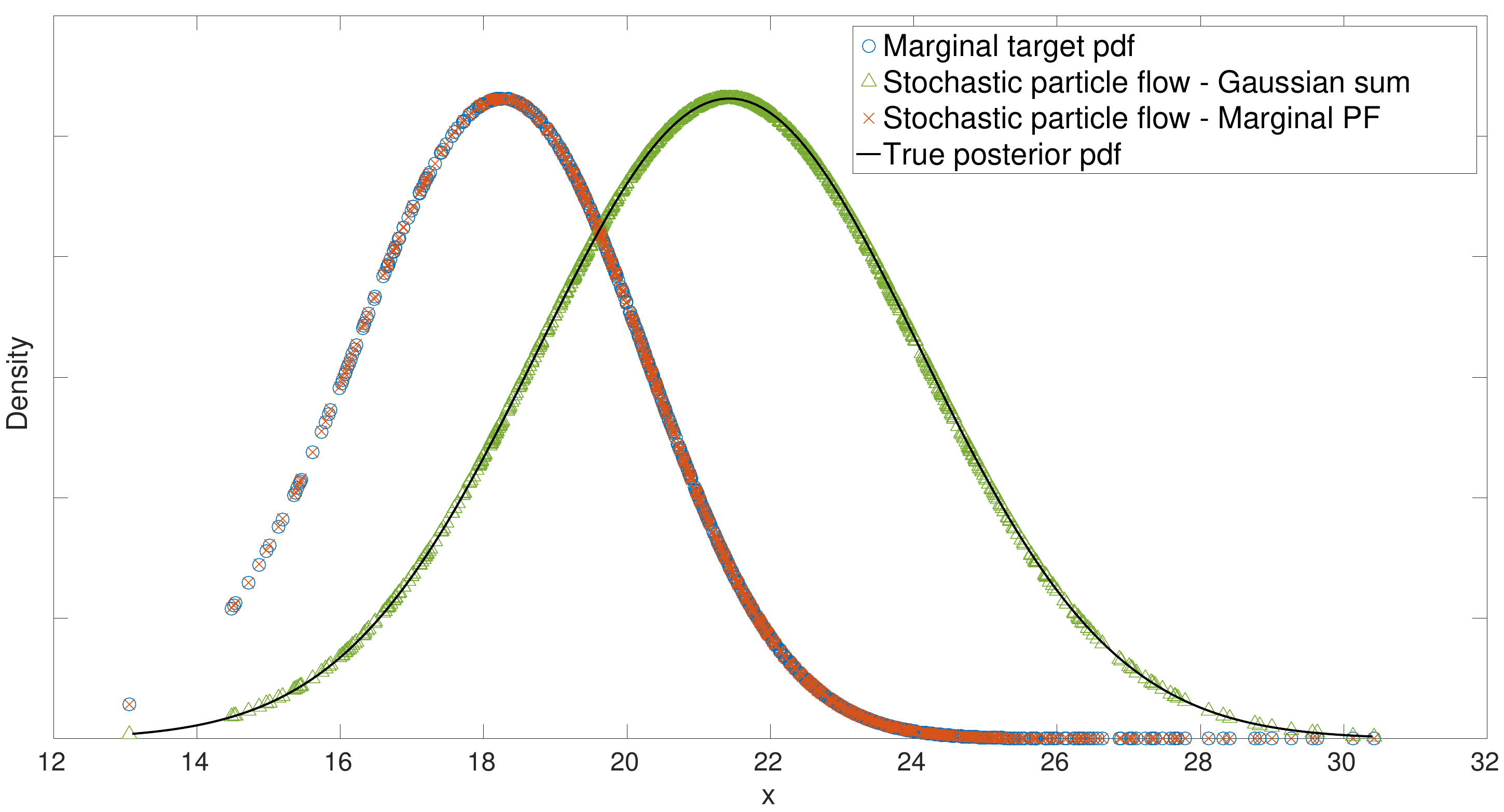}
\par\end{centering}
\caption{\label{fig:SPF-GS_versus_SPF-MPF}Densities generated by the SPF-GS
and SPF-MPF for the linear, univariate example}
\end{figure}

As depicted in Figure~\ref{fig:SPF-GS_versus_SPF-MPF}, the importance
density proposed by the SPF-MPF (red x's) is successful at aiming
the empirical marginal target (blue circles), generating a high effective
sample size. However, since the empirical target constitutes a poor
approximation to the true posterior pdf (black line), importance sampling
clearly fails and the SPF-MPF leads to a solution excessively biased.
In contrast, the direct filtering density generated by the SPF-GS
approximates the true posterior pdf accurately, generating a satisfactory
solution. These findings are quantified by the Jensen-Shannon divergences
averaged over 100 Monte Carlo runs and presented in Table~\ref{tab:Comparison-of-densities-for-the-univariate-examples}.
Table~\ref{tab:Comparison-of-densities-for-the-univariate-examples}
shows a neglible divergence between the density filtered by the SPF-GS
and the true posterior whereas the divergences computed for the target
density and for the proposal density constructed by the SPF-MPF are
significant.

\subsubsection{Quadratic, Univariate Model}

The quadratic, univariate model was tested with parameters set as
shown in the following table. This model is interesting because nonlinearity
of the observation process leads to bimodality of the filtered density.

\smallskip{}

\begin{center}
\begin{tabular}{|l|l|}
\hline 
\multicolumn{2}{|l|}{\emph{Parameters for the quadratic, univariate model}}\tabularnewline
\hline 
\hline 
Initial distribution & $\bar{\mathrm{x}}_{k-1}^{\vphantom{(i)}}=0$, $P_{k-1}=20$\tabularnewline
\hline 
Markov transition pdf & $f(\mathrm{x}_{k-1}^{\vphantom{(i)}})=\mathrm{x}_{k-1}^{\vphantom{(i)}}$,
$Q_{k}=20$\tabularnewline
\hline 
Likelihood function & $h(\mathrm{x}_{k}^{\vphantom{(i)}})=\mathrm{x}_{k}^{2}/20$, $R_{k}=50$\tabularnewline
\hline 
Observation & $\mathrm{y}_{k}=30$\tabularnewline
\hline 
\end{tabular}\smallskip{}
\par\end{center}

This nonlinear example was set to be favourable for marginal importance
sampling such that it would be possible to compare different marginal
particle filters against the SPF-MPF. The original particle flows,
GPF and SDPF, are compared to the SPF-MPF as well. The quantified
performances for this quadratic univariate model are shown in Table~\ref{tab:Comparison-of-densities-for-the-univariate-examples}.

\begin{figure}
\includegraphics[width=0.49\textwidth]{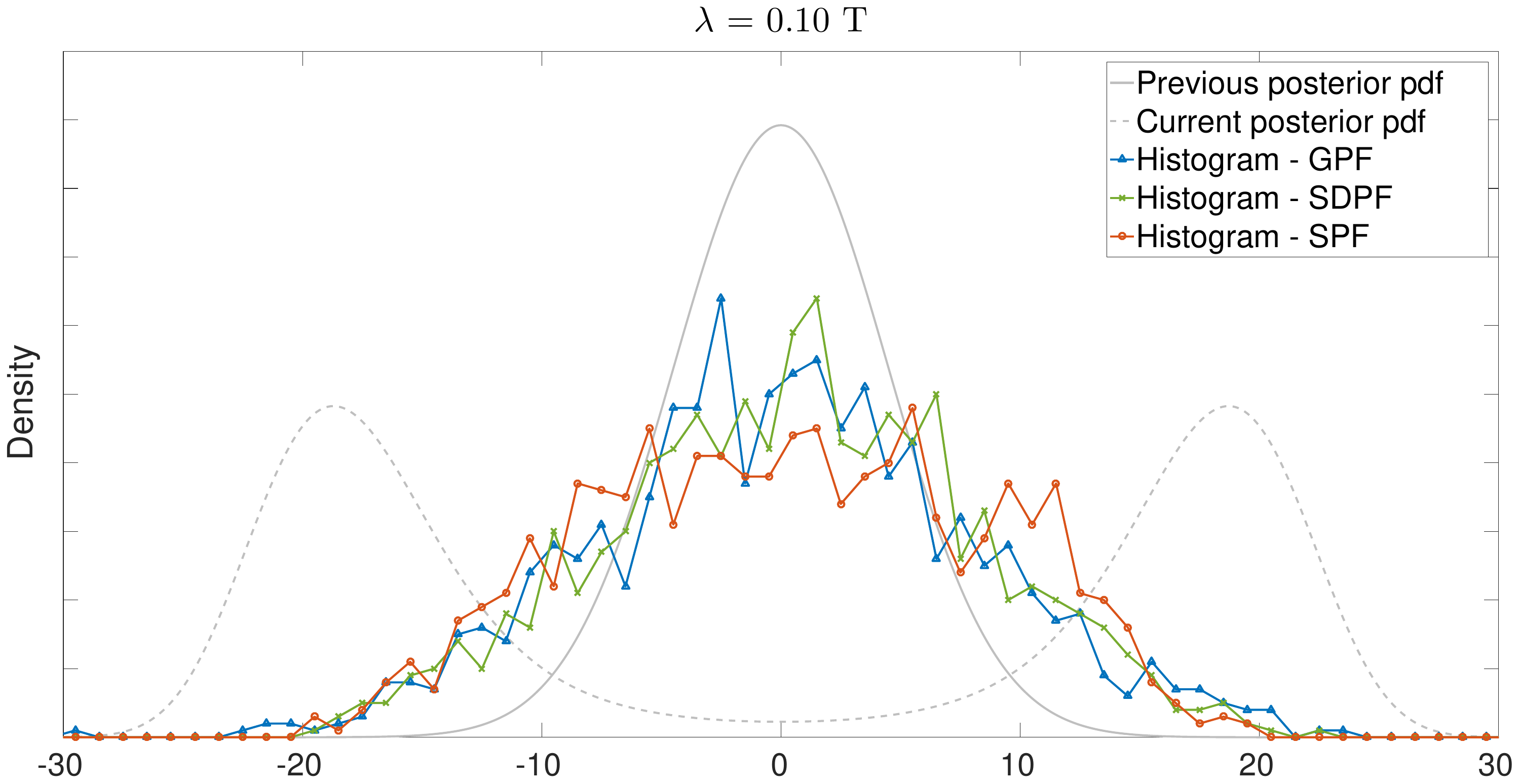}\hspace{0.02\textwidth}\includegraphics[width=0.49\textwidth]{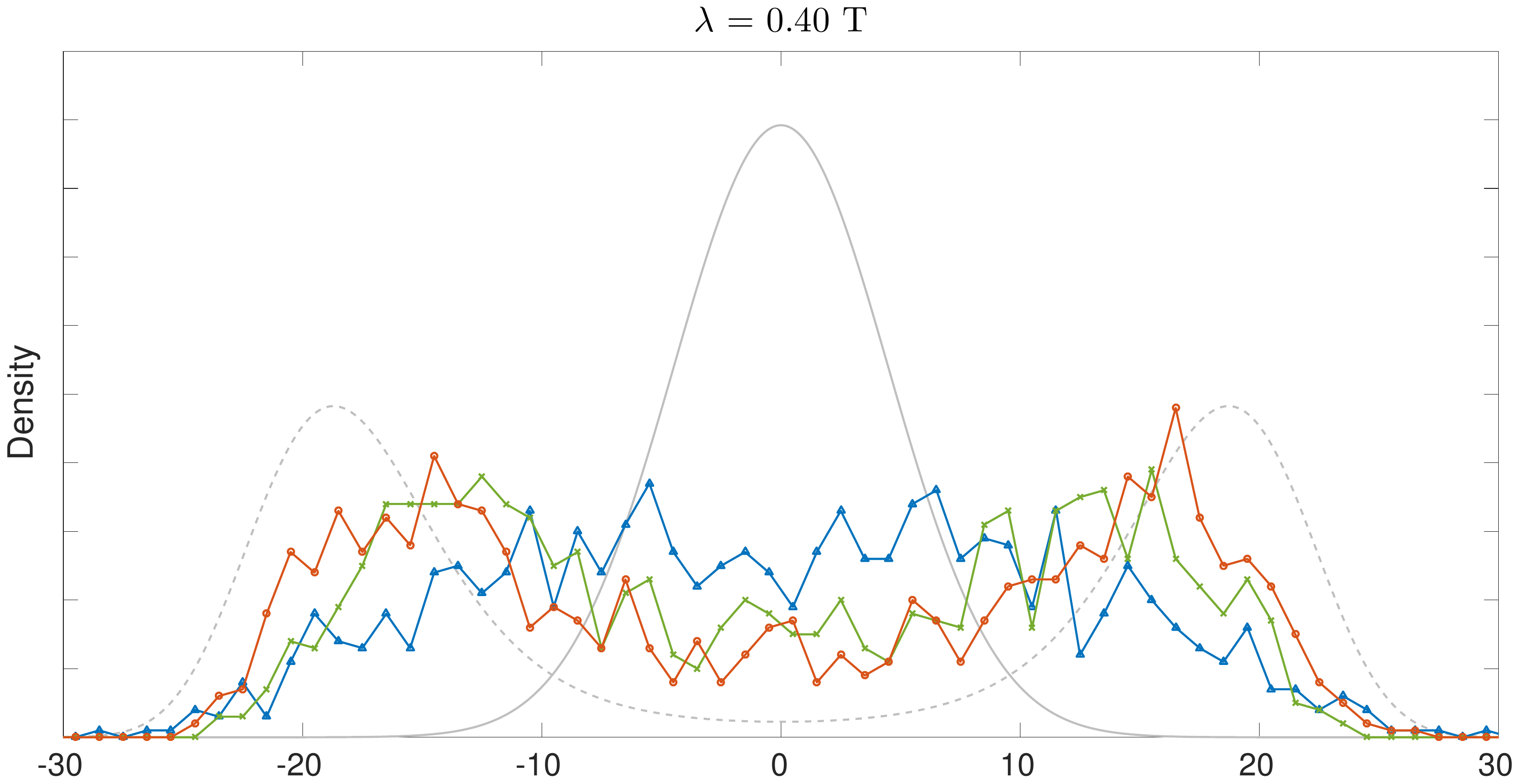}\medskip{}

\includegraphics[width=0.49\textwidth]{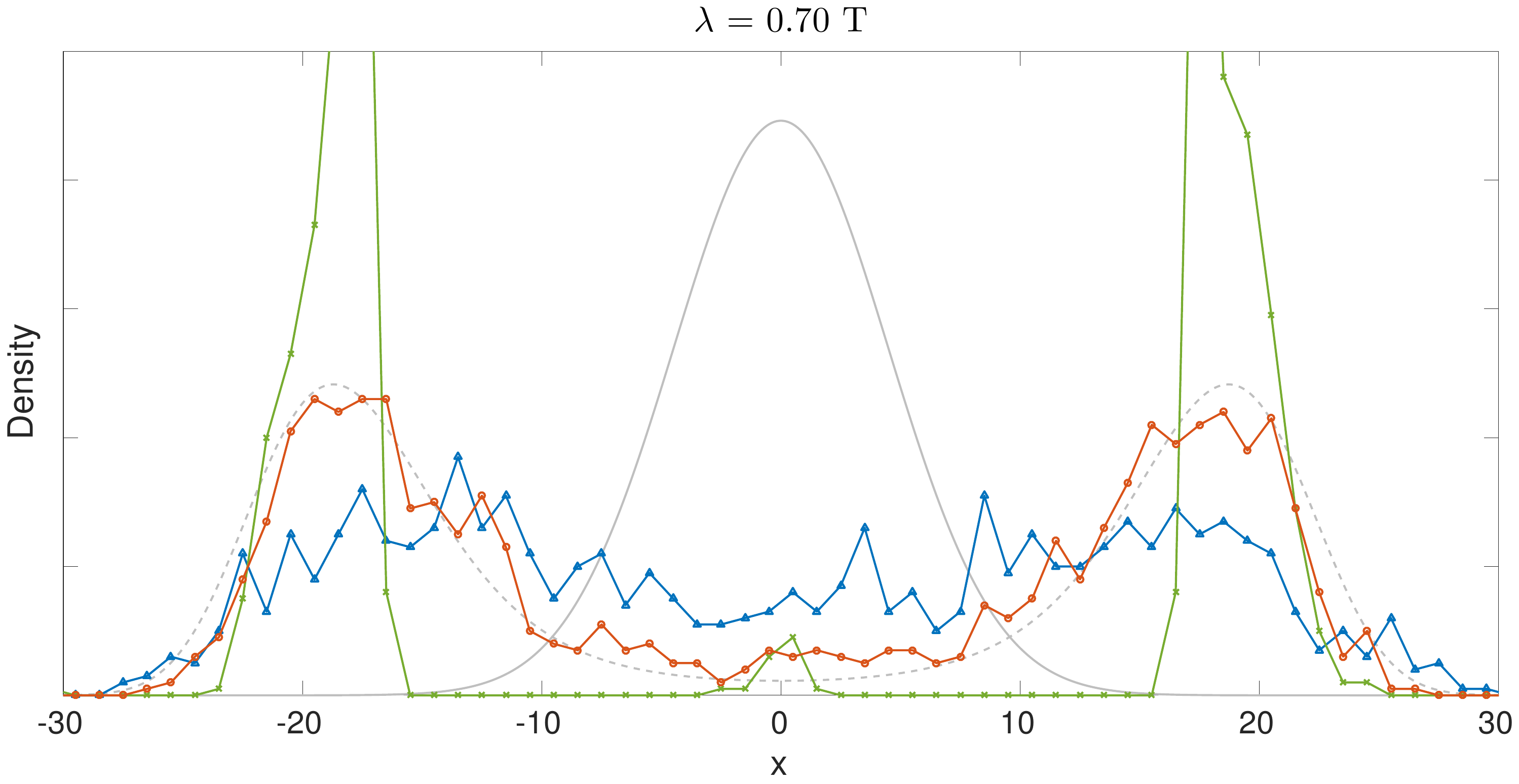}\hspace{0.02\textwidth}\includegraphics[width=0.49\textwidth]{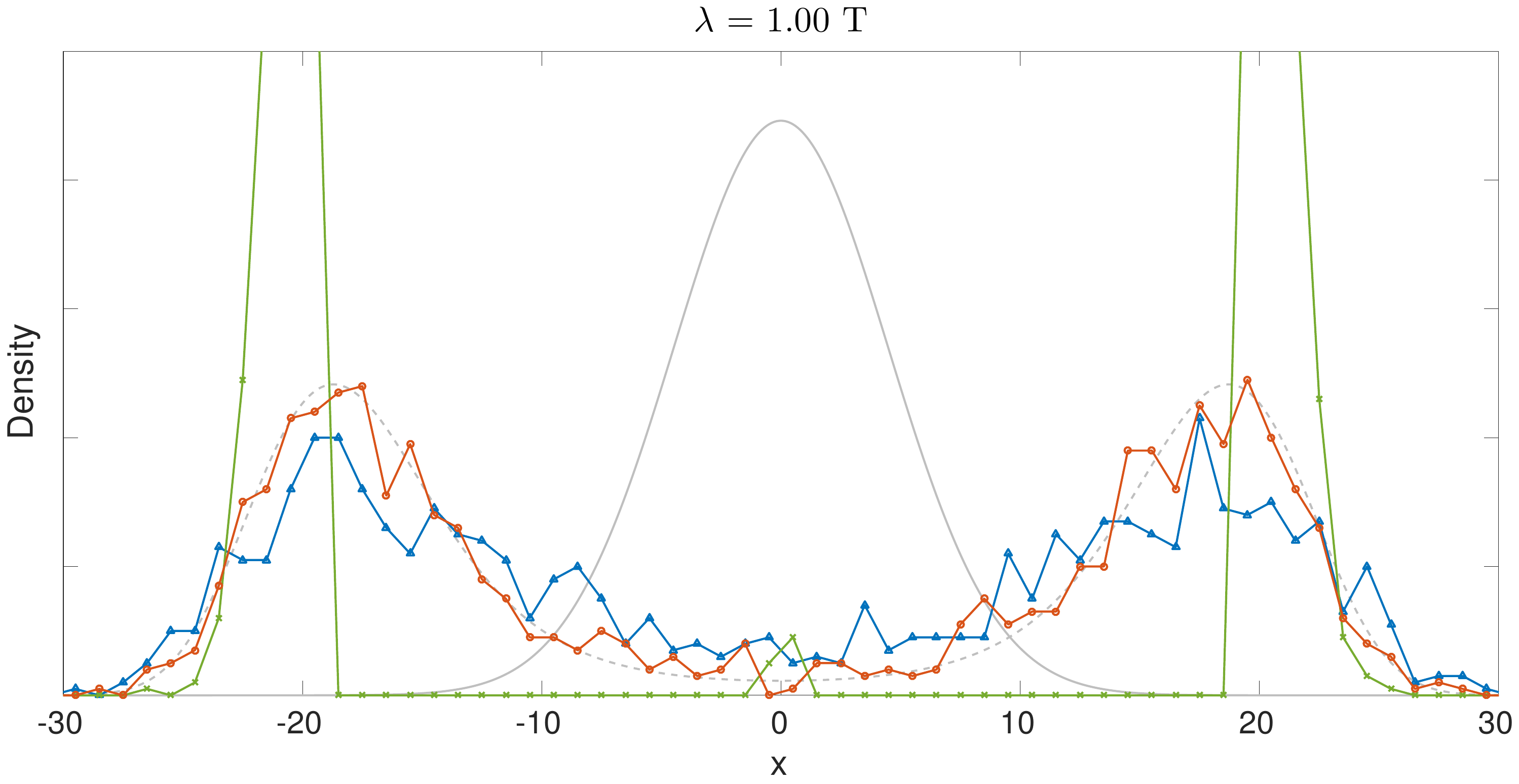}

\caption{Sequence of histograms achieved by propagating particles for the quadratic,
univariate example\label{fig:Sequence-of-histograms-1}}
\end{figure}

Firstly, we compare the sequence of histograms achieved when propagating
samples by the GPF, by the SDPF and by the SPF-GS. As it can be seen
in Figure~\ref{fig:Sequence-of-histograms-1}, for this example,
stochastic particle flow provides the best distribution of particles
to approximate the posterior density, denoting a higher level of accuracy
and regularity of the flow formulated as a diffusion.

\begin{figure}[H]
\begin{centering}
\includegraphics[width=1\textwidth]{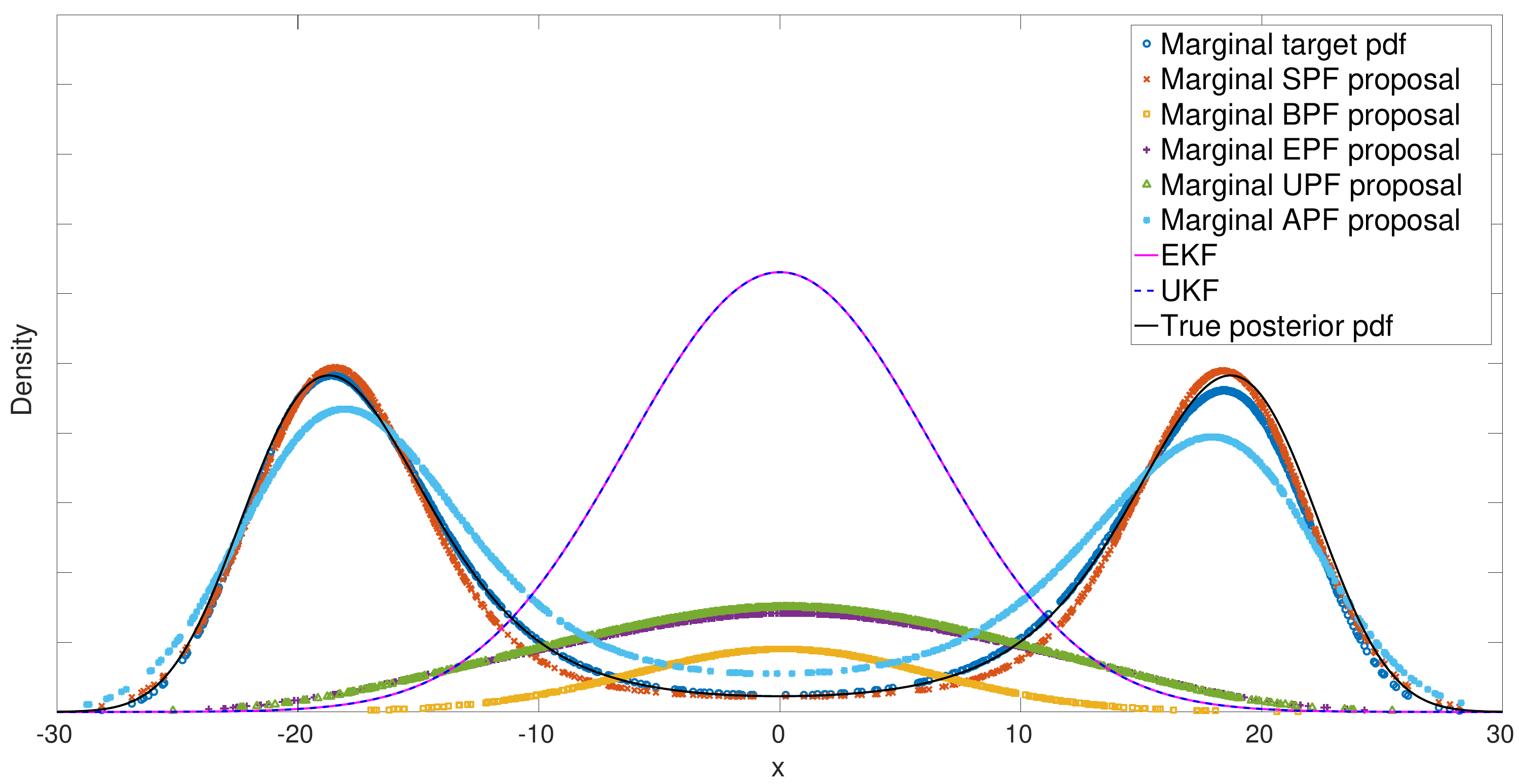}
\par\end{centering}
\caption{\label{fig:Comparison-of-proposals-1}Comparison of proposal densities
for the quadratic, univariate example}

\end{figure}

\begin{figure}[H]
\begin{centering}
\includegraphics[width=1\textwidth]{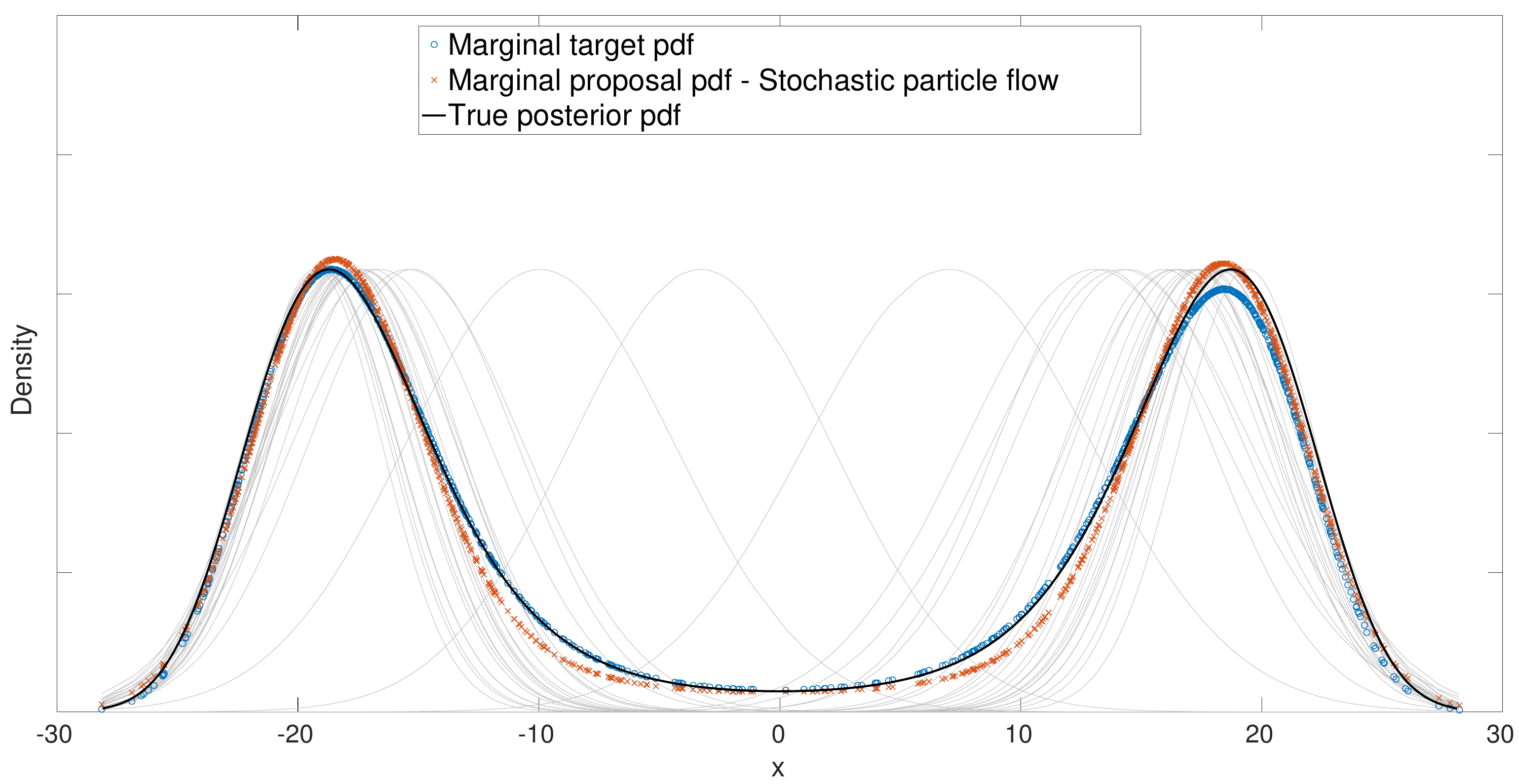}
\par\end{centering}
\caption{\label{fig:Marginal-SPF-density-1}Marginal proposal density based
on the stochastic particle flow}
\end{figure}

Regarding the marginal importance densities illustrated in Figure~\ref{fig:Comparison-of-proposals-1},
we observe a high degree of similarity of the SPF-MPF proposal density
to the marginal target density. In the same manner, the filtering
density achieved by the SPF-GS accurately approximates the true posterior
density, as evidenced in Table~\ref{tab:Comparison-of-densities-for-the-univariate-examples}.
In Figure~\ref{fig:Marginal-SPF-density-1} we can see in detail
the proximity of the SPF-MPF proposal density to both the marginal
target density and the true posterior density, along with some of
the proposal mixture components. The density proposed by the marginal
(optimal) auxiliary particle filter (MAPF) is also very similar to
the marginal target, providing an accurate solution, whereas all other
filters propose densities less effective for this example. These observations
are quantitatively captured by the performance data summarized in
Table~\ref{tab:Comparison-of-densities-for-the-univariate-examples}.

\subsubsection{Cubic, Univariate Model}

The cubic, univariate model was tested with parameters set as shown
in the following table.

\smallskip{}

\begin{center}
\begin{tabular}{|l|l|}
\hline 
\multicolumn{2}{|l|}{\emph{Parameters for the cubic, univariate model}}\tabularnewline
\hline 
\hline 
Initial distribution & $\bar{\mathrm{x}}_{k-1}^{\vphantom{(i)}}=0$, $P_{k-1}=20$\tabularnewline
\hline 
Markov transition pdf & $f(\mathrm{x}_{k-1}^{\vphantom{(i)}})=\mathrm{x}_{k-1}^{\vphantom{(i)}}$,
$Q_{k}=20$\tabularnewline
\hline 
Likelihood function & $h(\mathrm{x}_{k}^{\vphantom{(i)}})=\mathrm{x}_{k}^{3}/120$, $R_{k}=50$\tabularnewline
\hline 
Observation & $\mathrm{y}_{k}=20$\tabularnewline
\hline 
\end{tabular}\smallskip{}
\par\end{center}

\begin{figure}[H]
\begin{centering}
\includegraphics[width=1\textwidth]{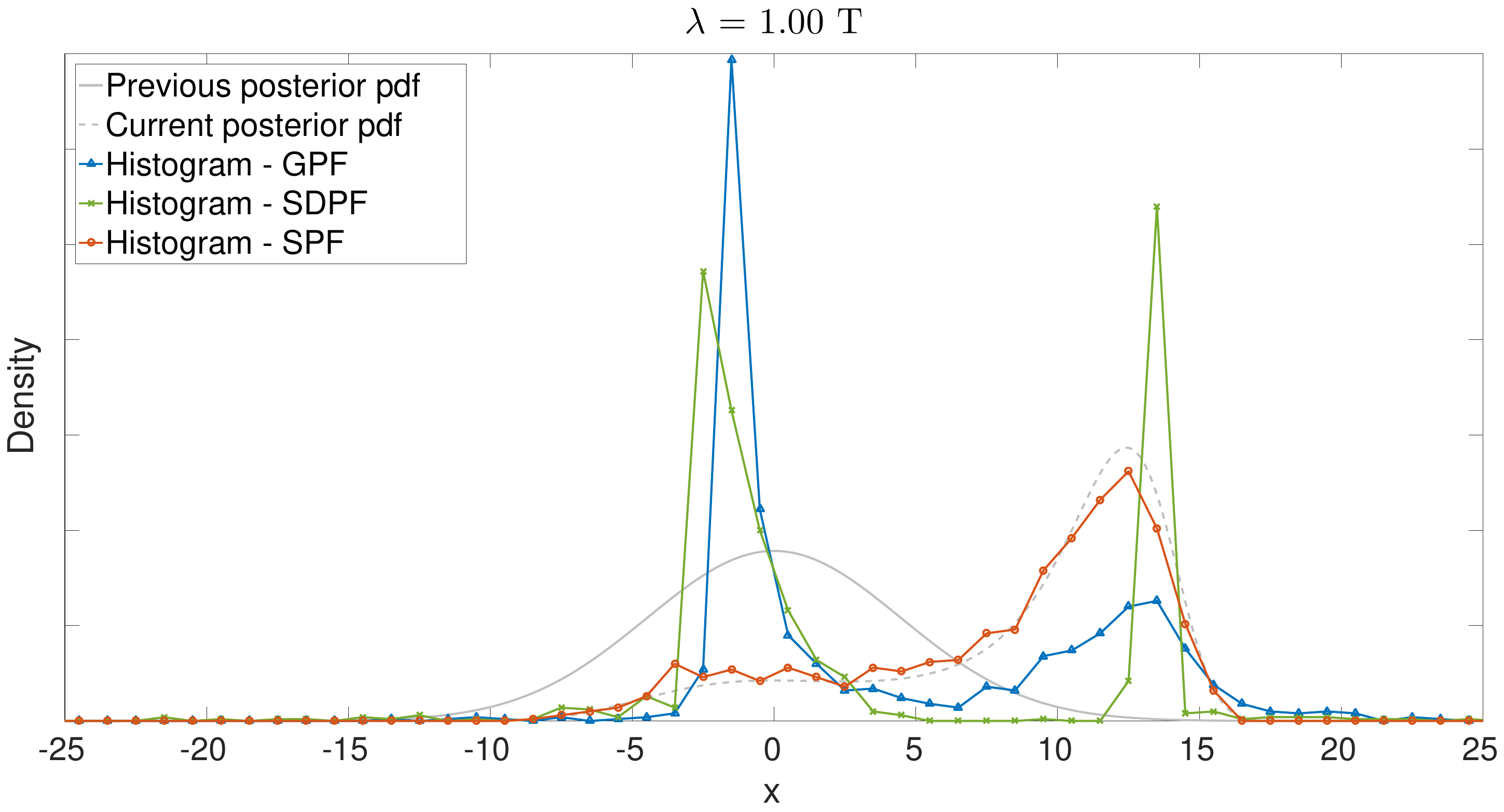}
\par\end{centering}
\caption{\label{fig:Histogram-2}Resulting histograms of particles for the
cubic, univariate example}
\end{figure}

\begin{figure}[H]
\begin{centering}
\includegraphics[width=1\textwidth]{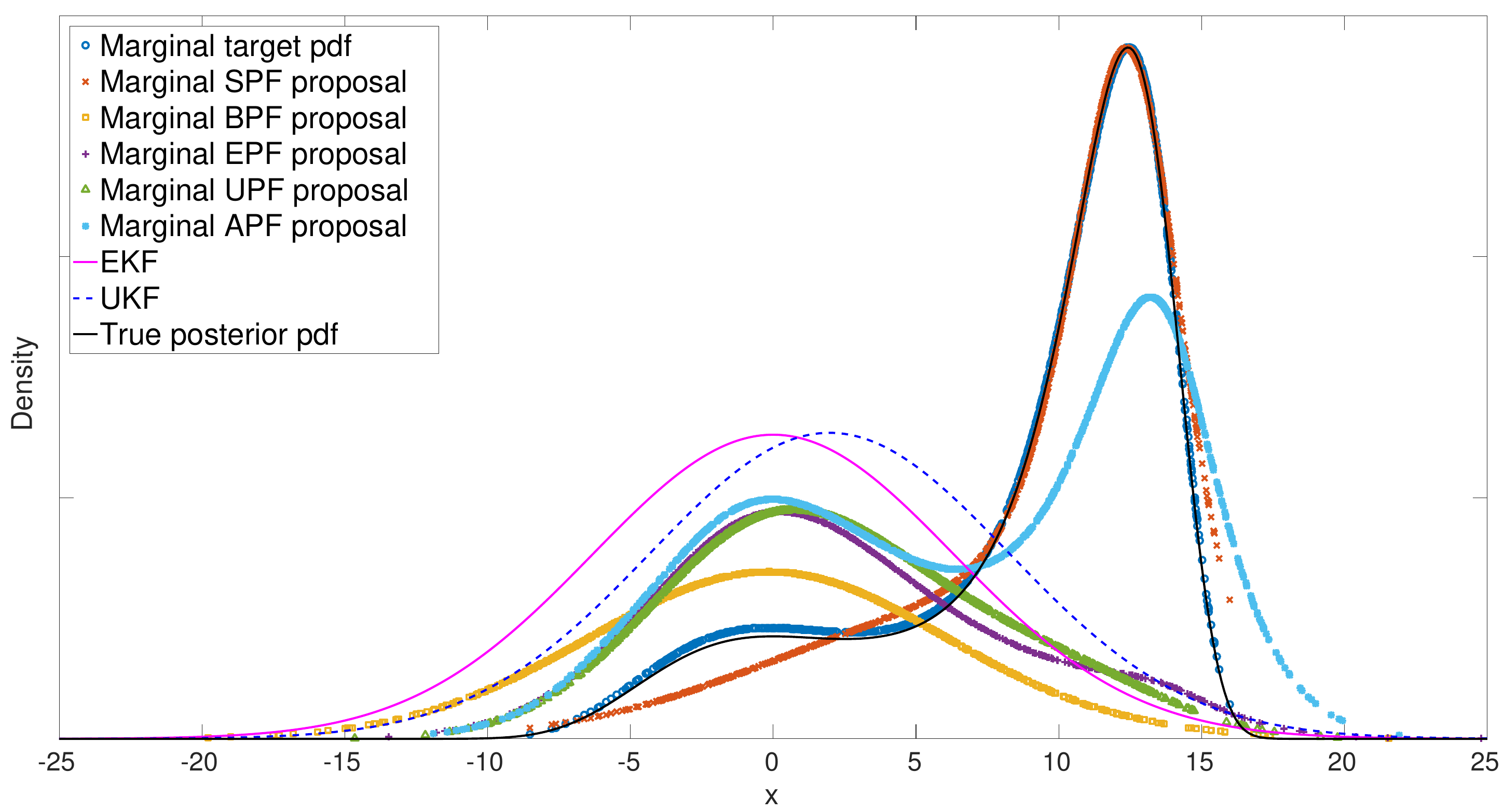}
\par\end{centering}
\caption{\label{fig:Comparison-of-proposals-2}Comparison of proposal densities
for the cubic, univariate example}
\end{figure}

This nonlinear example was also set to be favourable for marginal
importance sampling, i.e., avoiding the scenario described in the
first toy example where importance sampling fails. By comparing the
resulting histograms achieved when propagating samples by the GPF,
by the SDPF and by the stochastic particle flow, it is remarkable
in Figure~\ref{fig:Histogram-2} that the stochastic particle flow
provides a fairly superior distribution of particles to approximate
the posterior density. This superiority is incorporated in the importance
density proposed by the \mbox{SPF-MPF} as can be seen in Figure~\ref{fig:Comparison-of-proposals-2}.
The importance density proposed by the marginal auxiliary particle
filter (MAPF) also provides an accurate solution to the filtering
problem, but it is slightly less effective than the SPF-MPF. All the
other marginal particle filters present less effective solutions.
The comparison of all filters for this example is quantified in Table~\ref{tab:Comparison-of-densities-for-the-univariate-examples}.
\begin{table}
\begin{centering}
\begin{tabular}{|l|r@{\extracolsep{0pt}.}l|r@{\extracolsep{0pt}.}l|r@{\extracolsep{0pt}.}l|r@{\extracolsep{0pt}.}l|r@{\extracolsep{0pt}.}l|r@{\extracolsep{0pt}.}l|}
\hline 
\multirow{2}{*}{Density} & \multicolumn{4}{c|}{\emph{Linear}} & \multicolumn{4}{c|}{\emph{Quadratic}} & \multicolumn{4}{c|}{\emph{Cubic}}\tabularnewline
\cline{2-13} 
 & \multicolumn{2}{c|}{$\text{JSD}_{\text{avg}}$} & \multicolumn{2}{c|}{$\text{ESS}_{\text{avg}}$} & \multicolumn{2}{c|}{$\text{JSD}_{\text{avg}}$} & \multicolumn{2}{c|}{$\text{ESS}_{\text{avg}}$} & \multicolumn{2}{c|}{$\text{JSD}_{\text{avg}}$} & \multicolumn{2}{c|}{$\text{ESS}_{\text{avg}}$}\tabularnewline
\hline 
\hline 
\textcolor{gray}{\emph{Marginal target}} & \textcolor{gray}{0}&\textcolor{gray}{1572} & \multicolumn{2}{c|}{\textcolor{gray}{-}} & \textcolor{gray}{0}&\textcolor{gray}{0028} & \multicolumn{2}{c|}{\textcolor{gray}{-}} & \textcolor{gray}{0}&\textcolor{gray}{0001} & \multicolumn{2}{c|}{\textcolor{gray}{-}}\tabularnewline
\hline 
\emph{SPF-GS} & \textcolor{blue}{0}&\textcolor{blue}{0000} & \multicolumn{2}{c|}{-} & \textcolor{blue}{0}&\textcolor{blue}{0013} & \multicolumn{2}{c|}{-} & 0&0165 & \multicolumn{2}{c|}{-}\tabularnewline
\hline 
\emph{SPF-MPF} & 0&1574 & \textcolor{blue}{100}&\textcolor{blue}{00\%} & 0&0052 & \textcolor{blue}{97}&\textcolor{blue}{12\%} & \textcolor{blue}{0}&\textcolor{blue}{0071} & \textcolor{blue}{96}&\textcolor{blue}{74\%}\tabularnewline
\hline 
\emph{Marginal BPF} & 0&9876 & 0&21\% & 0&2641 & 1&79\% & 0&1723 & 12&60\%\tabularnewline
\hline 
\emph{Marginal EPF} & 0&7857 & 1&90\% & 0&3097 & 16&69\% & 0&1820 & 28&63\%\tabularnewline
\hline 
\emph{Marginal UPF} & 0&7870 & 2&04\% & 0&3112 & 14&46\% & 0&1674 & 25&32\%\tabularnewline
\hline 
\emph{Marginal APF} & 0&0670 & \textcolor{blue}{100}&\textcolor{blue}{00\%} & 0&0153 & 92&92\% & 0&0596 & 72&00\%\tabularnewline
\hline 
\end{tabular}\medskip{}
\par\end{centering}
\caption{Comparison of densities for the univariate examples\label{tab:Comparison-of-densities-for-the-univariate-examples}}

\end{table}

\subsubsection{Linear, Bimodal, Bivariate Model}

This example poses a bimodal model where the modes arise from two
different observations with a joint likelihood explicitly known. In
the algorithm for propagating particles, we implemented a scheme that
preselects samples to be filtered for either observation. This is
done according to a set of indexes that are sampled from a binomial
distribution $B\left(u_{1},u_{2};1,w_{l,(1)},w_{l,(2)}\right)\propto w_{l,(1)}^{u_{1}}w_{l,(2)}^{u_{2}}$
where $u_{1},u_{2}\in\left[0,1\right]$, $u_{1}+u_{2}=1$, such that
indexes are uniquely associated to either event $u_{1}$ or $u_{2}$,
with probability of either observation, $w_{l,(1)}$ or $w_{l,(2)}$
respectively. The linear, bimodal, bivariate model was tested with
parameters set as shown in Table~\ref{tab:Parameters-1}. These parameters
were chosen to result in quite distinct local properties of the two
modes.
\begin{table}[H]
\begin{centering}
\begin{tabular}{|l|l|}
\hline 
\multicolumn{2}{|l|}{\emph{Parameters for the linear, bimodal, bivariate model}}\tabularnewline
\hline 
\hline 
Initial distribution & $\bar{\mathrm{x}}_{k-1}^{\vphantom{(i)}}=\left(\begin{array}{c}
0\\
0
\end{array}\right)$, $P_{k-1}=\left(\begin{array}{cc}
9 & 0\\
0 & 9
\end{array}\right)_{\vphantom{2}}^{\vphantom{2}}$\tabularnewline
\hline 
Markov transition pdf & $f(\mathrm{x}_{k-1}^{\vphantom{(i)}})=\mathrm{x}_{k-1}^{\vphantom{(i)}}$,
$Q_{k}=\left(\begin{array}{cc}
16 & 0\\
0 & 16
\end{array}\right)_{\vphantom{}}^{\vphantom{2}}$\tabularnewline
\hline 
Likelihood function: & $h(\mathrm{x}_{k}^{\vphantom{(i)}})=\mathrm{x}_{k_{\vphantom{2}}}^{\vphantom{3}}$\tabularnewline
\cline{2-2} 
\emph{Mode 1} & $R_{k,(1)}=\left(\begin{array}{cc}
0.8 & 0\\
0 & 0.2
\end{array}\right)_{\vphantom{2}}^{\vphantom{2}}$, $w_{l,(1)}=0.2$\tabularnewline
\cline{2-2} 
\emph{Mode 2} & $R_{k,(2)}=\left(\begin{array}{cc}
4.0 & 0\\
0 & 1.0
\end{array}\right)_{\vphantom{2}}^{\vphantom{2}}$, $w_{l,(2)}=0.8$\tabularnewline
\hline 
Observations & $\mathrm{y}_{k,(1)}=\left(\begin{array}{c}
+10\\
+20
\end{array}\right)$, $\mathrm{y}_{k,(2)}=\left(\begin{array}{c}
+10\\
-20
\end{array}\right)_{\vphantom{2}}^{\vphantom{2}}$\tabularnewline
\hline 
\end{tabular}\smallskip{}
\par\end{centering}
\caption{\label{tab:Parameters-1}Parameters for the bimodal bivariate model}

\end{table}

\begin{figure}
\includegraphics[width=0.49\textwidth]{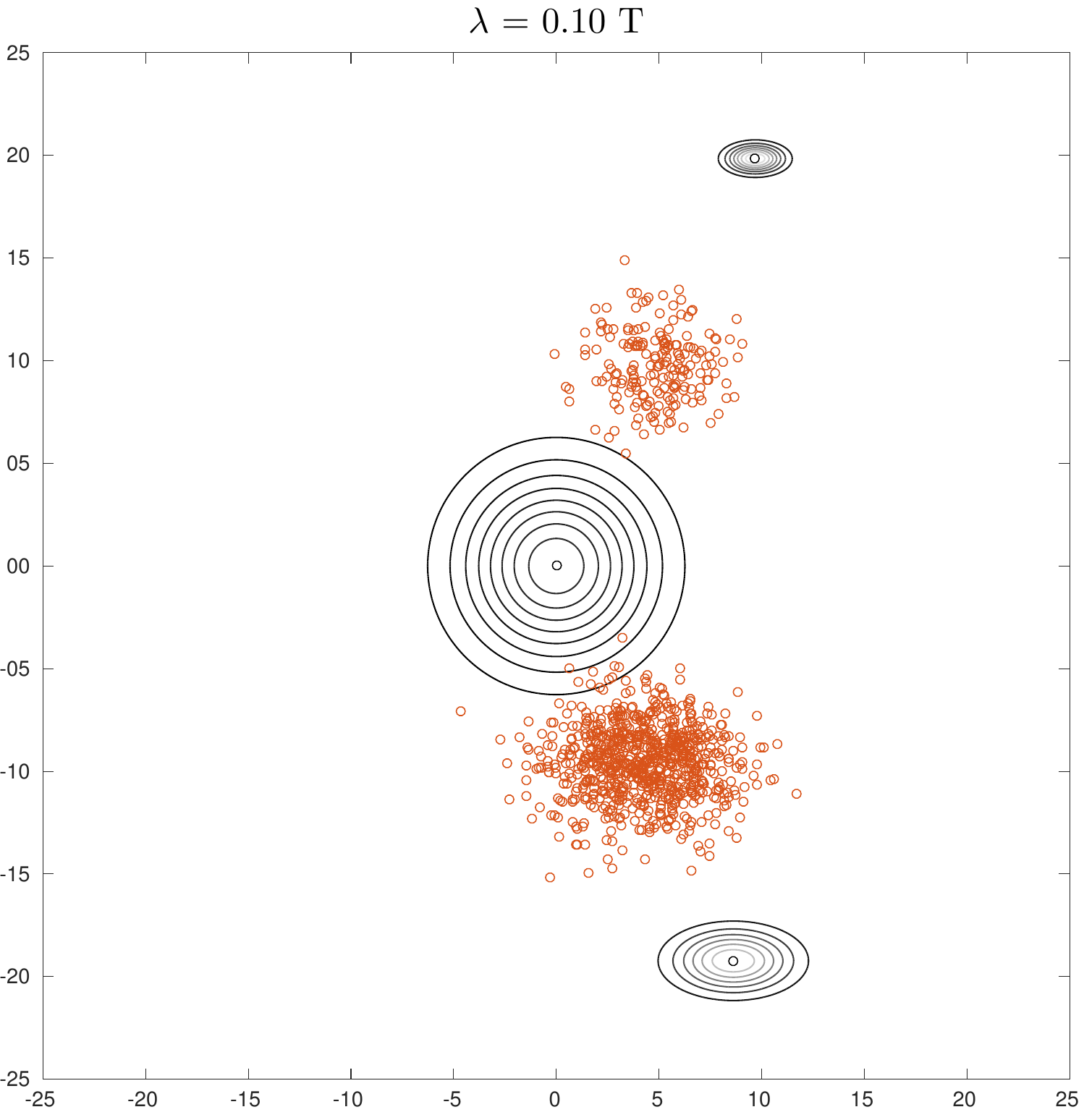}\hspace{0.02\textwidth}\includegraphics[width=0.49\textwidth]{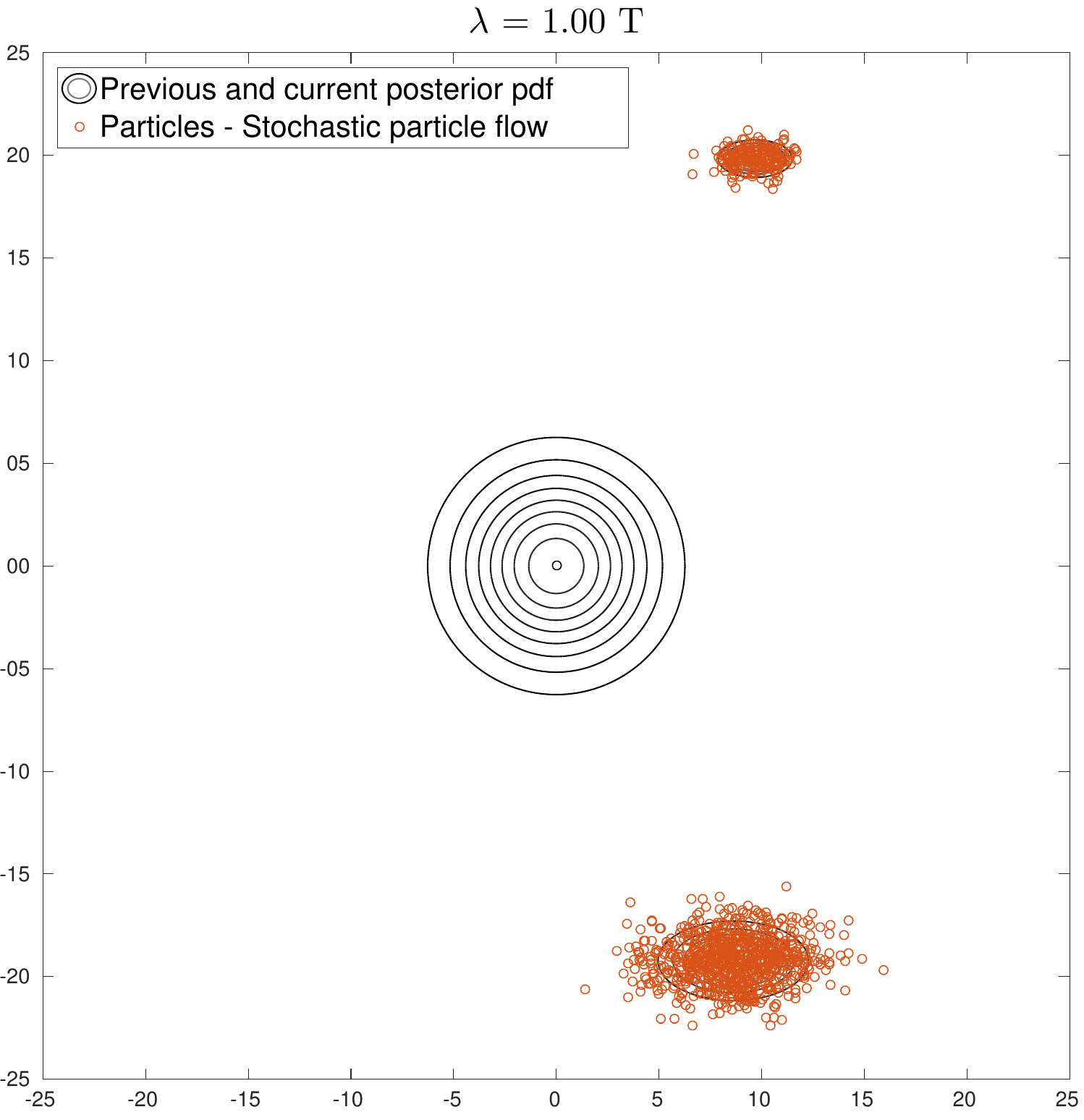}\medskip{}

\includegraphics[width=0.49\textwidth]{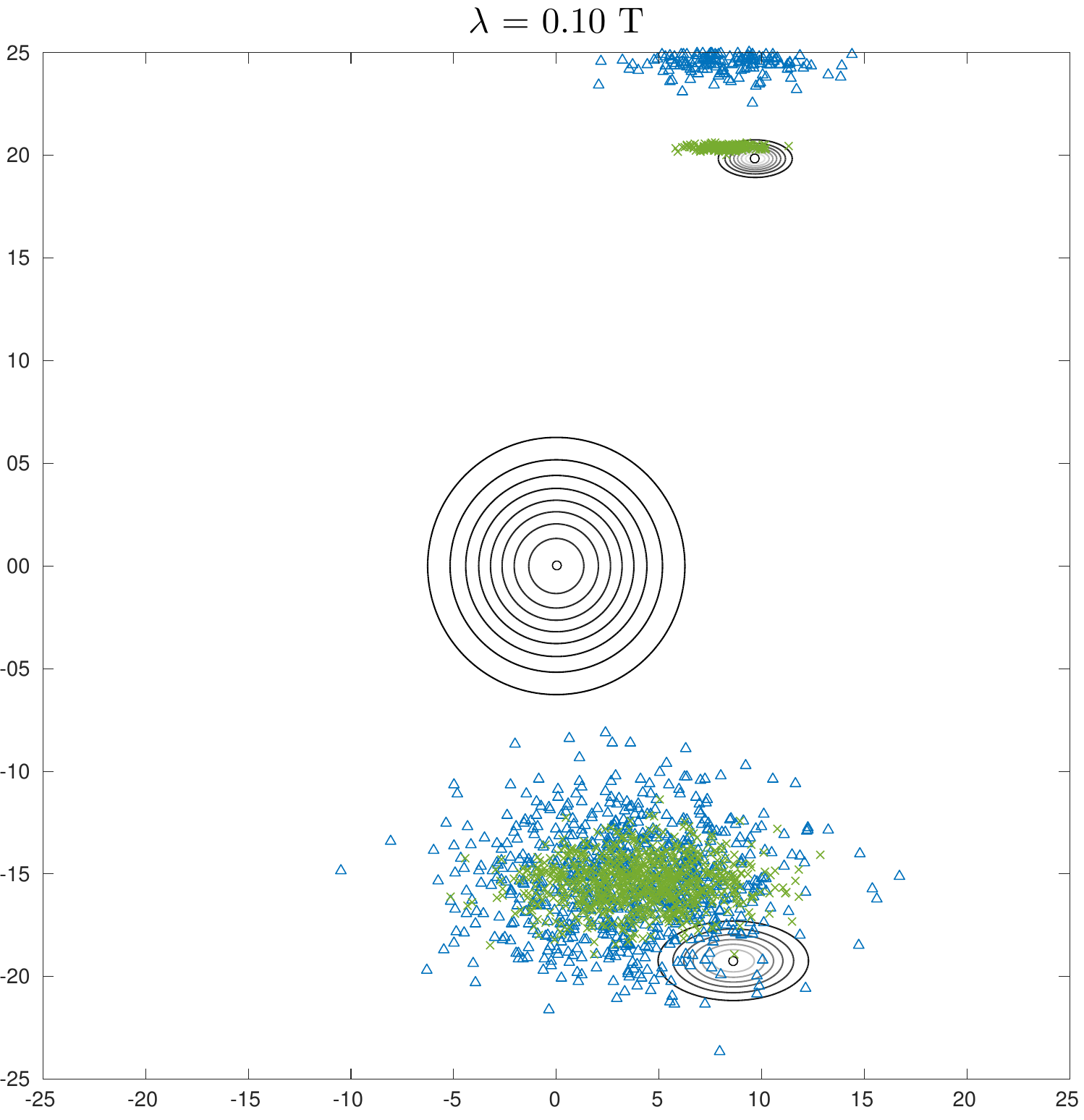}\hspace{0.02\textwidth}\includegraphics[width=0.49\textwidth]{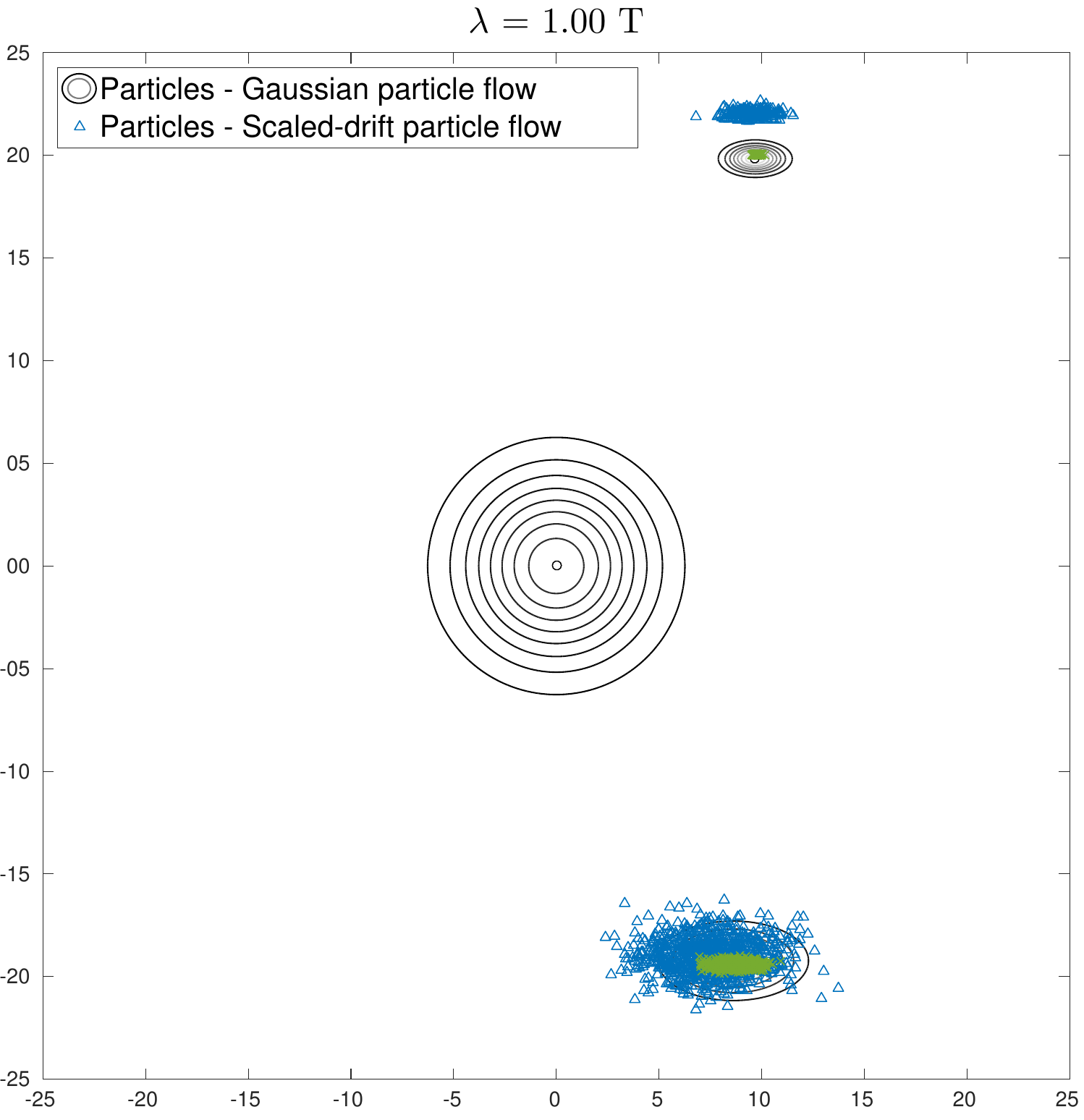}

\caption{Sequence of distributions achieved by propagating particles for the
bimodal, bivariate example\label{fig:Sequence-of-distributions3}}
\end{figure}

For this example, we analyze stochastic particle flow methods, \mbox{SPF-GS}
and \mbox{SPF-MPF}, against original particle flow methods only.
We exemplify the sequence of particles' distributions acquired by
the GPF, by the SDPF and by the stochastic particle flow in Figure~\ref{fig:Sequence-of-distributions3}.
It becomes clear that the final distribution generated by the stochastic
particle flow is closely similar to the true posterior density, precisely
describing the local moments of the two modes. In opposition, the
GPF generates a distribution that is excessively biased for the most
peaky mode whereas the SDPF generates a distribution that does not
describe correctly the covariances of each mode. 

These findings are quantified by the average Jensen-Shannon divergences
presented in Table~\ref{tab:Comparison-of-densities-for-the-bivariate-examples}.
Table~\ref{tab:Comparison-of-densities-for-the-bivariate-examples}
shows a small divergence of the density filtered by the SPF-GS with
respect to the true posterior, a small divergence of the SPF-MPF proposal
density as well as of the target density, whereas the divergences
of the original particle flows are fairly big. The SPF-MPF provides
a high effective sample size.

\subsubsection{Nonlinear, Unimodal, Bivariate Model}

The nonlinear bivariate model was tested in two cases: 
\begin{enumerate}
\item favourable for marginal particle filters, and 
\item unfavourable, i.e., emulating a scenario similar to that presented
in the first toy example where importance sampling fails. 
\end{enumerate}
The parameters used for cases\emph{~1} and \emph{2} are presented
in Table~\ref{tab:Parameters-2-1} and Table~\ref{tab:Parameters-2-2},
respectively. 

\begin{figure}
\includegraphics[width=0.49\textwidth]{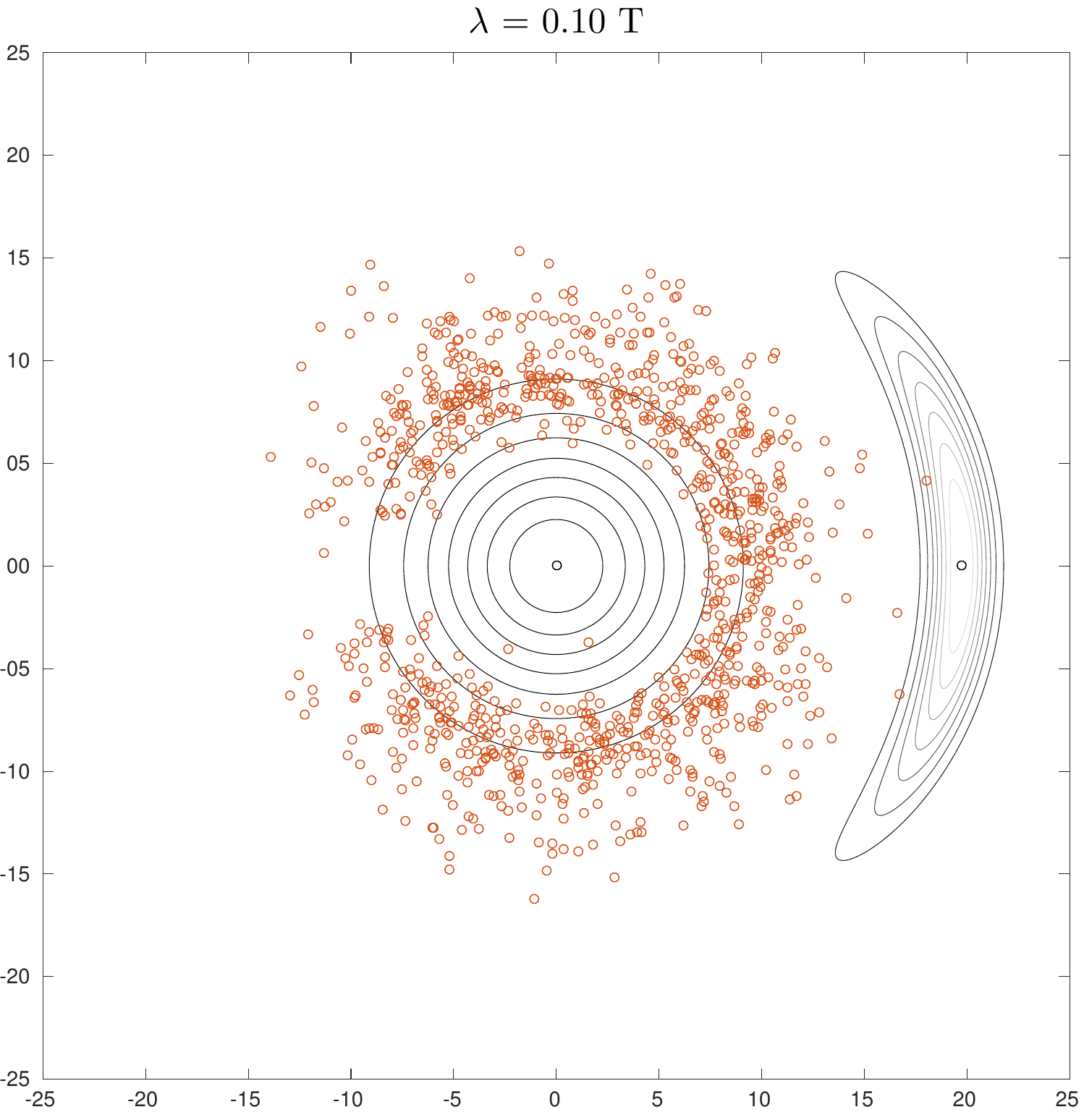}\hspace{0.02\textwidth}\includegraphics[width=0.49\textwidth]{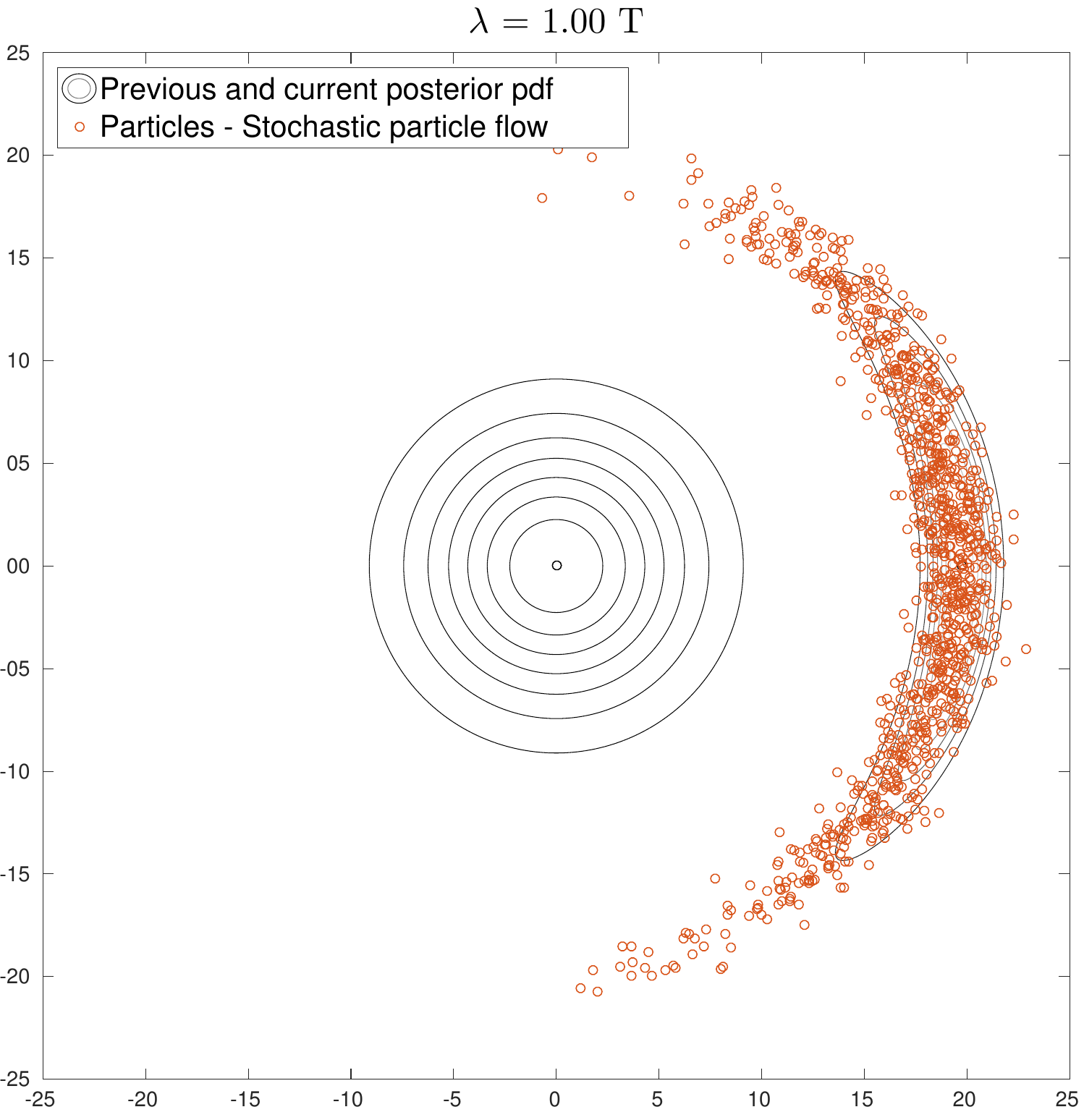}\medskip{}

\includegraphics[width=0.49\textwidth]{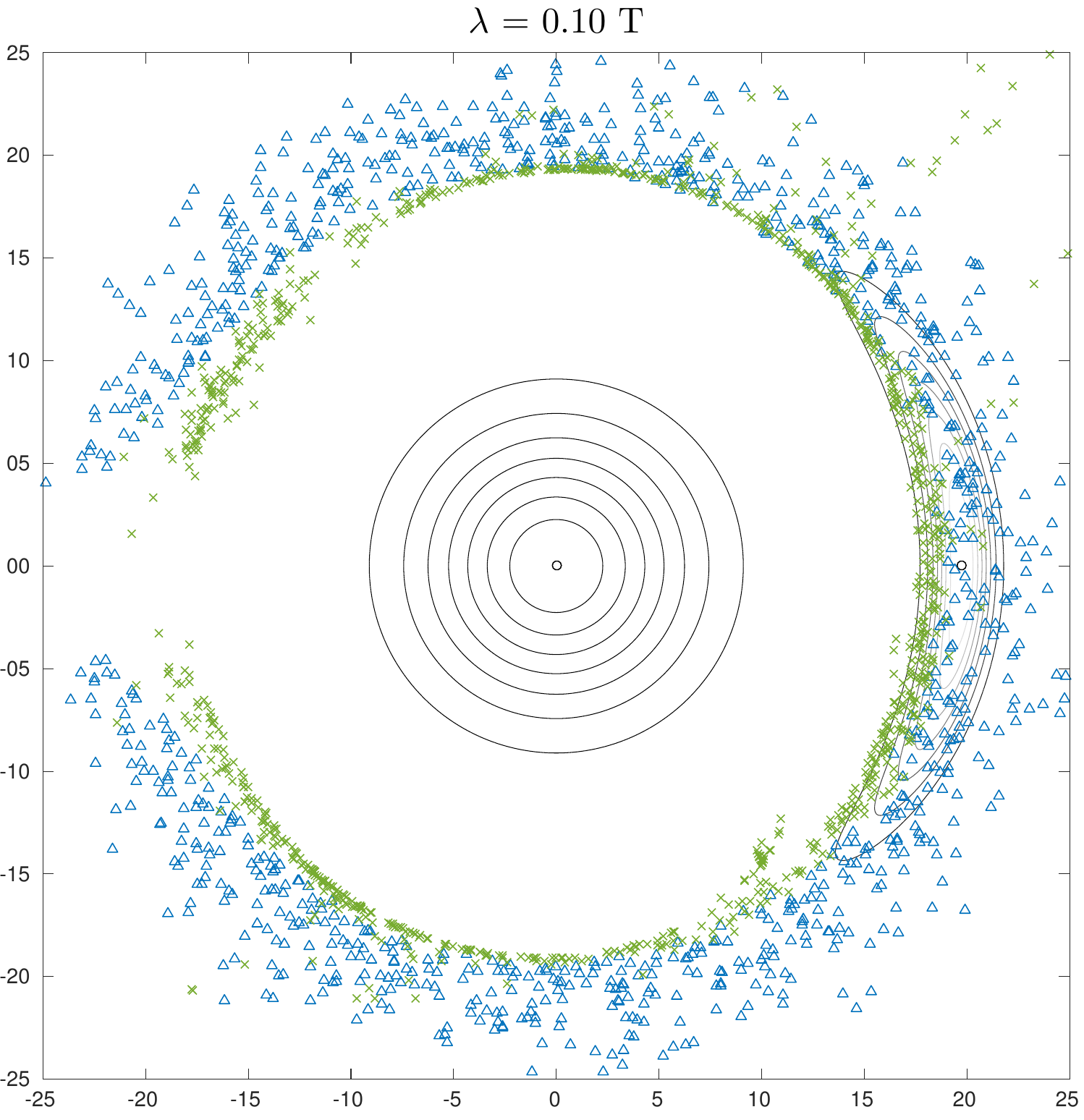}\hspace{0.02\textwidth}\includegraphics[width=0.49\textwidth]{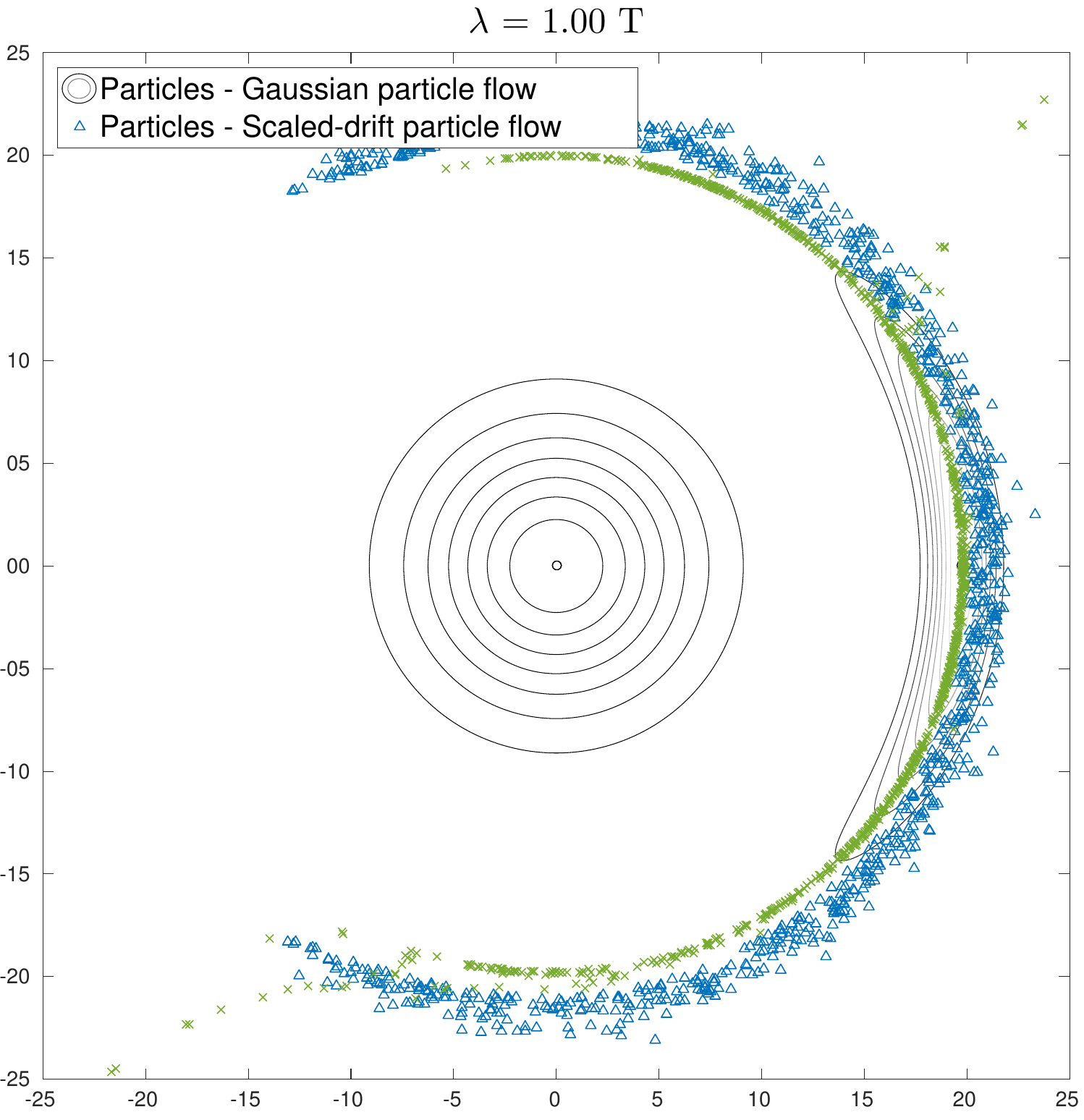}

\caption{Sequence of distributions achieved by propagating particles for the
nonlinear, bivariate example\label{fig:Sequence-of-distributions4}}
\end{figure}

In either cases the sequence of distributions generated by the original
particle flows and by the stochastic particle flow are as illustrated
in Figure~\ref{fig:Sequence-of-distributions4}. Once more it becomes
evident that the stochastic particle flow provides a superior distribution
of samples to approximate the posterior density, which demonstrates
its higher level of accuracy and regularity. Similarly to results
presented for previous examples, the GPF seems to generate substantially
biased distributions whereas the SDPF seems highly prone to regularity
problems. These aspects are well corroborated by the average Jensen-Shannon
divergences presented in Table~\ref{tab:Comparison-of-densities-for-the-bivariate-examples}.

\begin{figure}
\includegraphics[width=1\textwidth]{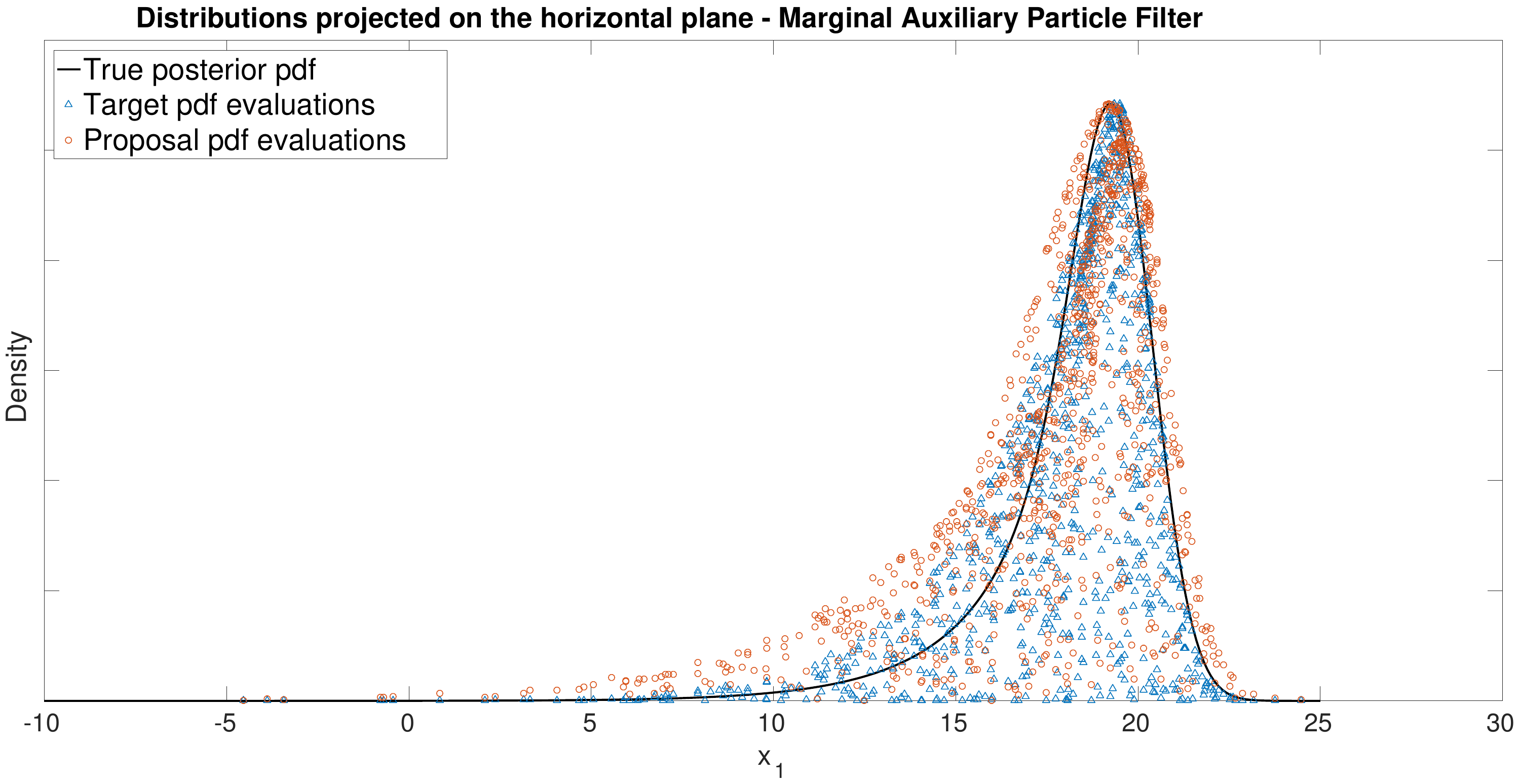}\medskip{}

\includegraphics[width=1\textwidth]{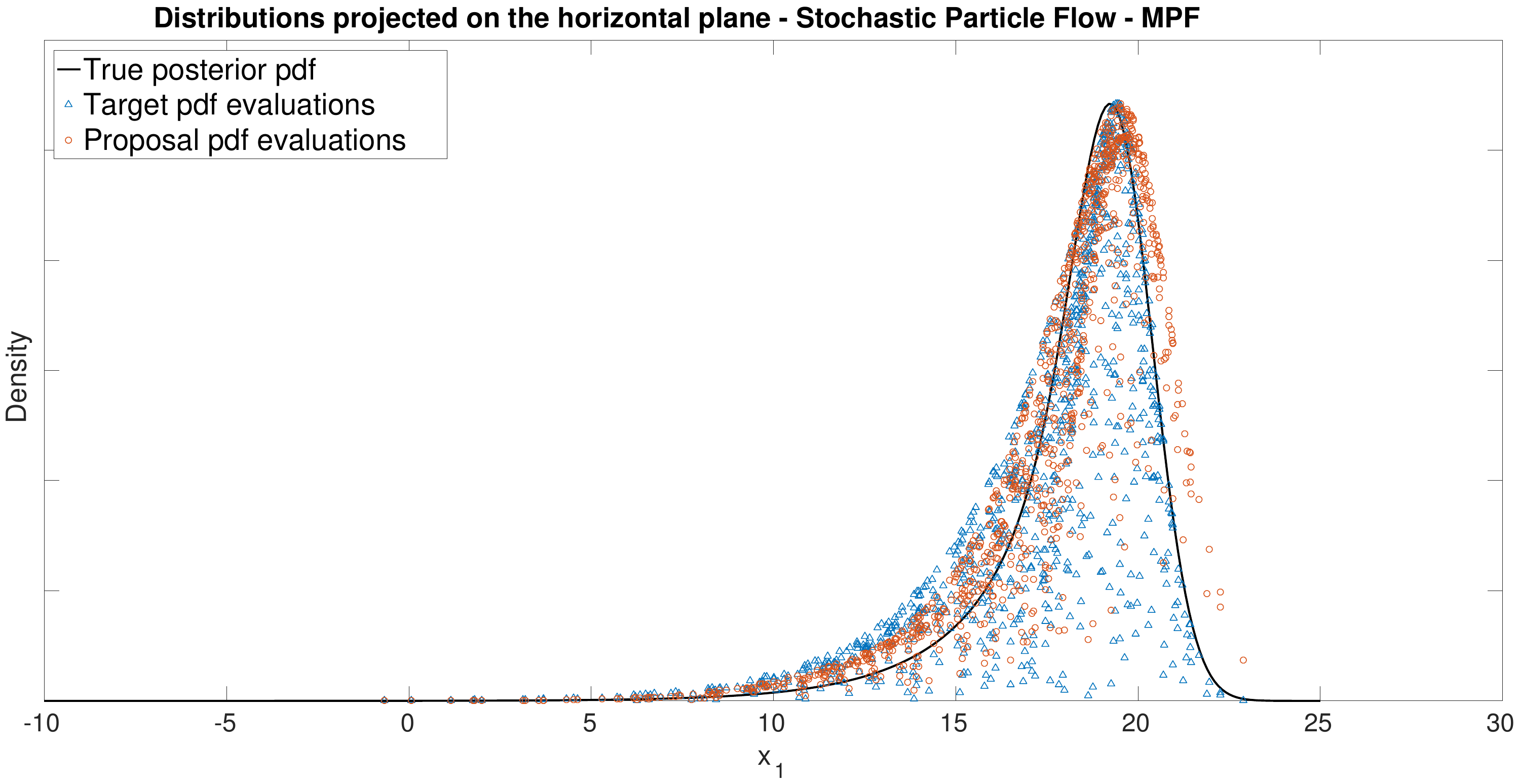}

\caption{Horizontal-plane projection of densities for the MAPF and SPF-MPF
(nonlinear, bivariate example)\label{fig:Projected-distributions4a}}
\end{figure}

\begin{figure}
\includegraphics[width=1\textwidth]{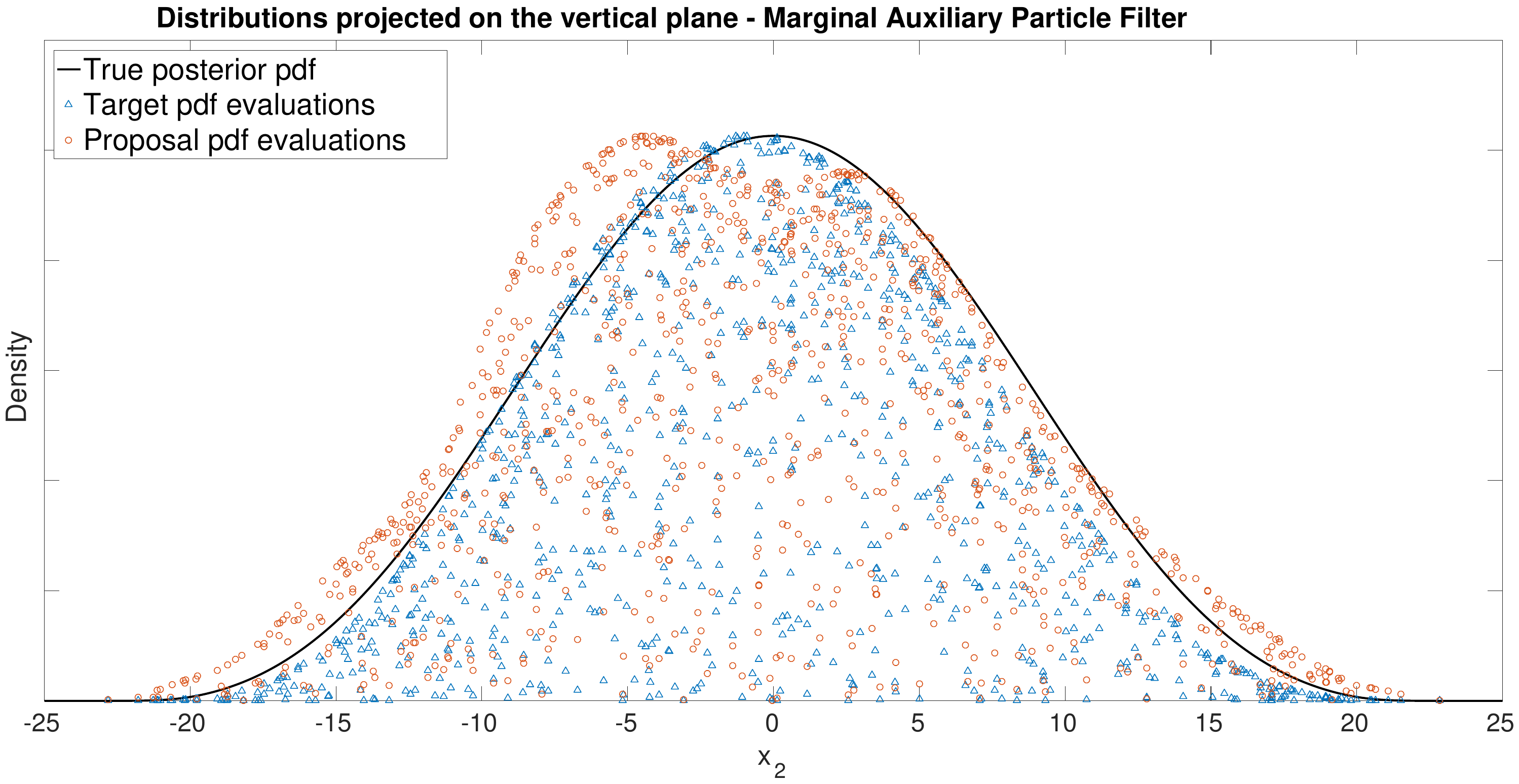}\medskip{}

\includegraphics[width=1\textwidth]{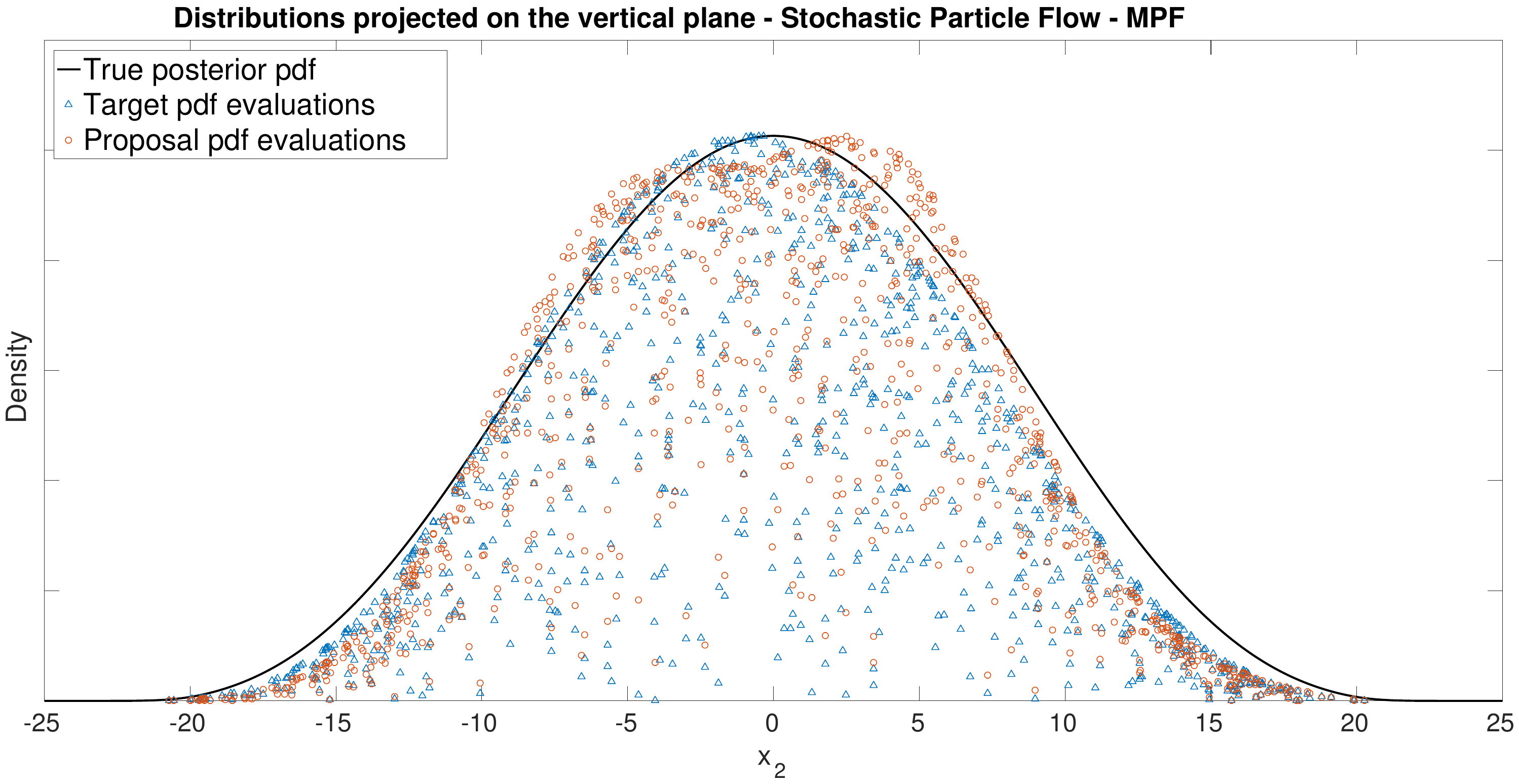}

\caption{Vertical-plane projection of densities for the MAPF and SPF-MPF (nonlinear,
bivariate example)\label{fig:Projected-distributions4b}}
\end{figure}

In the comparison we also included other marginal particle filters.
For \emph{case 1} (favourable), we illustrate in Figures~\ref{fig:Projected-distributions4a}
and \ref{fig:Projected-distributions4b} how the marginal importance
densities, projected (marginalized) onto the horizontal and vertical
planes, would look like as proposed by the marginal auxiliary particle
filter (MAPF) and by the SPF-MPF. It is clear that in this case both
MAPF and SPF-MPF generate proposal densities quite proximate of the
empirical marginal target, which in turn approximates well the true
posterior. Additionally, it is possible to visualize that the SPF-MPF
provides a slightly better proposal density in terms of similarity
to the target density, which is corroborated by a greater average
effective sample size as presented in Table~\ref{tab:Comparison-of-densities-for-the-bivariate-examples}.
All other marginal particle filters don't generate effective importance
densities in terms of approximating either the true posterior or the
target density.

\begin{figure}
\includegraphics[width=1\textwidth]{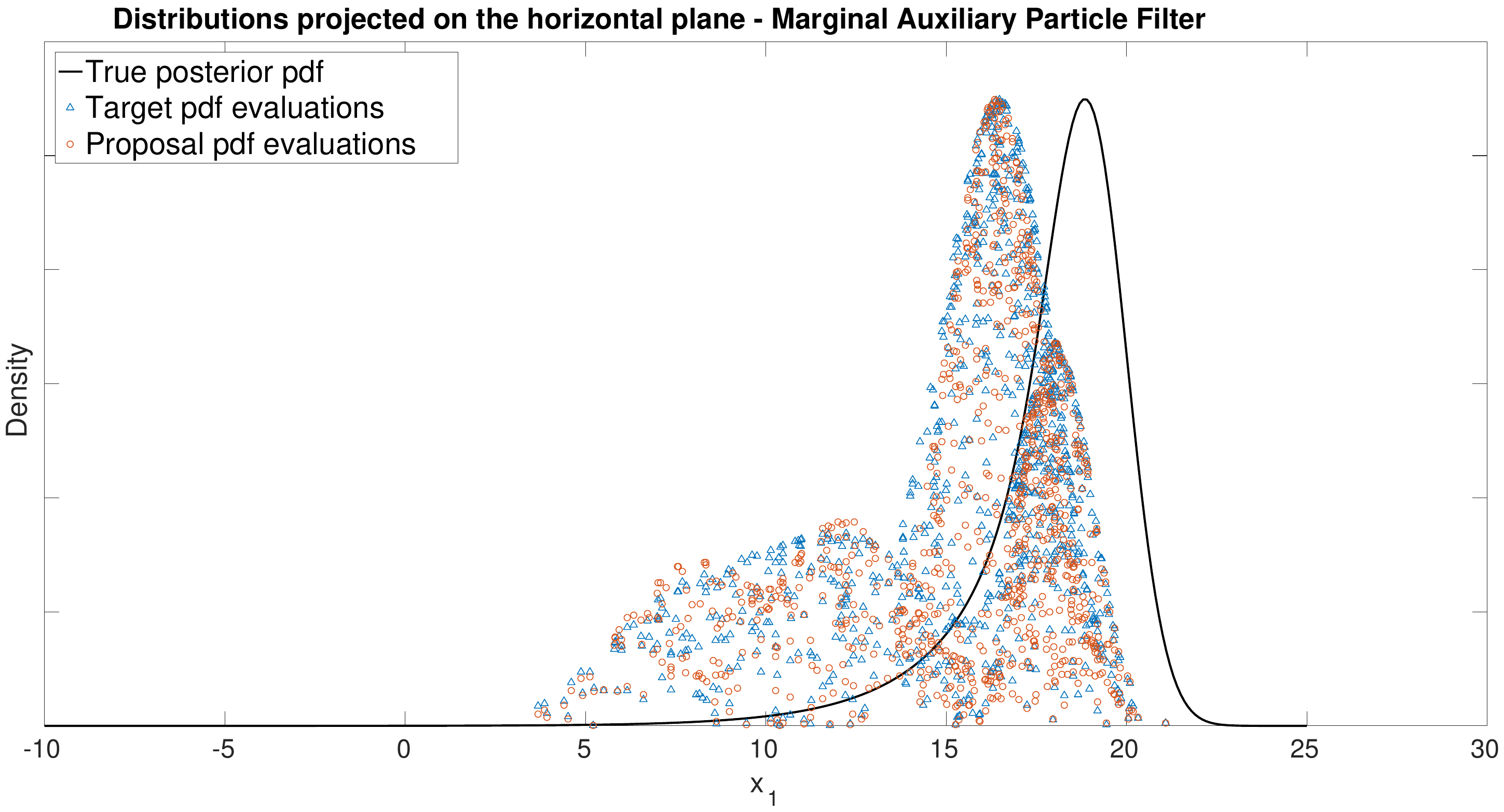}\medskip{}

\includegraphics[width=1\textwidth]{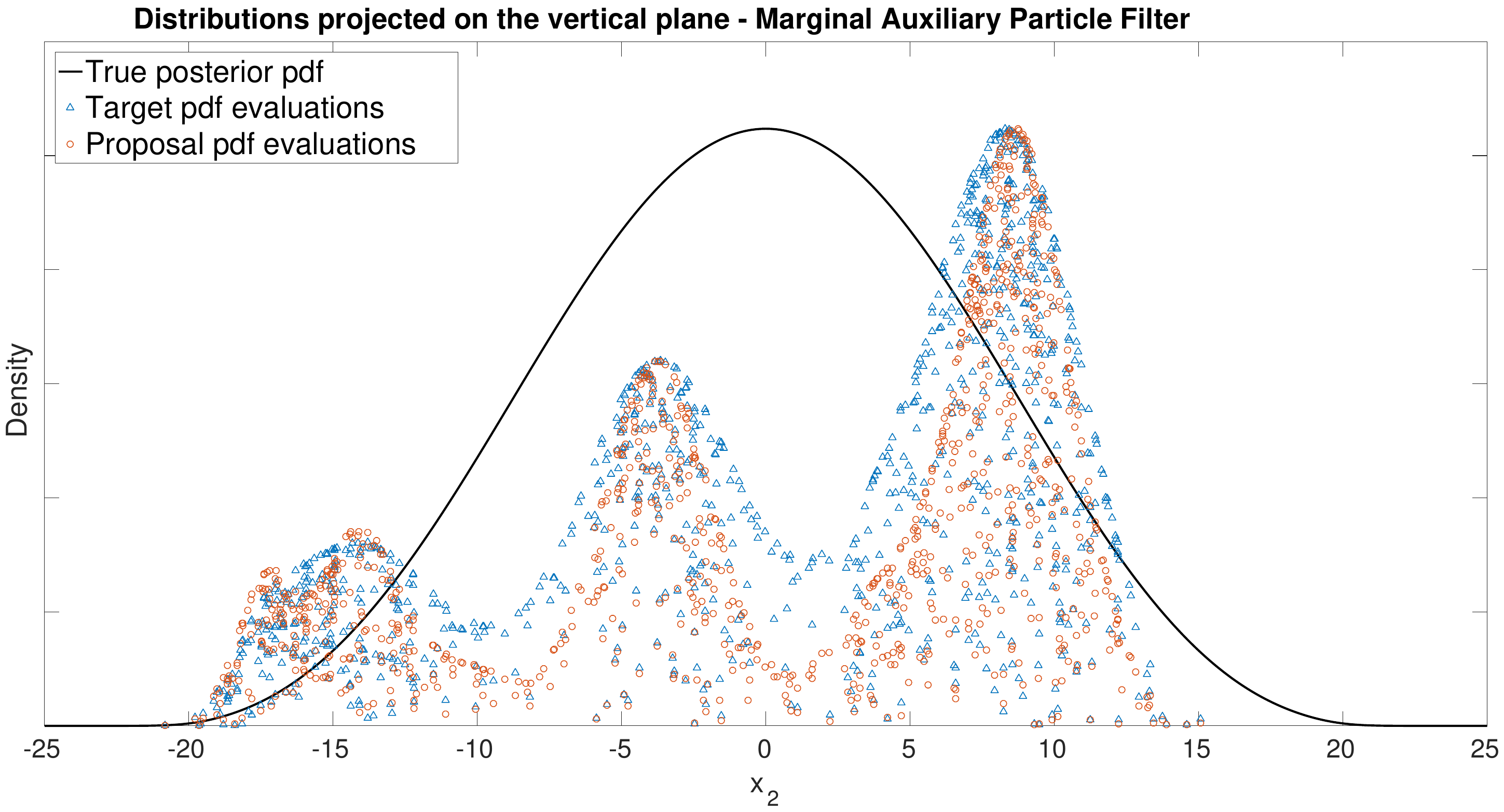}

\caption{Failure of marginal importance sampling for the nonlinear, bivariate
example\label{fig:Projected-distributions4c}}
\end{figure}

For \emph{case 2} (unfavourable), importance sampling fails as exemplified
by the projections of the importance density proposed by the MAPF
depicted in Figure~\ref{fig:Projected-distributions4c}. By the same
reason explained before, the importance sampling procedure fails to
provide a satisfactory filtering measure owing to the errors that
affect evaluations of both the marginal target density and the marginal
importance density. As a consequence, in this case, any marginal particle
filter generates a poor solution, although the MAPF provides a high
effective sample size. The SPF-MPF generates a remarkably poor solution
for \emph{case 2} because it distributes particles to approximate
the true posterior density by design, but must constrain the proposal
mixture components to match a very inaccurate empirical target density.

In contrast, in both cases\emph{ 1} and\emph{ 2} the SPF-GS proposes
a direct filtering density that accurately approximates the true posterior
density. The SPF-GS is demonstrated to be insensitive to the issues
caused by an observation located relatively far from the initial distribution.
These features are quantitatively captured by the performance indexes
summarized in Table~\ref{tab:Comparison-of-densities-for-the-bivariate-examples}.
\begin{table}
\begin{centering}
\begin{tabular}{|l|l|}
\hline 
\multicolumn{2}{|l|}{\emph{Parameters for the nonlinear bivariate model, case 1}}\tabularnewline
\hline 
\hline 
Initial distribution & $\bar{\mathrm{x}}_{k-1}^{\vphantom{(i)}}=\left(\begin{array}{c}
0\\
0
\end{array}\right)$, $P_{k-1}=\left(\begin{array}{cc}
20 & 0\\
0 & 20
\end{array}\right)_{\vphantom{2}}^{\vphantom{2}}$\tabularnewline
\hline 
Markov transition pdf & $f(\mathrm{x}_{k-1}^{\vphantom{(i)}})=\mathrm{x}_{k-1}^{\vphantom{(i)}}$,
$Q_{k}=\left(\begin{array}{cc}
20 & 0\\
0 & 20
\end{array}\right)_{\vphantom{}}^{\vphantom{2}}$\tabularnewline
\hline 
\multirow{2}{*}{Likelihood function} & $h(\mathrm{x}_{k}^{\vphantom{(i)}})=\left(\begin{array}{c}
\sqrt{\mathrm{x}_{k}\left(1\right)^{2}+\mathrm{x}_{k}\left(2\right)^{2}}\\
\text{atan}\left(\mathrm{x}_{k}\left(2\right)^{\vphantom{2}}/\mathrm{x}_{k}\left(1\right)^{\vphantom{2}}\right)
\end{array}\right)_{\vphantom{2}}^{\vphantom{2}}$, \tabularnewline
 & $R_{k}=\left(\begin{array}{cc}
1.00 & 0\\
0 & 0.16
\end{array}\right)_{\vphantom{2}}^{\vphantom{2}}$\tabularnewline
\hline 
Observation & $\mathrm{y}_{k}=\left(\begin{array}{c}
20\\
0^{\circ}
\end{array}\right)_{\vphantom{2}}^{\vphantom{2}}$\tabularnewline
\hline 
\end{tabular}\smallskip{}
\par\end{centering}
\caption{\label{tab:Parameters-2-1}Parameters for the nonlinear bivariate
model, case 1}
\end{table}
\begin{table}
\begin{centering}
\begin{tabular}{|l|l|}
\hline 
\multicolumn{2}{|l|}{\emph{Parameters for the nonlinear bivariate model, case 2}}\tabularnewline
\hline 
\hline 
Initial distribution & $\bar{\mathrm{x}}_{k-1}^{\vphantom{(i)}}=\left(\begin{array}{c}
0\\
0
\end{array}\right)$, $P_{k-1}=\left(\begin{array}{cc}
10 & 0\\
0 & 10
\end{array}\right)_{\vphantom{2}}^{\vphantom{2}}$\tabularnewline
\hline 
Markov transition pdf & $f(\mathrm{x}_{k-1}^{\vphantom{(i)}})=\mathrm{x}_{k-1}^{\vphantom{(i)}}$,
$Q_{k}=\left(\begin{array}{cc}
5 & 0\\
0 & 5
\end{array}\right)_{\vphantom{}}^{\vphantom{2}}$\tabularnewline
\hline 
\multirow{2}{*}{Likelihood function} & $h(\mathrm{x}_{k}^{\vphantom{(i)}})=\left(\begin{array}{c}
\sqrt{\mathrm{x}_{k}\left(1\right)^{2}+\mathrm{x}_{k}\left(2\right)^{2}}\\
\text{atan}\left(\mathrm{x}_{k}\left(2\right)^{\vphantom{2}}/\mathrm{x}_{k}\left(1\right)^{\vphantom{2}}\right)
\end{array}\right)_{\vphantom{2}}^{\vphantom{2}}$, \tabularnewline
 & $R_{k}=\left(\begin{array}{cc}
1.00 & 0\\
0 & 0.16
\end{array}\right)_{\vphantom{2}}^{\vphantom{2}}$\tabularnewline
\hline 
Observation & $\mathrm{y}_{k}=\left(\begin{array}{c}
20\\
0^{\circ}
\end{array}\right)_{\vphantom{2}}^{\vphantom{2}}$\tabularnewline
\hline 
\end{tabular}\smallskip{}
\par\end{centering}
\caption{\label{tab:Parameters-2-2}Parameters for the nonlinear bivariate
model, case 2}
\end{table}
\begin{table}
\begin{centering}
\begin{tabular}{|l|r@{\extracolsep{0pt}.}l|r@{\extracolsep{0pt}.}l|r@{\extracolsep{0pt}.}l|r@{\extracolsep{0pt}.}l|r@{\extracolsep{0pt}.}l|r@{\extracolsep{0pt}.}l|}
\hline 
\multirow{2}{*}{Density} & \multicolumn{4}{c|}{\emph{Multimodal, linear}} & \multicolumn{4}{c|}{\emph{Nonlinear - case 1}} & \multicolumn{4}{r|}{\emph{Nonlinear - case 2}}\tabularnewline
\cline{2-13} 
 & \multicolumn{2}{c|}{$\text{JSD}_{\text{avg}}$} & \multicolumn{2}{c|}{$\text{ESS}_{\text{avg}}$} & \multicolumn{2}{c|}{$\text{JSD}_{\text{avg}}$} & \multicolumn{2}{c|}{$\text{ESS}_{\text{avg}}$} & \multicolumn{2}{c|}{$\text{JSD}_{\text{avg}}$} & \multicolumn{2}{c|}{$\text{ESS}_{\text{avg}}$}\tabularnewline
\hline 
\hline 
\textcolor{gray}{\emph{Marginal target}} & \textcolor{gray}{0}&\textcolor{gray}{0118} & \multicolumn{2}{c|}{\textcolor{gray}{-}} & \textcolor{gray}{0}&\textcolor{gray}{0074} & \multicolumn{2}{c|}{\textcolor{gray}{-}} & \textcolor{gray}{0}&\textcolor{gray}{2444} & \multicolumn{2}{c|}{\textcolor{gray}{-}}\tabularnewline
\hline 
\emph{SPF-GS} & \textcolor{blue}{0}&\textcolor{blue}{0003} & \multicolumn{2}{c|}{-} & 0&0133 & \multicolumn{2}{c|}{-} & \textcolor{blue}{0}&\textcolor{blue}{0755} & \multicolumn{2}{c|}{-}\tabularnewline
\hline 
\emph{SPF-MPF} & 0&0217 & \textcolor{blue}{93}&\textcolor{blue}{00\%} & \textcolor{blue}{0}&\textcolor{blue}{0112} & 84&01\% & 0&2746 & 10&79\%\tabularnewline
\hline 
\emph{Gaussian particle flow} & 0&2647 & \multicolumn{2}{c|}{-} & 0&6563 & \multicolumn{2}{c|}{-} & 0&5279 & \multicolumn{2}{c|}{-}\tabularnewline
\hline 
\emph{Scaled-drift particle flow} & 0&3866 & \multicolumn{2}{c|}{-} & 0&4962 & \multicolumn{2}{c|}{-} & 0&5804 & \multicolumn{2}{c|}{-}\tabularnewline
\hline 
\emph{Marginal BPF} & \multicolumn{2}{c|}{-} & \multicolumn{2}{c|}{-} & 0&9969 & 0&37\% & 0&9998 & 0&13\%\tabularnewline
\hline 
\emph{Marginal EPF} & \multicolumn{2}{c|}{-} & \multicolumn{2}{c|}{-} & 0&3131 & 27&57\% & 0&5714 & 8&83\%\tabularnewline
\hline 
\emph{Marginal UPF} & \multicolumn{2}{c|}{-} & \multicolumn{2}{c|}{-} & 0&7753 & 4&44\% & 0&8136 & 2&31\%\tabularnewline
\hline 
\emph{Marginal APF} & \multicolumn{2}{c|}{-} & \multicolumn{2}{c|}{-} & 0&0119 & 81&32\% & 0&1467 & \textcolor{blue}{85}&\textcolor{blue}{11\%}\tabularnewline
\hline 
\end{tabular}\medskip{}
\par\end{centering}
\caption{Comparison of densities for the bivariate examples\label{tab:Comparison-of-densities-for-the-bivariate-examples}}
\end{table}

\pagebreak{}

\subsection{Multi-Sensor Bearings-Only Tracking}

Estimation in clutter of a target's position and velocity based solely
on angular measurements is a relevant problem that finds direct application
in airborne radar and sonar in passive listening mode. We propose
an example where a single target is observed by a circumferential
array of sensors. Each sensor measures the target's bearing with respect
to its own position.

In this example, we compare performances of the following filters:
\begin{itemize}
\item multi-sensor EKF that performs a series of centralized measurement
updates, considering each sensor in sequence; 
\item Information Matrix Fusion filter (IMF-EKF) \citep{Bar-Shalom2011}
that fuses distributed estimates (in parallel) into a global estimate
using the Information Matrix form of the EKF;
\item bootstrap particle filter (SIR);
\item Gaussian particle flow (GPF);
\item scaled-drift particle flow (SDPF); and
\item stochastic-particle-flow Gaussian sum (SPF-GS).
\end{itemize}
As mentioned before, in order to work properly, both the GPF and SDPF
are implemented based on a companion filter that estimates the state
covariance matrix. This is in accordance with implementations suggested
by Choi~\textit{et al.} \citep{Choi2011} and Ding \& Coates \citep{Ding2012}.
In this example we used the multi-sensor EKF as companion filter for
both the GPF and SDPF. In contrast, the stochastic particle flow does
not require a companion filter.

The bootstrap particle filter (SIR), the GPF and SDPF, and the SPF-GS
consider all measurements jointly according to a joint likelihood
function described in the next section. Performance is analyzed by
computing the root-mean-square error (RMSE) of estimates and the normalized-estimation
error squared (NEES) over 100 Monte Carlo runs. All particle-based
filters use 200 samples.\medskip{}

\subsubsection{A Multi-Sensor Bearings-Only Model}

When tracking in clutter based on multiple measurements, the usual
treatment rests on the probabilistic data association (PDA) \citep{Bar-Shalom2009}.
In the PDA model, a set of $m_{k}$ valid measurements is received
at each time step $k$ and assumed to be generated according to the
possibilities: (i) all measurements are false alarms (clutter), (ii)
one of the measurements is originated from the target and the remaining
are false alarms. Let $\theta_{k,i}$ be the association event that
the $i$th measurement is target-originated. The PDA filter computes
the association probabilities $p(\theta_{k,i}|\mathrm{y}_{1:k})$
conditional on the set of all received measurements up to time instant
$k$, and calculates the target state posterior density, $p(\mathrm{x}_{k}|\mathrm{y}_{1:k})$,
by marginalizing the joint density $p(\mathrm{x}_{k},\theta_{k,1:m_{k}}|\mathrm{y}_{1:k})$
over all possible associations.

In our example a single target is tracked by a set of $N_{s}$ sensors
located along a circumference that encloses the surveillance region,
at equally-spaced angular positions. As per the PDA model, one target
is known to exist a priori, detected with probability $P_{d,j}$ by
the $j$th sensor, and the number of clutter detections per sensor
is Poisson-distributed with mean $\lambda_{c}\cdot V$, where $\lambda_{c}$
is the clutter spatial density and $V$ is the surveillance region's
volume. For any given set of $N_{s}$ sensors, the expected likelihood
can be easily obtained by extending the procedure established by Marrs~\textit{et
al.} \citep{Marrs2002} to multiple sensors, to give
\begin{equation}
p\left(\mathrm{y}_{k}|\mathrm{x}_{k},\mathrm{y}_{1:k-1}\right)=\prod_{j=1}^{N_{s}}V^{-m_{k,j}}\frac{\left(\lambda_{c}V\right)^{m_{k,j}}e^{-\lambda_{c}V}}{m_{k,j}!}\left[\lambda_{c}\left(1-P_{d,j}\right)+\sum_{i=1}^{m_{k,j}}P_{d,j}\mathcal{N}\left(\mathrm{y}_{k,i(j)};h_{j}(\mathrm{x}_{k}),R_{k,j}\right)\right],\label{eq:multi-sensor-joint-likelihood}
\end{equation}
\noindent where $m_{k,j}$ is the total number of validated measurements
for the $j$th sensor, $\mathrm{y}_{k,i(j)}$ is the $i$th measurement
received by the $j$th sensor, $h_{j}(\cdot)$ and $R_{k,j}$ are
the observation function and the observation noise variance for the
$j$th sensor, respectively.

On a bidimensional state space, the bearing observations are modeled
by
\begin{equation}
h_{j}(\mathrm{x}_{k})=\text{atan}\left(\frac{\mathrm{x}_{k}\left(2\right)-p_{2,j}}{\mathrm{x}_{k}\left(1\right)-p_{1,j}}\right),\label{eq:bearing-observation-function}
\end{equation}
\noindent where $\mathrm{p}_{j}=\left(p_{1,j},p_{2,j}\right)^{T}$
are the position coordinates for the $j$th sensor. We assume a target
moving according to the nearly-constant velocity model, 
\begin{equation}
\mathrm{x}_{k}^{\vphantom{(i)}}=F\mathrm{x}_{k-1}^{\vphantom{(i)}}+\mathrm{u}_{k},\,\mathrm{u}_{k}\sim\mathcal{N}(\mathrm{u}_{k}^{\vphantom{(i)}};0_{n_{x}},Q_{k}),\label{eq:nearly-constant-velocity-model}
\end{equation}
\noindent where $\mathrm{x}_{k}=(p_{x_{1}},p_{x_{2}},v_{x_{1}},v_{x_{2}})_{k}^{T}$
is the state vector composed of position and velocity in Cartesian
coordinates $\left(x_{1},x_{2}\right)$, and
\begin{align}
F_{\hphantom{k}} & =\left(\begin{array}{cccc}
1 & 0 & \Delta t & 0\\
0 & 1 & 0 & \Delta t\\
0 & 0 & 1 & 0\\
0 & 0 & 0 & 1
\end{array}\right),\label{eq:state-transition-matrix}\\
Q_{k} & =\left(\begin{array}{cccc}
\nicefrac{\Delta t^{3}}{3} & 0 & \nicefrac{\Delta t^{2}}{2} & 0\\
0 & \nicefrac{\Delta t^{3}}{3} & 0 & \nicefrac{\Delta t^{2}}{2}\\
\nicefrac{\Delta t^{2}}{2} & 0 & \Delta t & 0\\
0 & \nicefrac{\Delta t^{2}}{2} & 0 & \Delta t
\end{array}\right)\sigma_{q}^{2}.\label{eq:process-noise-covariance}
\end{align}

The multi-sensor joint likelihood (\ref{eq:multi-sensor-joint-likelihood})
is incorporated in the bootstrap (SIR) filter, the GPF and SDPF, and
the SPF-GS by considering their filtering densities to target a posterior
density involving such joint likelihood. Regarding the implementation
of particle flows, specifically for this problem, the GPF and SDPF
reinterpret the filtered density empirically as a Gaussian pdf at
the end of each iteration in order to avoid exponential growth of
the number of mixture components over time. This practical aspect
does not affect the SPF-GS, whose filtered density is a mixture composed
invariably of $N$ local solutions to the actual posterior pdf, where
$N$ is the number of samples (and mixture components).

Generally speaking, multi-sensor bearings-only tracking is a difficult
problem to solve when the observation noise has high variance, the
probability of detection is relatively low and the probability of
having clutter in the surveillance region is not negligible. In this
scenario, the difficulty stems from the fact that the joint multi-sensor
likelihood (\ref{eq:multi-sensor-joint-likelihood}) is a product
of mixtures composed of several nonlinear and non-informative likelihood
terms: when nonlinearity is pronounced by a high-variance observation
noise, the resulting posterior density may not be well expressed by
simple parametric densities. In addition, this difficulty is modulated
by the amount of information available: the fewer the number of sensors
the more difficult to solve the problem. Another aspect that poses
additional concern is the system's observability. It is highly dependent
on the relative position of a sensor with respect to the target's
trajectory, i.e., trajectories radially aligned with a sensor's position
provide less information on the target's velocity. \medskip{}

\subsubsection{Results}

A challenging scenario was set for comparing the filters in order
to exacerbate differences of their performances to a noticeable level.
In this very difficult scenario, state process noise is assumed with
variance scaled by ${\sigma_{q}^{2}=25\,m^{2}}$, observation noise
variance is ${R_{k}=25\,deg^{2}}$, $P_{d}=0.50$ and $\lambda_{c}\times V=1.00$
false alarm/sensor/scan, for identical sensors. Even though estimation
errors generated for such a scenario might not be feasible as an Engineering
solution, it is certainly of practical interest to examine how the
estimates' errors scale to such extreme scenarios, which might happen
in real applications. This problem is particularly interesting because
the smaller the number of sensors the more difficult to achieve reasonable
estimates since, in this case, the low signal-to-noise ratio would
deteriorate inference. For this example, the SPF-GS has been set with
time horizon $T=10\,s$ and integration step size $\Delta\lambda=1\,s$.

\begin{figure}[H]
\begin{centering}
\includegraphics[width=0.9\textwidth]{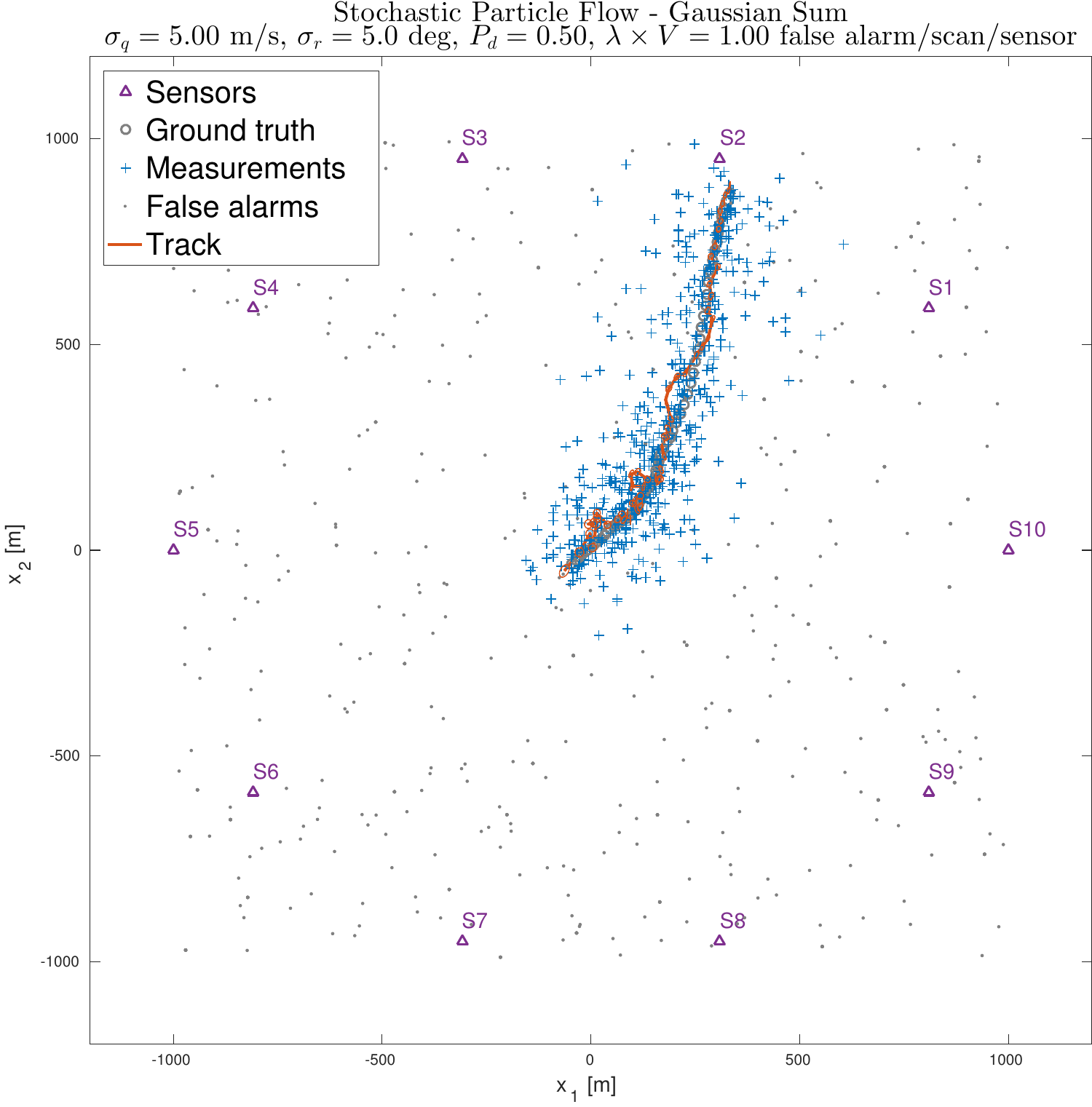}
\par\end{centering}
\caption{Illustration of bearing-only multi-sensor tracking\label{fig:Bearing-only-tracking-illustration}}
\end{figure}

The resulting track of an exemplar run is shown in Figure~\ref{fig:Bearing-only-tracking-illustration}.
The track initiation is based on an overdetermined triangulation of
measurements for the first two steps. No gating has been performed,
i.e., no preprocessing to discard measurements that fall outside a
high-confidence region of each sensor. Figure~\ref{fig:Bearing-only-tracking-illustration}
depicts a successful tracking of the target despite the difficult
scenario.

\begin{figure}[H]
\begin{centering}
\includegraphics[width=1\textwidth]{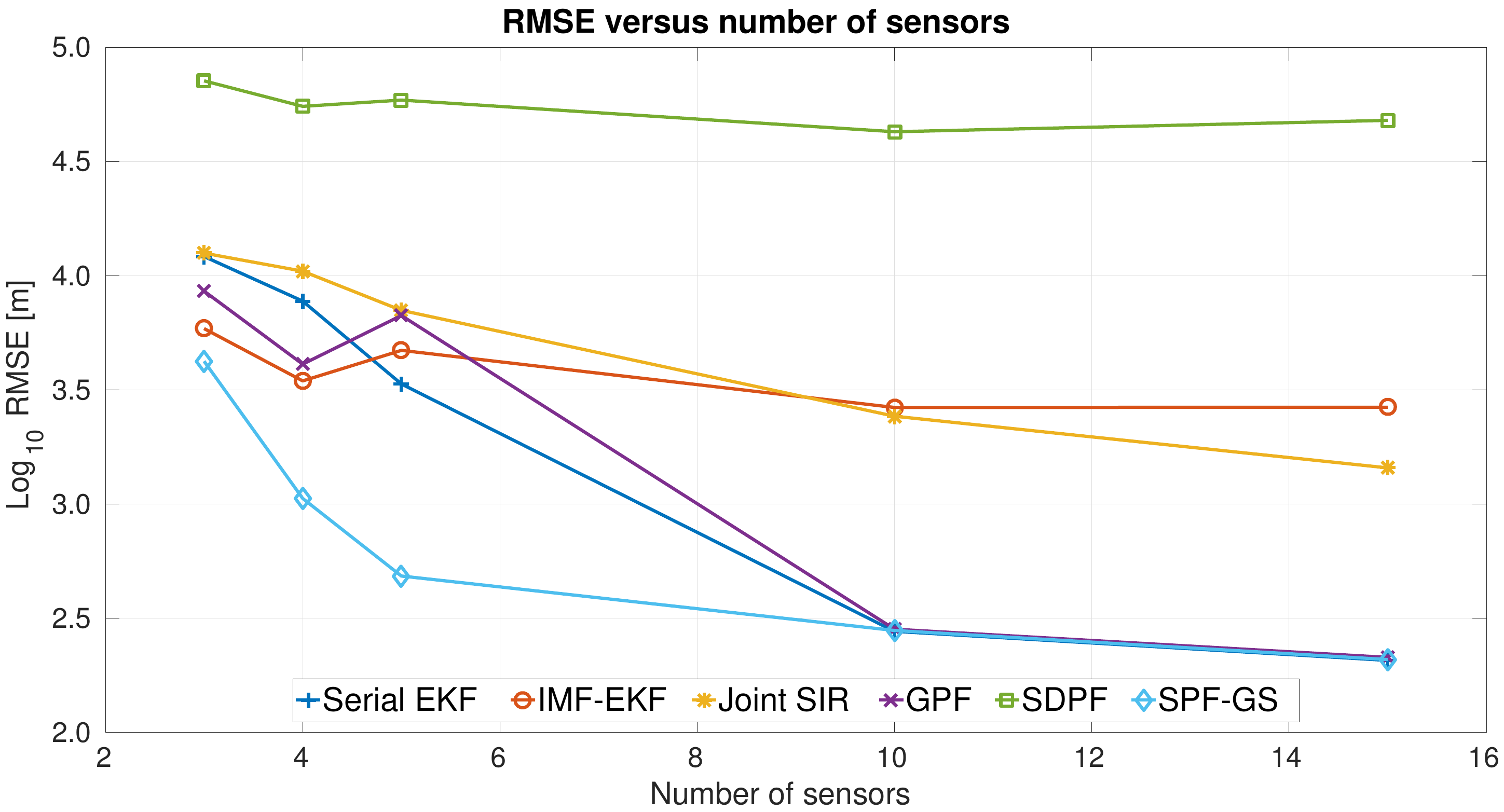}
\par\end{centering}
\caption{RMSE for the multi-sensor bearing-only tracking example\label{fig:RMSE-bearing-only-example}}
\end{figure}

\begin{figure}[H]
\begin{centering}
\includegraphics[width=1\textwidth]{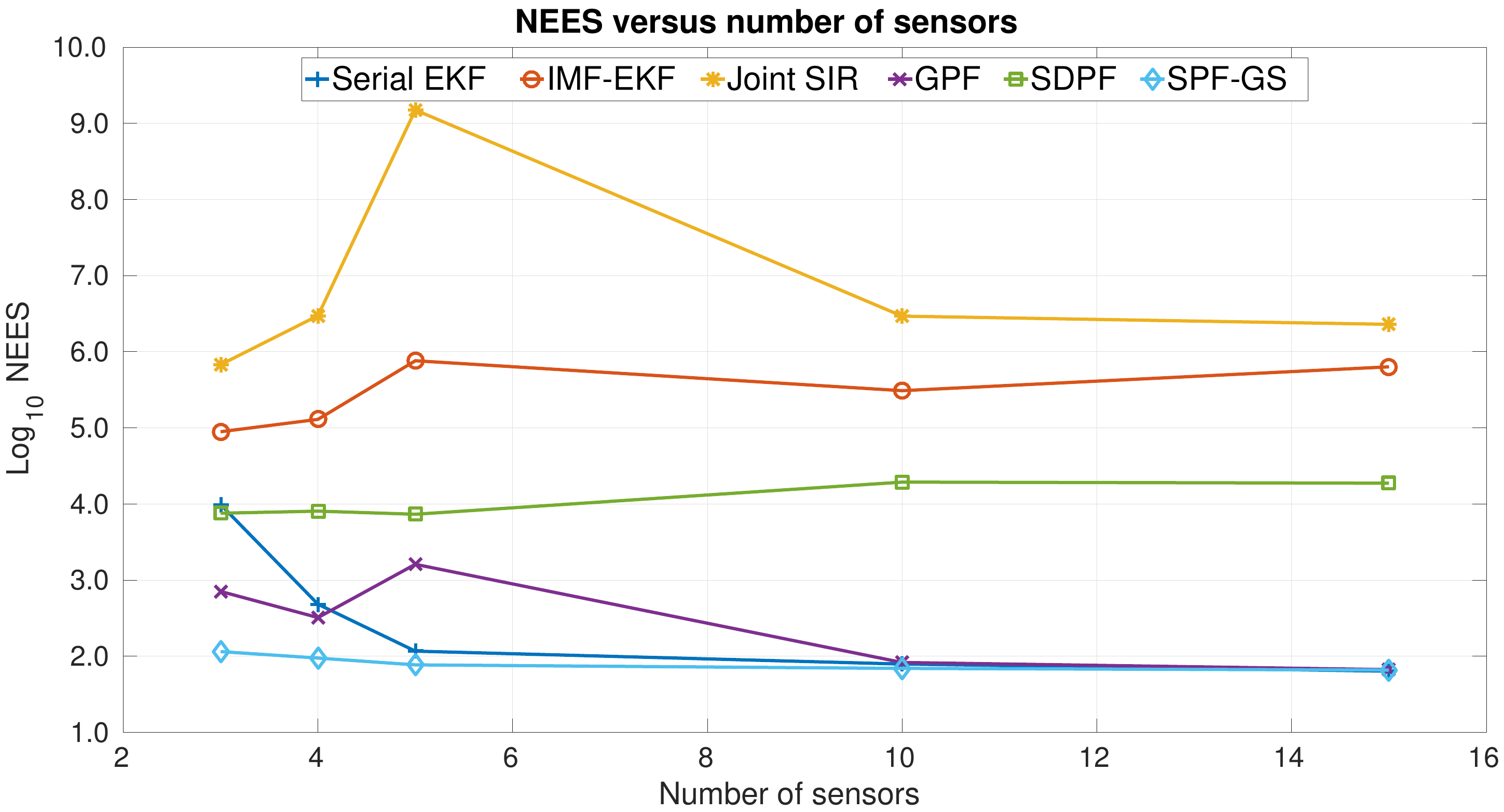}
\par\end{centering}
\caption{NEES for the multi-sensor bearing-only tracking example\label{fig:NEES-bearing-only-example}}
\end{figure}

\begin{figure}[H]
\begin{centering}
\includegraphics[width=1\textwidth]{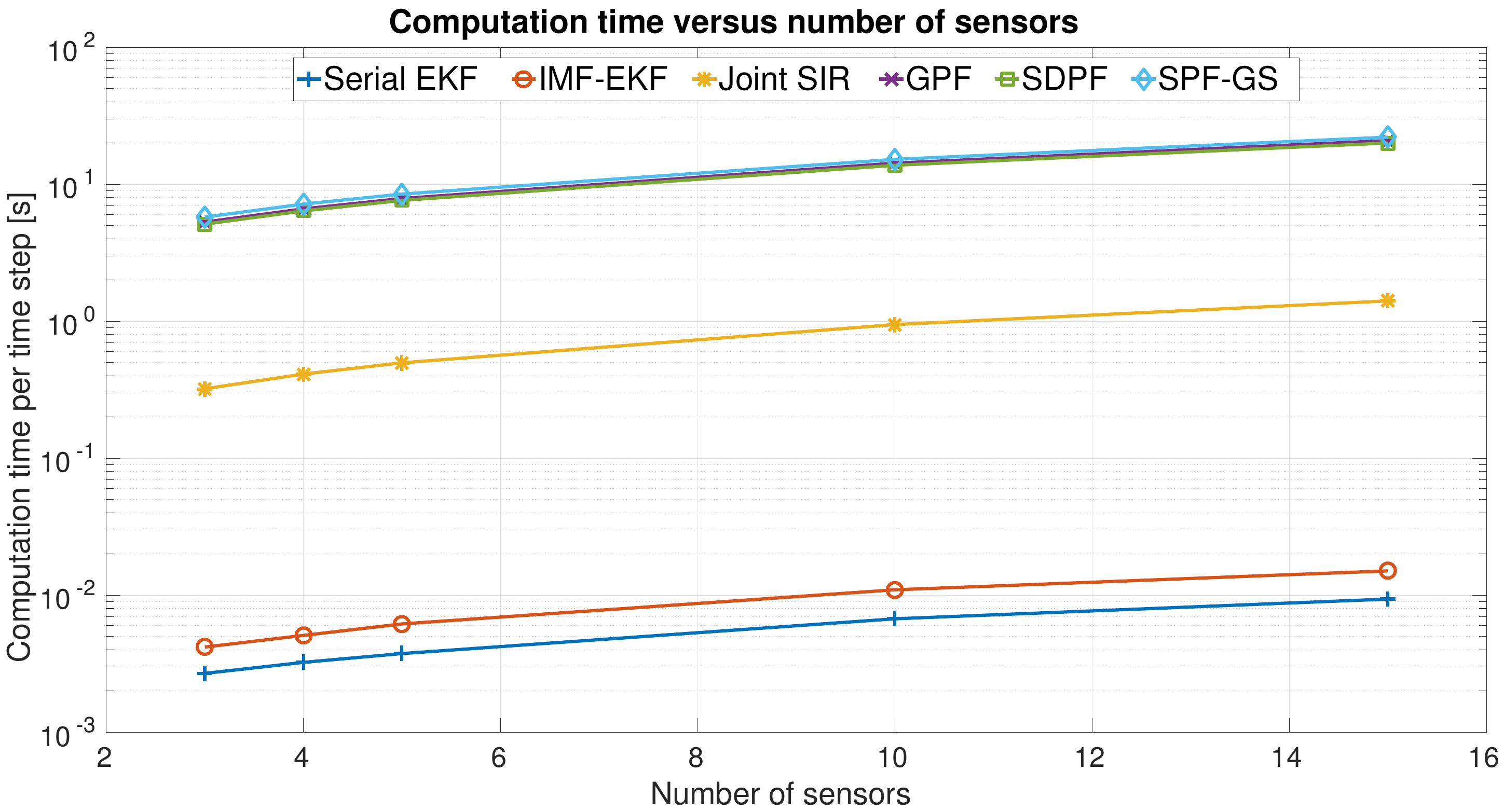}
\par\end{centering}
\caption{Average computation time for the multi-sensor bearing-only tracking
example\label{fig:Computation-time-bearing-only-example}}
\end{figure}

The resulting root-mean-square error (RMSE), normalized-estimation
error squared (NEES), and average computation time (per time step)
of all filters for different numbers of sensors are shown in Figures~\ref{fig:RMSE-bearing-only-example},
\ref{fig:NEES-bearing-only-example} and \ref{fig:Computation-time-bearing-only-example}
respectively. The following important aspects can be observed from
Figure~\ref{fig:RMSE-bearing-only-example}:
\begin{itemize}
\item Somewhat counterintuitively, the multi-sensor serial EKF provides
better estimates than that of Information Matrix Fusion EKF (IMF-EKF),
both in terms of precision (RMSE) and ``consistency'' (or credibility\footnote{As presented in \citep{Li2006}, the normalized-estimation error squared
(NEES) is the simplest metric that indicates an estimator's credibility.} as indicated by NEES).
\item The scaled-drift particle flow (SDPF) shows remarkably poor performance.
\item The bootstrap particle filter (SIR) provides mediocre performance,
eventually becoming better than the IMF-EKF as the number of sensors
increases.
\item Accuracies shown by the Gaussian particle flow (GPF) and the serial
EKF are commensurate and similar to that of stochastic particle flow
(SPF-GS) when the number of sensors is high.
\item As expected, the overall estimation accuracy is improved as the number
of sensors is increased.
\item The SPF-GS provides the most accurate estimates in difficult scenarios,
i.e., when the number of sensors is small.
\item Estimation by the SPF-GS is more consistent (or credible), which is
denoted by an NEES closer to one ($\log_{10}NEES\rightarrow0$) from
above.
\end{itemize}
It is worth commenting on the results comparing GPF and SPF-GS. Specifically
for this problem, when the number of sensors is sufficiently high,
the GPF provides estimates as accurate as those of SPF-GS at a slightly
lower computational cost. It is also remakable the successful synergy
between the GPF and its companion filter, a multi-sensor serial EKF
that provides the covariance estimates. However, it is difficult to
justify the calculated NEES for original particle flows since their
first and second moment estimates are underpinned by distinct filtering
methods. On the other hand, stochastic particle flow (SPF-GS) provides
fairly accurate state estimates and securely constitutes the most
credible estimator among all evaluated filters.

\subsection{Convoy Tracking}

Tracking multiple objects in clutter is as challenging as important
for real applications. In the multi-target tracking standard methods,
the most common treatment assumes the targets' states to be independent
so that the joint probability density is the product of their marginal
densities. While this assumption is fairly reasonable for applications
where objects are far apart most of the time, the same cannot be stated
for cases where objects are in proximity for a considerable part of
time. This latter cases elicit tracking all targets jointly in the
hope of implicitly capturing dependencies between targets. When targets
are tracked jointly the problem's dimensionality scales with the number
of targets.

To illustrate this situation, we propose an example of a convoy of
vehicles that are forced into mutual proximity when trafficking on
a road . The vehicles have explicit interactions as each driver aims
driving at the maximum allowed speed unless there is another vehicle
immediately in front at a slower speed. This scenario demands care
for a safety distance. The intent of the tracker is then to provide
the best estimate of each vehicle on a convoy, given a set of non-identified
measurements corrupted by noise and possible false alarms (clutter)
reported by a position sensor.

We compare performances of the following filters:
\begin{itemize}
\item Joint Probabilistic Data Association (JPDA) filter \citep{Bar-Shalom2009}; 
\item Global Nearest Neighbor Data Association (GNN) filter;
\item JPDA with a Gaussian mixture per target (JPDA-GM), applying mixture
reduction \citep{Pao1994};
\item Coupled Probabilistic Data Association (CPDA) filter~\citep{Blom2000};
\item multi-target bootstrap particle filter (joint SIR) based on the description
by Blom \& Bloem \citep{Blom2006};
\item Gaussian particle flow (GPF);
\item scaled-drift particle flow (SDPF); and
\item stochastic-particle-flow Gaussian sum (SPF-GS).
\end{itemize}
The GPF and the SDPF rely on a companion filter to estimate state
covariance matrices correctly, according to implementation guidelines
by Choi~\textit{et al.} \citep{Choi2011} and Ding \& Coates \citep{Ding2012}.
In this example, we used the CPDA \citep{Blom2000} as companion filter
for the original particle flows (GPF, SDPF). In contrast, the SPF-GS
does not require a companion filter.

The CPDA, the joint bootstrap particle filter (SIR), the GPF and SDPF,
and the SPF-GS consider all targets' states jointly, as a single high-dimensional
state. In contrast, the classical multi-target filters track targets
separately, where each target's state is described by the nearly-constant
velocity model. Performance is analyzed by computing the root-mean-square
error (RMSE) of estimates and the normalized-estimation error squared
(NEES) over 100 Monte Carlo runs. The particle filter (SIR), the original
particle flows and stochastic particle flow use 200 samples.\medskip{}

\subsubsection{The Intelligent Driver Model}

The Intelligent Driver Model (IDM) \citep{Treiber2013} is a model\footnote{In its simplest form, the IDM is focused on the interaction of vehicles
moving along a single-carriageway road. More complex variants exist
to model overtaking, for example, and consider factors such as the
politeness of the driver.} used in Traffic Engineering to simulate phenomena such as congestion
and to analyze the traffic behaviour as a response to changes in the
transport system. Because the interaction between vehicles is explicitly
taken into account by the IDM, tracking based on it involves consideration
of the joint state of multiple targets. Even though the IDM establishes
an empirical description of traffic for multiple vehicles, it has
not been previously used in the context of multi-target tracking.
We propose a stochastic version of the IDM and discretize it in order
to make it compatible with multi-target trackers formulated on the
joint state space.

The IDM describes the dynamics of vehicles in traffic, in terms of
positions and velocities, incorporating the interaction between each
vehicle and the vehicle directly in front. Provided a vehicle indexed
as $\alpha$ with length $l_{\alpha}$, the dynamics of its position
$x_{\alpha}$ and velocity $v_{\alpha}$ are given by the following
(continuous-time) stochastic differential equations:
\begin{align}
dx_{\alpha} & =v_{\alpha}dt,\label{eq:IDM-1}\\
dv_{\alpha} & =\underbrace{a\left[1-\left(\frac{v_{\alpha}}{v_{0}}\right)^{\delta}\right]^{\vphantom{2}}}_{\dot{v}_{\alpha}^{\text{free road\ensuremath{\hphantom{o}}}}}dt-\underbrace{a\left[\frac{\bar{s}}{s_{\alpha}}\vphantom{\left(\frac{v_{\alpha}}{v_{0}}\right)^{\delta}}\right]^{2}}_{\dot{v}_{\alpha}^{\text{interaction}}}dt+dw_{t},\label{eq:IDM-2}
\end{align}
\noindent where $\left\{ w_{t}\right\} _{t\ge0}$ is a Wiener process,
${s_{\alpha}=x_{\alpha-1}-x_{\alpha}-l_{\alpha}}$ is the net distance
between vehicles, the approaching rate is given by ${\Delta v_{\alpha}=v_{\alpha-1}-v_{\alpha}}$,
and ${\bar{s}=\bar{s}\left(v_{\alpha},\Delta v_{\alpha}\right)}$
is the expected distance defined as
\begin{equation}
\bar{s}\left(v_{\alpha},\Delta v_{\alpha}\right)=s_{0}+v_{\alpha}T_{h}+\frac{v_{\alpha}\Delta v_{\alpha}}{2\sqrt{a\cdot b}}.\label{eq:IDM-5}
\end{equation}

The model dynamics is such that when a vehicle is travelling on a
free road it will predominantly accelerate according to $\dot{v}_{\alpha}^{\text{free road}}$
up to the maximum allowed speed $v_{0}$, whereas when it approaches
another vehicle immediately in front, the decrement in acceleration
according to $\dot{v}_{\alpha}^{\text{interaction}}$ becomes relevant
to maintain a safe-time headway $T_{h}$ and to avoid approaching
closer than the minimum safe distance $s_{0}$. The IDM parameters
are summarized in the following table.

\begin{center}
\begin{tabular}{|l|l|}
\hline 
\textbf{Parameter} & \textbf{Description}\tabularnewline
\hline 
\hline 
$a$ & nominal maximum acceleration\tabularnewline
\hline 
$b$ & comfortable braking decceleration\tabularnewline
\hline 
$\delta$ & acceleration exponent (driver dynamics)\tabularnewline
\hline 
$v_{0}$ & free-road desired velocity\tabularnewline
\hline 
$s_{0}$ & minimum allowed distance between vehicles\tabularnewline
\hline 
$T_{h}$ & safe-time headway\tabularnewline
\hline 
$\alpha-1$ & index of the vehicle direcly in front\tabularnewline
\hline 
\end{tabular}\medskip{}
\par\end{center}

In order to use the stochastic IDM as the state process for a multi-target
tracker, its continuous-time equations are discretized by a first-order
approximation (Markov random field). This assumes that the state's
derivative with respect to time is linear in time between two subsequent
measurements, but the interactions between non-adjacent vehicles are
negligible when compared to the interactions between adjacent vehicles.
The discretized version of the stochastic IDM is presented in the
Appendix~\ref{sec:Discrete-time-stochastic-IDM}.

\medskip{}

\subsubsection{The Multi-Target Joint Likelihood Function}

The joint multi-target filters extend the joint probabilistic data
association (JPDA) \citep{Bar-Shalom2009} framework for situations
where the targets' states are not mutually independent conditioned
on the past observations. This formulation has been first proposed
as the JPDA Coupled filter (JPDAC) \citep{Bar-Shalom2009} and further
generalized by Blom \& Bloem \citep{Blom2000,Blom2006} who consider
the measurement-to-target associations implicitly.

In the JPDA model, a set of $N_{m}$ valid measurements is received
at each time step $k$ and assumed to be generated according to the
possibilities: (i) each of the measurements may be originated from
each target, considering all possible associations, (ii) a measurement
not originated from any target is due to a false alarm (clutter).
These possibilities are exhaustive such that a measurement can have
only one source, and at most one of the validated measurements can
originate from a target.

Let ${\phi_{k,i}\in\left\{ 0,1,\,\dots,\,N_{m}\right\} }$ be an association
event that maps each target $i\in\{1,\dots,N_{t}\}$ to the measurement
indexed as $\phi_{k,i}$, where $\phi_{k,i}=0$ means that no measurement
is associated to the $i$th target. The Coupled JPDA filter computes
the joint association probabilities $p(\phi_{k,1:N_{t}}|\mathrm{y}_{1:k})$
conditional on the set of all received measurements up to time instant
$k$, and calculates the joint state posterior density, $p(\mathrm{x}_{k,1:N_{t}}|\mathrm{y}_{1:k})$,
by marginalizing $p(\mathrm{x}_{k,1:N_{t}},\phi_{k,1:N_{t}}|\mathrm{y}_{1:k})$
over all possible joint associations.

In the JPDA framework, $N_{t}$ targets are known to exist a priori,
detected with probability $P_{d}$ by a single sensor; the number
of clutter detections is Poisson-distributed with mean $\lambda_{c}\cdot V$,
where $\lambda_{c}$ is the clutter spatial density and $V$ is the
surveillance region's volume; the location of each clutter detection
is independently distributed according to a spatial density $\eta_{c}(\mathrm{y})$;
and the likelihood function of the $j$th measurement being originated
from the $i$th detected target is $p(\mathrm{y}_{k,j}|\mathrm{x}_{k,i})$.
Denoting the joint multi-target state as $\mathrm{x}_{k,1:N_{t}}$
and the joint observation as $\mathrm{y}_{k,1:N_{m}},$ the joint
likelihood can be either obtained as in \citep{Blom2006} or by a
formulation equivalent to the Coupled JPDA as
\begin{multline}
p\left(\mathrm{y}_{k,1:N_{m}}|\mathrm{x}_{k,1:N_{t}}\right)\\
=\frac{\left[\prod_{j=1}^{N_{m}}\eta_{c}\left(\mathrm{y}_{k,j}\right)\right]}{N_{t}!}\sum_{N_{d}=0}^{N_{t}}\frac{\left(\lambda_{c}V\right)^{N_{m}-N_{d}}e^{-\lambda_{c}V}}{\left(N_{m}-N_{d}\right)!}P_{d}^{N_{d}}\left(1-P_{d}\right)^{N_{t}-N_{d}}\sum_{\phi_{k,1:N_{t}}|N_{d}}\prod_{i=1}^{N_{t}}\frac{p\left(\mathrm{y}_{k,\phi_{k,i}}|\mathrm{x}_{k,i}\right)}{\eta_{c}\left(\mathrm{y}_{k,\phi_{k,i}}\right)}.\label{eq:joint-multi-target-likelihood}
\end{multline}
The joint state vector $\mathrm{x}_{k,1:N_{t}}=(p_{1},\dots,p_{N_{t}},v_{1},\dots,v_{N_{t}})_{k}^{T}$
is composed of position and velocity of all vehicles in the convoy,
and the joint observation $\mathrm{y}_{k,1:N_{m}}=(\mathrm{y_{1},\dots,\mathrm{y}}_{N_{m}})_{k}^{T}$
contains position measurements of all targets and possible false alarms
obtained at a given time instant $k$.

The joint bootstrap particle filter, the GPF and SDPF, and the SPF-GS
consider their filtering densities to target a joint posterior density
incorporating the joint multi-target likelihood function (\ref{eq:joint-multi-target-likelihood}).
Regarding the implementation of original particle flows, the GPF and
SDPF reinterpret the filtered density empirically as a Gaussian pdf
at the end of each iteration in order to avoid exponential growth
of the number of mixture components over time. In contrast, this practical
aspect does not affect the stochastic particle flow, whose filtered
density is a mixture composed of a fixed number of local solutions
to the actual posterior pdf.

\medskip{}

\subsubsection{Results}

We simulated the trajectories of vehicles on a ring road by integrating
the continuous-time stochastic IDM over 60 seconds with the parameters
presented as follows. The convoy was set to start from rest with the
vehicles initially positioned apart, led by a truck so that the queue
of cars is slowed down and forced into mutual proximity. The minimum
allowed distance between vehicles was set to be exaggeratedly small
($s_{0}=0.5\,m$) to induce the model to control the distance between
cars mainly based on the safe-time headway $T_{h}$. In this case,
the safe-time headway indirectly determines the desired distance between
vehicles, which is denoted as \emph{target distance} in the table
below. At the final steady state, the net speed of the convoy is dominated
by the free-road speed of the truck, which motivates the safe-time
headway being computed based on $v_{0,\text{truck}}$.

\begin{center}
\begin{tabular}{|l|r|r|}
\hline 
\textbf{Parameter} & \textbf{car} & \textbf{truck}\tabularnewline
\hline 
\hline 
$a$ & $0.5\,m/s^{2}$ & $0.4\,m/s^{2}$\tabularnewline
\hline 
$b$ & $1.5\,m/s^{2}$ & $1.2\,m/s^{2}$\tabularnewline
\hline 
$\delta$ & $4$ & $4$\tabularnewline
\hline 
$v_{0}$ & $15\,m/s$ & $10\,m/s$\tabularnewline
\hline 
$s_{0}$ & $0.5\,m$ & $0.5\,m$\tabularnewline
\hline 
$T_{h}$ & $\frac{\vphantom{()^{2}}\text{\{\text{target distance (m)}\}}}{v_{0,\text{truck}}}$ & $\hphantom{\frac{\text{\{\text{target distance (m)}\}}}{v_{0,\text{truck}}}}$-\tabularnewline
\hline 
$l_{\alpha}$ & $5\,m$ & $20\,m$\tabularnewline
\hline 
\end{tabular}\smallskip{}
\par\end{center}

The joint state process covariance matrix is assumed as scaled by
${\sigma_{q}^{2}=0.0625\,\left(m/s\right)^{2}}$, each position observation
has variance ${\sigma_{r}^{2}=4\,m^{2}}$, ${P_{d}=0.80}$ and $\lambda_{c}\times V=0.01$
false alarm/scan, and the surveillance region's ``volume'', $V$,
is in fact the length covered by a confidence region ($\approx99.73\%$)
that contains all the vehicles. Proposing a method to effectively
initiate tracks was out of the example's scope, thus track initiation
was considered to be ideal, i.e., the initial position and velocity
of the targets are known with initial uncertainty scaled by the observation
noise. The stochastic particle flow has been set with time horizon
$T=15\,s$ and integration step size $\Delta\lambda=1\,s$.

\begin{figure}[H]
\begin{centering}
\includegraphics[width=1\textwidth]{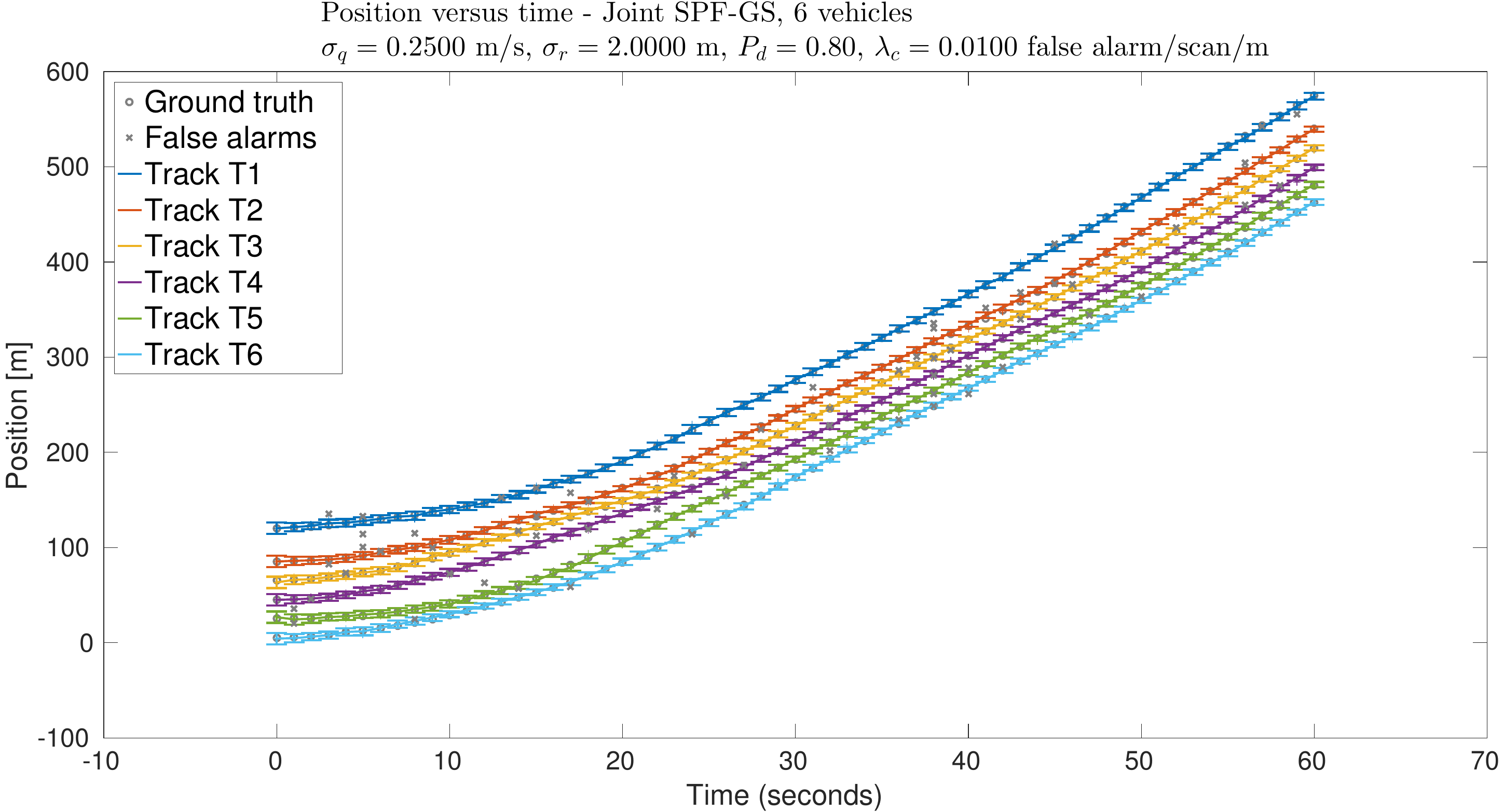}
\par\end{centering}
\caption{Position estimates for an exemplar run of the convoy tracking\label{fig:Convoy-tracking-illustration-2}}
\end{figure}

\begin{figure}[H]
\begin{centering}
\includegraphics[height=0.48\textheight]{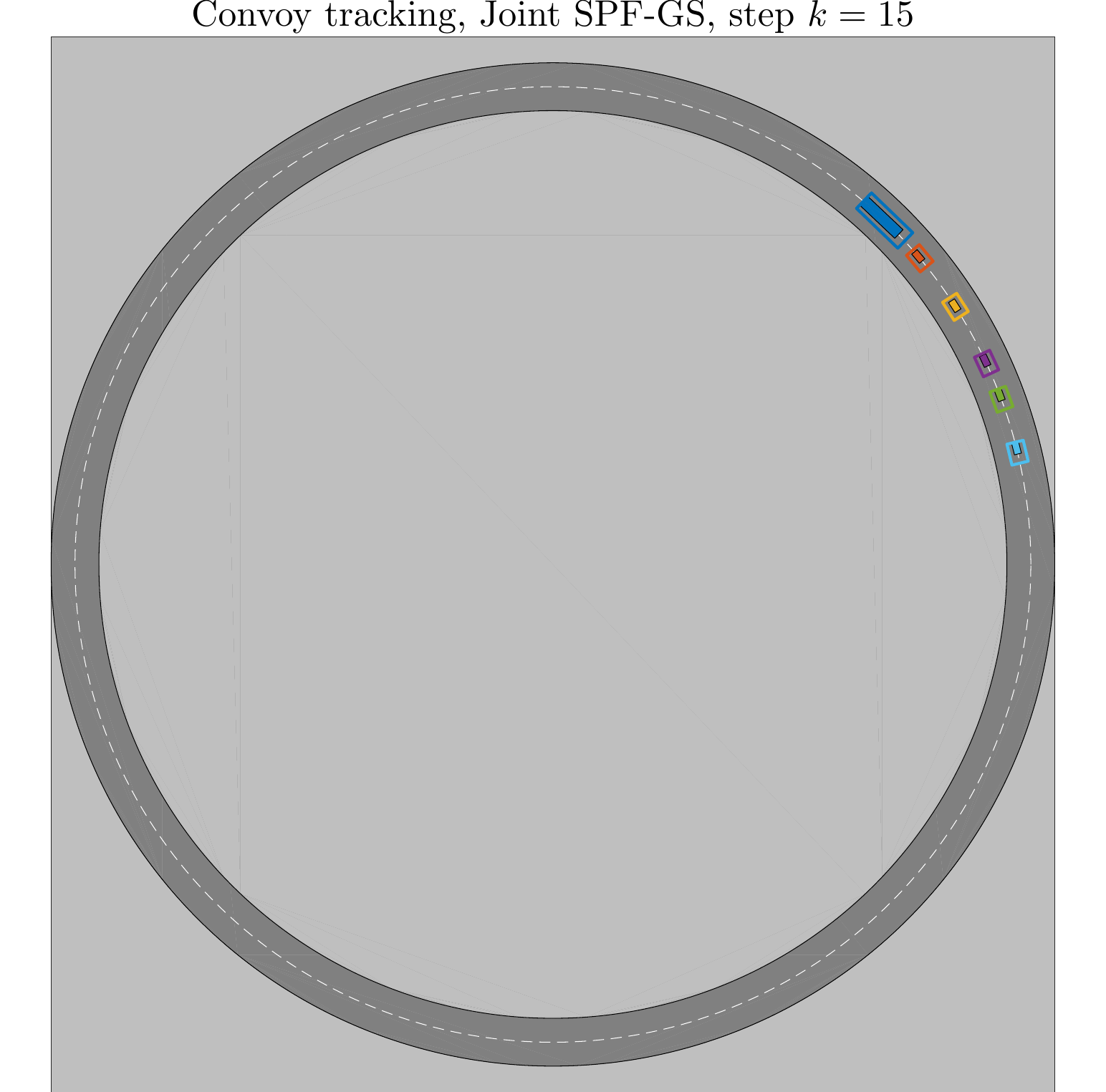}\medskip{}
\par\end{centering}
\begin{centering}
\includegraphics[height=0.48\textheight]{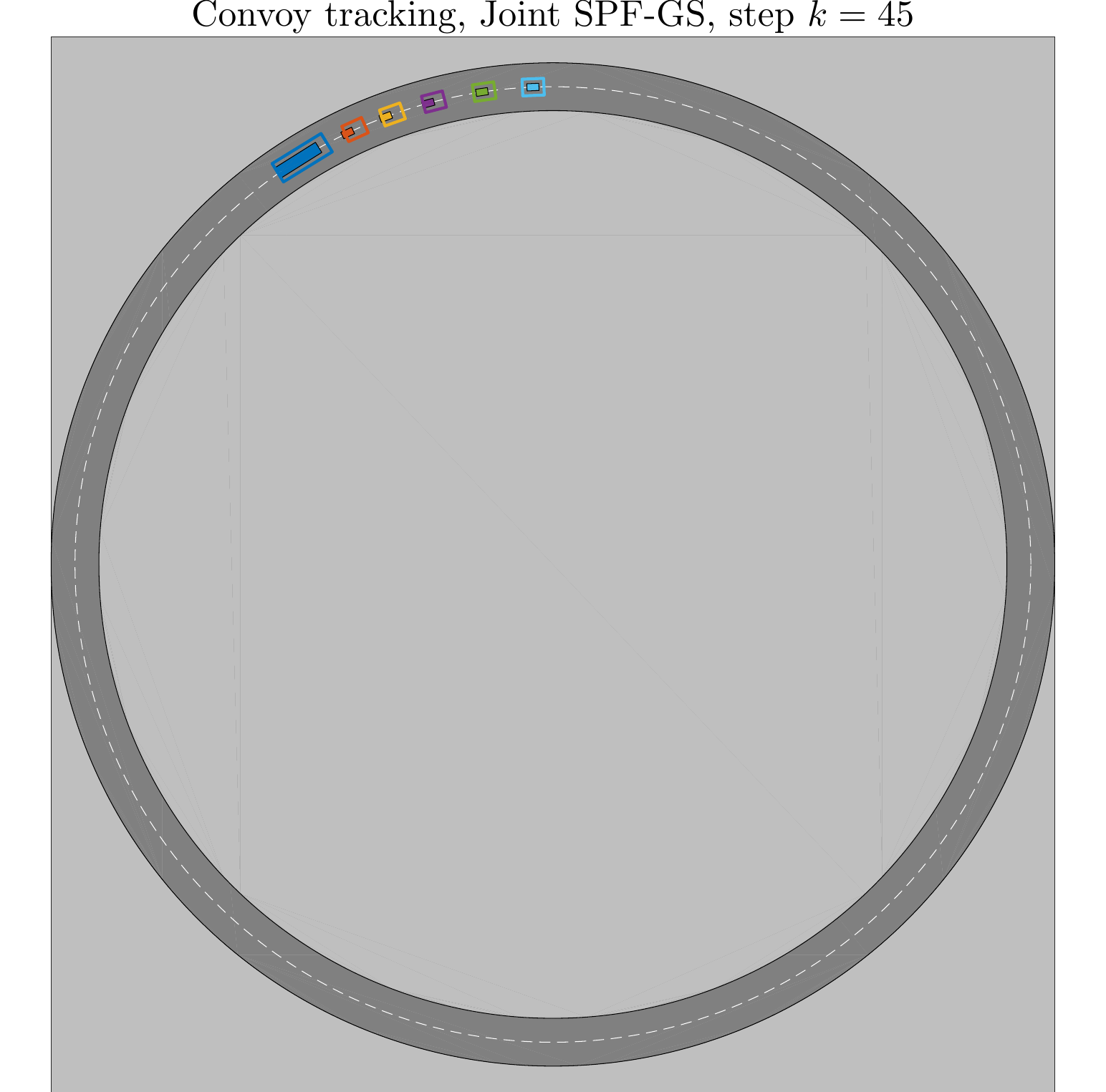}
\par\end{centering}
\caption{Illustration of convoy tracking on a ring road\label{fig:Convoy-tracking-illustration}}
\end{figure}

Figure~\ref{fig:Convoy-tracking-illustration} shows two frames of
an exemplar run, demonstrating the situation where a queue of cars
is slowed down by a truck, forcing them into proximity. The non-filled
rectangles depicted in Figure~\ref{fig:Convoy-tracking-illustration}
denote the position estimates provided by the filter applied for that
run. Interactions between vehicles in the convoy, due to proximity,
can be well perceived by position estimates of the exemplar run as
shown in Figure~\ref{fig:Convoy-tracking-illustration-2}.

\begin{figure}[H]
\begin{centering}
\includegraphics[width=1\textwidth]{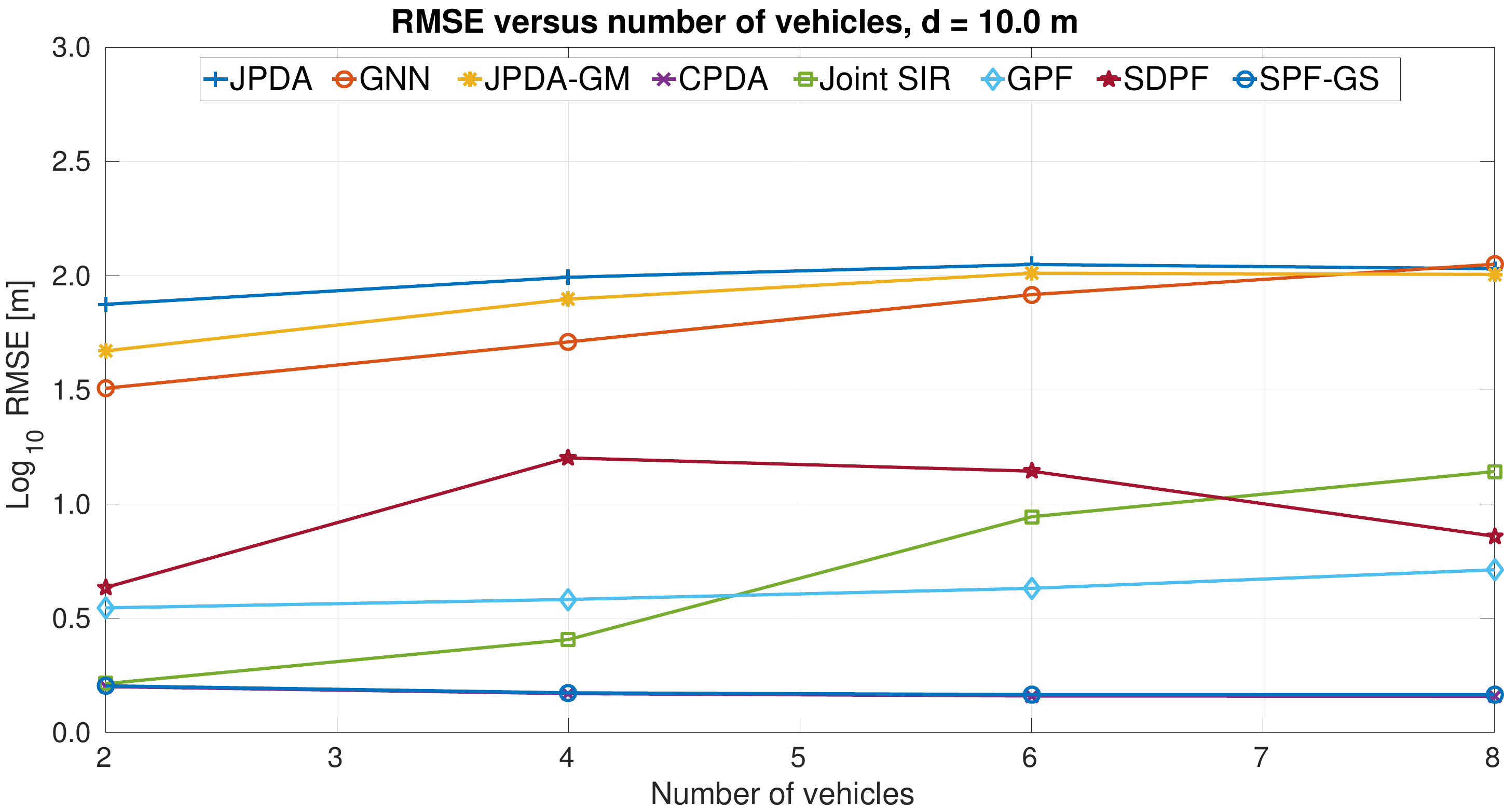}
\par\end{centering}
\caption{RMSE for the convoy tracking example\label{fig:RMSE-convoy-tracking}}
\end{figure}

\begin{figure}[H]
\begin{centering}
\includegraphics[width=1\textwidth]{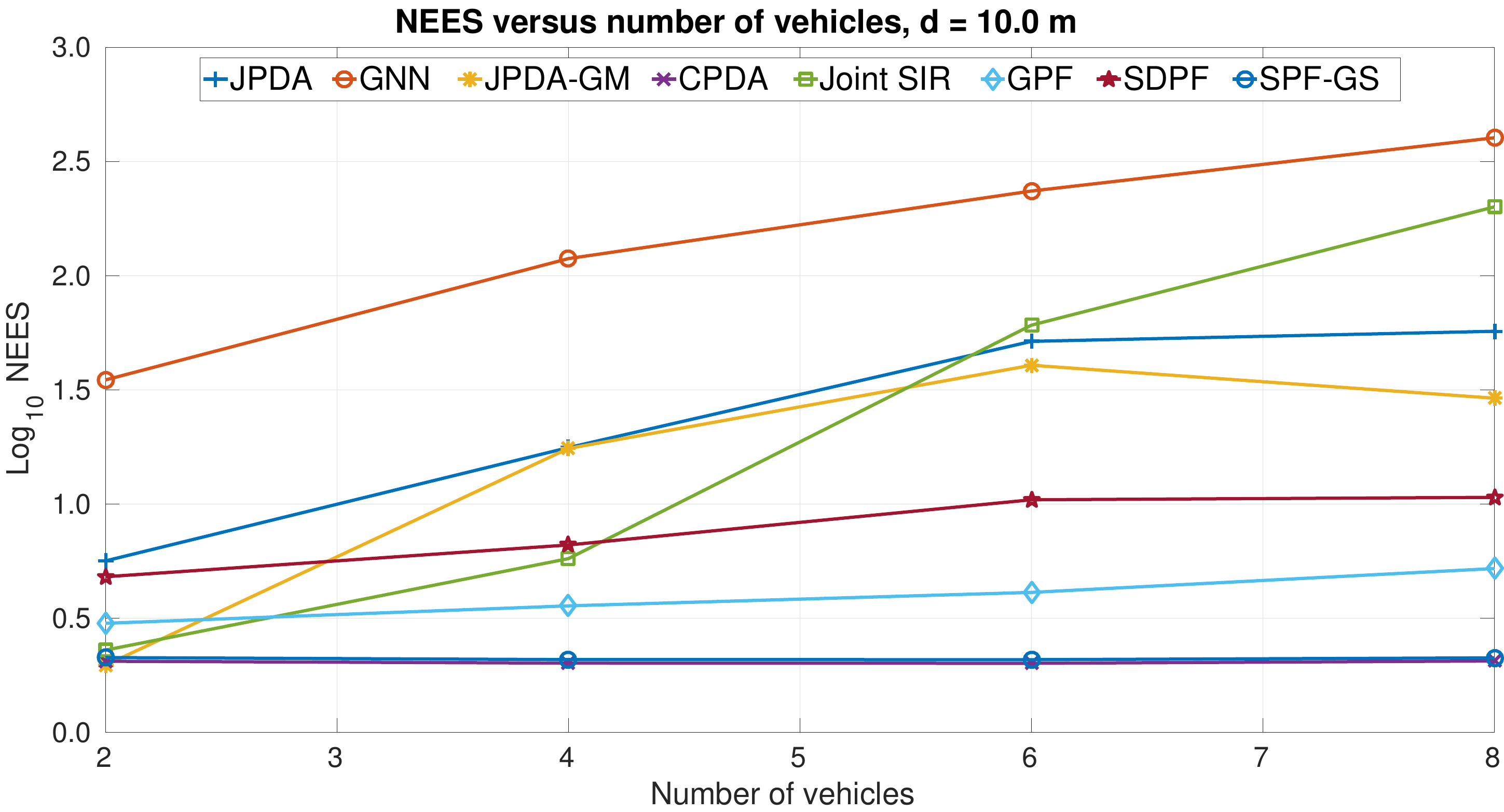}
\par\end{centering}
\caption{NEES for the convoy tracking example\label{fig:NEES-convoy-tracking}}
\end{figure}

\begin{figure}[H]
\begin{centering}
\includegraphics[width=1\textwidth]{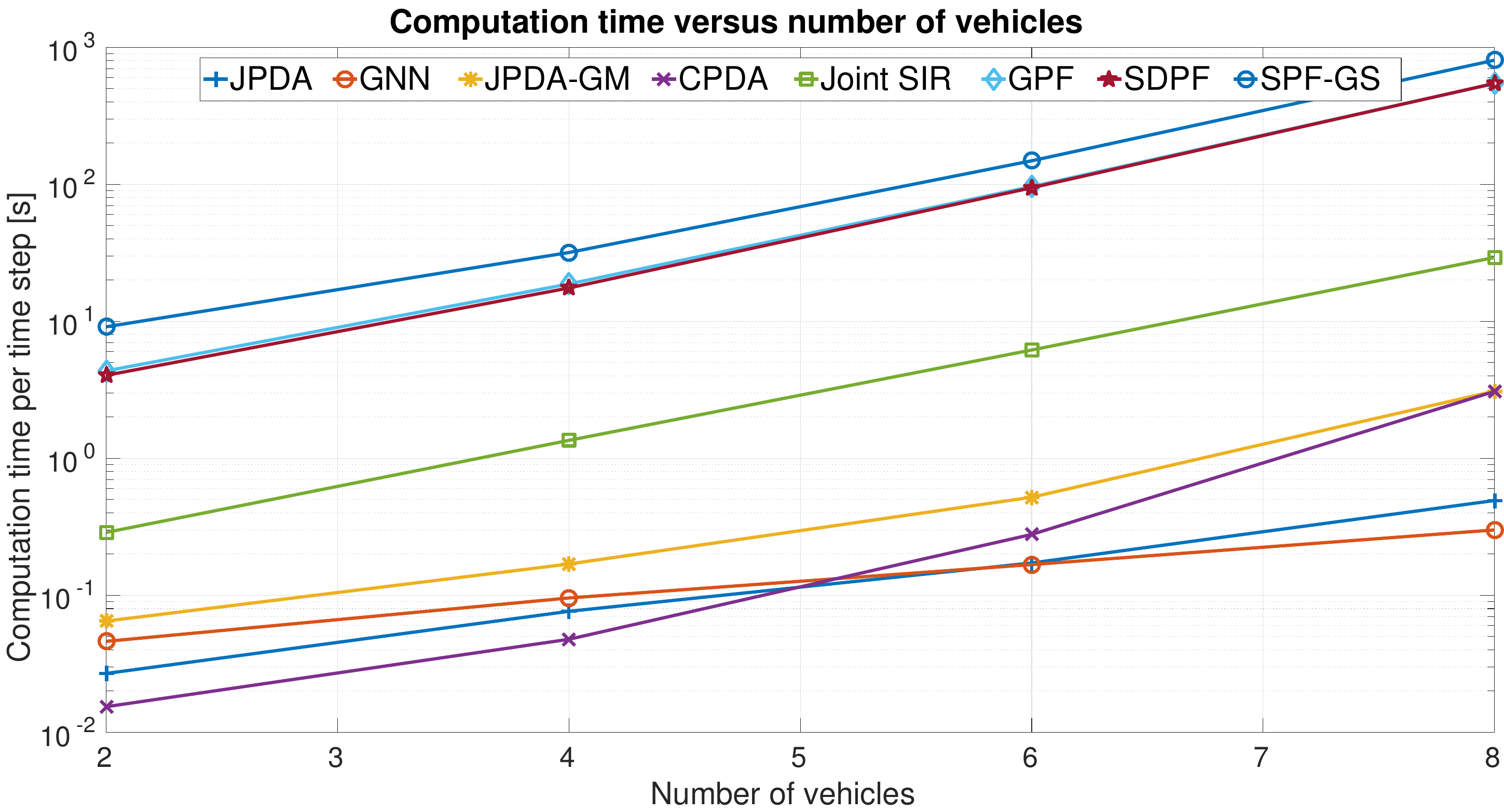}
\par\end{centering}
\caption{Average computation time for the convoy tracking example\label{fig:Computation-time-convoy-tracking}}
\end{figure}

The resulting root-mean-square error (RMSE), normalized-estimation
error squared (NEES), and average computation time (per time step)
of all filters, for different numbers of vehicles, and target distance
between vehicles $d=10\,m$, are shown in Figures~\ref{fig:RMSE-convoy-tracking},
\ref{fig:NEES-convoy-tracking} and \ref{fig:Computation-time-convoy-tracking}
respectively. RMSE and NEES were computed over 100 Monte Carlo runs,
with the particle-based filters using 200 samples. The following important
aspects can be noted from Figure~\ref{fig:RMSE-convoy-tracking}:
\begin{itemize}
\item In general, trackers that estimate on the joint $2N_{t}$-dimensional
state space clearly outperform the classical multi-target trackers
(JPDA, GNN and JPDA-GM), both in terms of precision (RMSE) and credibility
(NEES).
\item Among the classical multi-target trackers, the Global Nearest Neighbor
(GNN) association filter is the one that provides the most accurate
estimates. This can be explained by an increasingly detrimental effect
of the association uncertainty on estimation, which is more prominent
in the JPDA and less prominent on the GNN filter.
\item Estimation errors committed by the multi-target particle filter (joint
SIR) grow exponentially with the number of state-space dimensions
($2N_{t}$), as expected, due to the curse of dimensionality.
\item The Coupled PDA (CPDA) and the SPF-GS present commensurate root-mean-square
errors, suggesting that most of their accuracy gain originates from
tracking on the joint $2N_{t}$-dimensional state space and accounting
for inherent dependencies between targets.
\item The original particle flows (GPF, SDPF) show significant values of
NEES. Most likely this is because the evaluated implementation of
these filters cannot provide reliable estimates for state covariance
matrices and depend on a dissimilar companion filter to work around
it, which affects consistency of their estimates. 
\item The SPF-GS provides overall higher estimation accuracy (RMSE) and
consistency (NEES), with low sensitivity to increasing the problem's
number of dimensions.
\end{itemize}
Based on the results for this example, two important remarks are worth
making. Firstly, the results show a notable performance improvement
with adoption of filtering on the joint $2N_{t}$-dimensional state
space: about 30-fold improvement in estimation precision (RMSE) and
nearly 10-fold in credibility (NEES). The fact that the Coupled PDA
performs as well as the SPF-GS suggests that modeling inherent dependencies
between targets and filtering on the joint space provides most of
the performance gain. Secondly, the example not only illustrates well
the curse of dimensionality for the multi-target particle filter (joint
SIR), but also corroborates the success of principled choices made
in the SPF's formulation in order to avoid degeneracy in high-dimensional
problems. This latter observation becomes clear when we realize that
the performance indexes for stochastic particle flow scale gently
with the number of dimensions.

Additionally, it is also worth noting that the original particle flows
provide relatively accurate estimates, scaling well with the number
of dimensions. For the evaluated implementation, both particle flow
filters (GPF, SDPF) rely on covariance matrices estimated by the CPDA
as a companion filter. Due to this fact, calculated NEES for these
filters is not reliable since their first and second moment estimates
are underpinned by distinct filtering methods. This does not disqualify
the original particle filters per se since observed characteristics
are probably due to the implementation settings. Under these circumstances,
actual consistency (credibility) of their estimates cannot be quantified
and, ultimately, evokes the question about the extent to which the
success of the adopted implementation is due to the companion filter.

\subsection{Inference on Large Spatial Sensor Networks}

In this section we consider the problem proposed by Septier \& Peters
\citep{Septier2016}, posed to address inference of physical quantities
of complex phenomena from a collection of noisy measurements obtained
by a large network of spatially distributed sensors. According to
Septier \& Peters \citep{Septier2016}, a large number of applications
could adopt such sensor networks to make inferences related to complex
phenomena. Applications include environmental monitoring, weather
forecasting etc. In this framework, a fusion center would regularly
receive observations from sensors set up as a grid, which monitor
a time-varying physical phenomenon presenting spatially diverse attributes
such as pressure, temperature, concentrations of substance, radiation
levels, seismic activity etc. Upon fusing the observations, the solution
to the problem consists of estimating the phenomenon state at the
current time instant at each of the sensor's positions. The problem
becomes particularly challenging as the number of sensors in the grid
increases, since solving the problem then demands efficient algorithms
for inference in high-dimensions. 

The physical phenomenon is modeled as a time-varying spatially-dependent
continuous process defined over a two-dimensional space which is observed
sequentially in time by a 2D spatial grid of $n_{x}=N_{s}$ sensors,
where $n_{x}$ is the state dimension. At time instant $k$, each
sensor independently produces a noisy measurement of an attribute
of interest about the phenomenon at its specific location, giving
$\mathrm{y}_{k,j}|\mathrm{x}_{k}\sim p(\mathrm{y}_{k}^{(j)}|\mathrm{x}_{k}^{\vphantom{(j)}}),\,\forall j=1,\dots,N_{s}$.
Based on the historic set of observations $\mathrm{Y}_{1:k}\coloneqq\{\mathrm{Y}_{k^{\prime}}:k^{\prime}=1,\dots,k\}$,
where $Y_{k}\coloneqq\{\mathrm{y}_{k,j}:j=1,\dots,N_{s}\}$, one is
required to estimate, at time $k$, the state of the physical phenomenon
$\mathrm{x}_{k}\in\mathbb{R}^{n_{x}}$ across the locations of all
sensors in the grid. The state process that models the time-varying
physical phenomenon is considered to follow a transition multivariate
Generalized Hyperbolic (GH) density as 
\begin{equation}
p_{t}\left(\mathrm{x}_{k}|\mathrm{x}_{k-1}\right)\propto\frac{\mathbb{K}_{c_{1}-\nicefrac{n_{x}}{2}}\left[\sqrt{(c_{2}+Q(\mathrm{x}_{k},\mathrm{x}_{k-1}))(c_{3}+\mathrm{\gamma}^{T}\Sigma^{-1}\mathrm{\gamma})}\right]}{\left(\sqrt{(c_{2}+Q(\mathrm{x}_{k},\mathrm{x}_{k-1}))(c_{3}+\mathrm{\gamma}^{T}\Sigma^{-1}\mathrm{\gamma})}\right)^{\nicefrac{n_{x}}{2}-c_{1}}}\cdot e^{(\mathrm{x}_{k}-\alpha\mathrm{x}_{k-1})\Sigma^{-1}\mathrm{\gamma}}\label{eq:transition-pdf-sensor-networks}
\end{equation}
\noindent where $Q(\mathrm{x}_{k},\mathrm{x}_{k-1})=(\mathrm{x}_{k}-\alpha\mathrm{x}_{k-1})^{T}\Sigma^{-1}(\mathrm{x}_{k}-\alpha\mathrm{x}_{k-1})$,
$\alpha\in\mathbb{R}$ is the location constant, and $\mathbb{K}_{c_{1}}[\cdot]$
denotes the modified Bessel function of the second kind, of order
$c_{1}$. The parameters $c_{1}$, $c_{2}$, and $c_{3}$ are scalar
values that determine the shape of the distribution, $\Sigma\in\mathbb{R}^{n_{x}\times n_{x}}$
is the dispersion matrix, and the vector $\mathrm{\gamma}\in\mathbb{R}^{n_{x}}$
is the skewness parameter. The choice of transition density in \ref{eq:transition-pdf-sensor-networks}
can account for heavy-tailed and asymmetric data \citep{Septier2016},
which is beneficial when modeling physical process with extremal behavior.
In special cases, the transition density becomes the normal, normal
inverse Gaussian, skewed-t, and other densities. To generate the prior
distribution at the first time step, we take $p_{x}(\mathrm{x}_{0})=p_{t}(\mathrm{x}_{0}|\mathrm{x}_{-1}=0)$.

The dispersion matrix is positive definite and is defined such that
the degree of spatial correlation across sites of a physical phenomenon
is given in terms of the separation between locations as 
\begin{equation}
\left[\Sigma\right]_{ij}=\alpha_{0}\exp\left[-\beta^{-1}\Vert\mathcal{S}_{i}-\mathcal{S}_{j}\Vert_{2}^{2}\right]+\alpha_{1}\delta_{ij},\label{eq:spatial-correlation-sensor-networks}
\end{equation}
\noindent where $\Vert\cdot\Vert_{2}$ is the L2-norm, $\delta_{ij}$
the Kronecker symbol, $\alpha_{0},\alpha_{1}\in\mathbb{R}$, and $\mathcal{S}_{m}\in\mathbb{R}^{2}$
are the physical locations of the sensors for $m=1,\dots,N_{s}$.

For this example, we compare the performance of following filters:
\begin{itemize}
\item the Sequential Importance Resampling filter (SIR); 
\item the block SIR filter, which partitions the state space into separate
subspaces of smaller dimensions (blocks of 4 sensors each) and run
a particle filter on each subspace \citep{Rebeschini2015};
\item the Sequential manifold Metropolis-Adjusted Algorithm (SmMALA) filter~\citep{Girolami2011,Septier2016};
\item the Sequential manifold Hamiltonian Monte Carlo (SmHMC) filter~\citep{Girolami2011,Septier2016};
\item the Stochastic Particle Flow, Gaussian sum (SPF-GS).
\end{itemize}
These filters are compared for two cases:
\begin{itemize}
\item Gaussian state process and Gaussian likelihood;
\item Skewed-t state process and Poisson-distributed observations.
\end{itemize}
Note that SmMALA and SmHMC are chosen because we perceive they constitute
two of the best sequential MCMC filters that exist. It is essential
to justify why we have not included annealed importance sampling (AIS)
\citep{Neal2001} and SMC samplers \citep{DelMoral2006} in our comparisons.
Although these techniques are built on fast mixing Markov chains,
their filtering procedures operate on the joint space along the complete
path of samples. This makes them highly prone to the curse of dimensionality
for long-time horizons. A careful explanation of this issue can be
found in \citep{Klaas2012} (and Section \ref{subsec:Marginal-particle-filtering}),
but the key point is that as dimension increases, it becomes increasingly
important to avoid consideration of the path. SMC samplers could be
adapted to filter in the marginal space (at the cost of up to $\mathcal{O}(N^{3})$
evaluations) but this would require an ad-hoc approximation to the
target pdf. Developing the approximations that would be needed to
enable SMC samplers to be applied to the problem we consider is not
the focus of our paper. We have compared performance of SPF-GS with
state-of-the-art techniques (SmMALA and SmHMC) that consider the marginal
distribution as well as techniques (SIR and block SIR) that we perceive
to be good examples of the class of algorithms, which also includes
AIS and SMC samplers, that consider the joint space of the complete
sample path and can be applied, without modifications, to the problem
we are focused on solving. 

These filters are compared for two cases:
\begin{itemize}
\item Gaussian state process and Gaussian likelihood; 
\item Skewed-t state process and Poisson-distributed observations. 
\end{itemize}
The implementations of SmMALA and SmHMC used are exactly as made available
by Septier \& Peters \citep{Septier2016} on their web page\footnote{Code available at \href{http://pagesperso.telecom-lille.fr/septier/software.html}{http://pagesperso.telecom-lille.fr/septier/software.html}. }.
These algorithms make use of a refinement step of the state \citep{Septier2016}
at the current time, performed with a random partitioning of size
4, by using the empirical approximation of the previous posterior
distribution as proposal distribution. 

\subsubsection{Results}

\subsubsection*{Gaussian State Process and Gaussian Likelihood}

We first consider a trivial special case of the GH family as the transition
density, namely the multivariate normal distribution. In this setting,
each sensor measures the attribute of a physical process with some
Gaussian noise. The resulting model is given by
\begin{align}
p_{t}(\mathrm{x}_{k}|\mathrm{x}_{k-1}) & =\mathcal{N}(\mathrm{x}_{k};\,\alpha\mathrm{x}_{k-1},\Sigma),\nonumber \\
p(\mathrm{y}_{k}|\mathrm{x}_{k}) & =\mathcal{N}(\mathrm{y}_{k};\,\hphantom{\alpha}\mathrm{x}_{k},_{\hphantom{-1}}R),\label{eq:sensor-network-model-case-1}
\end{align}
\noindent where $R=\sigma_{y}^{2}\mathbb{I}_{n_{x}}$, and the following
model parameters are used: $\alpha=0.9$, $\sigma_{y}^{2}=2$, and
with the dispersion matrix constructed using $\alpha_{0}=3$, $\alpha_{1}=0.01$,
$\beta=20$. When performing the comparison between the filters, we
use as a reference the estimates provided by the Kalman filter (which
is optimal in this special case). For the SmMALA and SmHMC, the proposed
metric tensor is given by $G\left(\mathrm{x}_{k}\right)=R^{-1}+\Sigma^{-1}$. 

The methodology presented in \citep{Septier2016} for performance
evaluation was reproduced. Instead of presenting the baseline (provided
by the Kalman filter) performance explicitly, the accuracy of each
filter is evaluated with respect to the baseline performance, i.e.,
we observe the difference between the log mean square error for each
filter against the log mean square error for the Kalman filter. The
relative log mean square error, log normalized-estimation error squared
(NEES) and average computation time (per time step) of all filters,
for different numbers of sensors in the grid, are shown in Figures~\ref{fig:MSE-sensor-network-linear},
\ref{fig:NEES-sensor-network-linear}, \ref{fig:Computation-sensor-network-linear}
respectively. The mean square error of the estimates were computed
over 100 Monte Carlo runs, with the particle-based filters using 200
samples. The step size adopted for the sequential MCMC filters is
$\Delta\lambda=0.5$ and the number of steps obtained as $L=L_{0}+N$,
where $L_{0}=0.2N$ is the number of steps for the burn-in phase,
whereas for the SPF-GS we applied the empirical rules for time horizon
and step step size as presented in subsection \ref{subsec:Selection-of-time}.
From Figures~\ref{fig:MSE-sensor-network-linear}, \ref{fig:NEES-sensor-network-linear},
\ref{fig:Computation-sensor-network-linear}, we can note that for
this example:
\begin{itemize}
\item The performances of all sequential MCMC filters are in accordance
with the results shown in \citep{Septier2016}. 
\item The SPF-GS outperforms all other filters in terms of mean square error
and normalized-estimation error squared. 
\item The SPF-GS demands the highest computational effort when the number
of sensors is small, and its computing time scales better than that
of SmMALA and worse than that of SmHMC for higher dimensions. 
\end{itemize}
\begin{center}
\begin{figure}
\begin{centering}
\includegraphics[width=1\textwidth]{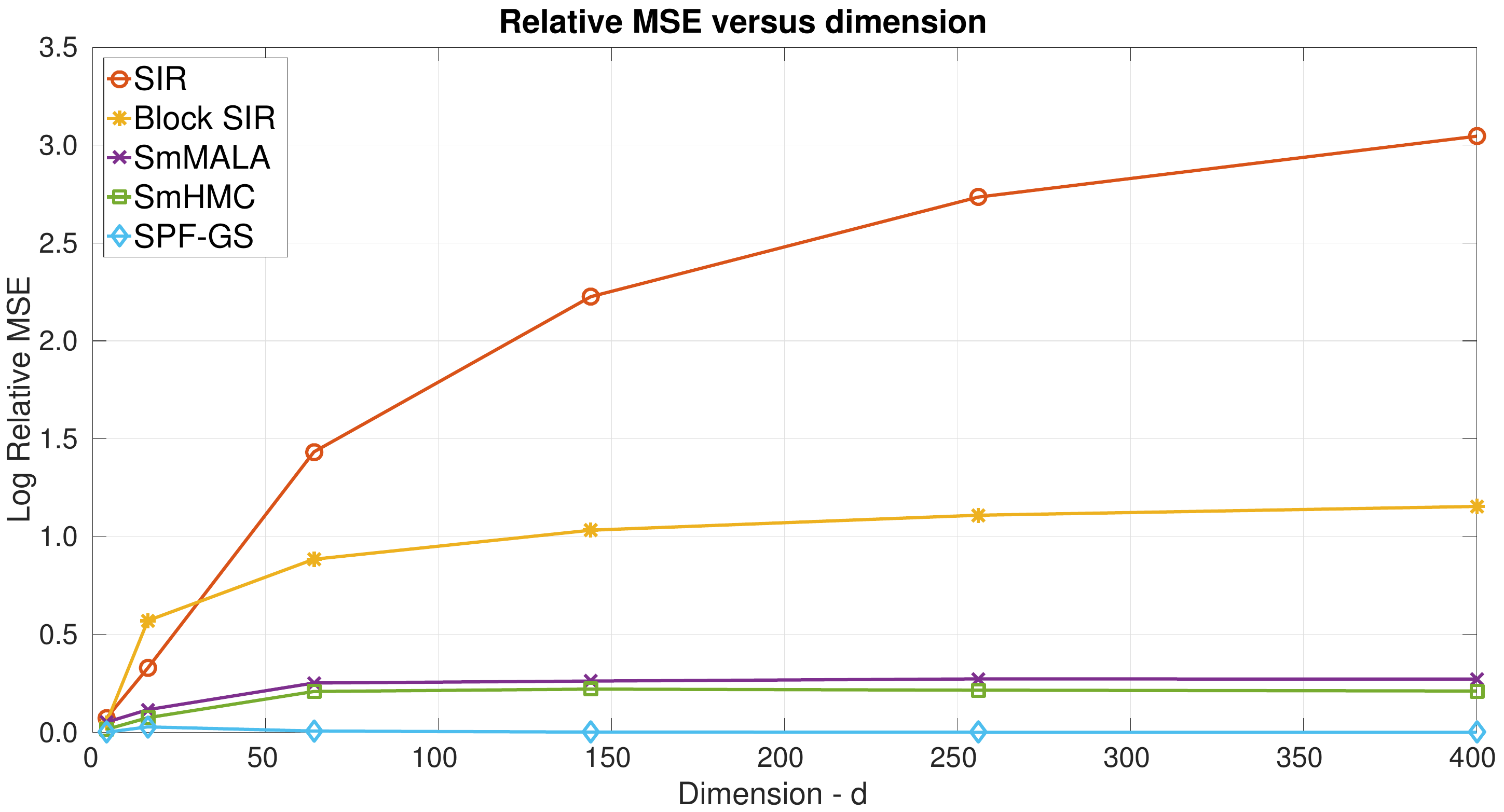}
\par\end{centering}
\caption{Relative MSE for the linear, Gaussian sensor network example\label{fig:MSE-sensor-network-linear}}
\end{figure}
\begin{figure}
\begin{centering}
\includegraphics[width=1\textwidth]{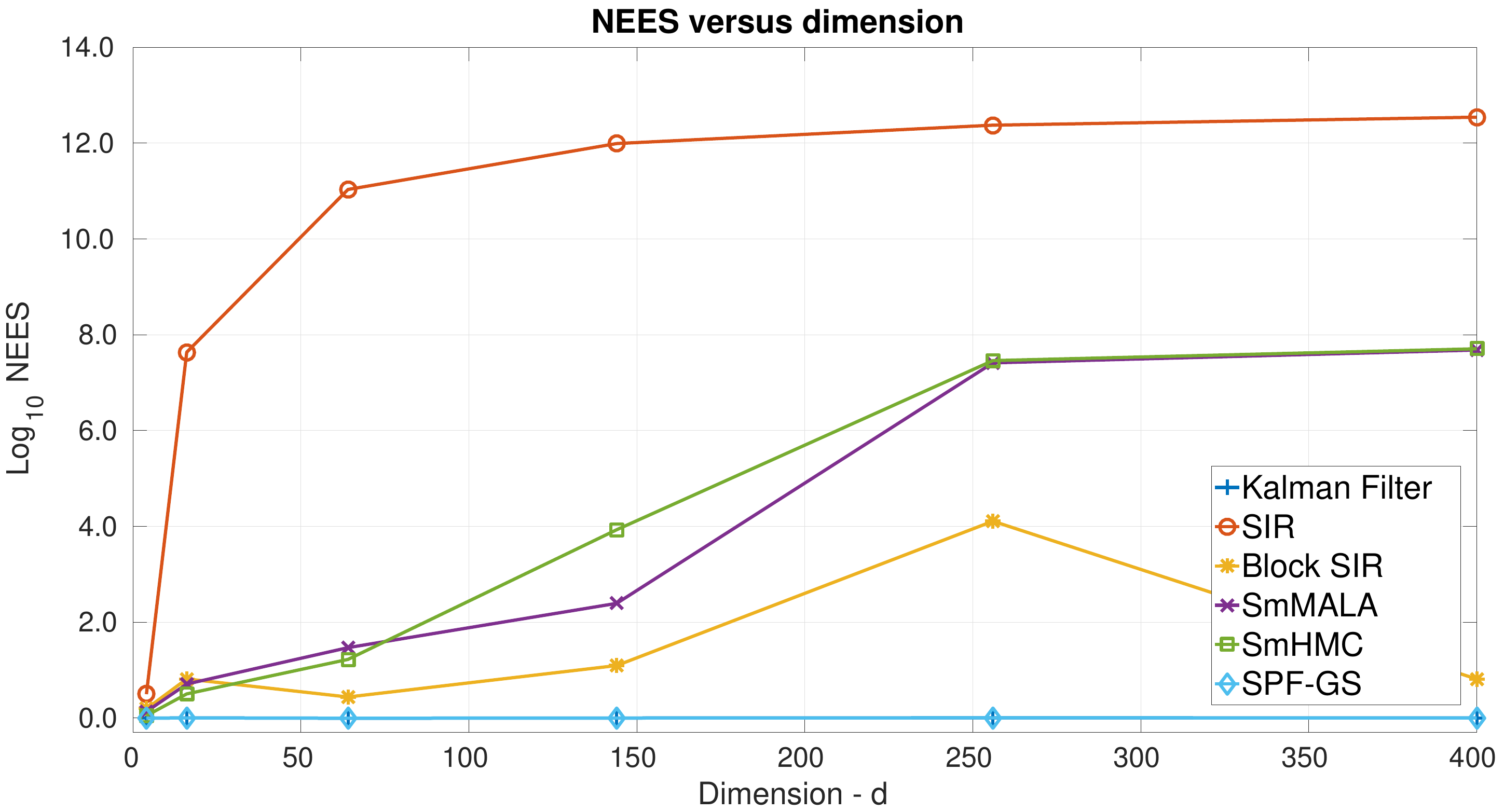}
\par\end{centering}
\caption{NEES for the linear, Gaussian sensor network example\label{fig:NEES-sensor-network-linear}}
\end{figure}
\begin{figure}
\begin{centering}
\includegraphics[width=1\textwidth]{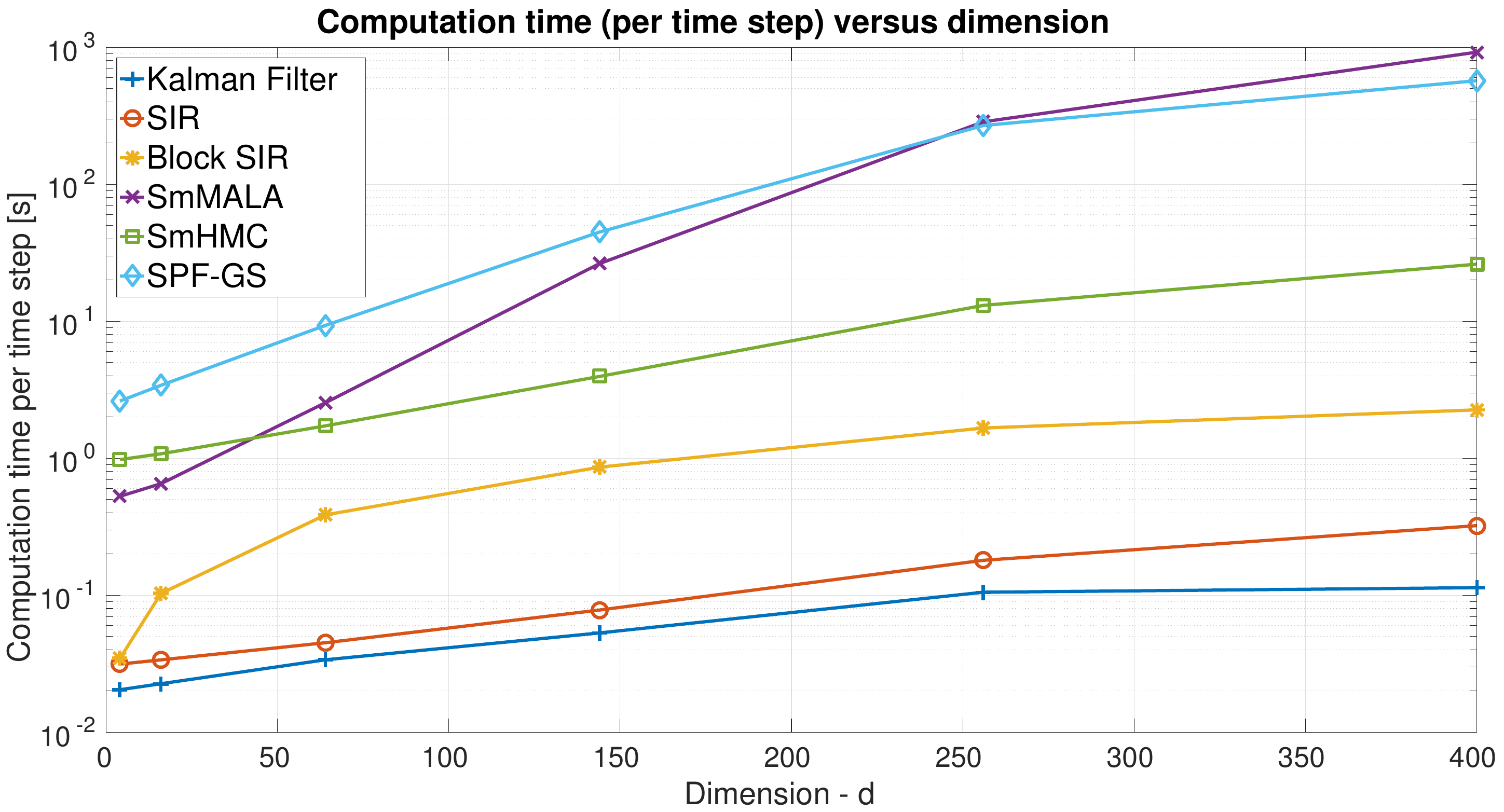}
\par\end{centering}
\caption{Average computation time for the linear, Gaussian sensor network example
\label{fig:Computation-sensor-network-linear}}
\end{figure}
\par\end{center}

The results suggest that SPF-GS is the most accurate among the compared
methods. It is unambiguous that the consistency of the estimates produced
by the SPF-GS is better than that by all other filters: the NEES for
SPF-GS is very close to one (from above) for all evaluated dimensions.
It may be well possible that better results could be achieved for
the sequential MCMC filters by carefully choosing the step size and
number of steps. This would slightly change the performance indexes.
However, we strongly believe that such changes would not be enough
to modify the conclusions. Additional tests with the SPF-GS demonstrate
the computational cost can be directly traded with estimation accuracy.
By allowing more steps, via the criteria of Section \ref{subsec:Selection-of-time},
SPF-GS is very close to optimal as depicted in Figure \ref{fig:MSE-sensor-network-linear}.
On the other hand, if one fixes the step size to $\Delta\lambda=0.5$
and number of steps to $L=20$, the results are as presented in Figures~\ref{fig:MSE-sensor-network-linear}
and \ref{fig:Computation-sensor-network-linear}, where it is becomes
clear computational cost alleviation at the expense of slightly degrading
mean square error.
\begin{figure}
\begin{centering}
\includegraphics[width=1\textwidth]{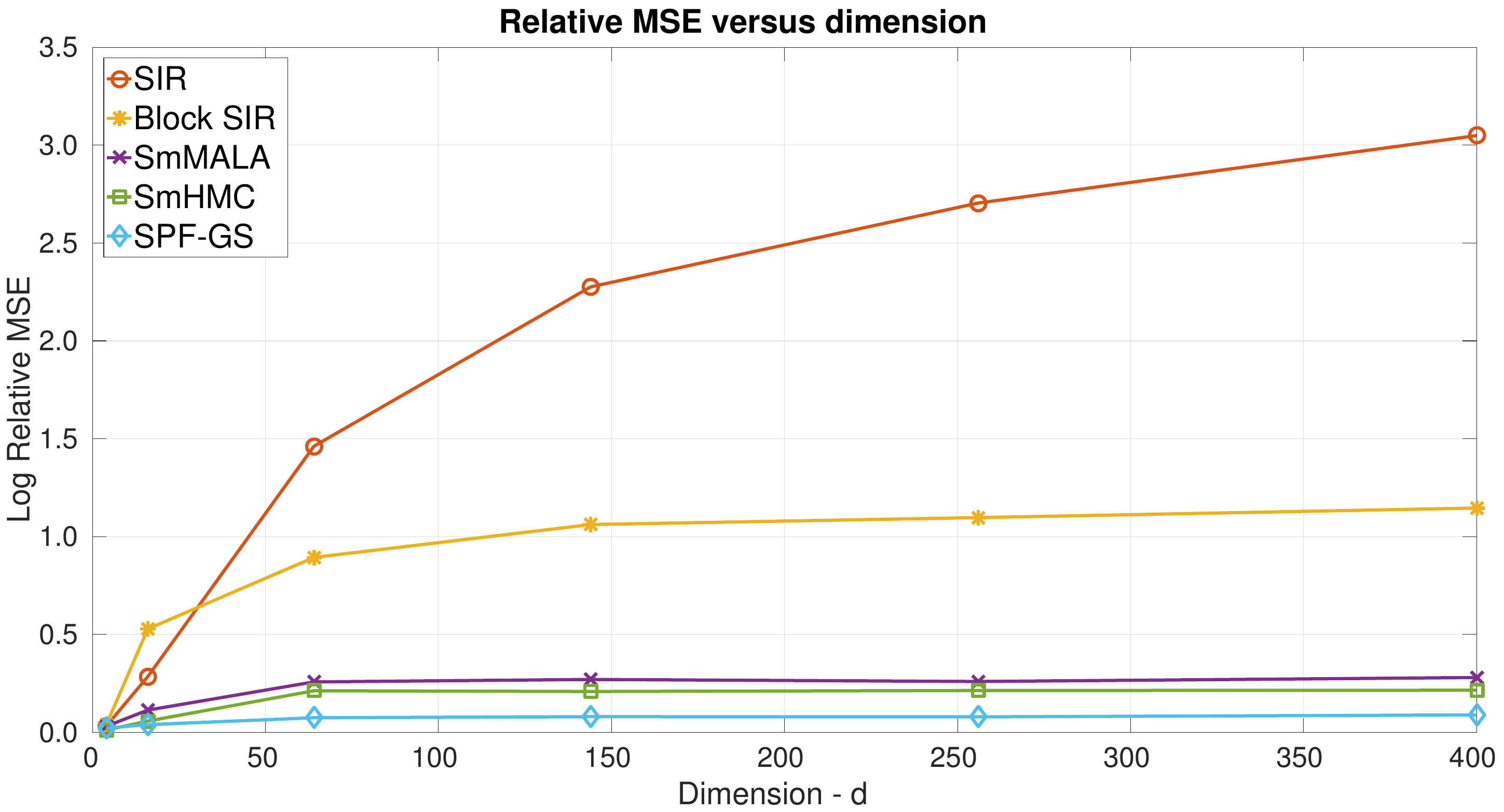}
\par\end{centering}
\caption{Relative MSE for the linear, Gaussian example (SPF-GS with $\Delta\lambda=0.5$,
$L=20$)\label{fig:MSE-sensor-network-linear-1}}
\end{figure}
\begin{figure}
\begin{centering}
\includegraphics[width=1\textwidth]{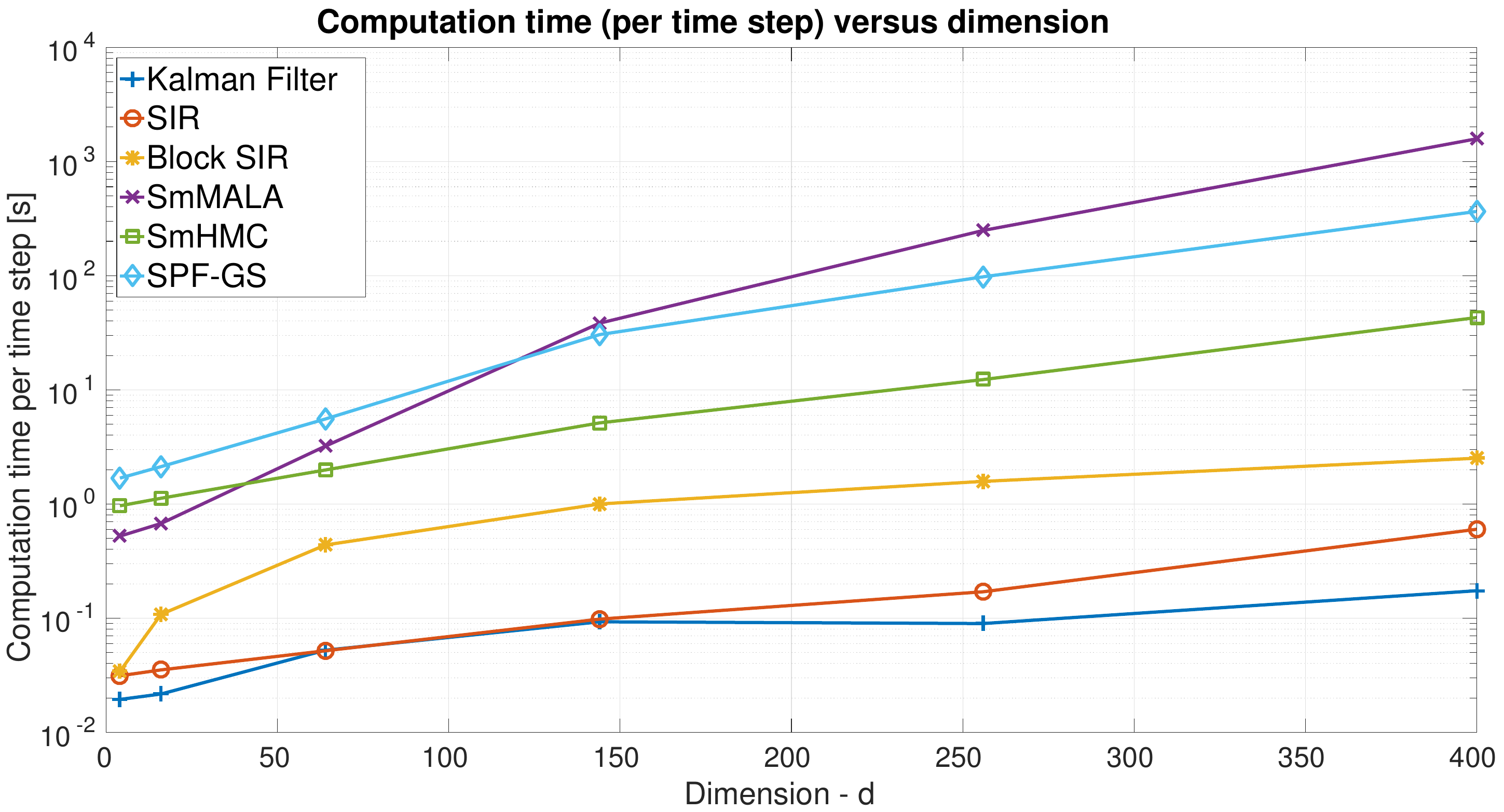}
\par\end{centering}
\caption{Computation time for the linear, Gaussian example (SPF-GS with $\Delta\lambda=0.5$,
$L=20$) \label{fig:Computation-sensor-network-linear-1}}
\end{figure}

\subsubsection*{Skewed-t State Process with Poisson-Distributed Observations}

A high-dimensional non-linear and non-Gaussian state-space model is
now studied. The transition kernel is proposed to be a multivariate
GH skewed-t density described by (\ref{eq:transition-pdf-sensor-networks})
with $c_{1}=-\nu/2$, $c_{2}=\nu$ and $c_{3}=0$. The likelihood
function is assumed to be a Poisson distribution, highly non-linear
on the state $\mathrm{x}_{k}$, given by
\begin{equation}
p(\mathrm{y}_{k}|\mathrm{x}_{k})=\prod_{j=1}^{n_{x}}\frac{\lambda_{j}(\mathrm{x}_{j,k})^{\mathrm{y}_{j,k}}}{\mathrm{y}_{j,k}!}e^{-\lambda_{j}(\mathrm{x}_{j,k})},\quad\lambda_{j}(\mathrm{x}_{j,k})=m_{1}e^{m_{2}\mathrm{x}_{j,k}},\label{eq:sensor-network-model-case-2}
\end{equation}
\noindent such that $\mathrm{x}_{k}=(\mathrm{x}_{1,k},\dots,\mathrm{x}_{n_{x},k})^{T}$
and $\mathrm{y}_{k}=(\mathrm{y}_{1,k},\dots,\mathrm{y}_{n_{x},k})^{T}$.
The model parameters are fixed as $m_{1}=1$, $m_{2}=1/3$, $\alpha=0.9$,
$\nu=7$, $\mathrm{\gamma}=0.3\mathbbm{1}_{n_{x}\times1}$ with the
dispersion matrix constructed using $\alpha_{0}=3$, $\alpha_{1}=0.01$
and $\beta=20$. The implied prior density is not log concave, and
thus the tensor metric that defines the diffusion coefficient for
the sequential MCMC algorithms is modified according to 
\begin{equation}
G(\mathrm{x}_{k})=\Lambda(\mathrm{x}_{k})+\tilde{\Sigma}^{-1},\label{eq:tensor-metric-skewed-t-poisson}
\end{equation}
\noindent where
\begin{align}
\Lambda(\mathrm{x}_{k}) & =m_{1}m_{2}^{2}\left(\begin{array}{ccc}
e^{m_{2}\mathrm{x}_{1,k}} &  & 0\\
 & \ddots\\
0 &  & e^{m_{2}\mathrm{x}_{n_{x},k}}
\end{array}\right),\label{eq:Hessian-likelihood}\\
\tilde{\Sigma}= & \frac{\nu}{\nu-2}\Sigma+\frac{\nu\text{\texttwosuperior}}{(2\nu-8)(\nicefrac{1}{2}-1)^{2}}\mathrm{\gamma\mathrm{\gamma^{T},\quad\nu>4.}}\label{eq:Hessian-prior}
\end{align}

For this problem, the local linearization of stochastic particle flow
around a probability mass $\mathrm{x}_{l}$, analogous to (\ref{eq:Jacobian-of-the-drift-term})
and (\ref{eq:offset-of-the-drift-term}), is given by 
\begin{align}
\frac{1}{2}D\nabla_{\mathrm{x}}\log\tilde{\pi}(\mathrm{x}) & \approx C(\mathrm{x}_{l})\cdot\mathrm{x}+c(\mathrm{x}_{l}),\nonumber \\
C(\mathrm{x}_{l}) & =\frac{1}{2}D\left(-\mathrm{V}(\mathrm{x}_{l})^{-1}-P_{k|k-1}^{-1}\right),\label{eq:sensor-network-local-linearization-C}\\
c(\mathrm{x}_{l}) & =\frac{1}{2}D\left(m_{2}\mathrm{y}_{k}-\mathrm{v}(\mathrm{x}_{l})+\mathrm{V}(\mathrm{x}_{l})^{-1}\mathrm{x}_{l}+P_{k|k-1}^{-1}\alpha\mu_{m,k-1}\right),\label{eq:sensor-network-local-linearization-c}
\end{align}
\noindent where $\mathrm{V}(\mathrm{x}_{l})^{-1}=\Lambda(\mathrm{x}_{l})$,
$\mathrm{v}(\mathrm{x}_{l})=m_{1}m_{2}e^{m_{2}\mathrm{x}_{l}}$, and
\begin{equation}
P_{k|k-1}=\mathbb{E}\left[(\mathrm{x}_{k|k-1}-\alpha\mu_{m,k-1})(\mathrm{x}_{k|k-1}-\alpha\mu_{m,k-1})^{T}\right].\label{eq:sensor-network-predicted-covariance}
\end{equation}

\begin{figure}
\begin{centering}
\includegraphics[width=1\textwidth]{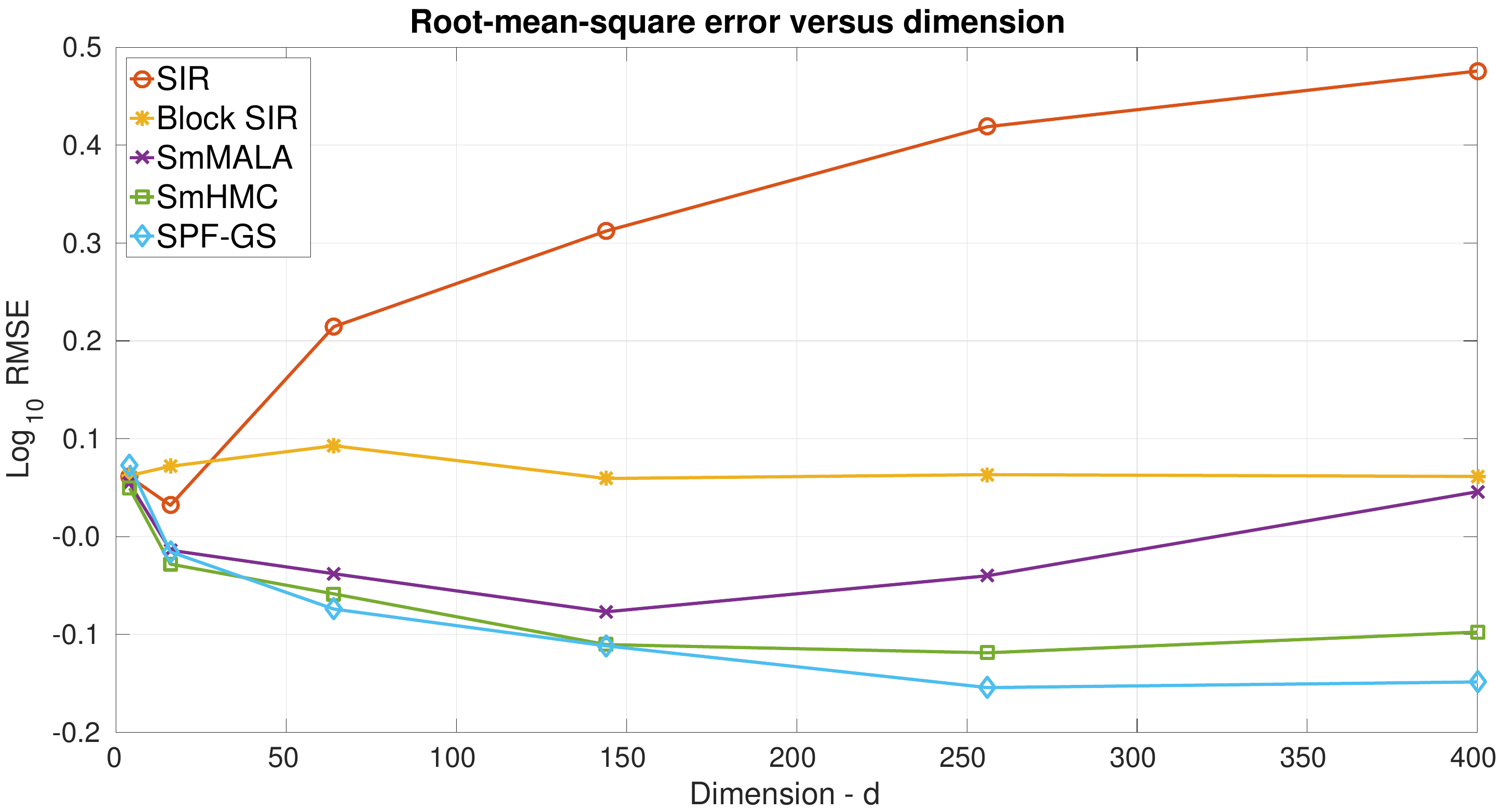}
\par\end{centering}
\caption{RMSE for the nonlinear, non-Gaussian sensor network example\label{fig:RMSE-sensor-network-nonlinear}}
\end{figure}

\begin{figure}
\begin{centering}
\includegraphics[width=1\textwidth]{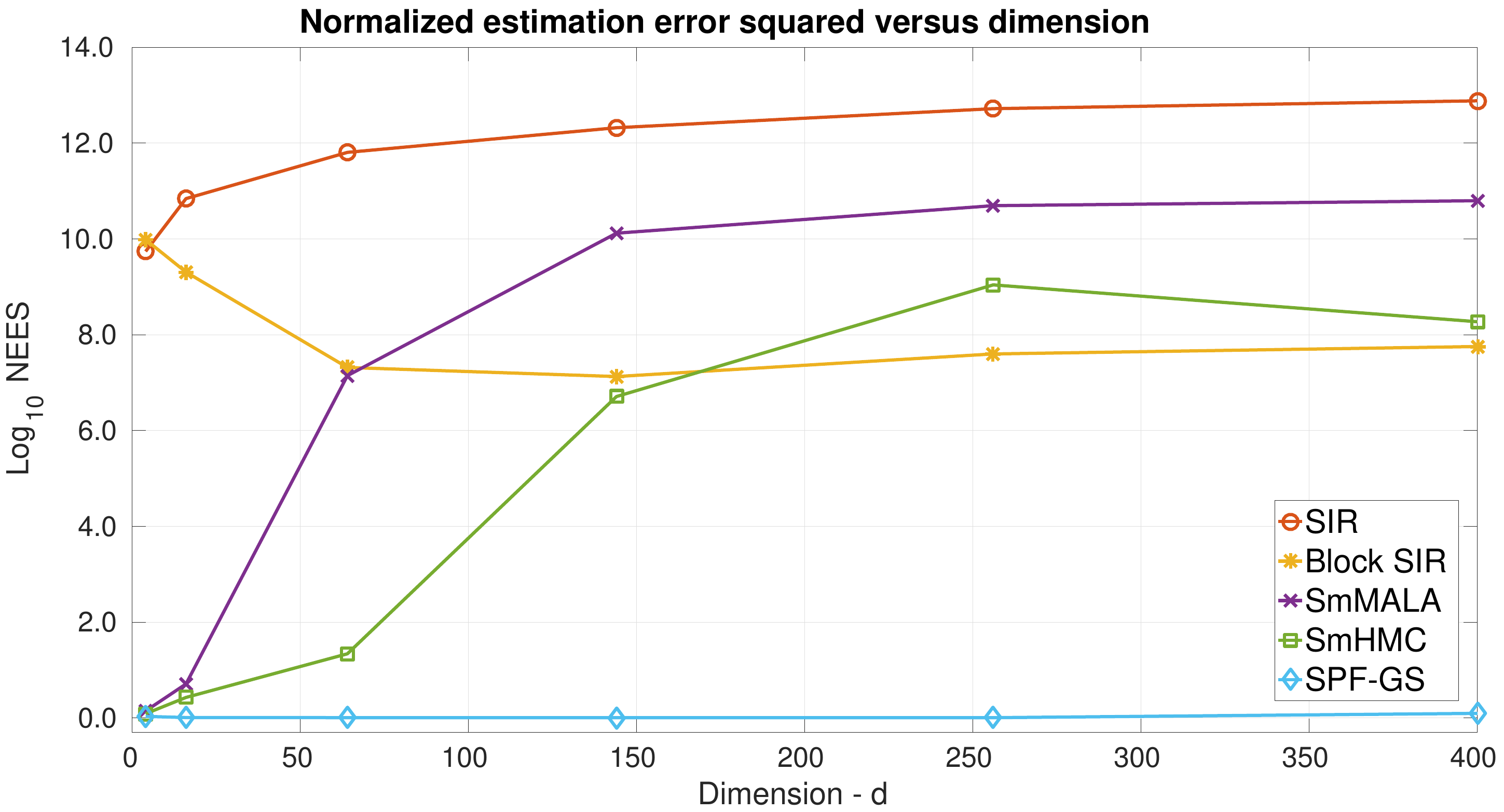}
\par\end{centering}
\caption{NEES for the nonlinear, non-Gaussian sensor network example\label{fig:NEES-sensor-network-nonlinear}}
\end{figure}

\begin{figure}
\begin{centering}
\includegraphics[width=1\textwidth]{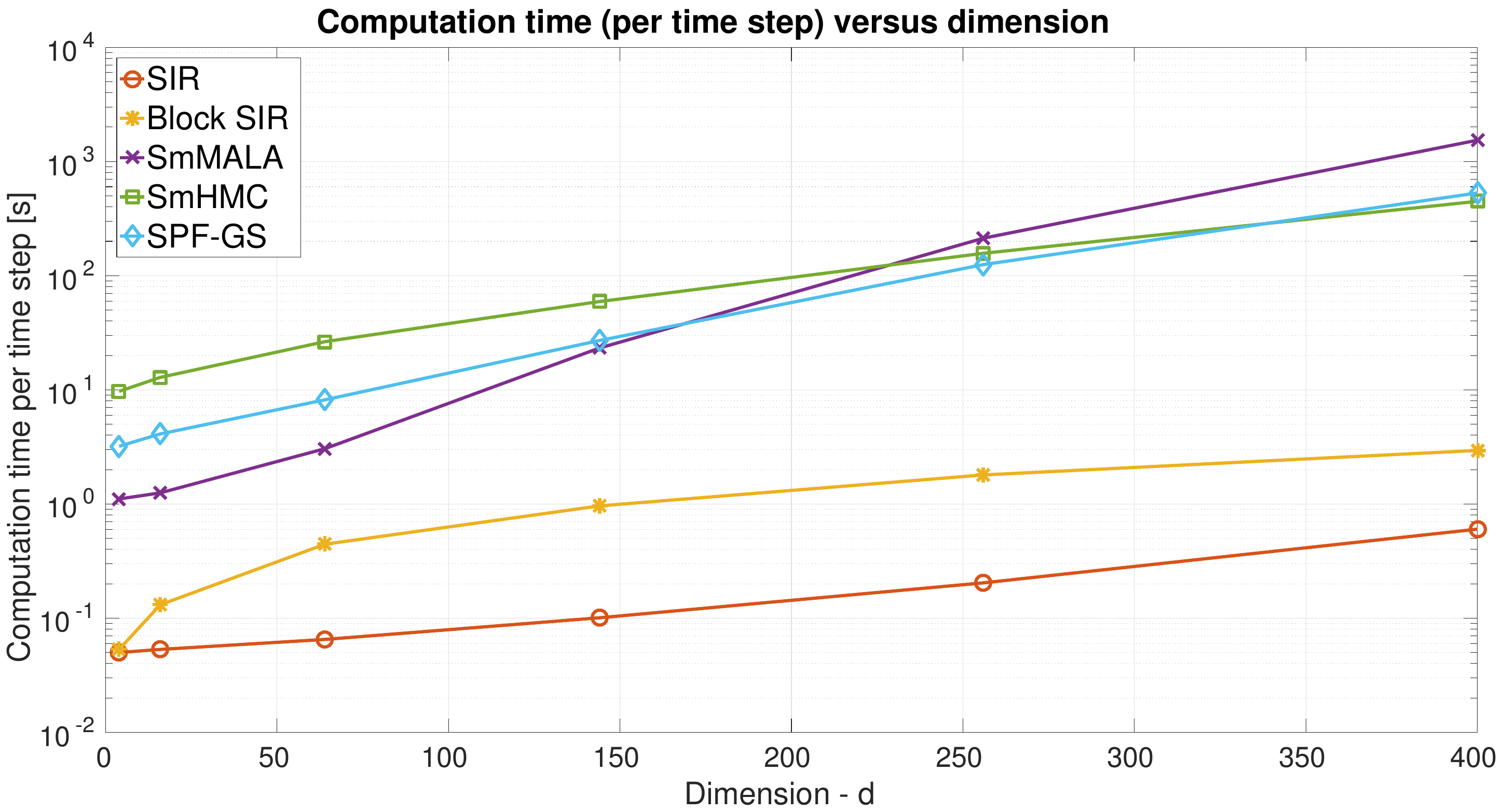}
\par\end{centering}
\caption{Average computation time for the nonlinear, non-Gaussian sensor network
example \label{fig:Computation-sensor-network-nonlinear}}
\end{figure}

\begin{figure}
\noindent \setlength\tabcolsep{2 pt}
\noindent \begin{centering}
\begin{tabular}{>{\centering}m{0.01\columnwidth}>{\centering}m{0.14\columnwidth}>{\centering}m{0.14\columnwidth}|>{\centering}m{0.14\columnwidth}>{\centering}m{0.14\columnwidth}|>{\centering}m{0.14\columnwidth}>{\centering}m{0.14\columnwidth}}
 & \multicolumn{2}{c|}{Time $k=2$} & \multicolumn{2}{c|}{Time $k=4$} & \multicolumn{2}{c}{Time $k=6$}\tabularnewline
 & State $x(2)$ & Obs. $y(2)$ & State $x(4)$ & Obs. $y(4)$ & State $x(6)$ & Obs. $y(6)$\tabularnewline
 & \includegraphics[clip,width=0.14\columnwidth]{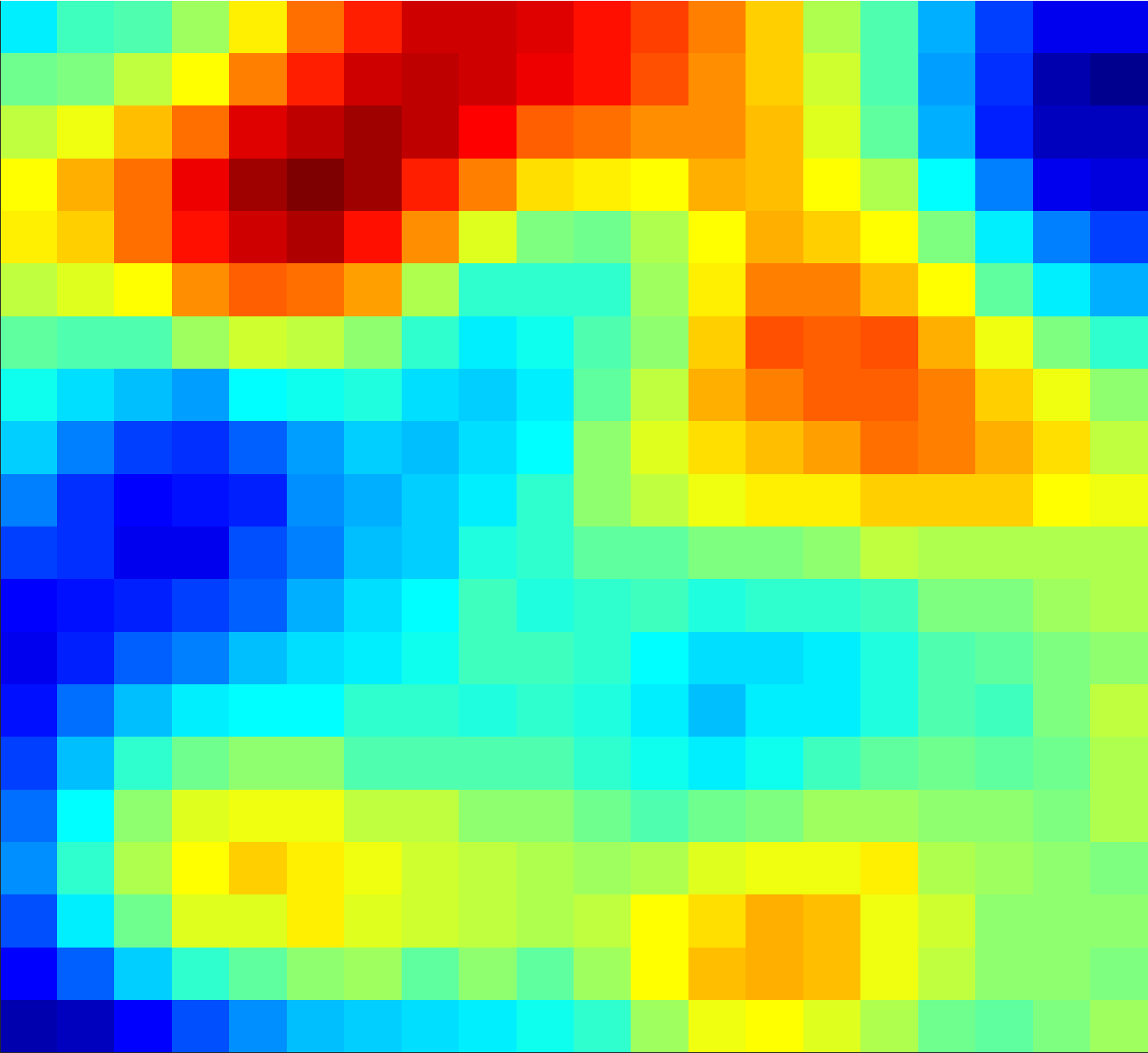} & \includegraphics[width=0.14\columnwidth]{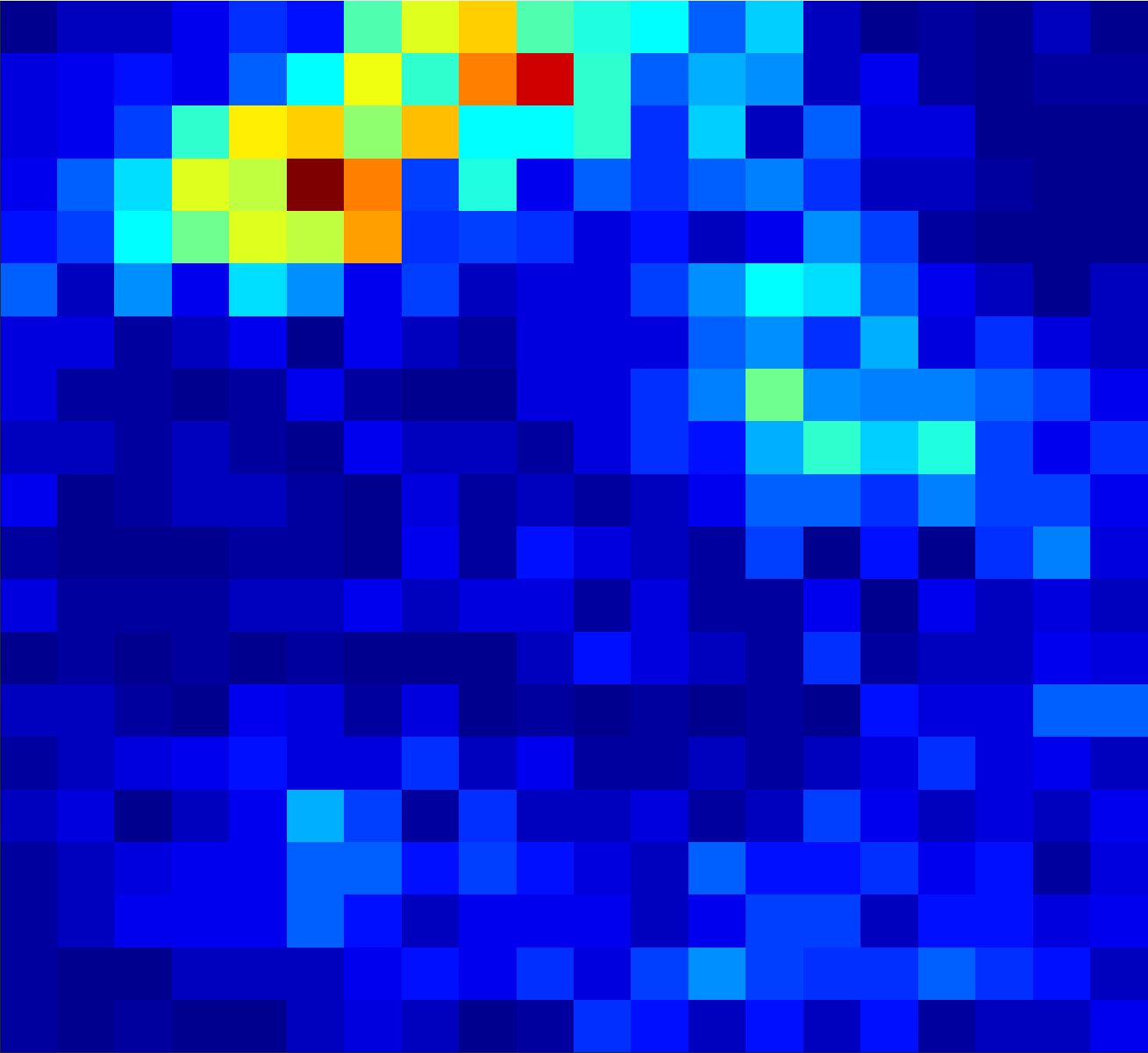} & \includegraphics[width=0.14\columnwidth]{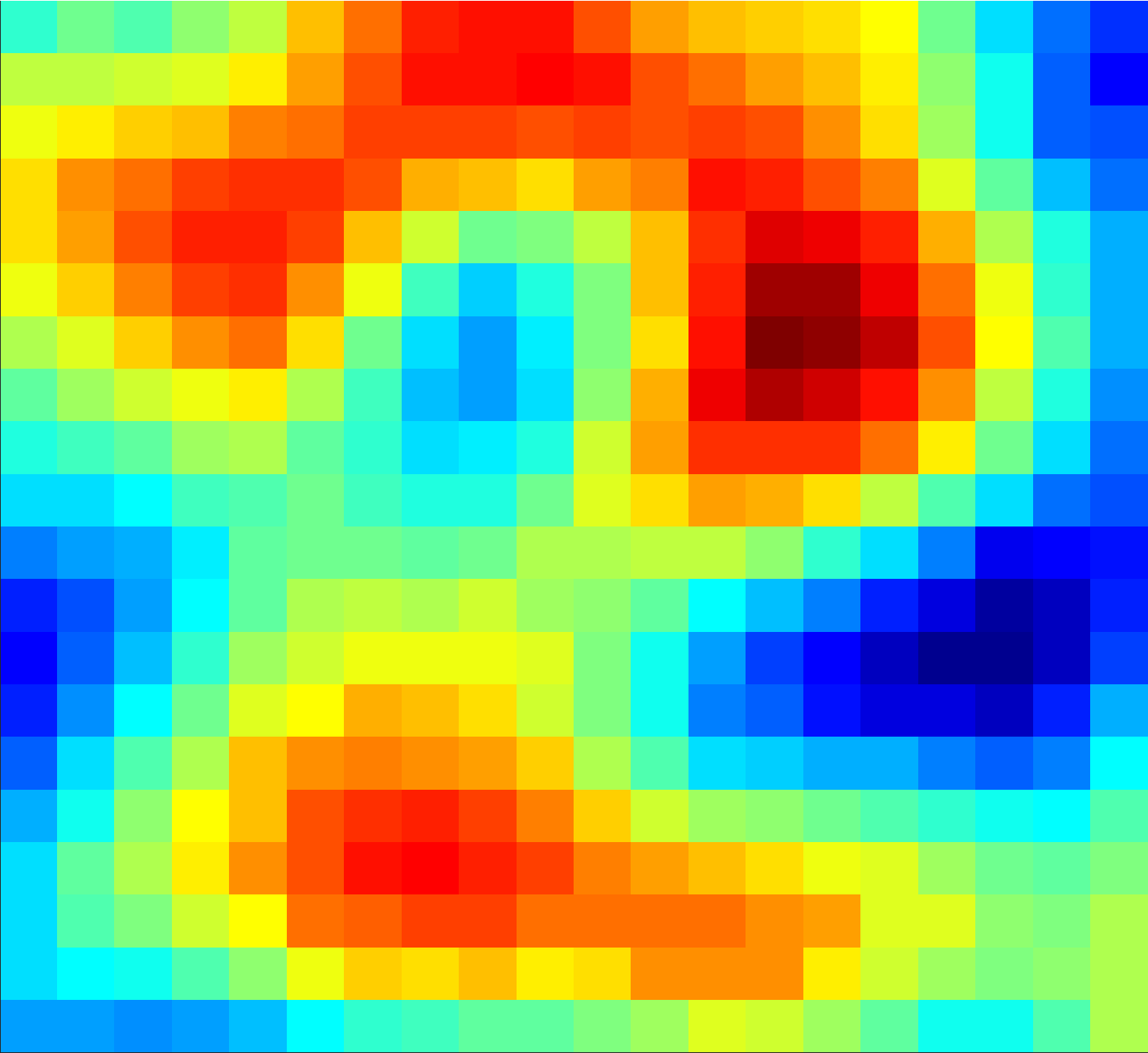} & \includegraphics[width=0.14\columnwidth]{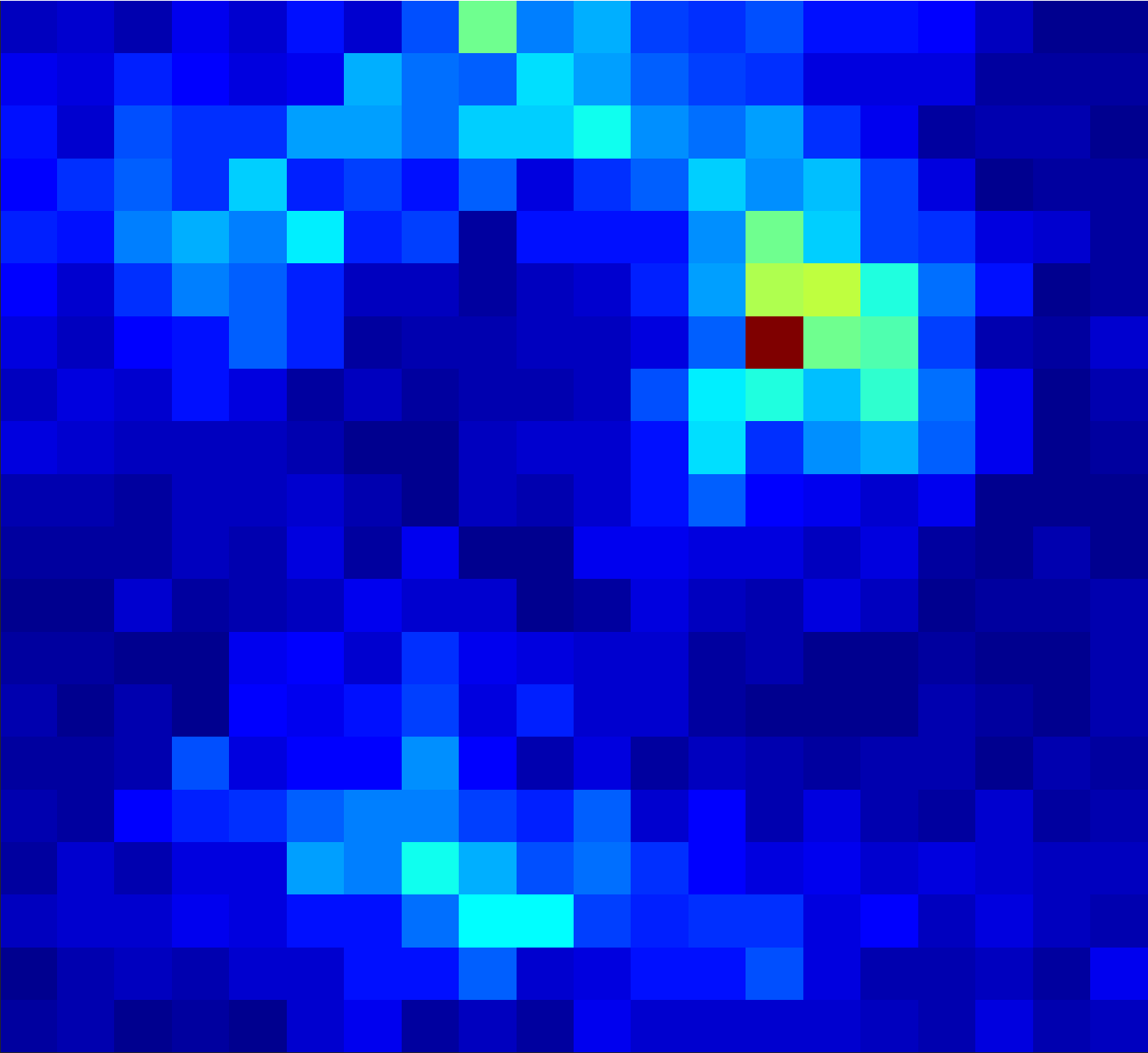} & \includegraphics[width=0.14\columnwidth]{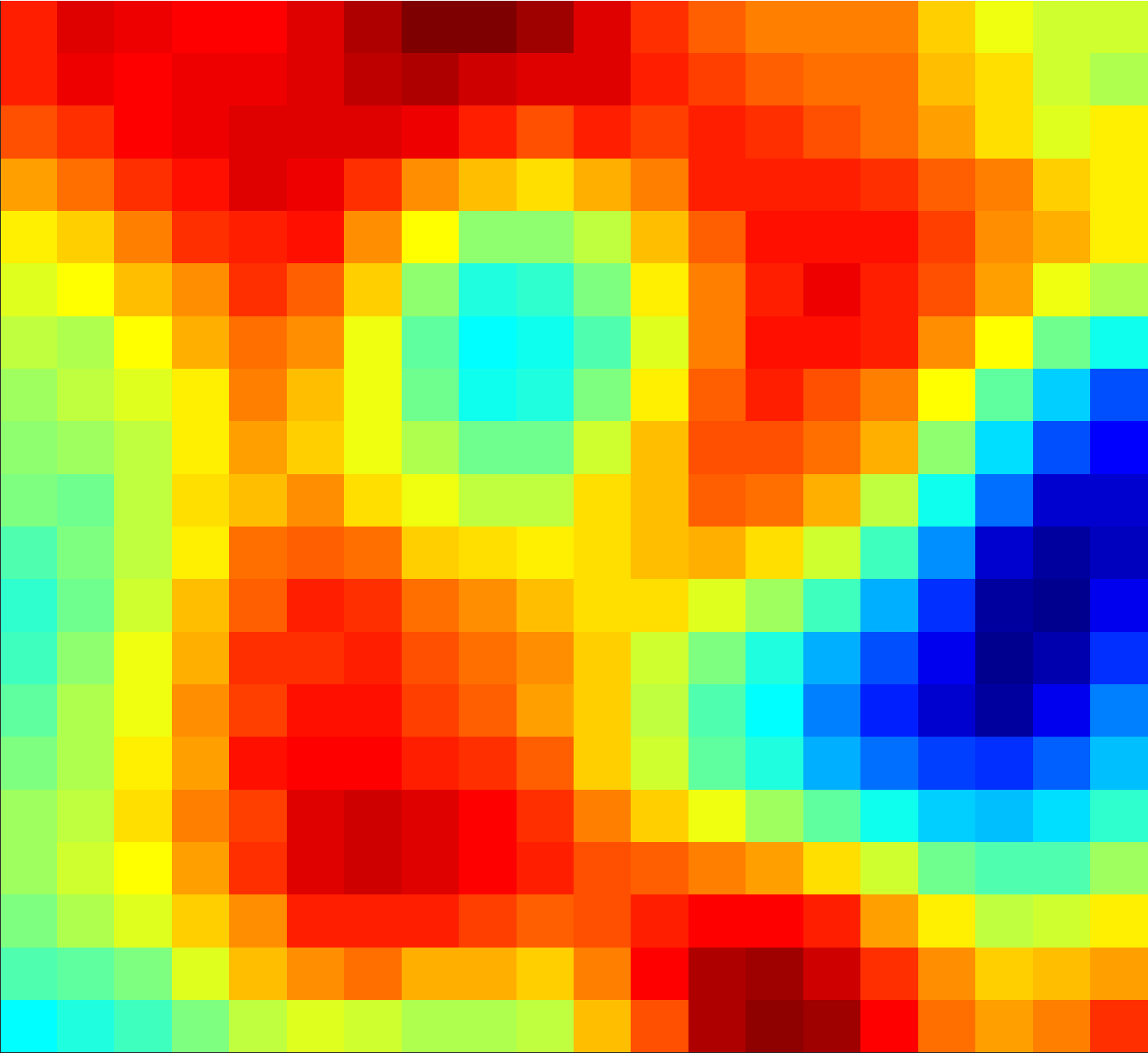} & \includegraphics[width=0.14\columnwidth]{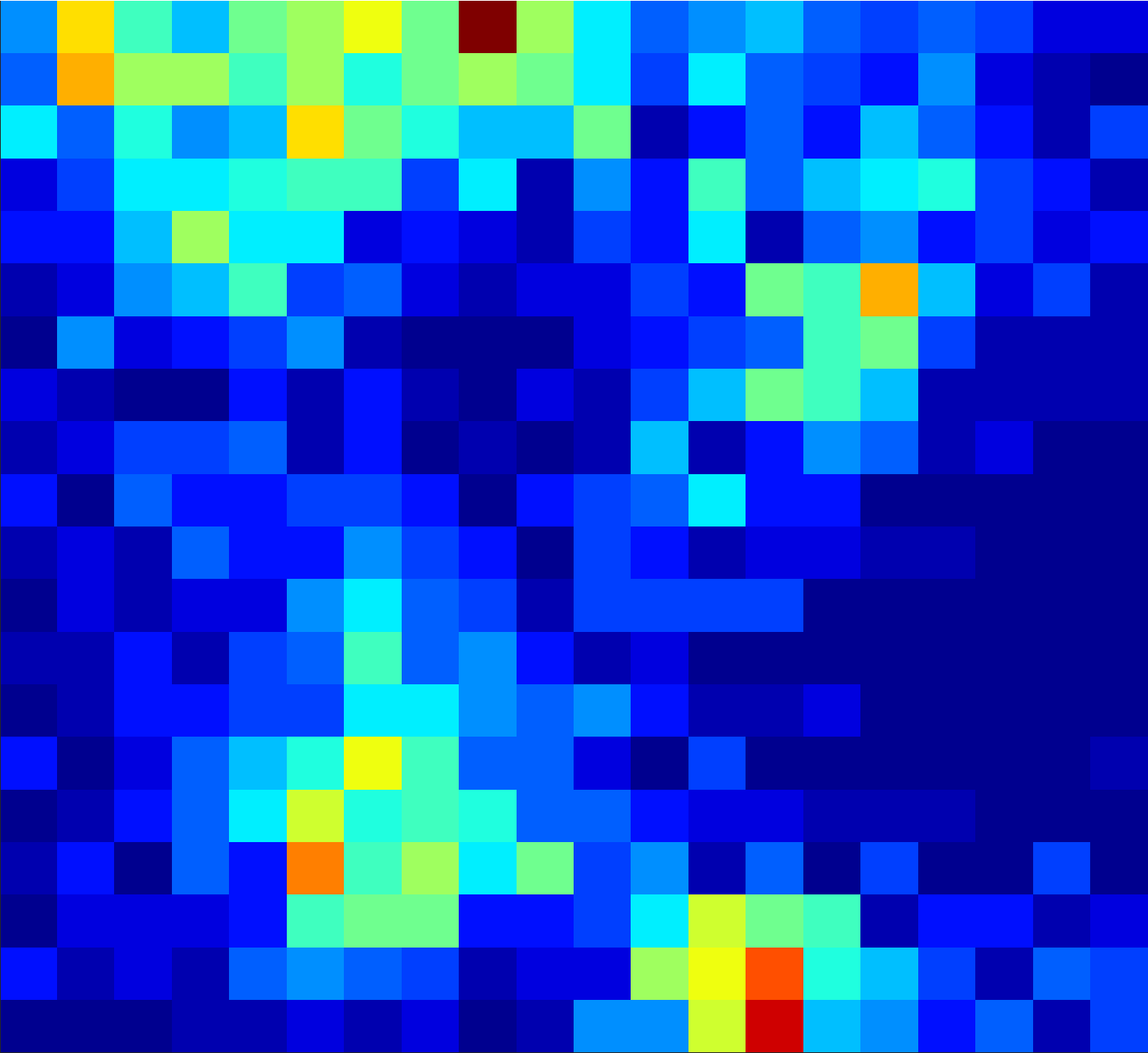}\tabularnewline
 & Post. Mean & Post. Var. & Post. Mean & Post. Var. & Post. Mean & Post. Var.\tabularnewline
\rotatebox[origin=c]{90}{SPF-GS} & \includegraphics[width=0.14\columnwidth]{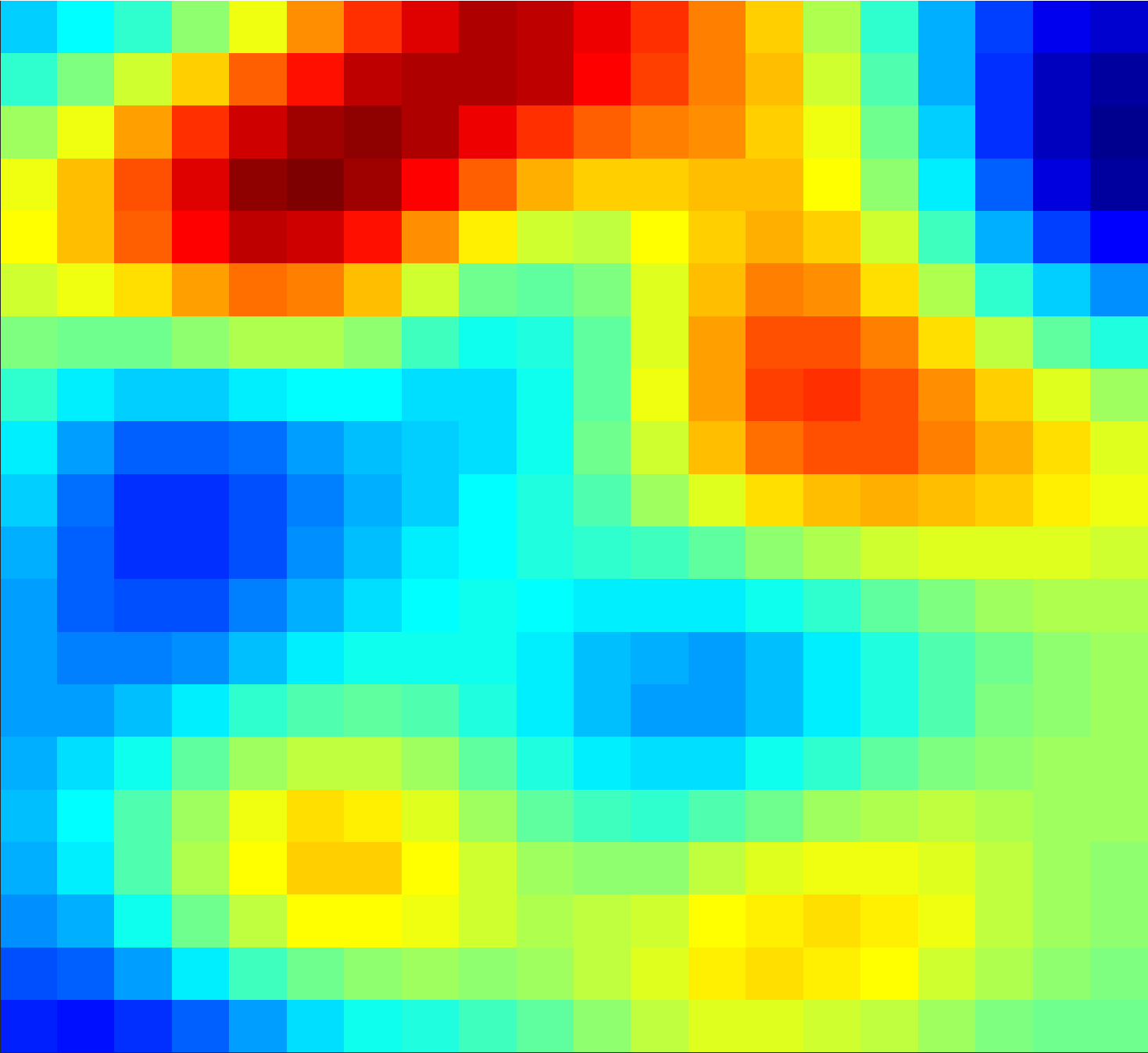} & \includegraphics[width=0.14\columnwidth]{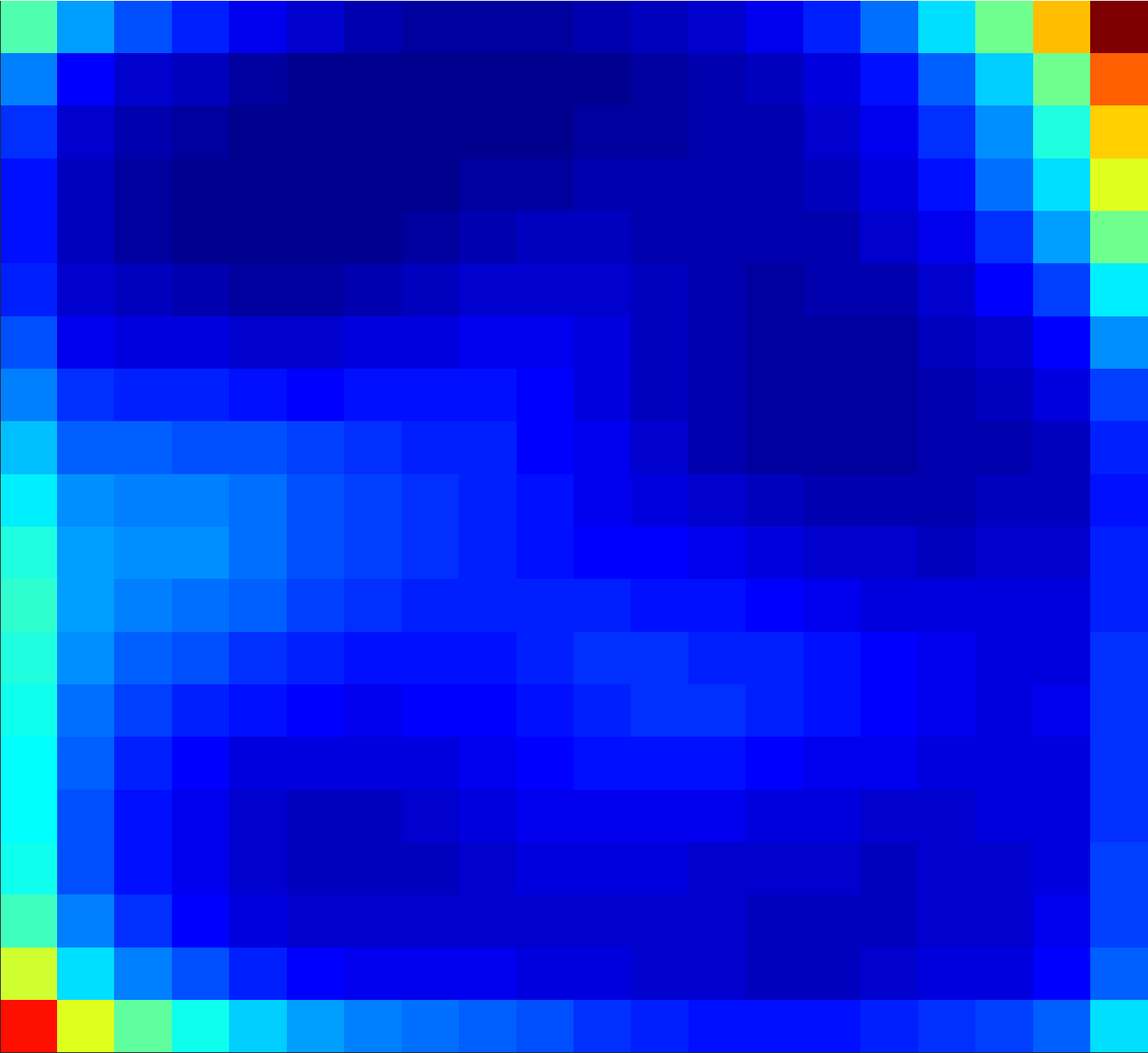} & \includegraphics[width=0.14\columnwidth]{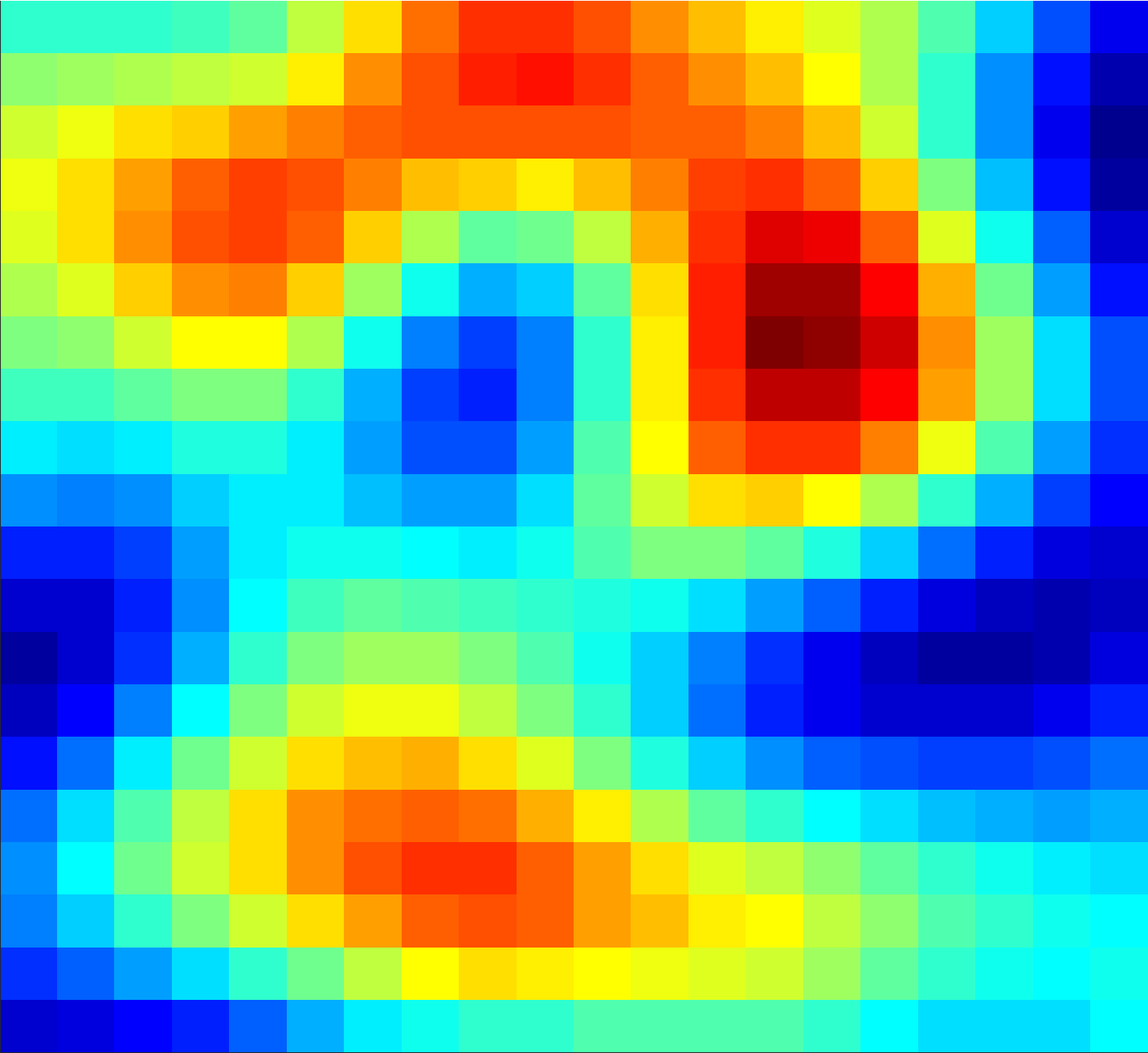} & \includegraphics[width=0.14\columnwidth]{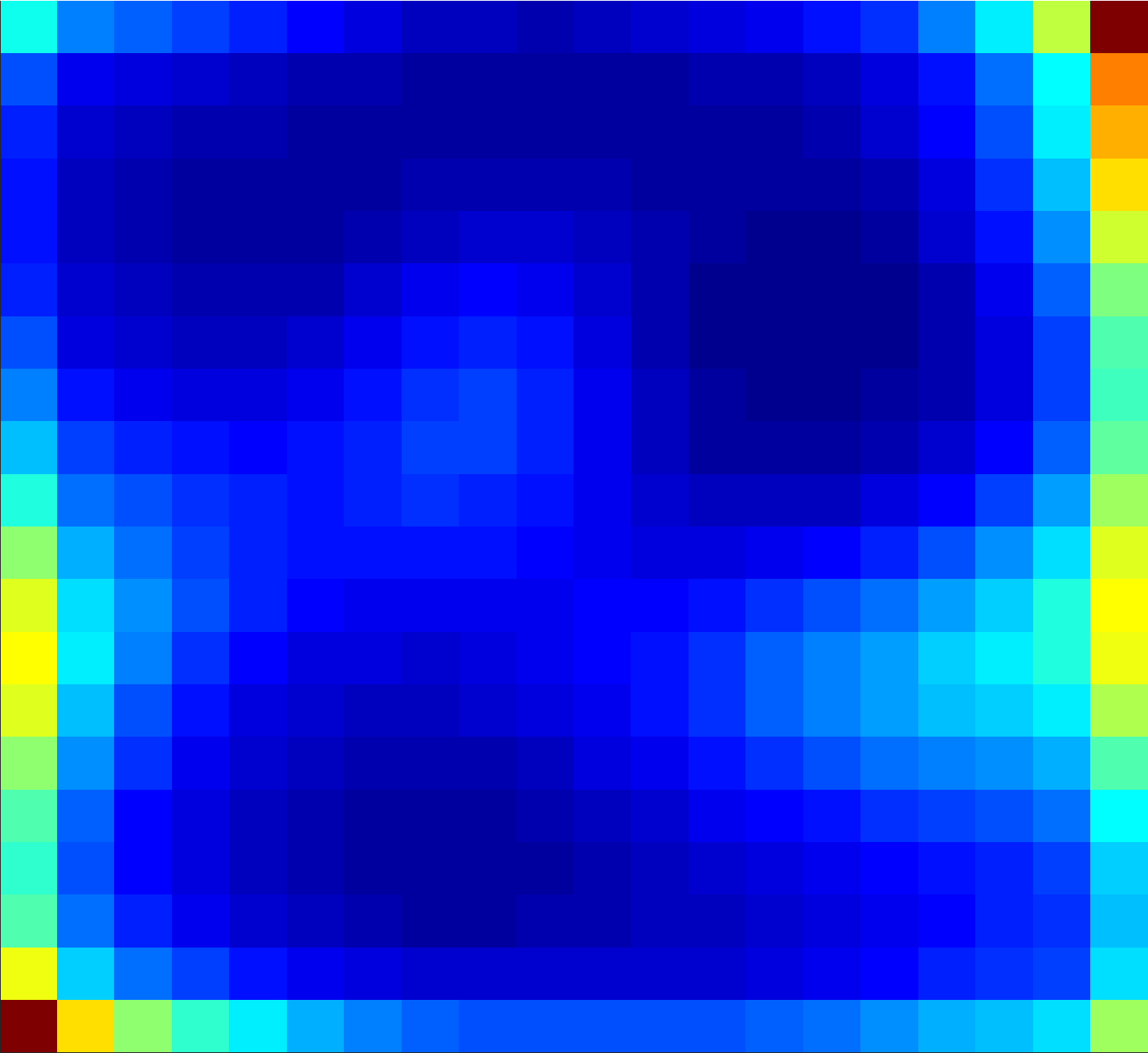} & \includegraphics[width=0.14\columnwidth]{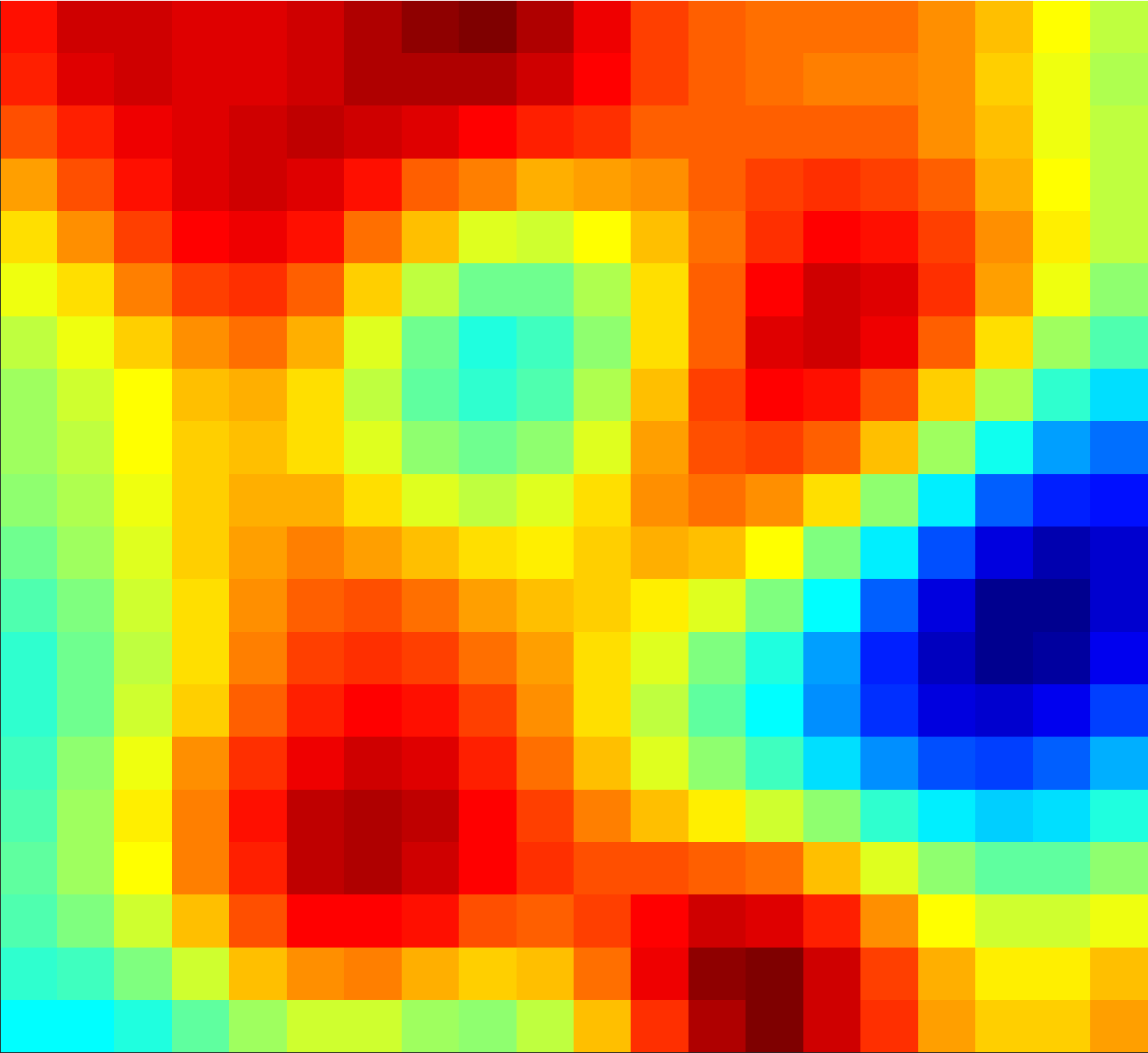} & \includegraphics[width=0.14\columnwidth]{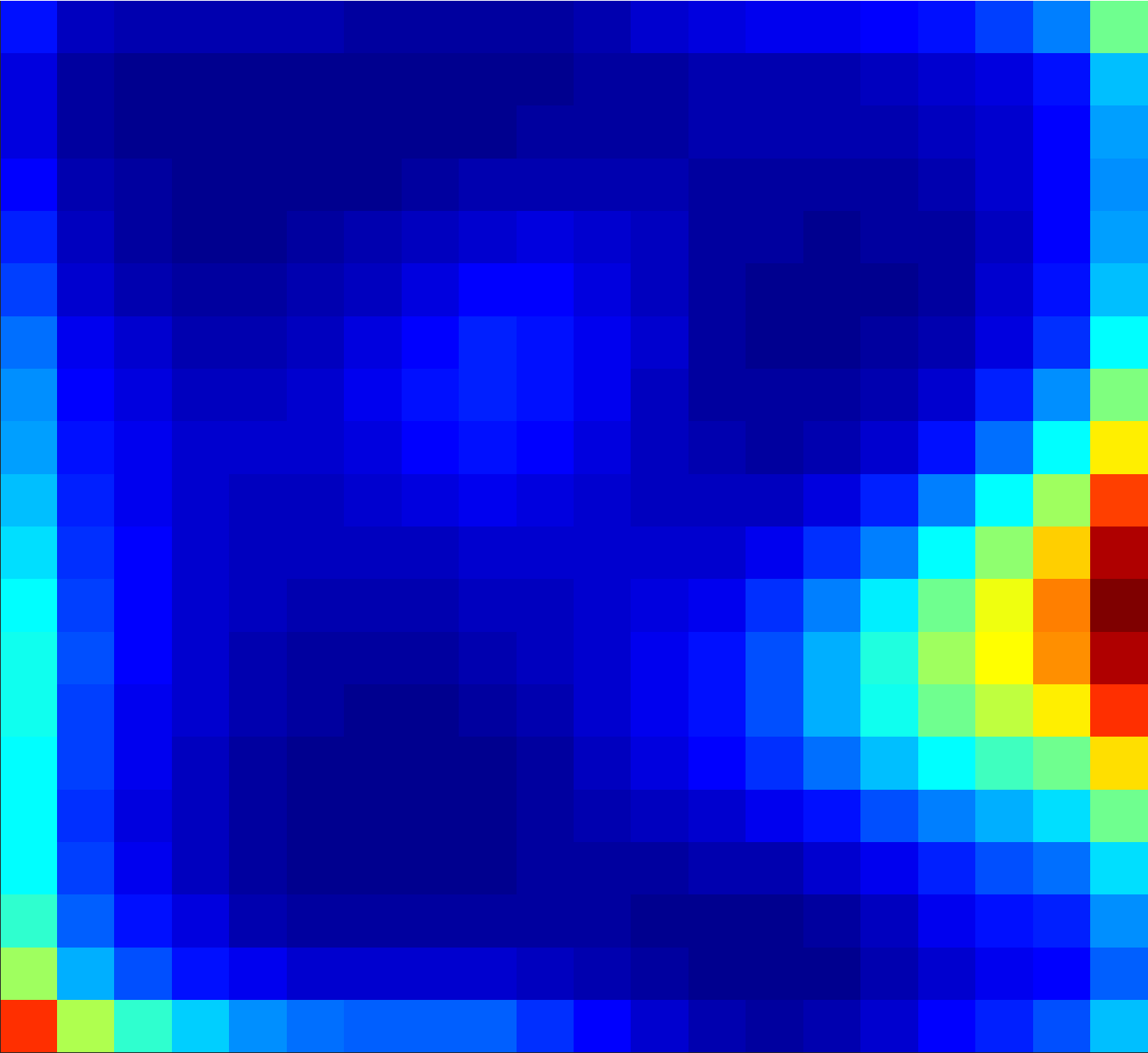}\tabularnewline
\rotatebox[origin=c]{90}{SmMALA} & \includegraphics[width=0.14\columnwidth]{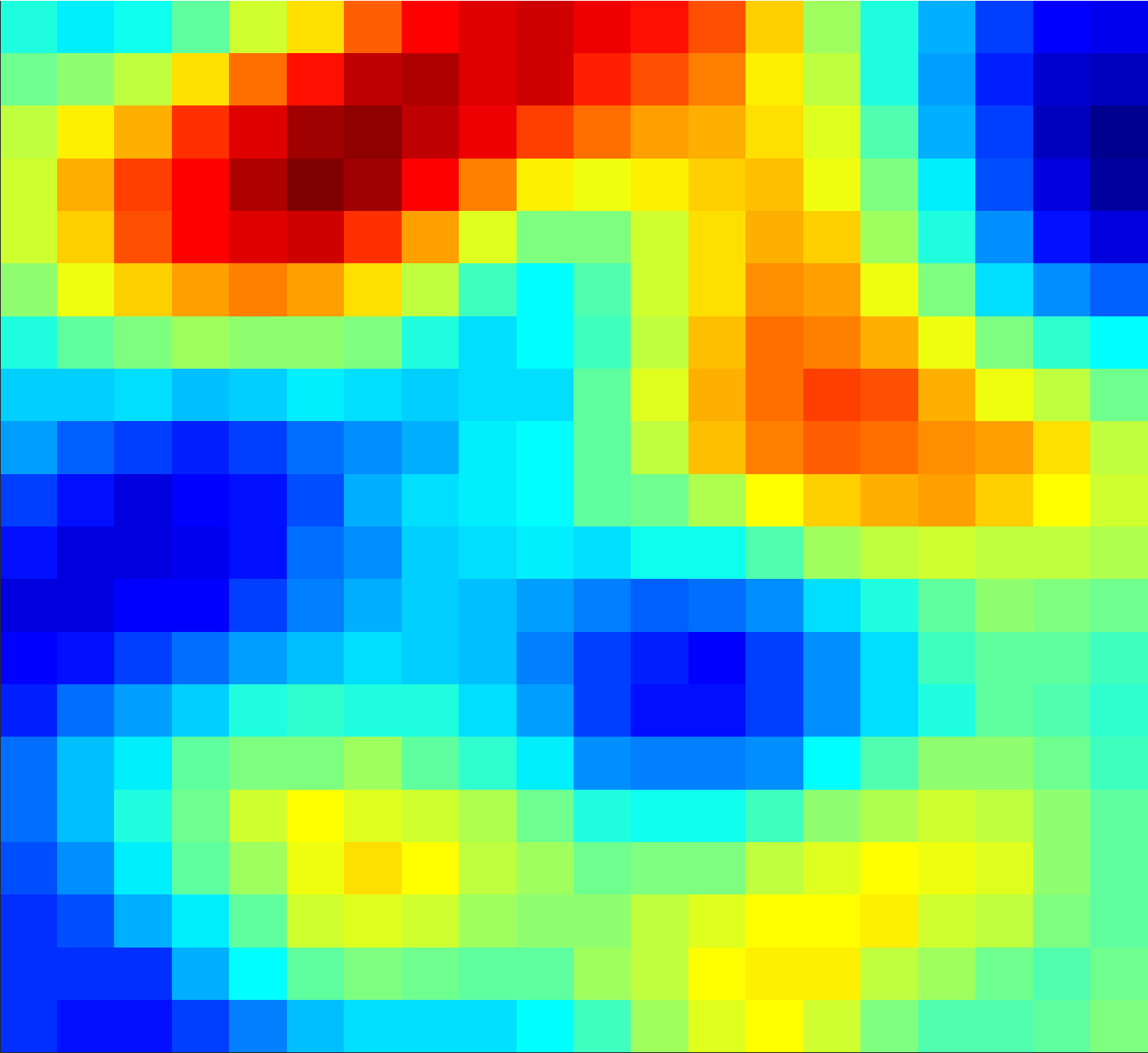} & \includegraphics[width=0.14\columnwidth]{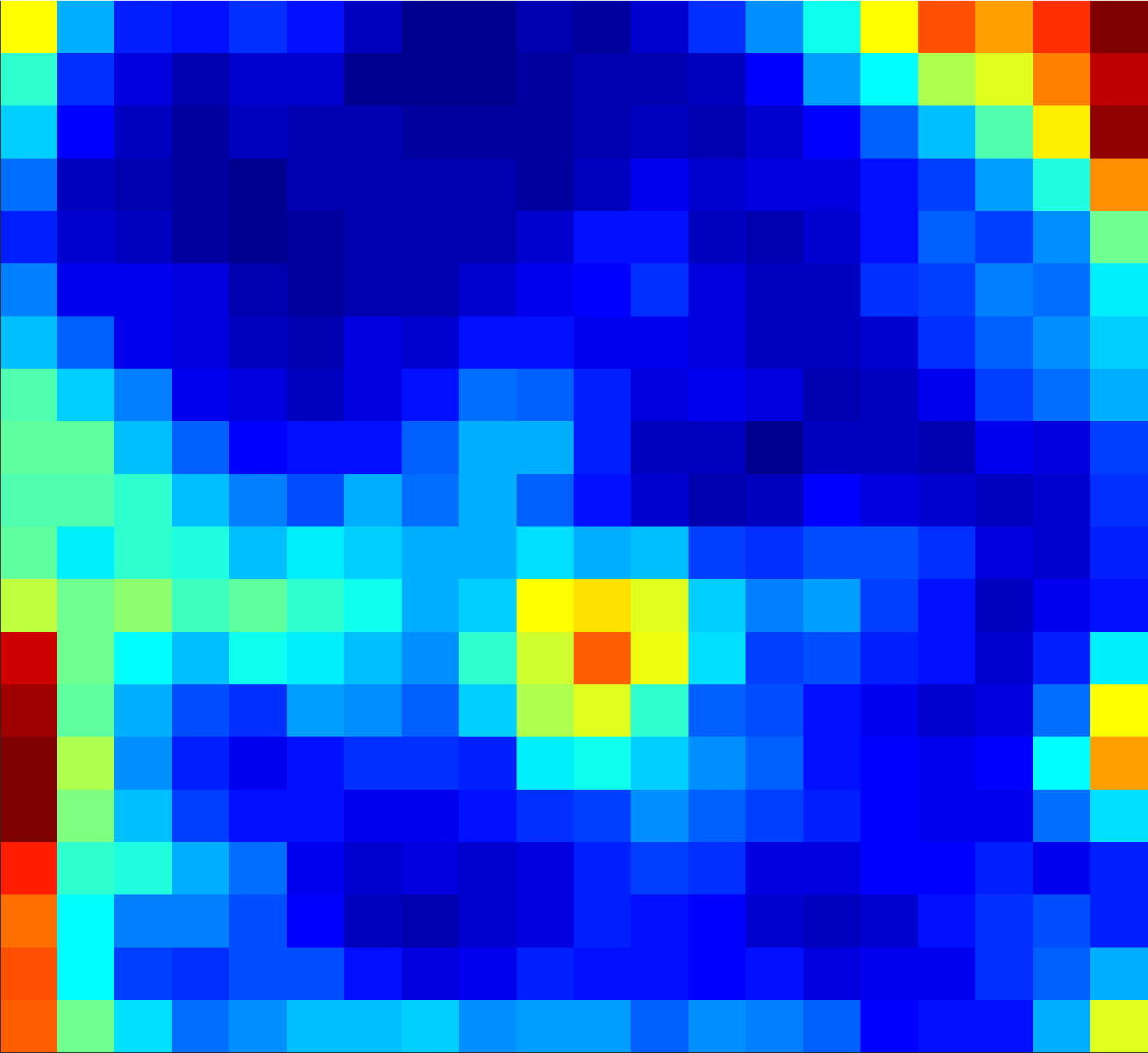} & \includegraphics[width=0.14\columnwidth]{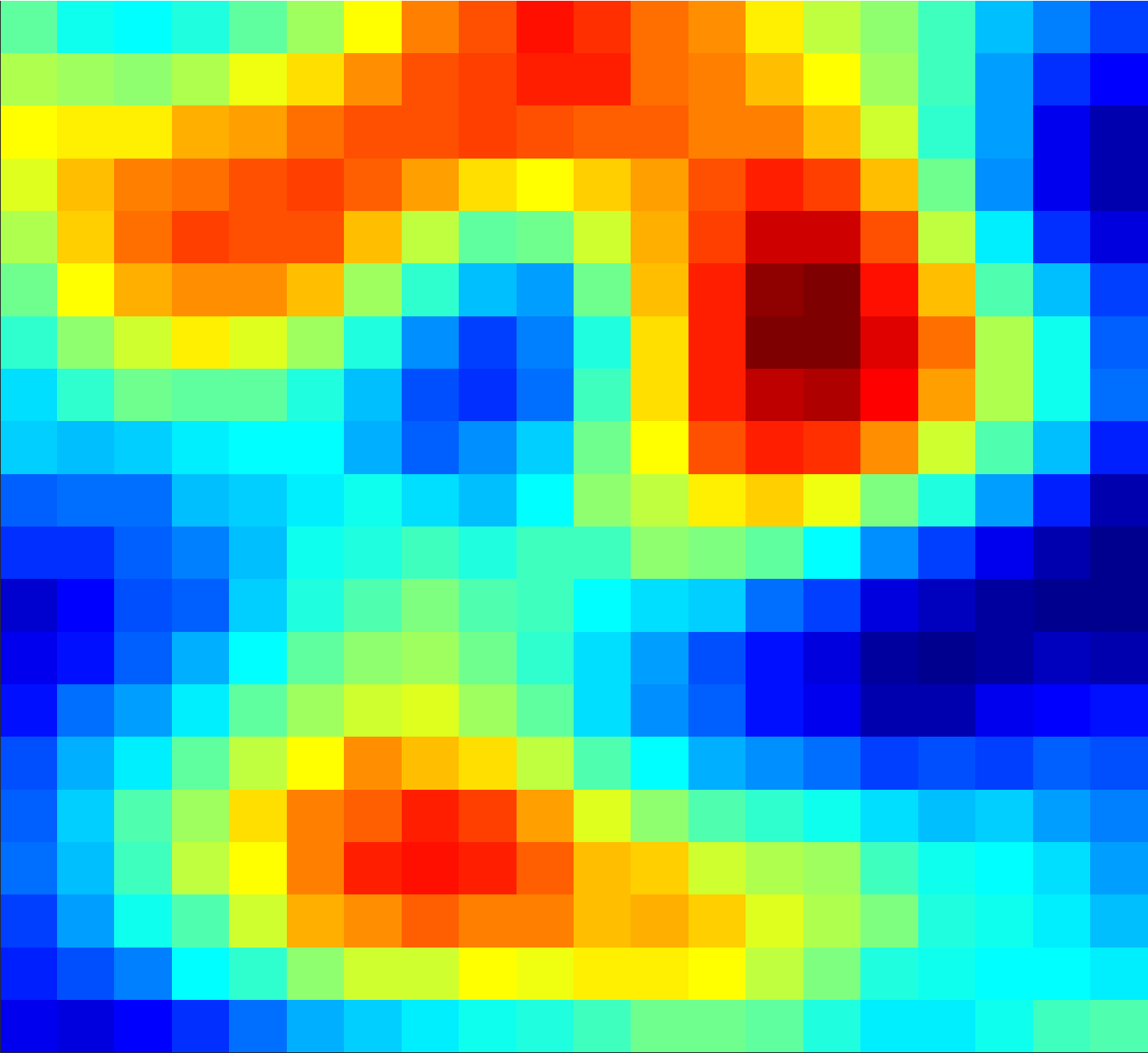} & \includegraphics[width=0.14\columnwidth]{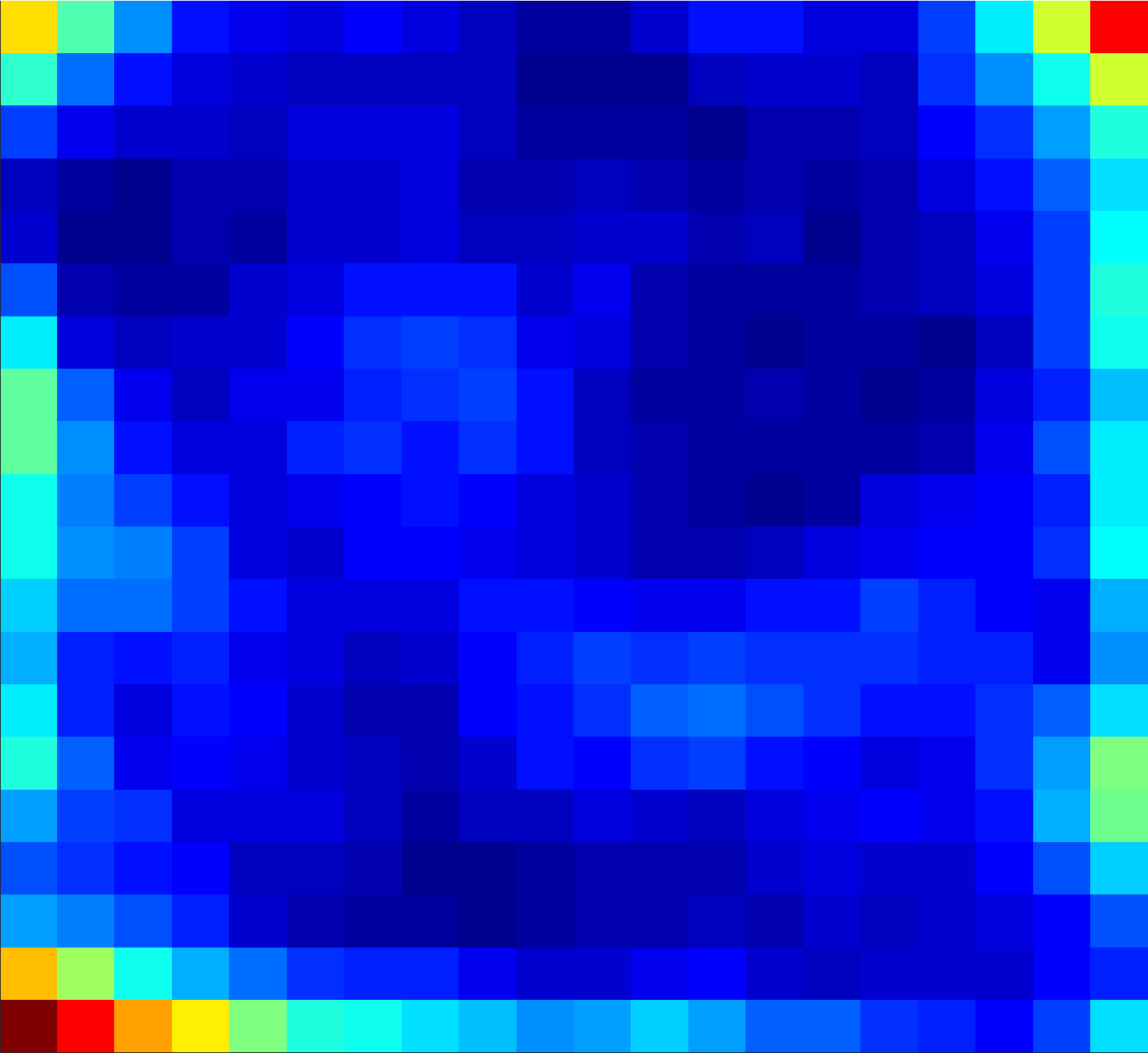} & \includegraphics[width=0.14\columnwidth]{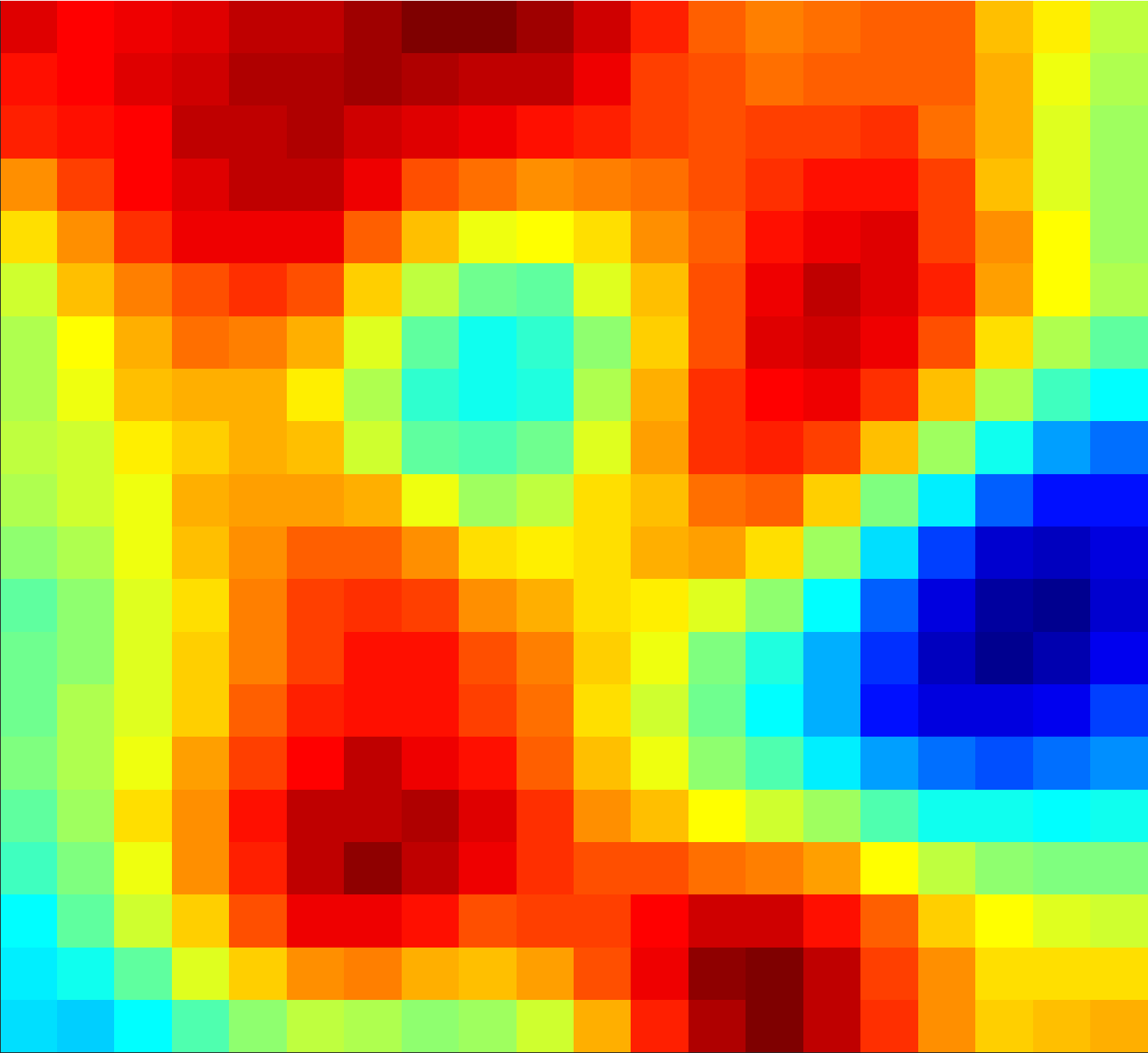} & \includegraphics[width=0.14\columnwidth]{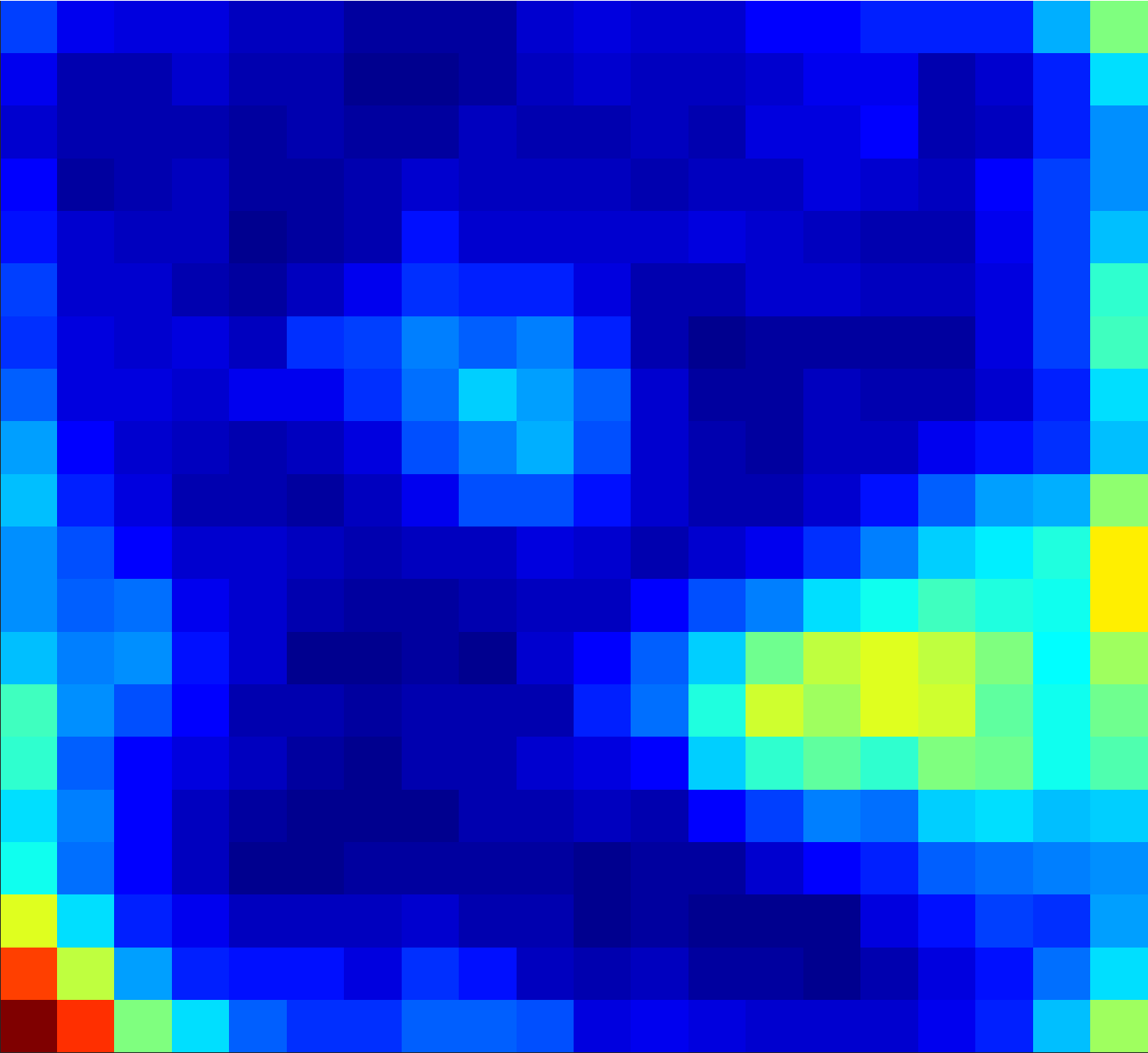}\tabularnewline
\rotatebox[origin=c]{90}{SmHMC} & \includegraphics[width=0.14\columnwidth]{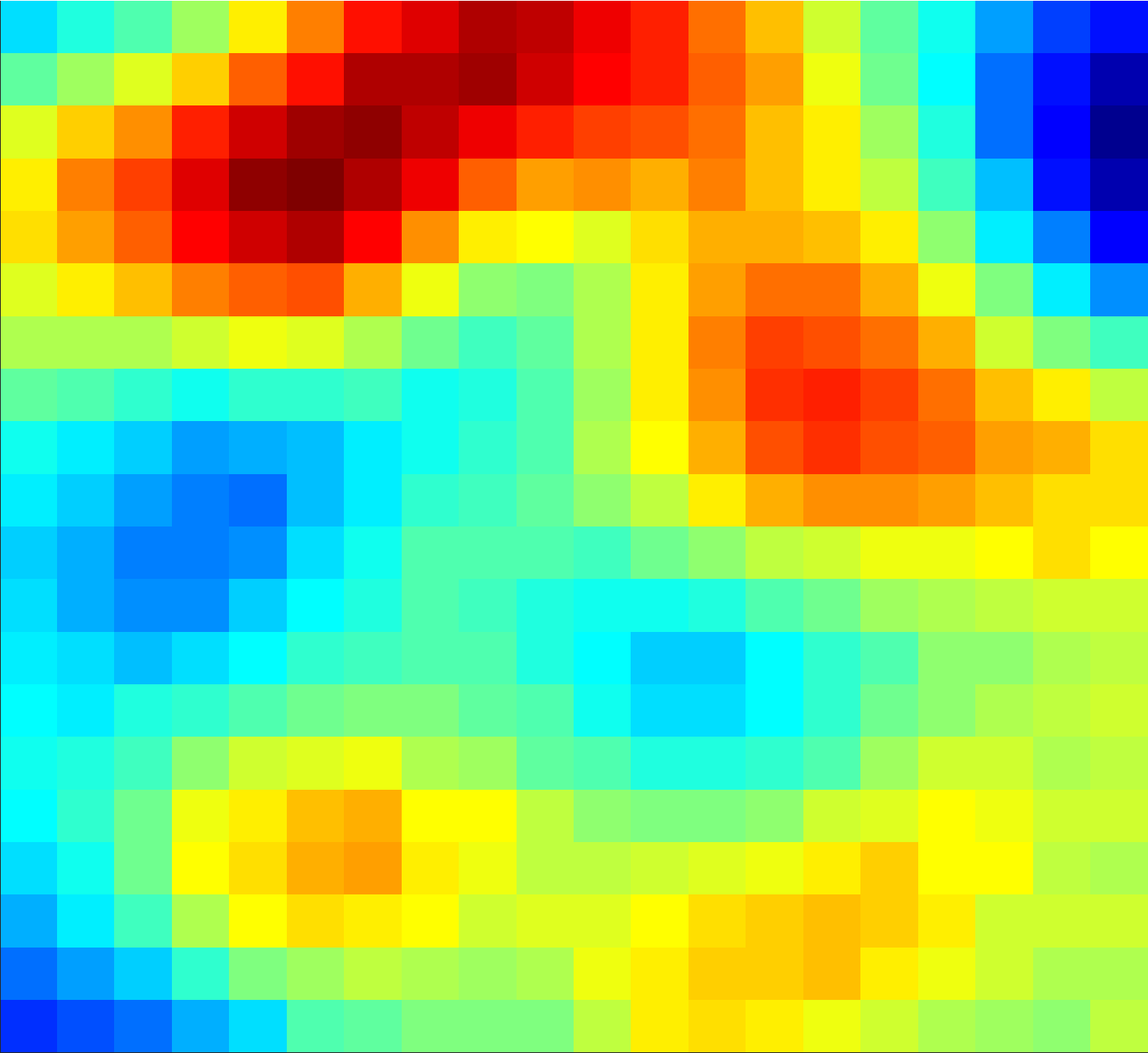} & \includegraphics[width=0.14\columnwidth]{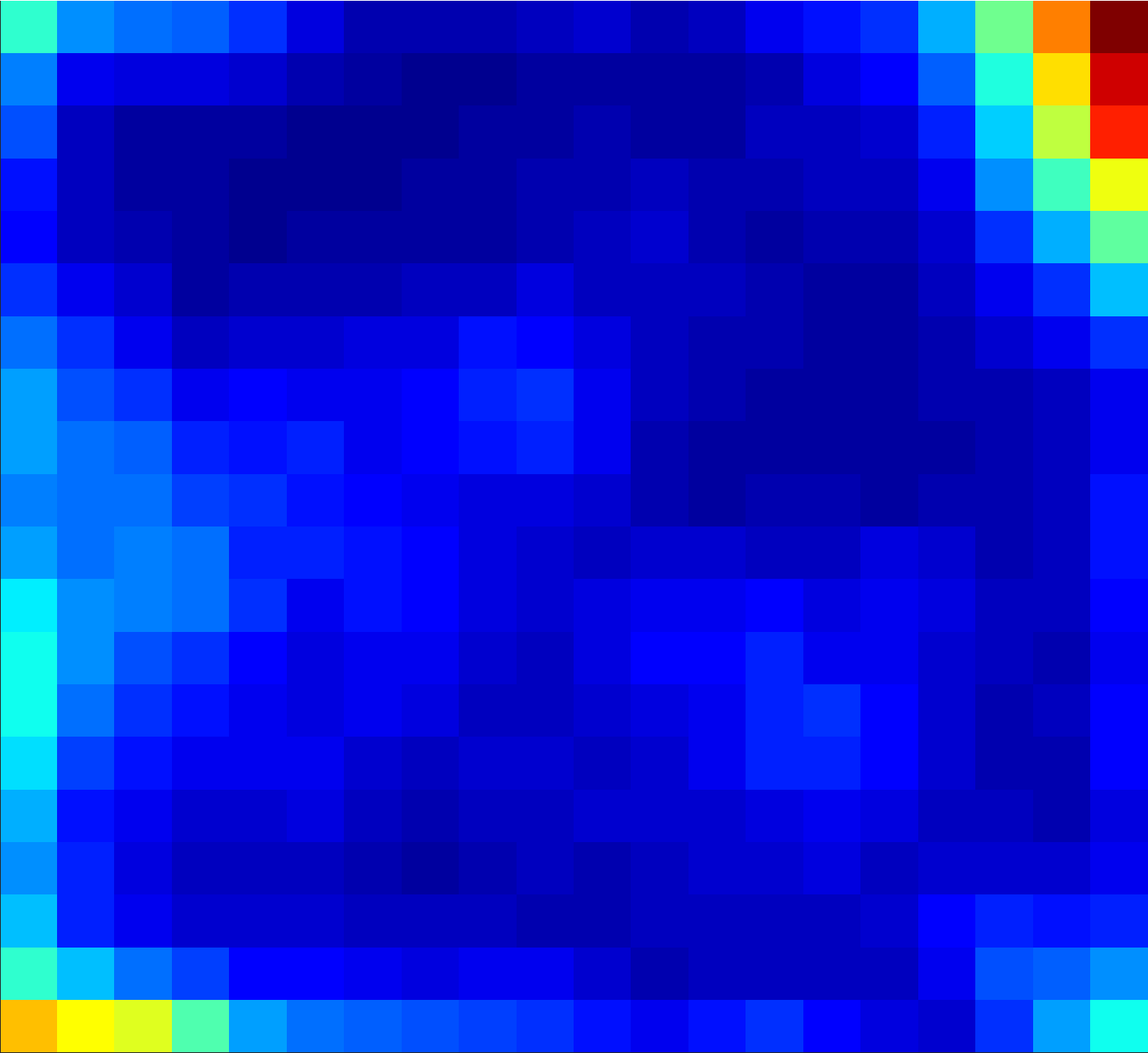} & \includegraphics[width=0.14\columnwidth]{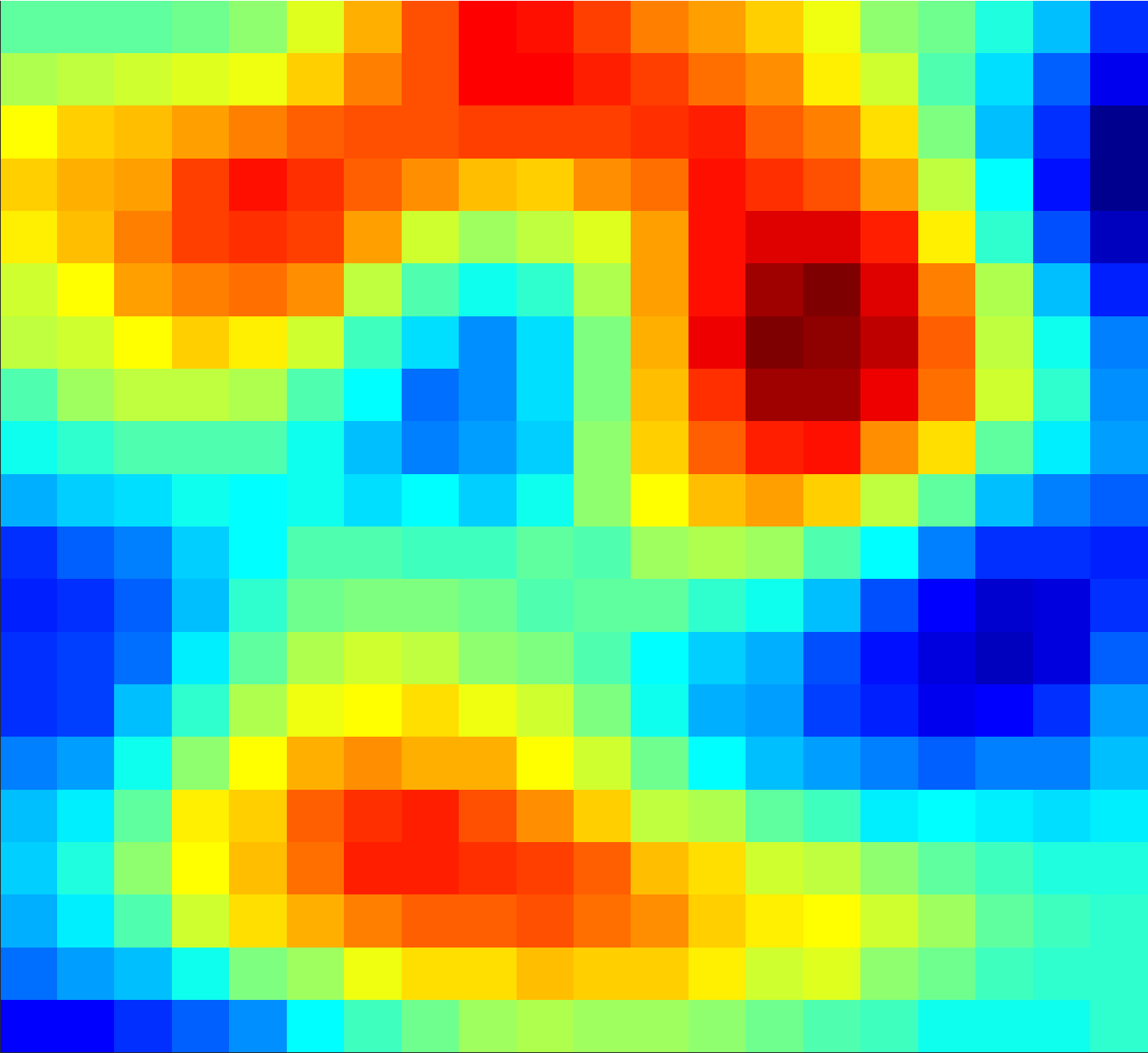} & \includegraphics[width=0.14\columnwidth]{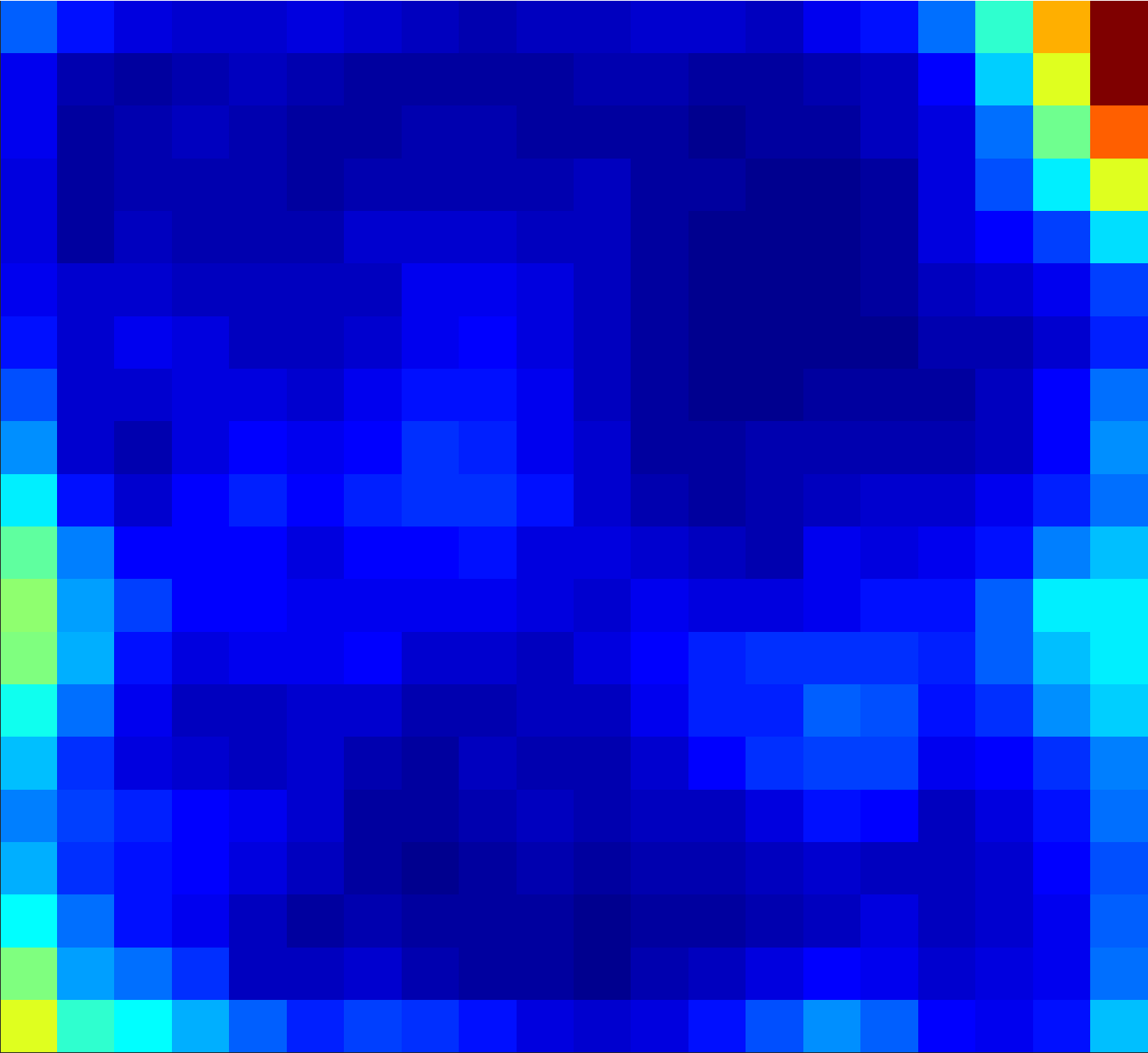} & \includegraphics[width=0.14\columnwidth]{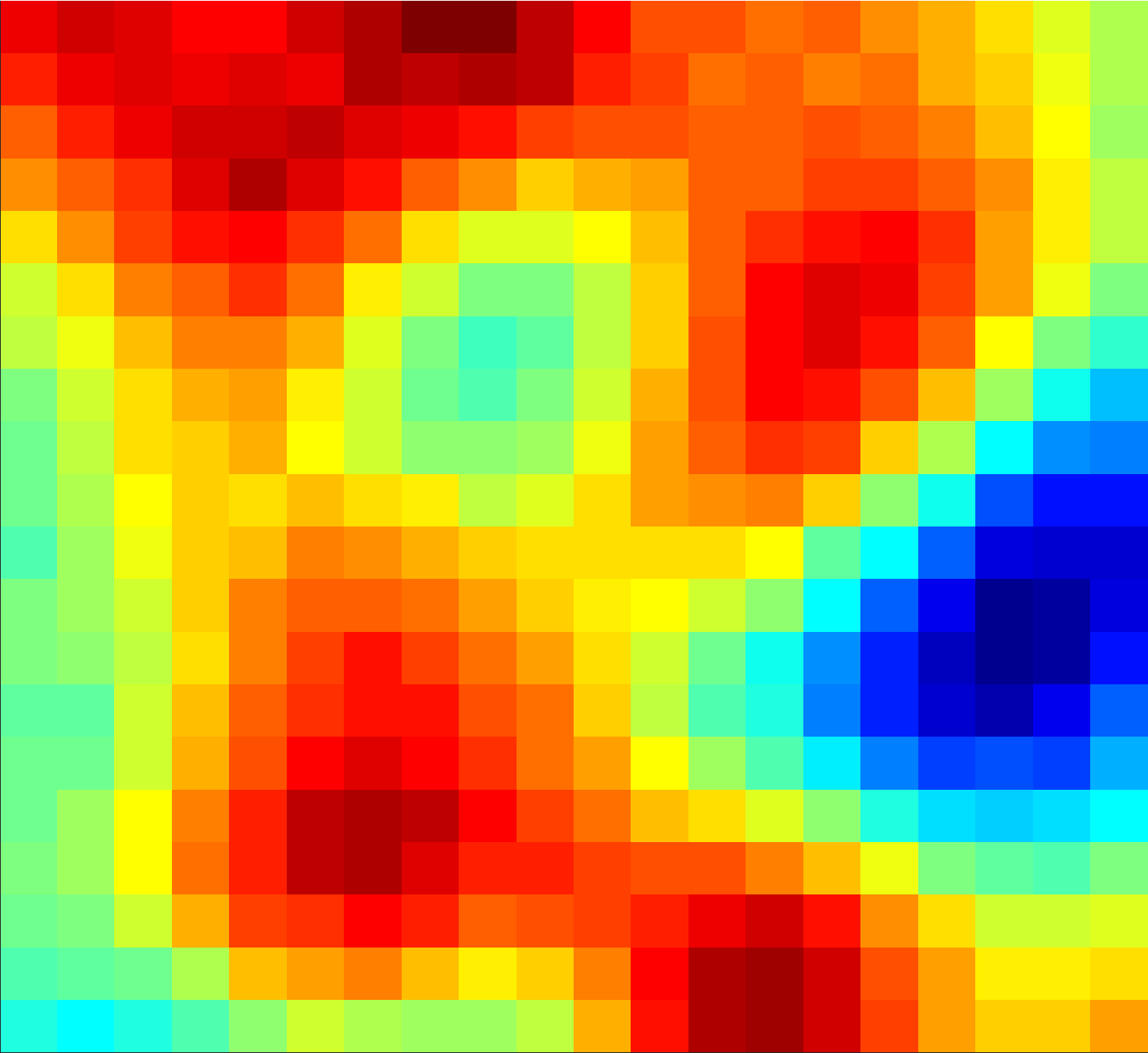} & \includegraphics[width=0.14\columnwidth]{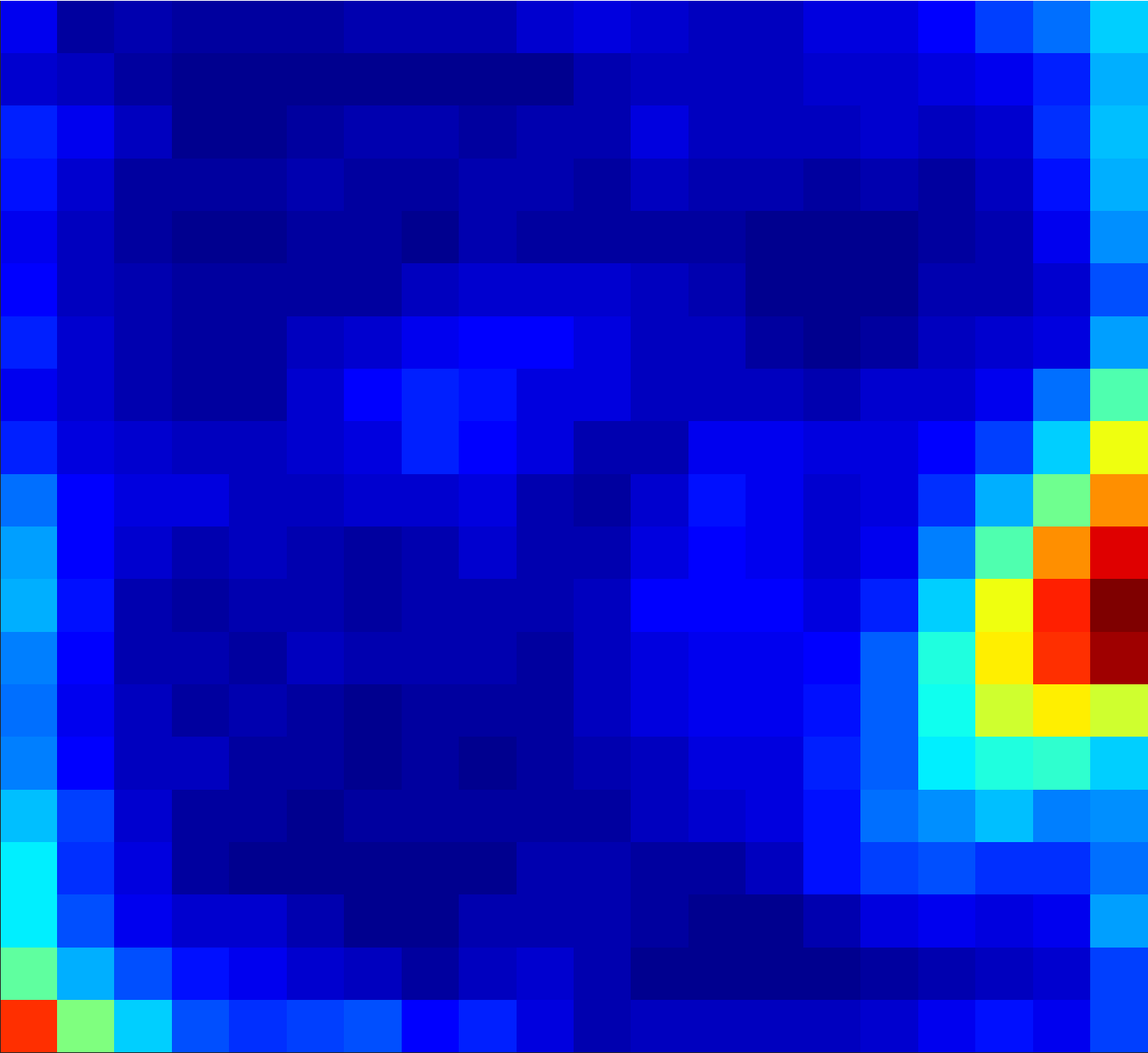}\tabularnewline
\rotatebox[origin=c]{90}{SIR} & \includegraphics[width=0.14\columnwidth]{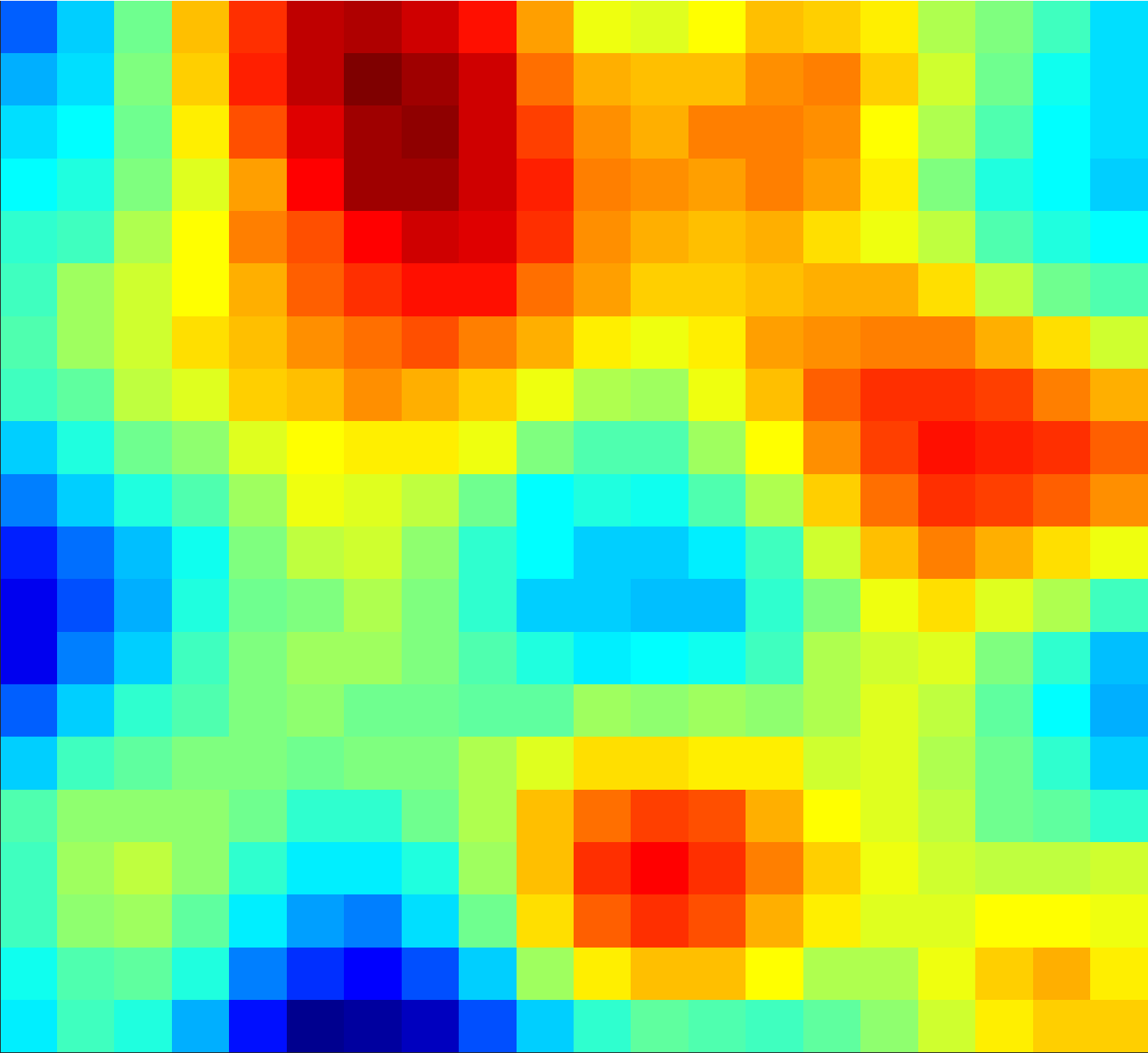} & \includegraphics[width=0.14\columnwidth]{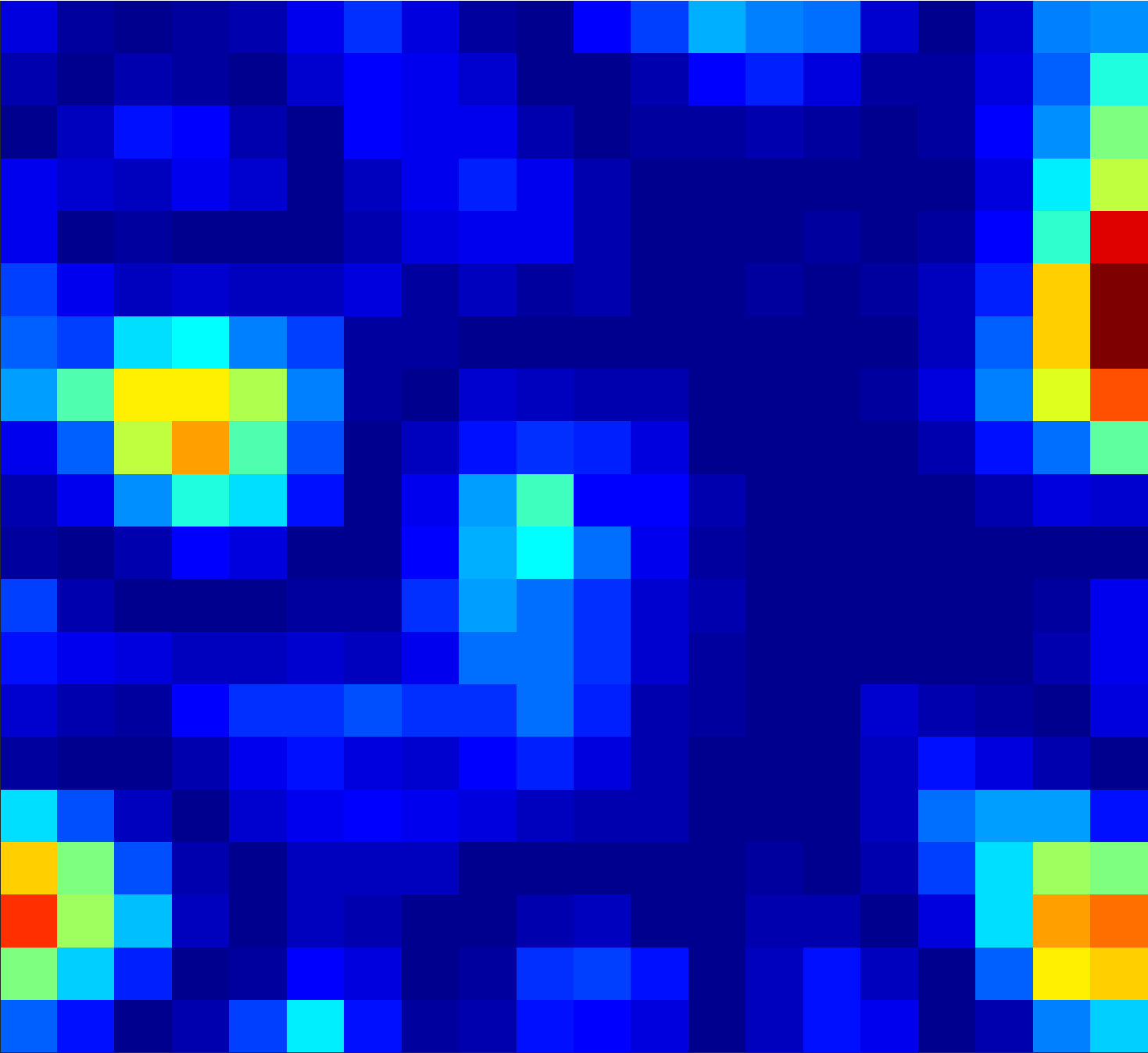} & \includegraphics[width=0.14\columnwidth]{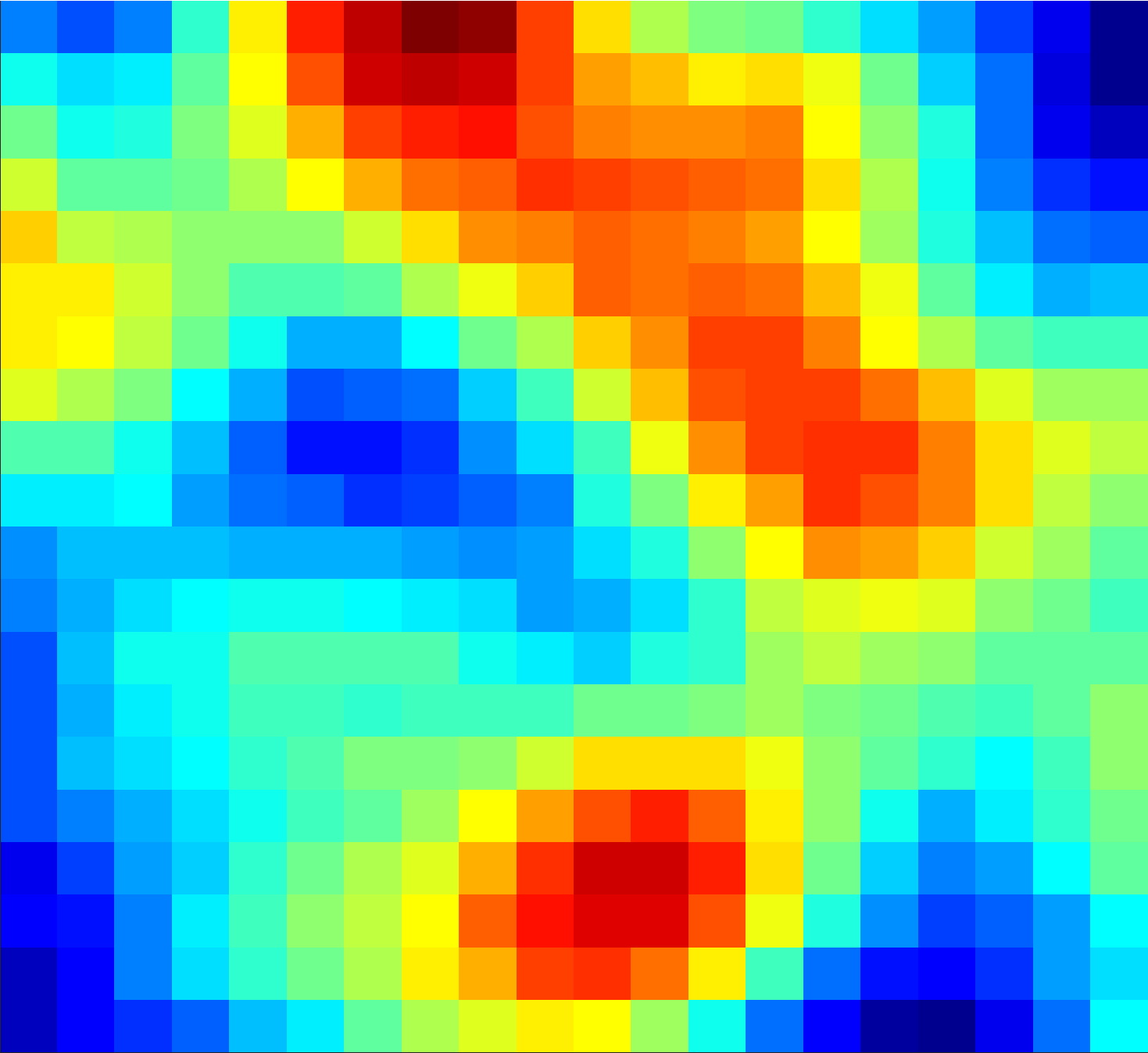} & \includegraphics[width=0.14\columnwidth]{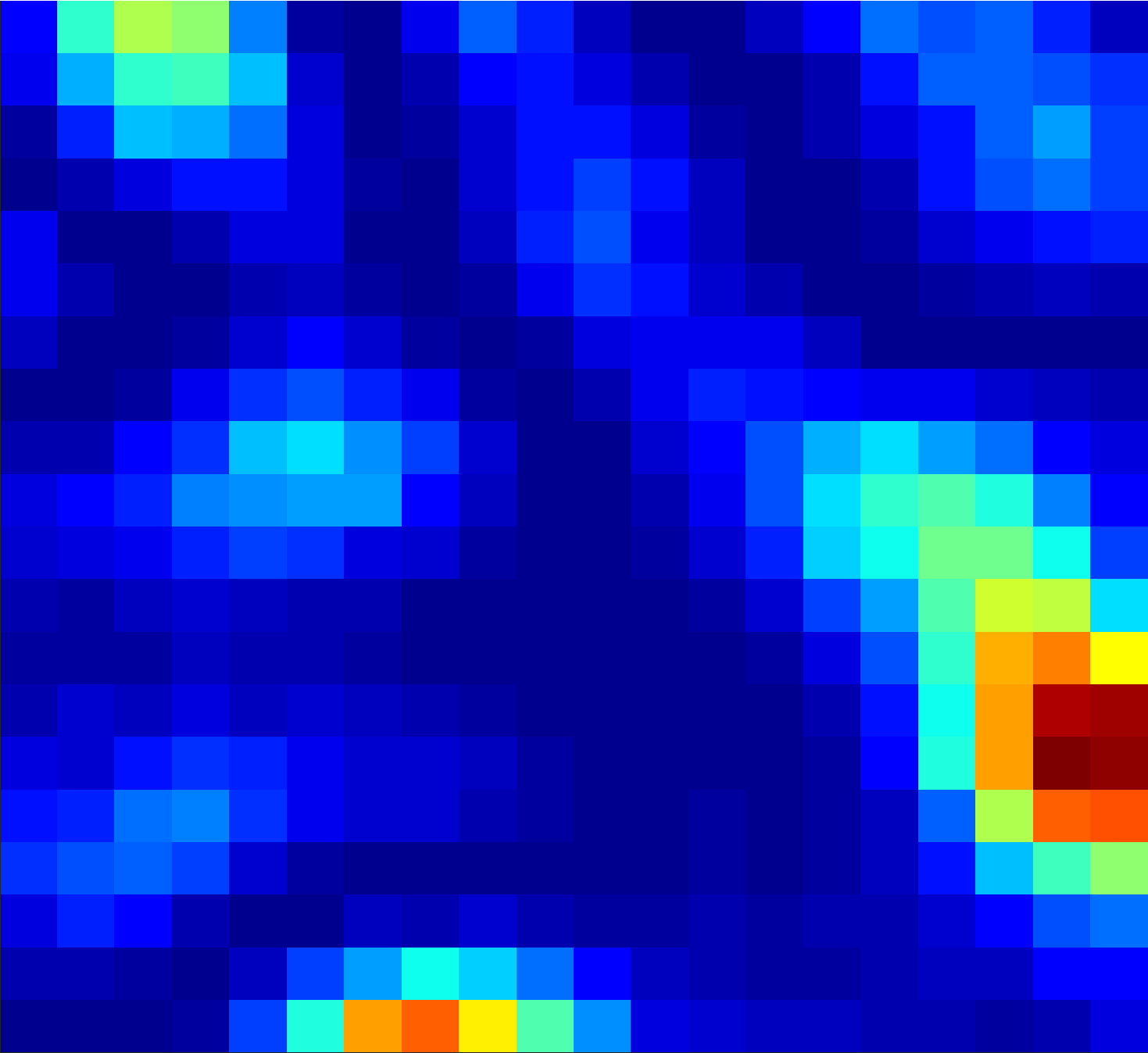} & \includegraphics[width=0.14\columnwidth]{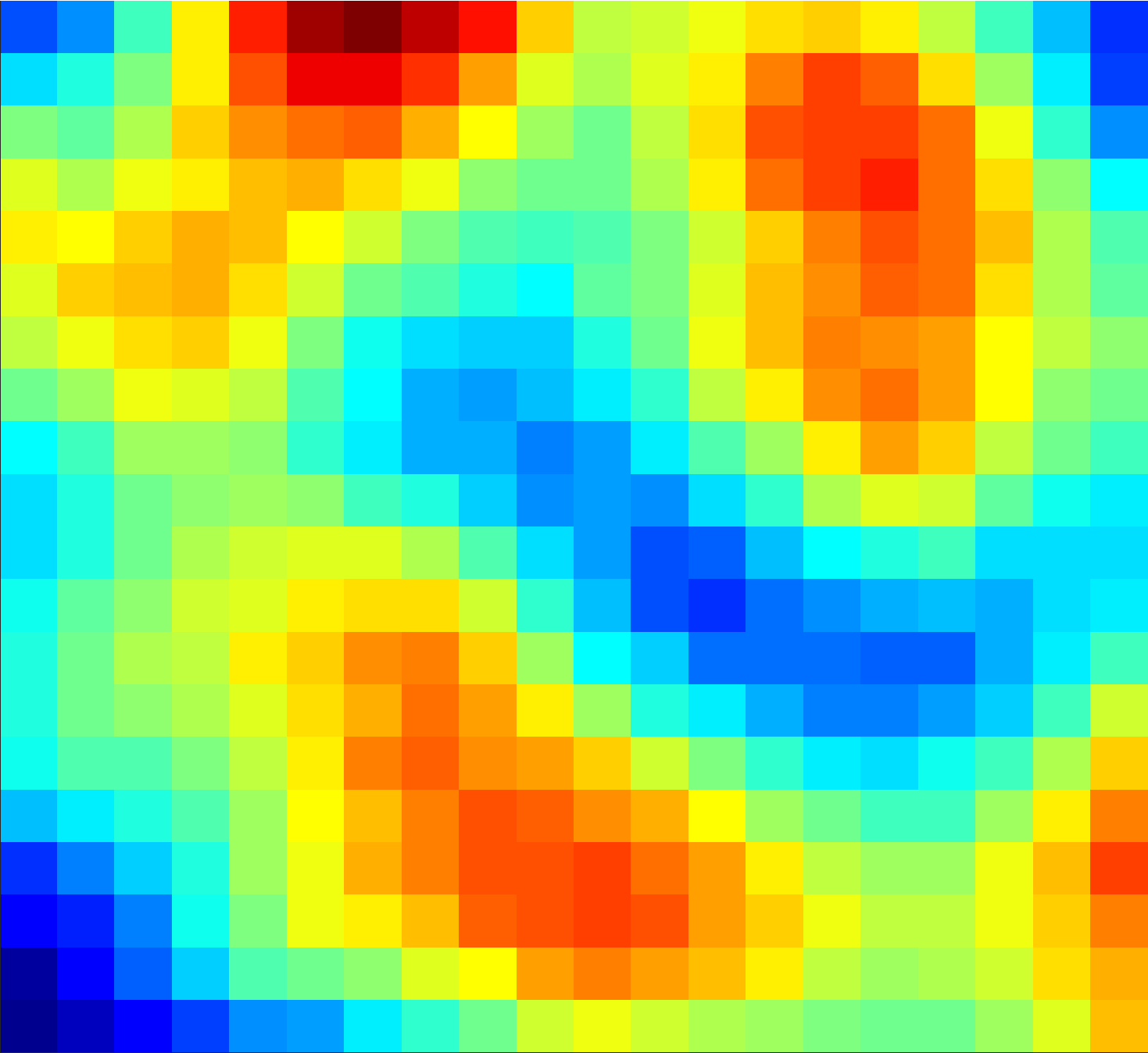} & \includegraphics[width=0.14\columnwidth]{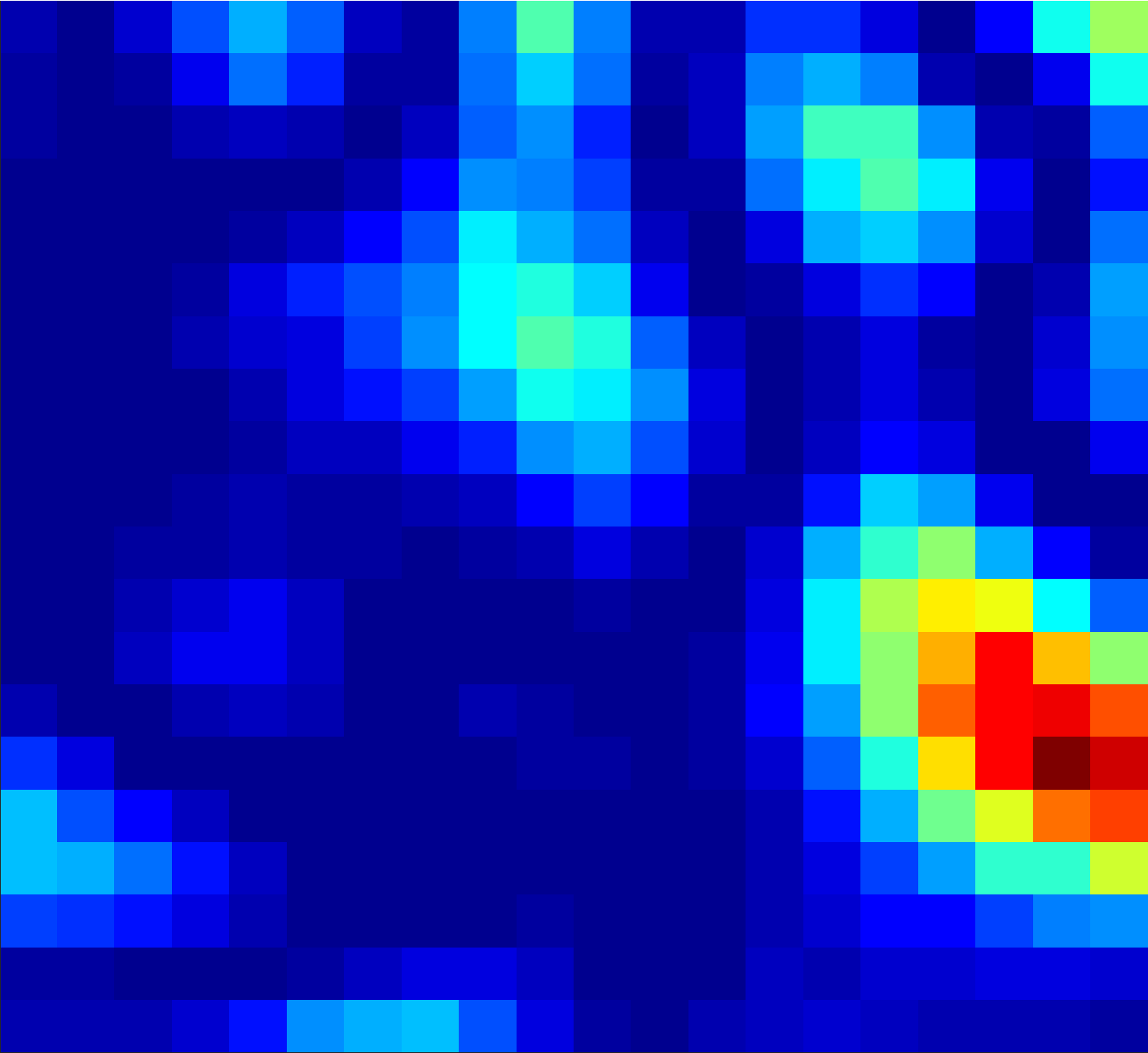}\tabularnewline
\rotatebox[origin=c]{90}{Block SIR} & \includegraphics[width=0.14\columnwidth]{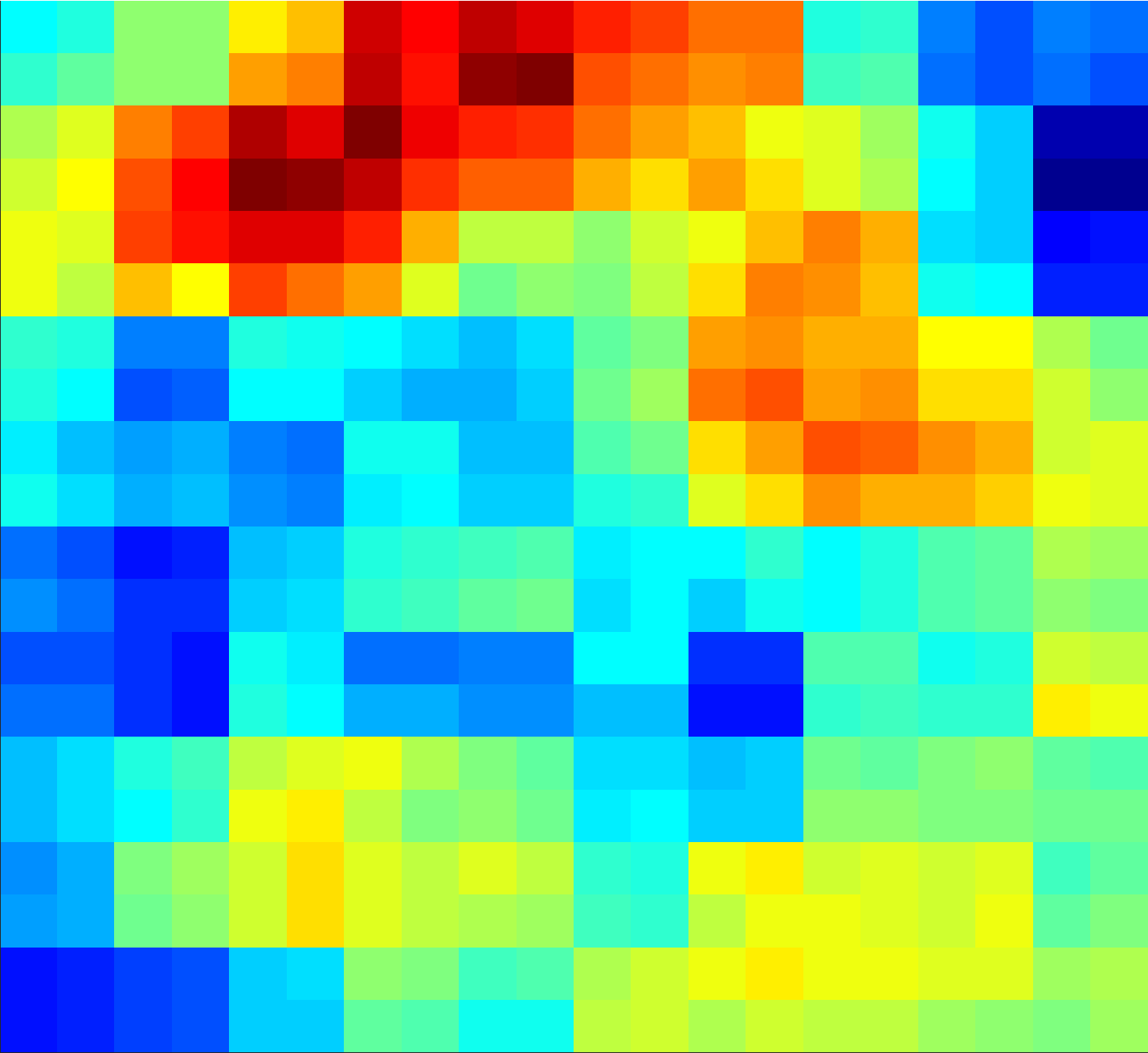} & \includegraphics[width=0.14\columnwidth]{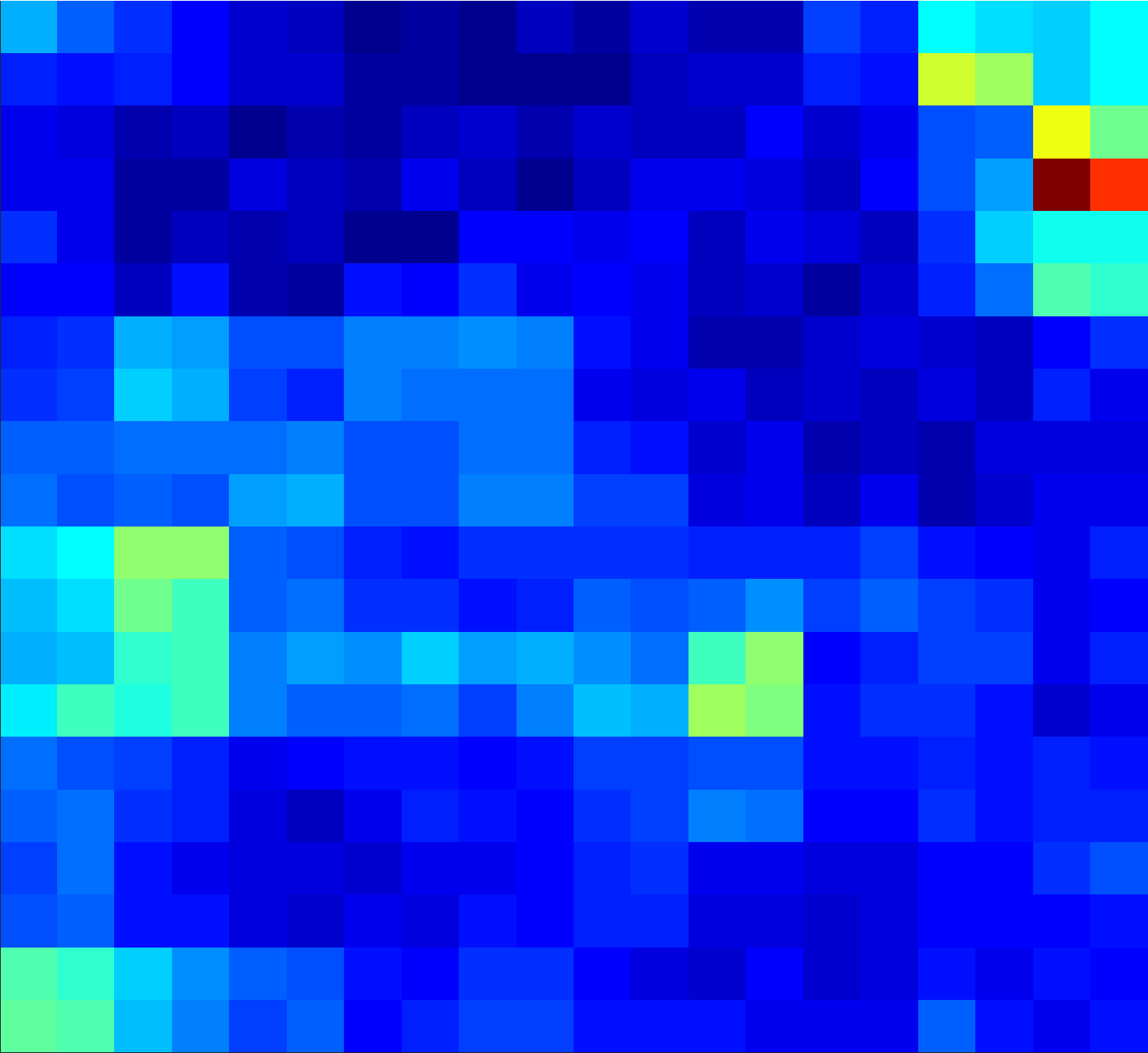} & \includegraphics[width=0.14\columnwidth]{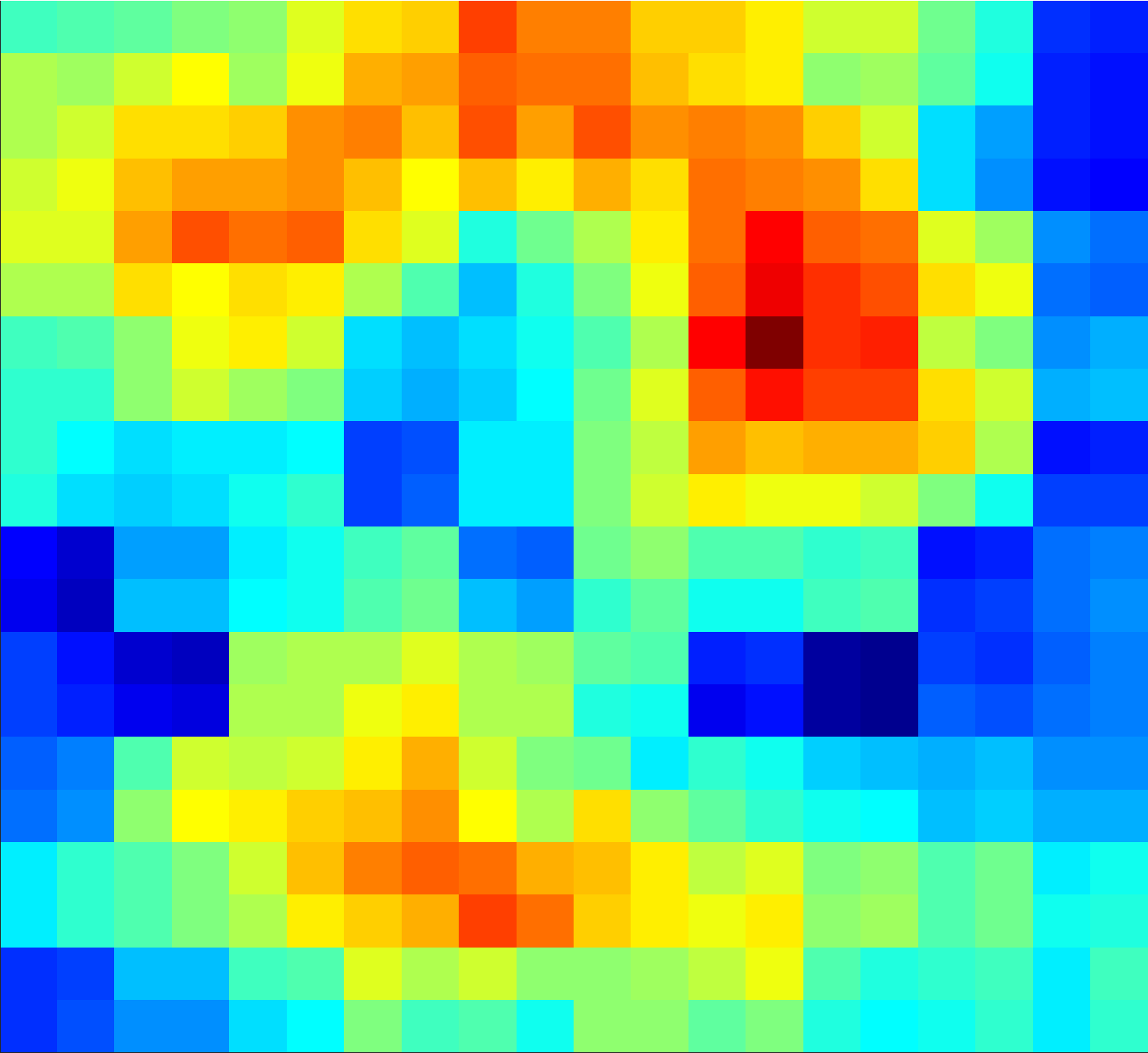} & \includegraphics[width=0.14\columnwidth]{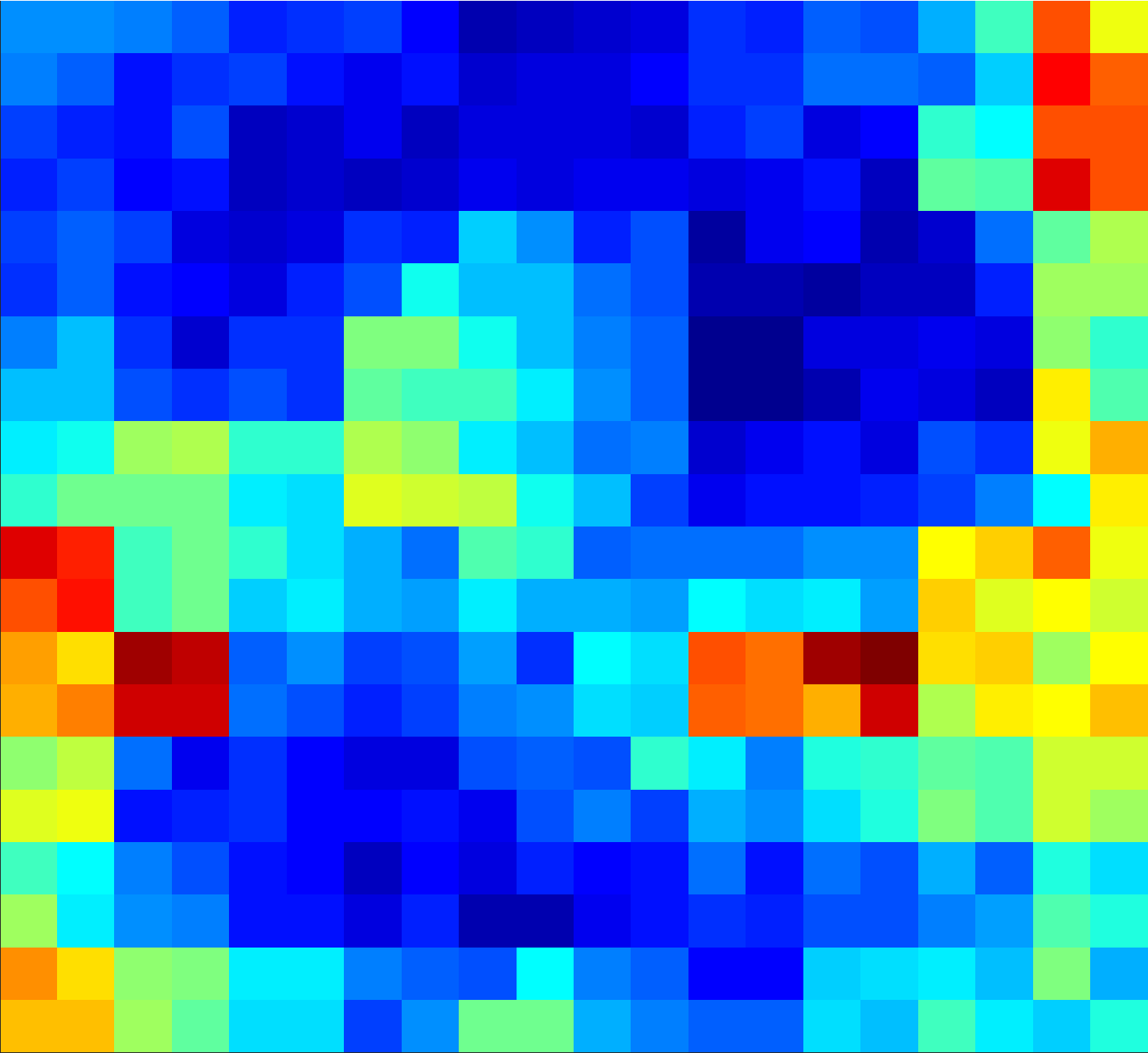} & \includegraphics[width=0.14\columnwidth]{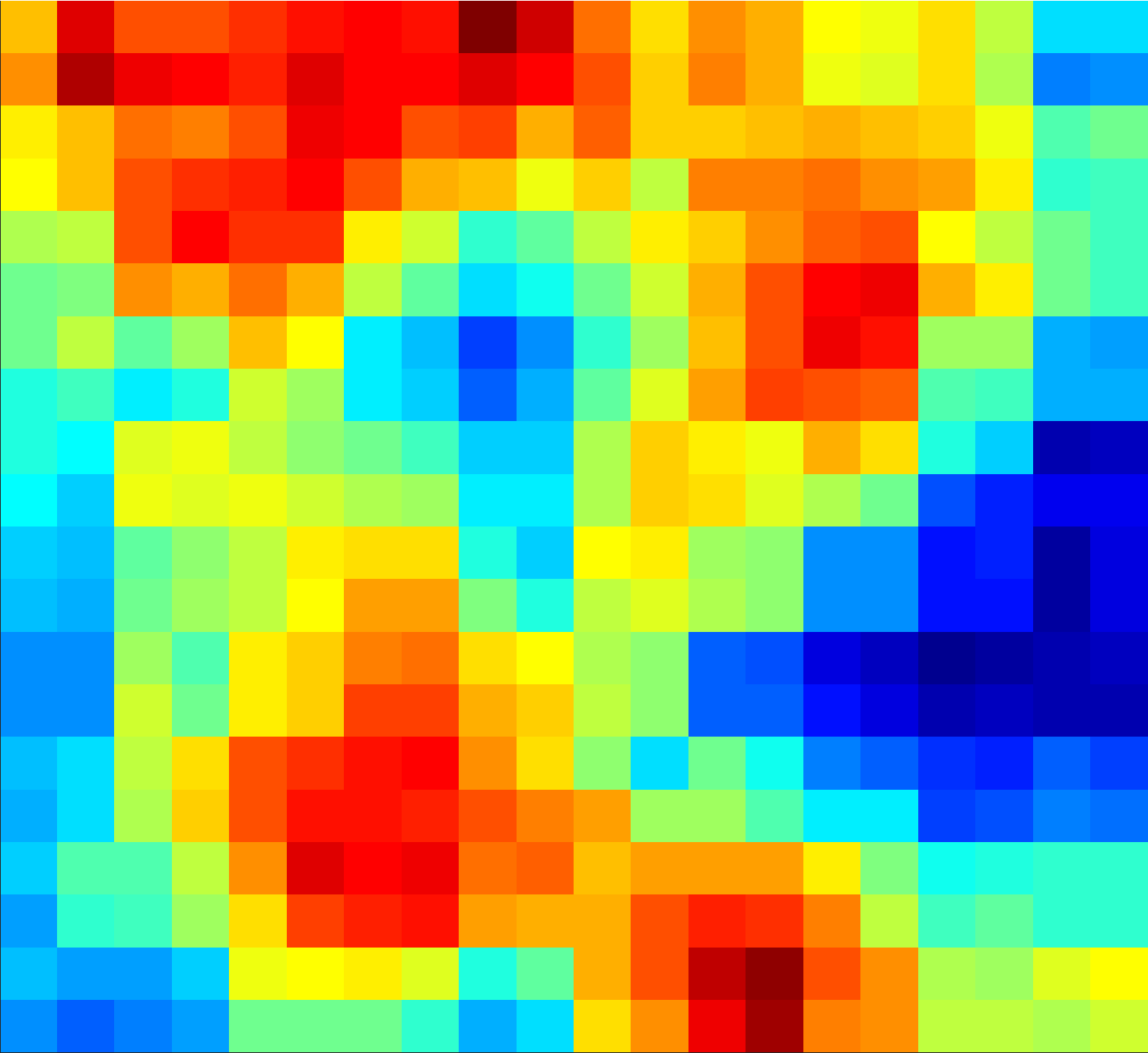} & \includegraphics[width=0.14\columnwidth]{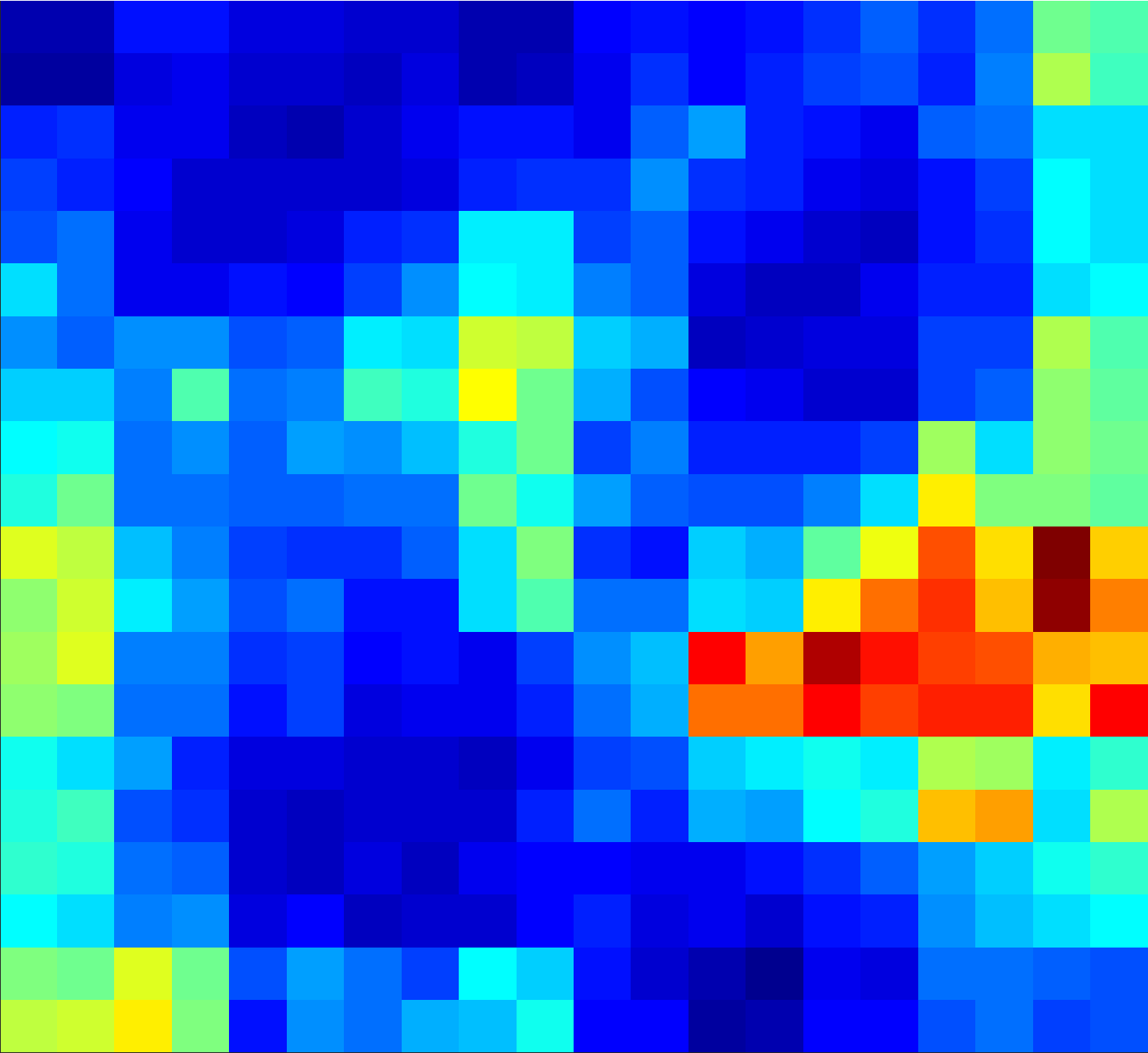}\tabularnewline
\end{tabular}
\par\end{centering}
\caption{Posterior statistics for the nonlinear, non-Gaussian sensor network
example ($n_{x}=400$)\label{fig:Posterior-sensor-network-nonlinear}}
\end{figure}

Once again we follow the methodology presented in \citep{Septier2016}
for performance evaluation. The log root-mean-square error, log normalized-estimation
error squared (NEES) and average computation time (per time step)
of all filters, for different numbers of sensors in the grid, are
shown in Figures~\ref{fig:RMSE-sensor-network-nonlinear}, \ref{fig:NEES-sensor-network-nonlinear},
and \ref{fig:Computation-sensor-network-nonlinear} respectively.
The mean square error of the estimates were computed over 100 Monte
Carlo runs, with the particle-based filters using 200 samples. The
step size adopted for the sequential MCMC filters is $\Delta\lambda=0.5$
and the number of steps computed as $L=L_{0}+N$, where $L_{0}=0.2N$
is the number of steps for the burn-in phase, whereas for the SPF-GS
we applied the empirical rules for time horizon and step-step size
as presented in Subsection \ref{subsec:Selection-of-time}. In Figure
\ref{fig:Posterior-sensor-network-nonlinear} we illustrate the posterior
means and variances of the state across the sensors grid ($n_{x}=400$)
at different time steps, for all evaluated filters. From Figures~\ref{fig:RMSE-sensor-network-nonlinear},
\ref{fig:NEES-sensor-network-nonlinear}, and \ref{fig:Computation-sensor-network-nonlinear},
we can note that
\begin{itemize}
\item The performances of all sequential MCMC filters are in accordance
with the results shown in \citep{Septier2016}.
\item The SPF-GS presents performance commensurate to that of the SmHMC
filter in terms of root-mean-square error, outperforming all other
filters.
\item The SPF-GS outperforms all other filters in terms normalized-estimation
error squared.
\item The SPF-GS demands a computational effort higher than that of SmMALA
when the number of sensors is small, but its computing time scales
better than that of SmMALA and similarly to that of SmHMC for higher
dimensions.
\end{itemize}
For this example, the results indicate that the SPF-GS is as accurate
as the SmHMC filter in general, and that SPF-GS is the most accurate
method in high dimensional problems. Once again it is clear that the
consistency (credibility) of estimates by the SPF-GS is higher than
that by all other filters: the NEES is very close to one (from above)
for all evaluated dimensions. It is worth noting that, to keep the
computational cost for the SPF-GS competitive, the total number of
steps was limited. This did result in some loss of accuracy but the
implication is perhaps that improved performance is possible if sufficient
computational resources are available. The associated trade-off is
sufficiently complex to form a hard obstacle against systematic solutions.
However, the techniques presented in Section \ref{subsec:Selection-of-time}
proved sufficient to generate the results presented here. 

\section{Conclusions and Future Work\label{sec:Conclusions}}

This paper builds on concepts, such as continuous-time filtering and
sequential Monte Carlo methods, being intensely studied by the research
community. The paper provides a description of these concepts that
aims to draw out some key insights from the theoretical research into
sequential Monte Carlo filtering while also responding to the significant
empirical challenges encountered when applying existing and emerging
tools to difficult real-world problems. More specifically, the paper
is part of a growing body of research which aims to apply concepts
from the sequential Monte Carlo community to problems involving high-dimensional
spaces. This broader body of research and this paper in particular
is motivated by the increasing demand for more statistically efficient
methods to solve difficult inference problems exemplified by those
involving large numbers of dimensions and relevant to a vast range
of applications. 

The paper capitalizes on some important findings that have been reported
recently \citep{Verge2015,Rebeschini2015,Septier2016} regarding how
the local properties of sequential Monte Carlo filtering measures
impact algorithms' abilities to solve high-dimensional problems. We
exploit the observation made by Rebeschini \& van Handel \citep{Rebeschini2015},
that, by using the decay of correlations property, it is possible
to develop particle filters based on local solutions whose approximation
error becomes less sensitive to augmenting the number of state dimensions. 

Within this context, we proposed a novel filter which aims to address
the well-known shortcomings of sequential Monte Carlo methods when
applied to nonlinear high-dimensional filtering problems. The novel
method uses a Monte Carlo procedure to generate a sequence of equally-weighted
samples that each guide a local solution of the Fokker-Planck equation.
Using these local approximations, a mixture is produced that approximates
the filtering density. The result is a statistically-sound general-purpose
class of algorithms. In the context of a simple, though not trivial,
high-dimensional inference problem and in comparison with state-of-the-art
algorithms, the proposed approach has been shown to offer significant
improvement in statistical consistency with commensurate computational
expense.

In its most computationally efficient form, SPF-GS, the proposed filter
has a complexity bounded by only $\mathcal{O}\left(NLn_{x}^{3}\right)$
evaluations\footnote{It is important to remark that the stochastic particle flow is much
more computationally complex than the original particle flows, which
are generally bounded by $\mathcal{O}\left(N\right)$ computations
both in theory and practice.} and it has the appealing property that its operations per sample
(and the associated mixture component) can be parallelized. When articulated
as a marginal particle filter, SPF-MPF, the complexity is bounded
by $\mathcal{O}\left(NLn_{x}^{3}+N+N^{2}\right)$ evaluations. That
said, our investigations indicate that further consideration is required
in order to explore fully the potential of stochastic particle flow.
Future work will focus on the computational cost of the algorithms
and guaranteeing certain properties of the diffusion matrix, (e.g.,
ensuring the matrix is positive definite and not singular). It should
be possible, at least in the context of some statistical models, to
reduce the computational complexity by exploiting or imposing sparsity
in the diffusion matrix. Girolami \& Calderhead \citep{Girolami2011}
suggest that the use of guiding Hamiltonians \citep{Duane1987} could
provide a way of reducing such computational cost, but it is currently
unclear how such a solution would be adopted in the context of stochastic
particle flow. Another potentially promising future direction would
be adopting the same approach as the well-known Broyden-Fletcher-Goldfarb-Shanno
(BFGS) algorithm and thereby work around the need to explicitly evaluate
the Hessian matrix at all simulation steps. We also strongly believe
that it would be possible for future work to result in further improvements
on the bounds used to compute the step size and number of simulation
steps. Such advances will likely improve the computational efficiency
of the algorithm and are the subject of ongoing development. 

\appendix

\section{Proofs\label{sec:Proofs}}

\subsection{Bounds for the Time Horizon and Step Size\label{subsec:Proofs-1}}

\subsubsection*{General Assumptions}

Let $\Phi:\mathbb{R}^{n_{x}}\rightarrow\mathbb{R}$ be a measurable
convex function, satisfying 
\begin{alignat}{1}
\int_{\mathbb{R}^{n_{x}}}\exp\{-\Phi\left(\mathrm{x}\right)\} & <\infty,\label{eq:G-Energy-condition}\\
\begin{gathered}\Phi\left(\mathrm{x}\right)-\Phi\left(\bar{\mathrm{x}}\right)-\nabla_{\mathrm{x}}\Phi\left(\bar{\mathrm{x}}\right)^{T}\left(\mathrm{x}-\bar{\mathrm{x}}\right)\end{gathered}
 & \ge\frac{1}{2}m\left\Vert \mathrm{x}-\bar{\mathrm{x}}\right\Vert _{2}^{2},\label{eq:G-Lipschitz-conditions-1}\\
\left\Vert \nabla_{\mathrm{x}}\Phi\left(\mathrm{x}\right)-\nabla_{\mathrm{x}}\Phi\left(\bar{\mathrm{x}}\right)\right\Vert _{2} & \le M\left\Vert \mathrm{x}-\bar{\mathrm{x}}\right\Vert _{2},\quad\forall\mathrm{x},\bar{\mathrm{x}}\in\mathbb{R}^{n_{x}},\label{eq:G-Lipschitz-conditions-2}
\end{alignat}
\noindent for two existing positive constants $m$ and $M$. Let
$\bar{\mathrm{x}}\in\mathbb{R}^{n_{x}}$ be the global minimum of
$\Phi\left(\mathrm{x}\right)$. We define the log-concave target density
for a Langevin algorithm as $\pi(\mathrm{x})=e^{-\Phi(\mathrm{x})}\left(\int_{\mathbb{R}^{n_{x}}}e^{-\Phi(\mathrm{x})}d\mathrm{x}\right)^{-1}$.
\begin{lem}
\label{lem:L3}Suppose a probability measure $\mathcal{P}_{\mathcal{L},T}$
produced by the exact integration, up to time horizon $T$, of the
Langevin diffusion SDE 
\begin{equation}
d\mathcal{L}_{\lambda}=-\frac{1}{2}D_{\lambda}\nabla\Phi(\mathcal{L}_{\lambda})d\lambda+D_{\lambda}^{1/2}dW_{\lambda},\,\lambda\ge0,\,\mathcal{L}_{0}=0,\label{eq:L3-st-1}
\end{equation}
\noindent departing from the initial density $\nu(\mathrm{x})$ and
targeting the invariant density $\pi(\mathrm{x})\propto\exp\{-\Phi(\mathrm{x})\}$
with measure $\mathcal{P}_{\pi}(d\mathrm{x})$. Process $\left\{ W_{\lambda}\right\} _{\lambda\ge0}$
is the standard Wiener process and $D_{\lambda}$ is the diffusion
matrix. Under assumptions (\ref{eq:G-Energy-condition}), (\ref{eq:G-Lipschitz-conditions-1}),
and (\ref{eq:G-Lipschitz-conditions-2}), 
\begin{equation}
\Vert\mathcal{P}_{\mathcal{L},T}-\mathcal{P}_{\vphantom{\mathcal{L}}\pi}\Vert_{\text{TV}}\le e^{-\frac{1}{2}Tm}\mathbb{E}^{\mathcal{P}_{\pi}}\left[\nu^{2}/\pi^{2}\right]^{1/2},\,T\ge0.\label{eq:L3-st-2}
\end{equation}
\end{lem}
\begin{proof}
By denoting the Markov transition kernel as $P_{t}(\mathrm{x},\cdot)$,
an argument similar to that given by Dalalyan \citep{Dalalyan2014},
invoking the Cauchy-Schwarz inequality and the spectral gap bound
for the transition operator, gives 
\begin{align}
\Vert\mathcal{P}_{\mathcal{L},T}-\mathcal{P}_{\vphantom{\mathcal{L}}\pi}\Vert_{\text{TV}} & =\sup_{\mathcal{A}\in\mathscr{B}(\mathbb{R}^{n_{x}})}\left|\int_{\mathbb{R}^{n_{x}}}P_{t}(\mathrm{x},\mathcal{A})\nu(\mathrm{x})d\mathrm{x}-\mathcal{P}_{\pi}(\mathcal{A})\right|\nonumber \\
 & =\sup_{A\in\mathscr{B}(\mathbb{R}^{n_{x}})}\left|\int_{\mathbb{R}^{n_{x}}}\left(P_{t}(\mathrm{x},\mathcal{A})-\mathcal{P}_{\pi}(\mathcal{A})\right)\nu(\mathrm{x})d\mathrm{x}\right|\nonumber \\
 & =\sup_{A\in\mathscr{B}(\mathbb{R}^{n_{x}})}\left|\int_{\mathbb{R}^{n_{x}}}\left(P_{t}(\mathrm{x},\mathcal{A})-\mathcal{P}_{\pi}(\mathcal{A})\right)\frac{\nu(\mathrm{x})}{\pi(\mathrm{x})}\pi(\mathrm{x})d\mathrm{x}\right|\nonumber \\
\text{} & \le\sup_{A\in\mathscr{B}(\mathbb{R}^{n_{x}})}\int_{\mathbb{R}^{n_{x}}}\left|P_{t}(\mathrm{x},\mathcal{A})-\mathcal{P}_{\pi}(\mathcal{A})\right|\left|\frac{\nu(\mathrm{x})}{\pi(\mathrm{x})}\right|\pi(\mathrm{x})d\mathrm{x}\nonumber \\
\text{} & \le\sup_{A\in\mathscr{B}(\mathbb{R}^{n_{x}})}\left[\int_{\mathbb{R}^{n_{x}}}\left|P_{t}(\mathrm{x},\mathcal{A})-\mathcal{P}_{\pi}(\mathcal{A})\right|^{2}\pi(\mathrm{x})d\mathrm{x}\right]^{1/2}\left(\int_{\mathbb{R}^{n_{x}}}\left|\nu(\mathrm{x})/\pi(\mathrm{x})\right|^{2}\pi(\mathrm{x})d\mathrm{x}\right)^{1/2}\nonumber \\
\text{\text{(Dalalyan, 2014)}} & \le\frac{1}{2}e^{-\frac{1}{2}mT}\mathbb{E}^{\mathcal{P}_{\pi}}\left[\nu^{2}/\pi^{2}\right]^{1/2},\label{eq:L3-result}
\end{align}
\noindent where $\mathscr{B}(\mathbb{R}^{n_{x}})$ is the $\sigma$-algebra
of Borel sets of $\mathbb{R}^{n_{x}}$.
\end{proof}
\begin{lem}
\label{lem:L4}Under conditions (\ref{eq:G-Energy-condition}), (\ref{eq:G-Lipschitz-conditions-1})
and (\ref{eq:G-Lipschitz-conditions-2}), and assumptions of Lemma
\ref{lem:L3}, given the initial probability density $\nu(\mathrm{x})=\mathcal{N}(\mathrm{x};\,\mu_{\nu},\sigma_{\nu}^{2}\mathbb{I}_{n_{x}})$,
for $\sigma_{\nu}^{2}<2M^{-1}$, then 
\begin{equation}
\Vert\mathcal{P}_{\mathcal{L},T}-\mathcal{P}_{\vphantom{\mathcal{L}}\pi}\Vert_{\text{TV}}\le\frac{1}{2}\exp\left\{ -\frac{1}{2}mT+\frac{n_{x}}{4}\log\left(\frac{1}{m\sigma_{\nu}^{2}(2-M\sigma_{\nu}^{2})}\right)+\frac{M}{2(2-M\sigma_{\nu}^{2})}\Vert\bar{\mathrm{x}}-\mu_{\nu}\Vert_{2}^{2}\right\} .\label{eq:L4-st}
\end{equation}
\end{lem}
\begin{proof}
According to \emph{Lemma 4} in \citep{Dalalyan2014}, if (\ref{eq:G-Lipschitz-conditions-2})
holds then 
\begin{equation}
\begin{gathered}\Phi\left(\mathrm{x}\right)-\Phi\left(\bar{\mathrm{x}}\right)-\nabla_{\mathrm{x}}\Phi\left(\bar{\mathrm{x}}\right)^{T}\left(\mathrm{x}-\bar{\mathrm{x}}\right)\end{gathered}
\le\frac{M}{2}\left\Vert \mathrm{x}-\bar{\mathrm{x}}\right\Vert _{2}^{2}.\label{eq:Lemma-4-result}
\end{equation}
Thus, 
\begin{align}
\pi\left(\mathrm{x}\right) & \ge e^{-\frac{M}{2}\left\Vert \mathrm{x}-\bar{\mathrm{x}}\right\Vert _{2}^{2}-\Phi\left(\bar{\mathrm{x}}\right)}\left(\int_{\mathbb{R}^{n_{x}}}e^{-\frac{m}{2}\left\Vert \mathrm{x}-\bar{\mathrm{x}}\right\Vert _{2}^{2}-\Phi\left(\bar{\mathrm{x}}\right)}d\mathrm{x}\right)^{-1}\nonumber \\
 & =e^{-\frac{M}{2}\left\Vert \mathrm{x}-\bar{\mathrm{x}}\right\Vert _{2}^{2}-\Phi\left(\bar{\mathrm{x}}\right)+\Phi\left(\bar{\mathrm{x}}\right)}\left(\int_{\mathbb{R}^{n_{x}}}e^{-\frac{m}{2}\left\Vert \mathrm{x}-\bar{\mathrm{x}}\right\Vert _{2}^{2}}d\mathrm{x}\right)^{-1},\nonumber \\
\pi\left(\mathrm{x}\right) & \ge\left(2\pi m^{-1}\right)^{-n_{x}/2}e^{-\frac{M}{2}\left\Vert \mathrm{x}-\bar{\mathrm{x}}\right\Vert _{2}^{2}}.\label{eq:L4-pf-1}
\end{align}
We use (\ref{eq:L4-pf-1}), $\sigma_{\nu}^{-2}>M/2$, and define $W=2\sigma_{\nu}^{-2}$
to compute 
\begin{align}
\mathbb{E}^{\mathcal{P}_{\pi}}\left[\nu^{2}/\pi^{2}\right] & =\int_{\mathbb{R}^{n_{x}}}\left(\frac{\nu}{\pi}\right)^{2}\pi(\mathrm{x})d\mathrm{x}=\int_{\mathbb{R}^{n_{x}}}\frac{\nu^{2}}{\pi}d\mathrm{x}\nonumber \\
 & \le\int_{\mathbb{R}^{n_{x}}}\frac{\left[\left(2\pi\sigma_{\nu}^{2}\right)^{-n_{x}/2}e^{-\frac{\sigma_{\nu}^{-2}}{2}\Vert\mathrm{x}-\mu_{\nu}\Vert_{2}^{2}}\right]^{2}}{\left(2\pi m^{-1}\right)^{-n_{x}/2}e^{-\frac{M}{2}\left\Vert \mathrm{x}-\bar{\mathrm{x}}\right\Vert _{2}^{2}}}d\mathrm{x}\nonumber \\
 & =\left(2\pi\sigma_{\nu}^{2}\right)^{-n_{x}}\left(2\pi m^{-1}\right)^{n_{x}/2}\int_{\mathbb{R}^{n_{x}}}\frac{e^{-\sigma_{\nu}^{-2}\Vert\mathrm{x}-\mu_{\nu}\Vert_{2}^{2}}}{e^{-\frac{M}{2}\left\Vert \mathrm{x}-\bar{\mathrm{x}}\right\Vert _{2}^{2}}}d\mathrm{x}\nonumber \\
 & =\left(2\pi\cdot2W^{-1}\right)^{-n_{x}}\left(2\pi m^{-1}\right)^{n_{x}/2}\int_{\mathbb{R}^{n_{x}}}e^{-\frac{W}{2}\Vert\mathrm{x}-\mu_{\nu}\Vert_{2}^{2}+\frac{M}{2}\Vert\mathrm{x}-\bar{\mathrm{x}}\Vert_{2}^{2}}d\mathrm{x}\nonumber \\
 & =\frac{\left(2\pi m^{-1}\right)^{n_{x}/2}}{\left(2\pi\cdot2W^{-1}\right)^{n_{x}}}e^{\frac{1}{2}(W-M)^{-1}WM\Vert\bar{\mathrm{x}}-\mu_{\nu}\Vert_{2}^{2}}\int_{\mathbb{R}^{n_{x}}}e^{-\frac{(W-M)}{2}\Vert\mathrm{x}-(W-M)^{-1}(W\mu_{\nu}-M\bar{\mathrm{x}})\Vert_{2}^{2}}d\mathrm{x}\nonumber \\
 & =\frac{\left(2\pi m^{-1}\right)^{n_{x}/2}}{\left(2\pi\cdot2W^{-1}\right)^{n_{x}}}e^{\frac{1}{2}\left(M^{-1}-W^{-1}\right)^{-1}\Vert\bar{\mathrm{x}}-\mu_{\nu}\Vert_{2}^{2}}\left(2\pi(W-M)^{-1}\right)^{n_{x}/2}\nonumber \\
 & =\left(\frac{W^{2}}{4m(W-M)}\right)^{n_{x}/2}e^{\frac{1}{2}\left(M^{-1}-W^{-1}\right)^{-1}\Vert\bar{\mathrm{x}}-\mu_{\nu}\Vert_{2}^{2}}\nonumber \\
 & =\left(\frac{\sigma_{\nu}^{-4}}{m(2\sigma_{\nu}^{-2}-M)}\right)^{n_{x}/2}e^{\frac{1}{2}\left(M^{-1}-\frac{1}{2}\sigma_{\nu}^{2}\right)^{-1}\Vert\bar{\mathrm{x}}-\mu_{\nu}\Vert_{2}^{2}}\nonumber \\
 & =\left(\frac{1}{m\sigma_{\nu}^{2}(2-M\sigma_{\nu}^{2})}\right)^{n_{x}/2}e^{M\left(2-M\sigma_{\nu}^{2}\right)^{-1}\Vert\bar{\mathrm{x}}-\mu_{\nu}\Vert_{2}^{2}}\nonumber \\
\therefore\mathbb{E}^{\mathcal{P}_{\pi}}\left[\nu^{2}/\pi^{2}\right]^{1/2} & \le\exp\left\{ \frac{n_{x}}{4}\log\left(\frac{1}{m\sigma_{\nu}^{2}(2-M\sigma_{\nu}^{2})}\right)+\frac{M}{2(2-M\sigma_{\nu}^{2})}\Vert\bar{\mathrm{x}}-\mu_{\nu}\Vert_{2}^{2}\right\} .\label{eq:L4-result}
\end{align}

The result is complete by incorporating (\ref{eq:L4-result}) into
the result (\ref{eq:L3-result}). 
\end{proof}
\begin{lem}
\label{lem:L5}For $d\nu(\mathrm{x})\sim\delta(\mathrm{x}-\mathrm{x}_{\nu})d\mathrm{x}$,
the Radon-Nikodym derivative 
\begin{equation}
\frac{d\nu}{d\mathcal{P}_{\pi}}=\frac{1}{\int_{\Omega(\mathrm{x}_{\nu})}\pi(\mathrm{x}^{\prime})d\mathrm{\mathrm{x}^{\prime}}}\label{eq:L5-st}
\end{equation}
\noindent is valid, where $d\mathcal{P}_{\pi}(\mathrm{x})=\pi(\mathrm{x})d\mathrm{x}$
and the integral is taken over $\Omega(\mathrm{x}_{\nu})=\left\{ \mathrm{x}^{\prime}\in\mathbb{R}^{n_{x}}:\,\Vert\mathrm{x}^{\prime}-\bar{\mathrm{x}}\Vert_{2}\ge\Vert\mathrm{x}_{\nu}-\bar{\mathrm{x}}\Vert_{2}\right\} $.
\end{lem}
\begin{proof}
The Dirac-delta measure is singular, i.e., not absolutely continuous
with respect to the Lebesgue measure, and hence its Radon-Nikodym
derivative may not be formally defined in general. However, by abusing
notation one can obtain a functional expression for the enunciated
derivative. Observe the definition of the Radon-Nikodym derivative
as 
\begin{equation}
\frac{d\mathcal{P_{\pi}}}{d\nu}=f\implies\mathcal{P}_{\pi}(\mathcal{D})=\int_{\mathcal{D}}f\,d\nu,\label{eq:L5-pf-1}
\end{equation}
\noindent for any measurable domain $\mathcal{D}\subseteq\mathbb{R}^{n_{x}}$.
Given $\nu(\mathrm{x}_{\nu})=\delta(\mathrm{x}_{\nu}-\mathrm{x})\equiv\delta(\mathrm{x}-\mathrm{x}_{\nu})=\nu(\mathrm{x})$
and setting $f(\mathrm{x})\coloneqq\int_{\Omega(\mathrm{x})}\pi(\mathrm{x}^{\prime})d\mathrm{\mathrm{x}^{\prime}}$,
we write 
\begin{equation}
\mathcal{P}_{\pi}(\mathrm{x})=\int_{\mathbb{R}^{n_{x}}}f(\mathrm{x}_{\nu})\,d\nu(\mathrm{x}_{\nu})=\int_{\mathbb{R}^{n_{x}}}f(\mathrm{x}_{\nu})\,\delta(\mathrm{x}_{\nu}-\mathrm{x})d\mathrm{x}_{\nu}=f(\mathrm{x})=\int_{\Omega(\mathrm{x})}\pi(\mathrm{x}^{\prime})d\mathrm{\mathrm{x}^{\prime}},\label{eq:L5-pf-2}
\end{equation}
\noindent where $\frac{d\mathcal{P}_{\pi}}{d\mathrm{x}}=\pi(\mathrm{x})$
accordingly, and $f(\mathrm{x}_{\nu})=\int_{\Omega(\mathrm{x}_{\nu})}\pi(\mathrm{x}^{\prime})d\mathrm{\mathrm{x}^{\prime}}$.
Notice that $f(\mathrm{x})$ plays the role of a cumulative distribution
function that is integrated over the ``tail'' of $\pi(\mathrm{x}^{\prime})$,
in the region defined by $\Omega(\mathrm{x})$. 

Thus, applying definition (\ref{eq:L5-pf-1}), we get 
\begin{align}
\frac{d\mathcal{P}_{\pi}}{d\nu} & \coloneqq f(\mathrm{x}_{\nu})=\int_{\Omega(\mathrm{x}_{\nu})}\pi(\mathrm{x}^{\prime})d\mathrm{\mathrm{x}^{\prime}},\nonumber \\
\therefore\frac{d\nu}{d\mathcal{P}_{\pi}} & =\frac{1}{\int_{\Omega(\mathrm{x}_{\nu})}\pi(\mathrm{x}^{\prime})d\mathrm{\mathrm{x}^{\prime}}}.\label{eq:L5-result}
\end{align}
\end{proof}
\begin{rem}
\label{rem:R6}Notice that if $\pi(\mathrm{x}^{\prime})=(2\pi P^{-1})^{-\frac{n_{x}}{2}}e^{-\frac{P}{2}\Vert\mathrm{x}^{\prime}-\bar{\mathrm{x}}\Vert_{2}^{2}}$,
for some $P\in\mathbb{R}_{+}$, we can compute the resulting integral
in \emph{Lemma \ref{lem:L5}} using the spherical symmetry of the
Gaussian function to give 
\begin{align*}
\int_{\Omega(\mathrm{x}_{\nu})}\pi(\mathrm{x}^{\prime})d\mathrm{\mathrm{x}^{\prime}} & =(2\pi P^{-1})^{-\frac{n_{x}}{2}}\int_{\Omega(\mathrm{x}_{\nu})}e^{-\frac{P}{2}\Vert\mathrm{x}^{\prime}-\bar{\mathrm{x}}\Vert_{2}^{2}}d\mathrm{\mathrm{x}^{\prime}}\\
 & =(2\pi)^{-\frac{n_{x}}{2}}\int_{\{\Vert u\Vert_{2}\ge\sqrt{P}\Vert\mathrm{x}_{\nu}-\bar{\mathrm{x}}\Vert_{2}\}}e^{-\frac{1}{2}\Vert u\Vert_{2}^{2}}du\\
 & =(2\pi)^{-\frac{n_{x}}{2}}\int_{r_{\nu}}^{\infty}\int_{\mathcal{S}^{n_{x}-1}(r)}e^{-\frac{1}{2}r^{2}}d\mathcal{A}\,dr\\
 & =(2\pi)^{-\frac{n_{x}}{2}}\int_{r_{\nu}}^{\infty}e^{-\frac{1}{2}r^{2}}\mathcal{A}_{n_{x}-1}(r)\,dr\\
 & =(2\pi)^{-\frac{n_{x}}{2}}\mathcal{A}_{n_{x}-1}(1)\int_{r_{\nu}}^{\infty}e^{-\frac{1}{2}r^{2}}r^{n_{x}-1}\,dr\\
 & =(2\pi)^{-\frac{n_{x}}{2}}\frac{2\pi^{\frac{n_{x}}{2}}}{\Gamma\left(\frac{n_{x}}{2}\right)}\int_{r_{\nu}}^{\infty}e^{-\frac{1}{2}r^{2}}r^{n_{x}-1}\,dr\\
 & =(2\pi)^{-\frac{n_{x}}{2}}\frac{2\pi^{\frac{n_{x}}{2}}}{\Gamma\left(\frac{n_{x}}{2}\right)}2^{\frac{n_{x}}{2}-1}\int_{r_{\nu}^{2}/2}^{\infty}e^{-t}\cdot t^{\frac{n_{x}}{2}-1}\,dt\\
 & =\frac{1}{\Gamma\left(\frac{n_{x}}{2}\right)}\int_{r_{\nu}^{2}/2}^{\infty}t^{\frac{n_{x}}{2}-1}e^{-t}\,dt\\
 & =\Gamma_{u}\left(\frac{n_{x}}{2},\frac{r_{\nu}^{2}}{2}\right)=\Gamma_{u}\left(\frac{n_{x}}{2},\frac{P\Vert\mathrm{x}_{\nu}-\bar{\mathrm{x}}\Vert_{2}^{2}}{2}\right),
\end{align*}
\noindent where $du=d\mathcal{A}\,dr$ for a volume element $du\in\mathbb{R}^{n_{x}}$,
area element $d\mathcal{A}\in\mathbb{R}^{n_{x}-1}$ and radius element
$dr\in\mathbb{R}_{+}$. In addition, $\mathcal{S}^{n_{x}-1}(r)$ denotes
the $(n_{x}-1)$-sphere of radius $r$, with a total surface area
of $\mathcal{A}_{n_{x}-1}(r)$, and the recursion $\mathcal{A}_{n_{x}-1}(r)=\mathcal{A}_{n_{x}-1}(1)r^{n_{x}-1}$
has been applied with $\mathcal{A}_{n_{x}-1}(1)=2\pi^{\frac{n_{x}}{2}}/\Gamma(n_{x}/2)$.
The lower integration extreme is taken as $r_{\nu}\coloneqq\sqrt{P}\Vert\mathrm{x}_{\nu}-\bar{\mathrm{x}}\Vert_{2}$,
and $\Gamma_{u}\left(s,x\right)=\Gamma(s)^{-1}\int_{x}^{\infty}t^{s-1}e^{-t}dt$
is the upper incomplete gamma function. Also, we recall that $\int_{\Omega(\mathrm{x}_{\nu})}\pi(\mathrm{x}^{\prime})d\mathrm{\mathrm{x}^{\prime}}=1-\int_{\mathbb{R}^{n_{x}}\setminus\Omega(\mathrm{x}_{\nu})}\pi(\mathrm{x}^{\prime})d\mathrm{\mathrm{x}^{\prime}}$.
\end{rem}
\begin{lem}
\label{lem:L6}Under conditions (\ref{eq:G-Energy-condition}), (\ref{eq:G-Lipschitz-conditions-1})
and (\ref{eq:G-Lipschitz-conditions-2}), and assumptions of Lemma
\ref{lem:L3}, given a initial probability mass located at $\mathrm{x}=\mathrm{x}_{\nu}$,
i.e., $\nu(\mathrm{x})=\delta(\mathrm{x}-\mathrm{x}_{\nu})$ (Dirac
delta), then 
\begin{equation}
\Vert\mathcal{P}_{\mathcal{L},T}-\mathcal{P}_{\vphantom{\mathcal{L}}\pi}\Vert_{\text{TV}}\le\frac{1}{2}\exp\left\{ -\frac{1}{2}mT+\frac{n_{x}}{2}\log\left(\frac{M}{m}\right)-\log\left[\Gamma_{u}\left(\frac{n_{x}}{2},\frac{M\Vert\bar{\mathrm{x}}-\mathrm{x}_{\nu}\Vert_{2}^{2}}{2}\right)\right]\right\} .\label{eq:L6-st}
\end{equation}
\end{lem}
\begin{proof}
Considering $\nu(\mathrm{x})=\delta(\mathrm{x}-\mathrm{x}_{\nu})$
on the definition of total variation distance, we use \emph{Lemma~\ref{lem:L5}}
and \emph{Remark~\ref{rem:R6}}. The procedure follows as 
\begin{align}
\Vert\mathcal{P}_{\mathcal{L},T}-\mathcal{P}_{\vphantom{\mathcal{L}}\pi}\Vert_{\text{TV}} & =\sup_{\mathcal{A}\in\mathscr{B}(\mathbb{R}^{n_{x}})}\left|\int_{\mathbb{R}^{n_{x}}}P_{t}(\mathrm{x},\mathcal{A})d\nu(\mathrm{x})-\mathcal{P}_{\pi}(\mathcal{A})\right|\nonumber \\
 & =\sup_{A\in\mathscr{B}(\mathbb{R}^{n_{x}})}\left|\int_{\mathbb{R}^{n_{x}}}\left(P_{t}(\mathrm{x},\mathcal{A})-\mathcal{P}_{\pi}(\mathcal{A})\right)d\nu(\mathrm{x})\right|\nonumber \\
 & =\sup_{A\in\mathscr{B}(\mathbb{R}^{n_{x}})}\left|\int_{\mathbb{R}^{n_{x}}}\left(P_{t}(\mathrm{x},\mathcal{A})-\mathcal{P}_{\pi}(\mathcal{A})\right)\frac{d\nu}{d\mathcal{P}_{\pi}}d\mathcal{P}_{\pi}(\mathrm{x})\right|\nonumber \\
\text{(Lemma 5)} & =\sup_{A\in\mathscr{B}(\mathbb{R}^{n_{x}})}\left|\int_{\mathbb{R}^{n_{x}}}\left(P_{t}(\mathrm{x},\mathcal{A})-\mathcal{P}_{\pi}(\mathcal{A})\right)\frac{1}{\int_{\Omega(\mathrm{x}_{\nu})}\pi(\mathrm{x}^{\prime})d\mathrm{\mathrm{x}^{\prime}}}d\mathcal{P}_{\pi}(\mathrm{x})\right|\nonumber \\
 & \le\frac{1}{\int_{\Omega(\mathrm{x}_{\nu})}\pi(\mathrm{x}^{\prime})d\mathrm{\mathrm{x}^{\prime}}}\cdot\sup_{A\in\mathscr{B}(\mathbb{R}^{n_{x}})}\int_{\mathbb{R}^{n_{x}}}\left|P_{t}(\mathrm{x},\mathcal{A})-\mathcal{P}_{\pi}(\mathcal{A})\right|\pi(\mathrm{x})d\mathrm{x}\nonumber \\
 & \le\frac{1}{\int_{\Omega(\mathrm{x}_{\nu})}\pi(\mathrm{x}^{\prime})d\mathrm{\mathrm{x}^{\prime}}}\cdot\sup_{A\in\mathscr{B}(\mathbb{R}^{n_{x}})}\left[\int_{\mathbb{R}^{n_{x}}}\left|P_{t}(\mathrm{x},\mathcal{A})-\mathcal{P}_{\pi}(\mathcal{A})\right|^{2}\pi(\mathrm{x})d\mathrm{x}\right]^{1/2}\nonumber \\
\text{} & \le\frac{\left(2\pi m^{-1}\right)^{n_{x}/2}}{\int_{\Omega(\mathrm{x}_{\nu})}e^{-\frac{M}{2}\left\Vert \mathrm{x}^{\prime}-\bar{\mathrm{x}}\right\Vert _{2}^{2}}d\mathrm{x}^{\prime}}\sup_{A\in\mathscr{B}(\mathbb{R}^{n_{x}})}\left[\int_{\mathbb{R}^{n_{x}}}\left|P_{t}(\mathrm{x},\mathcal{A})-\mathcal{P}_{\pi}(\mathcal{A})\right|^{2}\pi(\mathrm{x})d\mathrm{x}\right]^{1/2}\nonumber \\
 & =\frac{\left(2\pi m^{-1}\right)^{n_{x}/2}\left(2\pi M^{-1}\right)^{-n_{x}/2}}{\int_{\Omega(\mathrm{x}_{\nu})}\mathcal{N}(\mathrm{x}^{\prime};\,\bar{\mathrm{x}},M^{-1}\mathbb{I}_{n_{x}})d\mathrm{x}^{\prime}}\sup_{A\in\mathscr{B}(\mathbb{R}^{n_{x}})}\left[\int_{\mathbb{R}^{n_{x}}}\left|P_{t}(\mathrm{x},\mathcal{A})-\mathcal{P}_{\pi}(\mathcal{A})\right|^{2}\pi(\mathrm{x})d\mathrm{x}\right]^{1/2}\nonumber \\
\text{(Remark 6)} & =\frac{\left(Mm^{-1}\right)^{n_{x}/2}}{\Gamma_{u}\left(\frac{n_{x}}{2},\frac{M\Vert\bar{\mathrm{x}}-\mathrm{x}_{\nu}\Vert_{2}^{2}}{2}\right)}\sup_{A\in\mathscr{B}(\mathbb{R}^{n_{x}})}\left[\int_{\mathbb{R}^{n_{x}}}\left|P_{t}(\mathrm{x},\mathcal{A})-\mathcal{P}_{\pi}(\mathcal{A})\right|^{2}\pi(\mathrm{x})d\mathrm{x}\right]^{1/2}\nonumber \\
\text{\text{(Dalalyan, 2014)}} & \le\left(\frac{M}{m}\right)^{n_{x}/2}\cdot\Gamma_{u}\left(\frac{n_{x}}{2},\frac{M\Vert\mathrm{x}_{\nu}-\bar{\mathrm{x}}\Vert_{2}^{2}}{2}\right)^{-1}\cdot\frac{1}{2}e^{-\frac{1}{2}mT}\nonumber \\
 & =\frac{1}{2}\exp\left\{ -\frac{1}{2}mT+\frac{n_{x}}{2}\log\left(\frac{M}{m}\right)-\log\left[\Gamma_{u}\left(\frac{n_{x}}{2},\frac{M\Vert\bar{\mathrm{x}}-\mathrm{x}_{\nu}\Vert_{2}^{2}}{2}\right)\right]\right\} .\label{eq:L6-result}
\end{align}
\end{proof}
\begin{lem}
\label{lem:L7-1}Suppose a probability measure $\mathcal{P}_{\tilde{\mathcal{L}}[\Delta\lambda],T}$
produced by numerical integration with step size, $\Delta\lambda$,
up to time horizon $T=L\Delta\lambda$, of the Langevin diffusion
SDE according to 
\begin{equation}
d\tilde{\mathcal{L}}_{\lambda}=-\frac{1}{2}\sum_{l=0}^{L-1}D_{l\Delta\lambda}\nabla\tilde{\Phi}(\tilde{\mathcal{L}}_{l\Delta\lambda})\mathbbm{1}_{[l\Delta\lambda,(l+1)\Delta\lambda)}d\lambda+\sum_{l=0}^{L-1}D_{l\Delta\lambda}^{1/2}dW_{\lambda},\,0\le\lambda\le L\Delta\lambda,\,\tilde{\mathcal{L}}_{0}=\mathrm{0}.\label{eq:L8-st-1}
\end{equation}

The resulting Markov chain is assumed to depart from the initial density
$\nu(\mathrm{x}_{0})=\mathcal{N}(\mathrm{x}_{0};\,\mu_{\nu},\sigma_{\nu}^{2}\mathbb{I}_{n_{x}})$
and targets the invariant density $\pi(\mathrm{x})\propto\exp\{-\Phi(\mathrm{x})\}$
with measure $\mathcal{P}_{\pi}(d\mathrm{x})$. The process $\left\{ W_{\lambda}\right\} _{\lambda\ge0}$
is the standard Wiener process and the diffusion matrix $D_{\lambda}=D(\mathrm{x}_{\lambda})$
is bounded by $K=\sup_{\mathrm{x}}\Vert D(\mathrm{x})\Vert_{2}$.
Under assumptions (\ref{eq:G-Energy-condition}), (\ref{eq:G-Lipschitz-conditions-1}),
and (\ref{eq:G-Lipschitz-conditions-2}), then 
\begin{multline}
\Vert\mathcal{P}_{\mathcal{L\vphantom{\tilde{\mathcal{L}}}},T}-\mathcal{P}_{\tilde{\mathcal{L}}[\Delta\lambda],T}\Vert_{\text{TV}}\le\frac{1}{2}\\
-\frac{1}{2}\exp\left\{ -\frac{n_{x}}{2}\frac{M^{3}K^{4}\gamma}{48(2\gamma-1)}\left(\sigma_{\nu}^{2}+\frac{1}{n_{x}}\Vert\mathrm{\bar{x}}-\mu_{\nu}\Vert_{2}^{2}+2T\right)\Delta\lambda^{2}-\frac{n_{x}M^{2}K^{3}T}{16}\Delta\lambda\right\} ,\,T\ge0;\label{eq:L7-st-2}
\end{multline}
\noindent where $\mathcal{P}_{\mathcal{L},T}$ is the probability
measure produced by the exact integration of the Langevin diffusion
SDE.
\end{lem}
\begin{proof}
We take a different approach as that proposed by Dalalyan \citep{Dalalyan2014}
for this proof. Instead of bounding the total variance distance $\Vert\mathcal{P}_{\mathcal{L\vphantom{\tilde{\mathcal{L}}}},T}-\mathcal{P}_{\tilde{\mathcal{L}}[\Delta\lambda],T}\Vert_{\text{TV}}$
by the Pinsker inequality, we treat it directly. First, identify that
\begin{align}
\Vert\mathcal{P}_{\mathcal{L\vphantom{\tilde{\mathcal{L}}}},T}-\mathcal{P}_{\tilde{\mathcal{L}}[\Delta\lambda],T}\Vert_{\text{TV}} & =\frac{1}{2}\int_{\mathbb{R}^{n_{x}}}\left|d\mathcal{P}_{\mathcal{L\vphantom{\tilde{\mathcal{L}}}},T}-d\mathcal{P}_{\tilde{\mathcal{L}}[\Delta\lambda],T}\right|\nonumber \\
 & =\frac{1}{2}\int_{\mathbb{R}^{n_{x}}}\left|1-\frac{d\mathcal{P}_{\tilde{\mathcal{L}}[\Delta\lambda],T}}{d\mathcal{P}_{\mathcal{L},T}}\right|d\mathcal{P}_{\mathcal{L\vphantom{\tilde{\mathcal{L}}}},T}\nonumber \\
 & =\frac{1}{2}\int_{\mathbb{R}^{n_{x}}}\left|1-\frac{\nu(\mathrm{x}_{0})dP_{t}^{\tilde{\mathcal{L}}}(\mathrm{x}_{0},\mathrm{x})}{\nu(\mathrm{x}_{0})dP_{t}^{\mathcal{L}}(\mathrm{x}_{0},\mathrm{x})}\right|d\mathcal{P}_{\mathcal{L},T}\nonumber \\
 & =\frac{1}{2}\int_{\mathbb{R}^{n_{x}}}\left|1-\frac{dP_{t}^{\tilde{\mathcal{L}}}(\mathrm{x}_{0},\mathrm{x})}{dP_{t}^{\mathcal{L}}(\mathrm{x}_{0},\mathrm{x})}\right|d\mathcal{P}_{\mathcal{L},T}(\mathrm{x})\nonumber \\
 & =\frac{1}{2}\int_{\mathbb{R}^{n_{x}}}\mathbb{E}^{P_{t}^{\mathcal{L}}}\left[\left|1-\frac{dP_{t}^{\tilde{\mathcal{L}}}(\mathrm{x}_{0},\mathrm{x})}{dP_{t}^{\mathcal{L}}(\mathrm{x}_{0},\mathrm{x})}\right||\mathrm{x}_{0}\right]d\nu(\mathrm{x}_{0}),\label{eq:L7-pf-1}
\end{align}
\noindent where $\mathcal{P}_{\mathcal{L},T}(d\mathrm{x})=\int_{\mathbb{R}^{n_{x}}}P_{t}^{\mathcal{L}}(\mathrm{x}_{0},d\mathrm{x})\nu(\mathrm{x}_{0})d\mathrm{x}_{0}$,
and $P_{t}^{\tilde{\mathcal{L}}}(\mathrm{x}_{0},\cdot)$ and $P_{t}^{\mathcal{L}}(\mathrm{x}_{0},\cdot)$
are the Markov transition kernels for the discrete-time and continuous-time
Langevin dynamics respectively. By applying Girsanov's theorem to
change the measure from $\mathcal{P}_{\mathcal{L\vphantom{\tilde{\mathcal{L}}}},T}$
to $\mathcal{P}_{\tilde{\mathcal{L}}[\Delta\lambda],T}$, one obtains
(see step 3 of the proof of \emph{Proposition 2} in \citep{Dalalyan2012})
\begin{multline}
\frac{dP_{t}^{\tilde{\mathcal{L}}}}{dP_{t}^{\mathcal{L}}}(\mathrm{x})\\
=\exp\left\{ \int_{0}^{T}\frac{1}{2}\left(-\sum_{l=0}^{L-1}D_{l\Delta\lambda}\nabla\tilde{\Phi}(\mathrm{x}_{l\Delta\lambda})\mathbbm{1}_{[l\Delta\lambda,(l+1)\Delta\lambda)}+D_{\lambda}\nabla\Phi(\mathrm{x}_{\lambda})\right)^{T}D_{\lambda}^{-1}\left(d\mathcal{L}_{\lambda}+\frac{1}{2}D_{\lambda}\nabla\Phi(\mathcal{L}_{\lambda})d\lambda\right)\right\} \\
\times\exp\left\{ -\frac{1}{2}\int_{0}^{T}\left\Vert -\frac{1}{2}\sum_{l=0}^{L-1}D_{l\Delta\lambda}\nabla\tilde{\Phi}(\mathrm{x}_{l\Delta\lambda})\mathbbm{1}_{[l\Delta\lambda,(l+1)\Delta\lambda)}+\frac{1}{2}D_{\lambda}\nabla\Phi(\mathrm{x}_{\lambda})\right\Vert _{2}^{2}d\lambda\right\} .\label{eq:L7-pf-2}
\end{multline}
Thus, since $\mathbb{E}\left[e^{z}\right]\ge e^{\mathbb{E}\left[z\right]}$
by Jensen's inequality, $D_{\lambda}=D(\mathrm{x}_{\lambda})\le K$,
and $\nabla\Phi(\mathrm{x})$ is Lipschitz-continuous with constant
$M$ (\ref{eq:G-Lipschitz-conditions-2}), we obtain 
\begin{align}
\mathbb{E}^{P_{t}^{\mathcal{L}}}\left[\frac{dP_{t}^{\tilde{\mathcal{L}}}}{dP_{t}^{\mathcal{L}}}(\mathrm{x})|\mathrm{x}_{0}\right] & \ge\exp\left\{ -\mathbb{E}^{P_{t}^{\mathcal{L}}}\left[\frac{1}{2}\sum_{l=0}^{L-1}\int_{l\Delta\lambda}^{(l+1)\Delta\lambda}\left\Vert \frac{1}{2}D_{\lambda}\nabla\Phi(\mathrm{x}_{\lambda})-\frac{1}{2}D_{l\Delta\lambda}\nabla\tilde{\Phi}(\mathrm{x}_{l\Delta\lambda})\right\Vert _{2}^{2}d\lambda|\mathrm{x}_{0}\right]\right\} \nonumber \\
 & =\exp\left\{ -\frac{1}{2}\sum_{l=0}^{L-1}\int_{l\Delta\lambda}^{(l+1)\Delta\lambda}\mathbb{E}^{P_{t}^{\mathcal{L}}}\left[\left\Vert \frac{1}{2}D_{\lambda}\nabla\Phi(\mathrm{x}_{\lambda})-\frac{1}{2}D_{l\Delta\lambda}\nabla\tilde{\Phi}(\mathrm{x}_{l\Delta\lambda})\right\Vert _{2}^{2}|\mathrm{x}_{0}\right]d\lambda\right\} \nonumber \\
 & \ge\exp\left\{ -\frac{K^{2}}{8}\sum_{l=0}^{L-1}\int_{l\Delta\lambda}^{(l+1)\Delta\lambda}\mathbb{E}^{P_{t}^{\mathcal{L}}}\left[\left\Vert \nabla\Phi(\mathrm{x}_{\lambda})-\nabla\tilde{\Phi}(\mathrm{x}_{l\Delta\lambda})\right\Vert _{2}^{2}|\mathrm{x}_{0}\right]d\lambda\right\} \nonumber \\
 & \ge\exp\left\{ -\frac{K^{2}M^{2}}{8}\sum_{l=0}^{L-1}\int_{l\Delta\lambda}^{(l+1)\Delta\lambda}\mathbb{E}^{P_{t}^{\mathcal{L}}}\left[\left\Vert \mathrm{x}_{\lambda}-\mathrm{x}_{l\Delta\lambda}\right\Vert _{2}^{2}|\mathrm{x}_{0}\right]d\lambda\right\} .\label{eq:L7-pf-3}
\end{align}
For each step $\mathrm{x}_{\lambda}-\mathrm{x}_{l\Delta\lambda}=-\frac{1}{2}D_{l\Delta\lambda}\nabla\Phi(\mathrm{x}_{l\Delta\lambda})(\lambda-l\Delta\lambda)+D_{l\Delta\lambda}^{1/2}(W_{\lambda}-W_{l\Delta\lambda})$,
hence 
\begin{align}
\mathbb{E}^{P_{t}^{\mathcal{L}}}\left[\frac{dP_{t}^{\tilde{\mathcal{L}}}}{dP_{t}^{\mathcal{L}}}(\mathrm{x})|\mathrm{x}_{0}\right] & \ge\exp\left\{ -\frac{K^{2}M^{2}}{8}\sum_{l=0}^{L-1}\int_{l\Delta\lambda}^{(l+1)\Delta\lambda}\mathbb{E}^{P_{t}^{\mathcal{L}}}\left[\left\Vert \frac{1}{2}D(\mathrm{x}_{l\Delta\lambda})\nabla\Phi(\mathrm{x}_{l\Delta\lambda})\right\Vert _{2}^{2}(\lambda-l\Delta\lambda)^{2}|\mathrm{x}_{0}\right]d\lambda\right\} \nonumber \\
 & \times\exp\left\{ -\frac{K^{2}M^{2}}{8}\sum_{l=0}^{L-1}\int_{l\Delta\lambda}^{(l+1)\Delta\lambda}\mathbb{E}^{P_{t}^{\mathcal{L}}}\left[\left\Vert D(\mathrm{x}_{l\Delta\lambda})^{1/2}(W_{\lambda}-W_{l\Delta\lambda})\right\Vert _{2}^{2}|\mathrm{x}_{0}\right]d\lambda\right\} \nonumber \\
 & \ge\exp\left\{ -\frac{K^{2}M^{2}}{8}\sum_{l=0}^{L-1}\int_{l\Delta\lambda}^{(l+1)\Delta\lambda}\frac{K^{2}}{4}\mathbb{E}^{P_{t}^{\mathcal{L}}}\left[\Vert\nabla\Phi(\mathrm{x}_{l\Delta\lambda})\Vert_{2}^{2}|\mathrm{x}_{0}\right](\lambda-l\Delta\lambda)^{2}d\lambda\right\} \nonumber \\
 & \times\exp\left\{ -\frac{K^{2}M^{2}}{8}\sum_{l=0}^{L-1}\int_{l\Delta\lambda}^{(l+1)\Delta\lambda}K(\lambda-l\Delta\lambda)n_{x}d\lambda\right\} \nonumber \\
 & =\exp\left\{ -\frac{M^{2}K^{4}\Delta\lambda^{3}}{96}\sum_{l=0}^{L-1}\mathbb{E}^{P_{t}^{\mathcal{L}}}\left[\Vert\nabla\Phi(\mathrm{x}_{l\Delta\lambda})\Vert_{2}^{2}|\mathrm{x}_{0}\right]-\frac{n_{x}M^{2}K^{3}L\Delta\lambda^{2}}{16}\right\} .\label{eq:L7-pf-4}
\end{align}

Now we invoke a result from \emph{Corollary 4} in \citep{Dalalyan2014},
that gives 
\begin{equation}
\Delta\lambda\sum_{l=0}^{L-1}\mathbb{E}\left[\Vert\nabla\Phi(\mathrm{x}_{l\Delta\lambda})\Vert_{2}^{2}\right]\le\frac{M\gamma}{2\gamma-1}\mathbb{E}\left[\Vert\mathrm{x}_{0}-\bar{\mathrm{x}}\Vert_{2}^{2}\right]+\frac{2\gamma MTn_{x}}{2\gamma-1},\label{eq:L7-pf-5}
\end{equation}
\noindent for some $\gamma>1$, $\Delta\lambda\le(\gamma M)^{-1}$
and $L>1$. Substituting $L\Delta\lambda=T$ and incorporating (\ref{eq:L7-pf-5}):
\begin{align}
\mathbb{E}^{P_{t}^{\mathcal{L}}}\left[\frac{dP_{t}^{\tilde{\mathcal{L}}}}{dP_{t}^{\mathcal{L}}}(\mathrm{x})|\mathrm{x}_{0}\right] & \ge\exp\left\{ -\frac{M^{2}K^{4}\Delta\lambda^{2}}{48}\frac{\gamma}{2\gamma-1}\left(\frac{M}{2}\mathbb{E}^{P_{t}^{\mathcal{L}}}\left[\Vert\mathrm{x}_{0}-\bar{\mathrm{x}}\Vert_{2}^{2}|\mathrm{x}_{0}\right]+n_{x}MT\right)-\frac{n_{x}M^{2}K^{3}T\Delta\lambda}{16}\right\} \nonumber \\
 & =\exp\left\{ -\frac{M^{2}K^{4}\Delta\lambda^{2}}{48}\frac{\gamma}{2\gamma-1}\left(\frac{M}{2}\Vert\mathrm{x}_{0}-\bar{\mathrm{x}}\Vert_{2}^{2}+n_{x}MT\right)-\frac{n_{x}M^{2}K^{3}T\Delta\lambda}{16}\right\} .\label{eq:L7-pf-6}
\end{align}
Since 
\[
0\le\mathbb{E}^{P_{t}^{\mathcal{L}}}\left[\frac{dP_{t}^{\tilde{\mathcal{L}}}}{dP_{t}^{\mathcal{L}}}(\mathrm{x})|\mathrm{x}_{0}\right]\le1,
\]
\noindent and applying expression (\ref{eq:L7-pf-6}) in (\ref{eq:L7-pf-1}),
we have 
\begin{align}
\Vert\mathcal{P}_{\mathcal{L\vphantom{\tilde{\mathcal{L}}}},T}-\mathcal{P}_{\tilde{\mathcal{L}}[\Delta\lambda],T}\Vert_{\text{TV}} & =\frac{1}{2}\int_{\mathbb{R}^{n_{x}}}\left|1-\frac{dP_{t}^{\tilde{\mathcal{L}}}(\mathrm{x}_{0},\mathrm{x})}{dP_{t}^{\mathcal{L}}(\mathrm{x}_{0},\mathrm{x})}\right|d\mathcal{P}_{\mathcal{L},T}(\mathrm{x})\nonumber \\
 & =\frac{1}{2}\int_{\mathbb{R}^{n_{x}}}d\mathcal{P}_{\mathcal{L},T}(\mathrm{x})-\frac{1}{2}\int_{\mathbb{R}^{n_{x}}}\mathbb{E}^{P_{t}^{\mathcal{L}}}\left[\frac{dP_{t}^{\tilde{\mathcal{L}}}(\mathrm{x}_{0},\mathrm{x})}{dP_{t}^{\mathcal{L}}(\mathrm{x}_{0},\mathrm{x})}|\mathrm{x}_{0}\right]d\nu(\mathrm{x}_{0})\nonumber \\
 & \le\frac{1}{2}-\frac{1}{2}\int_{\mathbb{R}^{n_{x}}}e^{-\frac{M^{2}K^{4}\Delta\lambda^{2}}{48}\frac{\gamma}{2\gamma-1}\left(\frac{M}{2}\Vert\mathrm{x}_{0}-\bar{\mathrm{x}}\Vert_{2}^{2}+n_{x}MT\right)-\frac{n_{x}M^{2}K^{3}T\Delta\lambda}{16}}\nu(\mathrm{x}_{0})d\mathrm{x}_{0}.\label{eq:L7-pf-7}
\end{align}

Finally, using the substitutions $A_{1}=\frac{M^{3}K^{4}}{96}\frac{\gamma}{2\gamma-1}$,
$A_{2}=\frac{n_{x}M^{3}TK^{4}}{48}\frac{\gamma}{2\gamma-1}$, $B=\frac{n_{x}M^{2}K^{3}T}{16}$,
$\mathbb{E}\left[e^{-z}\right]\ge e^{-\mathbb{E}\left[z\right]}$
by Jensen's inequality, and $\nu(\mathrm{x}_{0})=\mathcal{N}(\mathrm{x}_{0};\,\mu_{\nu},\sigma_{\nu}^{2}\mathbb{I}_{n_{x}})$,
we get 
\begin{align}
\Vert\mathcal{P}_{\mathcal{L\vphantom{\tilde{\mathcal{L}}}},T}-\mathcal{P}_{\tilde{\mathcal{L}}[\Delta\lambda],T}\Vert_{\text{TV}} & \le\frac{1}{2}-\frac{1}{2}e^{-A_{2}\Delta\lambda^{2}-B\Delta\lambda}\int_{\mathbb{R}^{n_{x}}}e^{-A_{1}\Delta\lambda^{2}\Vert\mathrm{x}_{0}-\bar{\mathrm{x}}\Vert_{2}^{2}}\nu(\mathrm{x}_{0})d\mathrm{x}_{0}\nonumber \\
 & \le\frac{1}{2}-\frac{1}{2}e^{-A_{2}\Delta\lambda^{2}-B\Delta\lambda}e^{-A_{1}\Delta\lambda^{2}\int_{\mathbb{R}^{n_{x}}}\Vert\mathrm{x}_{0}-\bar{\mathrm{x}}\Vert_{2}^{2}\nu(\mathrm{x}_{0})d\mathrm{x}_{0}}\nonumber \\
 & =\frac{1}{2}-\frac{1}{2}e^{-A_{2}\Delta\lambda^{2}-B\Delta\lambda-A_{1}\Delta\lambda^{2}\left(n_{x}\sigma_{\nu}^{2}+\Vert\mathrm{\bar{x}}-\mu_{\nu}\Vert_{2}^{2}\right)}\nonumber \\
 & =\frac{1}{2}-\frac{1}{2}e^{-\left(A_{1}\left(n_{x}\sigma_{\nu}^{2}+\Vert\mathrm{\bar{x}}-\mu_{\nu}\Vert_{2}^{2}\right)+A_{2}\right)\Delta\lambda^{2}-B\Delta\lambda}\nonumber \\
 & =\frac{1}{2}-\frac{1}{2}e^{-\frac{n_{x}}{2}\frac{M^{3}K^{4}\gamma}{48(2\gamma-1)}\left(\sigma_{\nu}^{2}+\frac{1}{n_{x}}\Vert\mathrm{\bar{x}}-\mu_{\nu}\Vert_{2}^{2}+2T\right)\Delta\lambda^{2}-\frac{n_{x}M^{2}K^{3}T}{16}\Delta\lambda}.\label{eq:L7-result}
\end{align}
\end{proof}
\begin{lem}
\label{lem:L8}Under the same assumptions as of Lemma \ref{lem:L7-1},
except for a initial density $\nu(\mathrm{x}_{0})=\delta(\mathrm{x}_{0}-\mathrm{x}_{\nu})$
(Dirac delta), i.e., a probability mass initially located at $\mathrm{x}_{0}=\mathrm{x}_{\nu}$,
then 
\begin{multline}
\Vert\mathcal{P}_{\mathcal{L\vphantom{\tilde{\mathcal{L}}}},T}-\mathcal{P}_{\tilde{\mathcal{L}}[\Delta\lambda],T}\Vert_{\text{TV}}\le\frac{1}{2}\\
-\frac{1}{2}\exp\left\{ -\frac{n_{x}}{2}\frac{M^{3}K^{4}\gamma}{48(2\gamma-1)}\left(\frac{1}{n_{x}}\Vert\mathrm{\bar{x}}-\mathrm{x}_{\nu}\Vert_{2}^{2}+2T\right)\Delta\lambda^{2}-\frac{n_{x}M^{2}K^{3}T}{16}\Delta\lambda\right\} ,\,T\ge0;\label{eq:L8-st}
\end{multline}
\end{lem}
\begin{proof}
The proof follows straightforwardly by substituting $\nu(\mathrm{x}_{0})=\delta(\mathrm{x}_{0}-\mathrm{x}_{\nu})$
into (\ref{eq:L7-pf-7}).
\end{proof}
\begin{thm}
\label{thm:T9} Let a convex function $\Phi$ satisfy the general
assumptions (\ref{eq:G-Energy-condition}), (\ref{eq:G-Lipschitz-conditions-1})
and (\ref{eq:G-Lipschitz-conditions-2}). Suppose a discrete-time
Langevin Monte Carlo algorithm integrates (\ref{eq:state-flow-equation}),
targeting the invariant density $\pi(\mathrm{x})\propto\exp\{-\Phi(\mathrm{x})\}$
with measure $\mathcal{P}_{\pi}(d\mathrm{x})$. In addition, assume
that for some $\gamma\ge1$ we have $\Delta\lambda\le\left(\gamma M\right)^{-1}$,
and $K=\sup_{\mathrm{x}}\Vert D(\mathrm{x})\Vert_{2}$ where $D_{\lambda}=D(\mathrm{x}_{\lambda})$
is the diffusion matrix. Then, for a time horizon, $T$, and step
size, $\Delta\lambda$, the total-variation distance between the target
measure $\mathcal{P}_{\pi}$ and the approximated measure $\mathcal{P}_{\mathcal{\tilde{\mathcal{L}}}(\Delta\lambda),T}$
furnished by a discrete-time Langevin Monte Carlo algorithm with initial
density $\nu\left(\mathrm{x}\right)=\mathcal{N}(\mathrm{x};\mu_{\nu},\sigma_{\nu}^{2}\mathbb{I}_{n_{x}})$,
for $\sigma_{\nu}^{2}<2M^{-1}$, satisfies 
\begin{multline}
\Vert\mathcal{P}_{\mathcal{\tilde{\mathcal{L}}}[\Delta\lambda],T}-\mathcal{P}_{\vphantom{\tilde{\mathcal{L}}}\pi}\Vert_{\text{TV}}\le\frac{1}{2}\exp\left\{ -\frac{1}{2}mT+\frac{n_{x}}{4}\log\left(\frac{1}{m\sigma_{\nu}^{2}(2-M\sigma_{\nu}^{2})}\right)+\frac{M}{2(2-M\sigma_{\nu}^{2})}\Vert\bar{\mathrm{x}}-\mu_{\nu}\Vert_{2}^{2}\right\} \\
+\frac{1}{2}-\frac{1}{2}\exp\left\{ -\frac{n_{x}}{2}\frac{M^{3}K^{4}\gamma}{48(2\gamma-1)}\left(\sigma_{\nu}^{2}+\frac{1}{n_{x}}\Vert\mathrm{\bar{x}}-\mu_{\nu}\Vert_{2}^{2}+2T\right)\Delta\lambda^{2}-\frac{n_{x}M^{2}K^{3}T}{16}\Delta\lambda\right\} .\label{eq:T9-st}
\end{multline}
\end{thm}
\begin{proof}
The proof follows from the triangle inequality 
\begin{equation}
\Vert\mathcal{P}_{\tilde{\mathcal{L}}[\Delta\lambda],T}-\mathcal{P}_{\vphantom{\tilde{\mathcal{L}}}\pi}\Vert_{\text{TV}}\le\Vert\mathcal{P}_{\mathcal{L\vphantom{\tilde{\mathcal{L}}}},T}-\mathcal{P}_{\vphantom{\tilde{\mathcal{L}}}\pi}\Vert_{\text{TV}}+\Vert\mathcal{P}_{\mathcal{L\vphantom{\tilde{\mathcal{L}}}},T}-\mathcal{P}_{\tilde{\mathcal{L}}[\Delta\lambda],T}\Vert_{\text{TV}},\label{eq:T9-pf-1}
\end{equation}
\noindent on which we subsitute the results of \emph{Lemmas \ref{lem:L4}
}and\emph{ \ref{lem:L7-1}} to give the final result (\ref{eq:T9-st}).
\end{proof}
\begin{cor}
\label{cor:C10} Let $n_{x}\ge2$, $\Phi$ satisfy (\ref{eq:G-Energy-condition}),
(\ref{eq:G-Lipschitz-conditions-1}) and (\ref{eq:G-Lipschitz-conditions-2}),
and $\varepsilon\in(0,\nicefrac{1}{2})$ be a desired precision level.
Let the time horizon, $T$, and the step size, $\Delta\lambda$, be
defined by 
\begin{alignat}{1}
T & \ge\frac{2\log\left(1/\varepsilon\right)+\frac{n_{x}}{2}\log\left(\frac{1}{m\sigma_{\nu}^{2}(2-M\sigma_{\nu}^{2})}\right)+\frac{M}{(2-M\sigma_{\nu}^{2})}\Vert\bar{\mathrm{x}}-\mu_{\nu}\Vert_{2}^{2}}{m},\label{eq:C10-st-1}\\
\Delta\lambda & \le\frac{-\frac{T}{16}+\sqrt{\left(\frac{T}{16}\right)^{2}+\frac{\gamma}{48(2\gamma-1)}\left(\sigma_{\nu}^{2}+\frac{1}{n_{x}}\Vert\mathrm{\bar{x}}-\mu_{\nu}\Vert_{2}^{2}+2T\right)M^{-1}K^{-2}\left[\frac{2}{n_{x}}\log\left(\frac{1}{1-\varepsilon}\right)\right]}}{\frac{\gamma}{48(2\gamma-1)}\left(\sigma_{\nu}^{2}+\frac{1}{n_{x}}\Vert\mathrm{\bar{x}}-\mu_{\nu}\Vert_{2}^{2}+2T\right)MK},\label{eq:C10-st-2}
\end{alignat}
\noindent where $\gamma\ge1$. Then the resulting probability distribution
of a Langevin Monte Carlo algorithm that integrates (\ref{eq:state-flow-equation})
after $L=\left\lceil T/\Delta\lambda\right\rceil $ steps, satisfies
$\Vert\mathcal{P}_{\mathcal{\tilde{\mathcal{L}}}[\Delta\lambda],T}-\mathcal{P}_{\vphantom{\tilde{\mathcal{L}}}\pi}\Vert_{\text{TV}}\le\varepsilon$.
\end{cor}
\begin{proof}
Bound each term on the right-hand side of (\ref{eq:T9-pf-1}) by half
of the required precision $\varepsilon$, i.e., 
\[
\Vert\mathcal{P}_{\mathcal{L\vphantom{\tilde{\mathcal{L}}}},T}-\mathcal{P}_{\vphantom{\tilde{\mathcal{L}}}\pi}\Vert_{\text{TV}}\le\varepsilon/2,\,\Vert\mathcal{P}_{\mathcal{L\vphantom{\tilde{\mathcal{L}}}},T}-\mathcal{P}_{\tilde{\mathcal{L}}[\Delta\lambda],T}\Vert_{\text{TV}}\le\varepsilon/2.
\]
For the first term: 
\begin{multline}
\frac{1}{2}\exp\left\{ -\frac{1}{2}mT+\frac{n_{x}}{4}\log\left(\frac{1}{m\sigma_{\nu}^{2}(2-M\sigma_{\nu}^{2})}\right)+\frac{M}{2(2-M\sigma_{\nu}^{2})}\Vert\bar{\mathrm{x}}-\mu_{\nu}\Vert_{2}^{2}\right\} \le\frac{\varepsilon}{2},\\
-\frac{1}{2}mT+\frac{n_{x}}{4}\log\left(\frac{1}{m\sigma_{\nu}^{2}(2-M\sigma_{\nu}^{2})}\right)+\frac{M}{2(2-M\sigma_{\nu}^{2})}\Vert\bar{\mathrm{x}}-\mu_{\nu}\Vert_{2}^{2}\le\log\varepsilon,\\
\frac{1}{2}mT\ge-\log\varepsilon+\frac{n_{x}}{4}\log\left(\frac{1}{m\sigma_{\nu}^{2}(2-M\sigma_{\nu}^{2})}\right)+\frac{M}{2(2-M\sigma_{\nu}^{2})}\Vert\bar{\mathrm{x}}-\mu_{\nu}\Vert_{2}^{2},\\
T\ge\frac{2\log\left(1/\varepsilon\right)+\frac{n_{x}}{2}\log\left(\frac{1}{m\sigma_{\nu}^{2}(2-M\sigma_{\nu}^{2})}\right)+\frac{M}{(2-M\sigma_{\nu}^{2})}\Vert\bar{\mathrm{x}}-\mu_{\nu}\Vert_{2}^{2}}{m}.\label{eq:C10-result-1}
\end{multline}
And for the second term: 
\begin{multline}
\frac{1}{2}-\frac{1}{2}\exp\left\{ \vphantom{\frac{n_{x}}{2}\frac{M^{3}K^{4}\gamma}{48(2\gamma-1)}\left(\sigma_{\nu}^{2}+\frac{1}{n_{x}}\Vert\mathrm{\bar{x}}-\mu_{\nu}\Vert_{2}^{2}+2T\right)}\smash{-\overbrace{\frac{n_{x}}{2}\frac{M^{3}K^{4}\gamma}{48(2\gamma-1)}\left(\sigma_{\nu}^{2}+\frac{1}{n_{x}}\Vert\mathrm{\bar{x}}-\mu_{\nu}\Vert_{2}^{2}+2T\right)}^{a\,(>0)}\Delta\lambda^{2}-\overbrace{\frac{n_{x}M^{2}K^{3}T}{16}}^{b\,(>0)}\Delta\lambda}\right\} \le\frac{\varepsilon}{2},\\
\exp\left\{ -a\Delta\lambda^{2}-b\Delta\lambda\right\} \ge1-\varepsilon,\\
-a\Delta\lambda^{2}-b\Delta\lambda\ge\log\left(1-\varepsilon\right),\\
a\Delta\lambda^{2}+b\Delta\lambda\le-\log\left(1-\varepsilon\right),\\
a\Delta\lambda^{2}+b\Delta\lambda+c\le0,\,c=\log\left(1-\varepsilon\right)=-\log\left(\frac{1}{1-\varepsilon}\right),\\
0<\Delta\lambda\le\frac{-b+\sqrt{b^{2}-4ac}}{2a},\\
\Delta\lambda\le\frac{-\frac{n_{x}M^{2}K^{3}T}{16}+\sqrt{\left(\frac{n_{x}M^{2}K^{3}T}{16}\right)^{2}+4\frac{n_{x}}{2}\frac{M^{3}K^{4}\gamma}{48(2\gamma-1)}\left(\sigma_{\nu}^{2}+\frac{1}{n_{x}}\Vert\mathrm{\bar{x}}-\mu_{\nu}\Vert_{2}^{2}+2T\right)\log\left(\frac{1}{1-\varepsilon}\right)}}{2\frac{n_{x}}{2}\frac{M^{3}K^{4}\gamma}{48(2\gamma-1)}\left(\sigma_{\nu}^{2}+\frac{1}{n_{x}}\Vert\mathrm{\bar{x}}-\mu_{\nu}\Vert_{2}^{2}+2T\right)},\\
\Delta\lambda\le\frac{-\frac{T}{16}+\sqrt{\left(\frac{T}{16}\right)^{2}+\frac{\gamma}{48(2\gamma-1)}\left(\sigma_{\nu}^{2}+\frac{1}{n_{x}}\Vert\mathrm{\bar{x}}-\mu_{\nu}\Vert_{2}^{2}+2T\right)M^{-1}K^{-2}\left[\frac{2}{n_{x}}\log\left(\frac{1}{1-\varepsilon}\right)\right]}}{\frac{\gamma}{48(2\gamma-1)}\left(\sigma_{\nu}^{2}+\frac{1}{n_{x}}\Vert\mathrm{\bar{x}}-\mu_{\nu}\Vert_{2}^{2}+2T\right)MK}.\label{eq:C10-result-2}
\end{multline}
\end{proof}

\subsubsection*{Proof of Theorem 1}
\begin{proof}
We follow the same procedure established for proving \emph{Theorem
\ref{thm:T9}}, subsituting the results of \emph{Lemmas \ref{lem:L6}
}and\emph{ \ref{lem:L8}} into (\ref{eq:T9-pf-1}) to give 
\begin{multline}
\Vert\mathcal{P}_{\tilde{\mathcal{L}}[\Delta\lambda],T}-\mathcal{P}_{\pi}\Vert_{\text{TV}}\le\frac{1}{2}\exp\left\{ -\frac{1}{2}mT+\frac{n_{x}}{2}\log\left(\frac{M}{m}\right)-\log\left[\Gamma_{u}\left(\frac{n_{x}}{2},\frac{M\Vert\bar{\mathrm{x}}-\mathrm{x}_{\nu}\Vert_{2}^{2}}{2}\right)\right]\right\} \\
+\frac{1}{2}-\frac{1}{2}\exp\left\{ -\frac{n_{x}}{2}\frac{M^{3}K^{4}\gamma}{48(2\gamma-1)}\left(\frac{1}{n_{x}}\Vert\mathrm{\bar{x}}-\mathrm{x}_{\nu}\Vert_{2}^{2}+2T\right)\Delta\lambda^{2}-\frac{n_{x}M^{2}K^{3}T}{16}\Delta\lambda\right\} .\label{eq:T1-result}
\end{multline}
\end{proof}

\subsubsection*{Proof of Corollary 2}
\begin{proof}
We follow the same procedure established for proving \emph{Corollary
\ref{cor:C10}}, bounding each term on the right-hand side of (\ref{eq:T9-pf-1})
in view of (\ref{eq:T1-result}) to obtain 
\begin{align}
T & \ge\frac{2\log\left(1/\varepsilon\right)+n_{x}\log\left(\frac{M}{m}\right)-2\log\left[\Gamma_{u}\left(\frac{n_{x}}{2},\frac{M\Vert\bar{\mathrm{x}}-\mathrm{x}_{\nu}\Vert_{2}^{2}}{2}\right)\right]}{m},\label{eq:C2-result-1}\\
\Delta\lambda & \le\frac{-\frac{T}{16}+\sqrt{\left(\frac{T}{16}\right)^{2}+\frac{\gamma}{48(2\gamma-1)}\left(\frac{1}{n_{x}}\Vert\mathrm{\bar{x}}-\mathrm{x}_{\nu}\Vert_{2}^{2}+2T\right)M^{-1}K^{-2}\left[\frac{2}{n_{x}}\log\left(\frac{1}{1-\varepsilon}\right)\right]}}{\frac{\gamma}{48(2\gamma-1)}\left(\frac{1}{n_{x}}\Vert\mathrm{\bar{x}}-\mathrm{x}_{\nu}\Vert_{2}^{2}+2T\right)MK}.\label{eq:C2-result-2}
\end{align}
\end{proof}

\subsection{On the Filtering Properties of the Stochastic Particle Flow\label{subsec:Proofs-2}}
\begin{thm}
\label{thm:T6}Define $\mathrm{x}\in\mathbb{R}^{n_{x}}$ to describe
an $n_{x}$-dimensional vector state. Let the vector field $\mu:\mathbb{R}^{n_{x}}\rightarrow\mathbb{R}^{n_{x}}$,
$\mu\left(\mathrm{x}\right)\in\mathcal{C}^{1}(\mathbb{R}^{n_{x}})$,
be a conservative field, i.e., there exists a scalar potential function
$\psi:\mathbb{R}^{n_{x}}\rightarrow\mathbb{R}$, $\psi\left(\mathrm{x}\right)\in\mathcal{C}^{2}(\mathbb{R}^{n_{x}}),$
such that 
\begin{equation}
\mu\left(\mathrm{x}\right)=-\nabla_{\mathrm{x}}\psi\left(\mathrm{x}\right).\label{eq:statement-conservative-field}
\end{equation}
Let $p\left(\mathrm{x},\lambda\right)$ be the density of an ensemble
of particles and, without loss of generality, can be assumed to be
a continuous probability density function on $\mathbb{R}^{n_{x}}$
that depends on the pseudo-time variable $\lambda\in\mathbb{R}$,
$\lambda\ge0$. Set $\pi\left(\mathrm{x}\right)\propto e^{-\psi\left(\mathrm{x}\right)}$
to be an invariant, locally log-concave probability density to which
the density $p\left(\mathrm{x},\lambda\right)$ is expected to converge
weakly at a stationary state in a finite time horizon $\lambda\ge T$,
$T\in\mathbb{R}_{+}$, i.e., 
\begin{equation}
\mathbb{E}_{p}\left[\varphi\left(\mathrm{x}\right)\right]\rightarrow\mathbb{E}_{\pi}\left[\varphi\left(\mathrm{x}\right)\right],\quad\lambda\rightarrow T;\label{eq:convergence-assumption}
\end{equation}
\noindent for all bounded, continuous functions $\varphi$, and where
$\mathbb{E}_{p}\left[.\right]$ is the expectation with respect to
the probability density $p\left(\mathrm{x},\lambda\right)$. If the
probability density $p\left(\mathrm{x},\lambda\right)$ satisfies
the continuity equation (Liouville's equation) 
\begin{equation}
\frac{\partial p}{\partial\lambda}=-\nabla_{\mathrm{x}}\cdot\left(p\,\mu\right),\quad\lambda\ge0;\label{eq:continuity-equation-proof}
\end{equation}
\noindent with the initial condition 
\begin{equation}
p\left(\mathrm{x},\lambda\right)=p_{0}\left(\mathrm{x}\right),\quad\lambda=0;\label{eq:conditions-proof}
\end{equation}
\noindent then any probability mass (particle) $\mathrm{x}_{m}\left(0\right)\sim p_{0}\left(\mathrm{x}\right)$,
when evolved according to the associated state equation 
\begin{equation}
d\mathrm{x}_{m}(\lambda)=\mu\left(\mathrm{x}_{m}(\lambda)\right)d\lambda,\quad\lambda\ge0;\label{eq:state-equation-proof}
\end{equation}
\noindent converges to 
\begin{equation}
\mathrm{x}_{m}\left(T\right)=\text{argmax}\left[\pi\left(\mathrm{x}\right)\right],\quad\lambda\ge T,\label{eq:final-condition}
\end{equation}
\noindent at a stable equilibrium.
\end{thm}
\begin{proof}
The general solution of the continuity equation without sources (\ref{eq:continuity-equation-proof})
assumes the form (see for example \citep{Lamb1895}) 
\begin{align}
p\left(\mathrm{x},\lambda\right) & =p_{0}\left(\mathrm{x}_{m}(\mathrm{x},\lambda)\right)\left|\frac{\partial\mathrm{x}_{m}}{\partial\mathrm{x}}\right|\nonumber \\
 & =p_{0}\left(\mathrm{x}_{m}(\mathrm{x},\lambda)\right)\left|\mathcal{J}_{\mathrm{x}}\left[\mathrm{x}_{m}(\mathrm{x},\lambda)\right]\right|,\label{eq:general-solution-of-continuity-equation}
\end{align}
\noindent where $\mathrm{x}_{m}(\mathrm{x},\lambda)$ is an arbitrary
element of mass that is regarded as a function of the pseudo-time
$\lambda$ and of the state $\mathrm{x}$ that it can possibly reach.
The matrix $\mathcal{J}_{\mathrm{x}}\left[\mathrm{x}_{m}(\mathrm{x},\lambda)\right]$
is the Jacobian matrix of $\mathrm{x}_{m}(\mathrm{x},\lambda)$ with
respect to $\mathrm{x}$. Conceptually, at the stationary state $\mathrm{x}_{m}\left(\mathrm{x}_{T},T\right)=\mathrm{x}_{T}$
the continuity equation (\ref{eq:continuity-equation-proof}) reads
\begin{equation}
\frac{\partial p}{\partial\lambda}=0,\quad\lambda\ge T.\label{eq:stationary-state}
\end{equation}

Using the general solution (\ref{eq:general-solution-of-continuity-equation})
to verify the stationary condition (\ref{eq:stationary-state}), we
conclude that 
\[
p_{0}\left(\mathrm{x}_{m}(\mathrm{x}_{T},T)\right)\left|\mathcal{J}_{\mathrm{x}}\left[\mathrm{x}_{m}(\mathrm{x}_{T},T)\right]\right|
\]
 \noindent must be constant with respect to the pseudo-time, thus
\begin{equation}
\frac{d\mathrm{x}_{m}(\lambda)}{d\lambda}=\mu\left(\mathrm{x}_{m}(\lambda)\right)=0,\quad\lambda\ge T.\label{eq:stationary-state-2}
\end{equation}

Following the assumption of conservative field, $\mu\left(\mathrm{x}_{T}\right)=-\nabla_{\mathrm{x}}\psi\left(\mathrm{x}_{T}\right)=0$
implies that the stationary state $\mathrm{x}_{T}$ is an equilibrium
point, i.e., an extreme of the potential function $\psi$. In addition,
since the potential function is assumed to be related to the stationary
distribution as ${\psi\left(\mathrm{x}\right)\propto-\log\pi\left(\mathrm{x}\right)}$,
the stationary state $\mathrm{x}_{T}$ is an extreme of the stationary
density.

A valid Lyapunov function of the flow is $V\left(\mathrm{x}\right)=\psi\left(\mathrm{x}\right)$,
which is positive semi-definite ($\psi\left(\mathrm{x}\right)\ge0$)
in the neighbourhood of the equilibrium point due to the local log-concavity
of the invariant density $\pi\left(\mathrm{x}\right)$. Analysing
the (Lie) time derivative of the Lyapunov function in the neighbourhood
of the equilibrium point, $\left\Vert \mathrm{x}-\mathrm{x}_{T}\right\Vert <\varepsilon$
for a sufficiently small $\varepsilon\in\mathbb{R}_{+}$, we have
\begin{align}
\frac{dV\left(\mathrm{x}\right)}{d\lambda} & =\nabla_{\mathrm{x}}V\left(\mathrm{x}\right)^{T}\cdot\frac{d\mathrm{x}}{d\lambda}=\nabla_{\mathrm{x}}V\left(\mathrm{x}\right)^{T}\cdot\mu\left(\mathrm{x}\right),\nonumber \\
\dot{V}\left(\mathrm{x}\right) & =\nabla_{\mathrm{x}}\psi\left(\mathrm{x}\right)^{T}\cdot\left(-\nabla_{\mathrm{x}}\psi\left(\mathrm{x}\right)\right),\nonumber \\
\dot{V}\left(\mathrm{x}\right) & =-\left\Vert \nabla_{\mathrm{x}}\psi\left(\mathrm{x}\right)\right\Vert ^{2}\le0,\,\left\Vert \mathrm{x}-\mathrm{x}_{T}\right\Vert <\varepsilon;\label{eq:stationary-state-3}
\end{align}
\noindent from which we conclude that $\mathrm{x}_{T}$ is a point
of (uniformily) stable equilibrium. Therefore, under the established
hypotheses, any arbitrary probability mass $\mathrm{x}_{m}\left(\lambda\right)$
evolved according to (\ref{eq:state-equation-proof}) converges to
\begin{align*}
\mathrm{x}_{m}\left(T\right) & =\text{argmin}\left[\psi\left(\mathrm{x}\right)\right]=\text{argmin}\left[-\log\pi\left(\mathrm{x}\right)\right],\\
\mathrm{x}_{m}\left(T\right) & =\text{argmax}\left[\pi\left(\mathrm{x}\right)\right],\quad\lambda\ge T;
\end{align*}
\noindent at a stable equilibrium.
\end{proof}
\begin{lem}
\label{lem:L7}Let $\left\{ \mathrm{X}_{\lambda}:t\le\lambda\le T\right\} $
be a diffusion process in $\mathbb{R}^{n_{x}}$ (hence a Markov process),
solution of 
\begin{align}
d\mathrm{X}_{\lambda} & =\mu_{f}\left(\mathrm{X}_{\lambda},\lambda\right)d\lambda+D_{f}\left(\lambda\right)^{\nicefrac{1}{2}}dW_{\lambda},\label{eq:proof-state-process-forward}\\
 & \hphantom{=\ }\mathrm{X}_{\lambda}\left(t\right)=\mathrm{x}_{t},t\le\lambda\le T;\nonumber 
\end{align}
\noindent where $\left\{ W_{\lambda}:t\le\lambda\le T\right\} $
is a standard Wiener process in $\mathbb{R}^{n_{x}}$ under the probability
measure $\mathcal{P}$, $\mu_{f}:\mathbb{R}^{n_{x}}\times[t,T]\rightarrow\mathbb{R}^{n_{x}}$
is the drift and $D_{f}:[t,T]\rightarrow\mathbb{R}^{n_{x}\times n_{x}}$
is a diffusion coefficient invariant over the space at any time instant.
There exists an equivalent process $\left\{ \bar{\mathrm{X}}_{\tau},\mathcal{V}_{\tau}:t\le\tau\le T\right\} $,
which is probabilistically the same as the original process, called
\emph{reverse process} on the interval $[t,T]$ (see \citep{Milstein2004}),
that provides the solution of the stochastic system 
\begin{alignat}{1}
d\bar{\mathrm{X}}_{\tau} & =\mu_{r}\left(\bar{\mathrm{X}}_{\tau},\tau\right)d\tau+D_{r}\left(\tau\right)^{\nicefrac{1}{2}}d\bar{W}_{\tau},\:\bar{\mathrm{X}}_{\tau}\left(t\right)=\bar{x}_{t};\label{eq:lemma-state-process-backward}\\
d\mathcal{V}_{\tau} & =v_{r}\left(\bar{\mathrm{X}}_{\tau},\tau\right)\mathcal{V}_{\tau}d\tau,\hphantom{D_{r}\left(\tau\right)^{\nicefrac{1}{2}}d\bar{W}_{\tau}^{P}\,\,\,\,}\mathcal{V}_{\tau}\left(t\right)=1;\label{eq:lemma-scalar-field-backward}
\end{alignat}
\noindent for a standard Wiener process $\left\{ \bar{W}_{\tau}:t\le\tau\le T\right\} $
in $\mathbb{R}^{n_{x}}$ under the measure $\mathcal{P}$, with the
reverse drift and diffusion coefficients given, respectively, by 
\begin{align}
\mu_{r}\left(\bar{\mathrm{X}}_{\tau},\tau\right) & =-\mu_{f}\left(\bar{\mathrm{X}}_{\tau},T+t-\lambda\right),\label{eq:drift-correspondence}\\
D_{r}\left(\tau\right) & =\hphantom{-}D_{f}\left(T+t-\lambda\right).\label{eq:diffusion-correspondence}
\end{align}
\end{lem}
\begin{proof}
The Markov process $\left\{ \mathrm{X}_{\lambda}\right\} $, as an
existing solution to the SDE (\ref{eq:proof-state-process-forward}),
has an associated probability density $p\left(\mathrm{x}_{\lambda},\lambda\right)$
that must satisfy the Kolmogorov forward equation (Fokker-Planck equation):
\begin{alignat*}{1}
\frac{\partial}{\partial\lambda}p & =-\nabla_{\mathrm{x}}\cdot\left(\mu_{f}p\right)+\frac{1}{2}\nabla_{\mathrm{x}}\cdot\left(D_{f}\nabla_{\mathrm{x}}p\right),\,\lambda\ge t,\\
 & \hphantom{=-\ }p\left(\mathrm{x}_{\lambda},t\right)=p_{t}\left(\mathrm{x}_{t}\right),\mathrm{x}_{\lambda}\in\mathbb{R}^{n_{x}}.
\end{alignat*}
The Fokker-Planck equation can be written in the non-divergence form
as 
\begin{equation}
\frac{\partial}{\partial\lambda}p=\hat{\mu}^{T}\nabla_{\mathrm{x}}p+\frac{1}{2}\nabla_{\mathrm{x}}\cdot\left(D_{f}\nabla_{\mathrm{x}}p\right)+\hat{v}\cdot p,\label{eq:FPE-non-divergence-form}
\end{equation}
\noindent where 
\begin{alignat*}{1}
\hat{\mu}\left(\mathrm{x}_{\lambda}\right) & =-\mu_{f}\left(\mathrm{x}_{\lambda}\right),\\
\hat{v}\left(\mathrm{x}_{\lambda}\right) & =-\nabla_{\mathrm{x}}\cdot\mu_{f}\left(\mathrm{x}_{\lambda}\right).
\end{alignat*}
We introduce the reverse time variable $\tau=T+t-\lambda$, so that
\[
p\left(\mathrm{x}_{\lambda}(T+t-\lambda),T+t-\lambda\right)\equiv\hat{p}\left(\bar{\mathrm{x}}_{\tau},\tau\right),
\]
 \noindent and hence $-\partial_{\lambda}p=\partial_{\tau}\hat{p}$.
Thus, rewritting (\ref{eq:FPE-non-divergence-form}) with respect
to $\hat{p}\left(\bar{\mathrm{x}}_{\tau},\tau\right)$ for $\tau\le T$,
$\bar{\mathrm{x}}_{\tau}\in\mathbb{R}^{n_{x}}$, and performing the
substitutions 
\begin{align*}
\mu_{r}\left(\bar{\mathrm{x}}_{\tau},\tau\right) & =-\mu_{f}\left(\mathrm{x}_{\lambda}(T+t-\lambda),T+t-\lambda\right),\\
D_{r}\left(\tau\right) & =\hphantom{-}D_{f}\left(T+t-\lambda\right),
\end{align*}
\noindent we obtain 
\begin{align}
\frac{\partial}{\partial\lambda}p & =-\mu_{f}^{T}\nabla_{\mathrm{x}}p+\frac{1}{2}\nabla_{\mathrm{x}}\cdot\left(D_{f}\nabla_{\mathrm{x}}p\right)+\left(-\nabla_{\mathrm{x}}\cdot\mu_{f}\right)\cdot p,\nonumber \\
-\frac{\partial}{\partial\tau}\hat{p} & =\hphantom{-}\mu_{r}^{T}\nabla_{\bar{\mathrm{x}}}\hat{p}+\frac{1}{2}\nabla_{\bar{\mathrm{x}}}\cdot\left(D_{r}\nabla_{\bar{\mathrm{x}}}\hat{p}\right)-\left(-\nabla_{\bar{\mathrm{x}}}\cdot\mu_{r}\right)\cdot\hat{p},\nonumber \\
-\frac{\partial}{\partial\tau}\hat{p} & =\hphantom{-}\mu_{r}^{T}\nabla_{\bar{\mathrm{x}}}\hat{p}+\frac{1}{2}\nabla_{\bar{\mathrm{x}}}\cdot\left(D_{r}\nabla_{\bar{\mathrm{x}}}\hat{p}\right)-v_{r}\cdot\hat{p},\,\tau\le T,\label{eq:Kolmogorov-backward-eq-proof}\\
 & \hphantom{=-\ }\hat{p}\left(\bar{\mathrm{x}}_{\tau},T\right)=\hat{p}_{T}\left(\bar{\mathrm{x}}_{T}\right)=p_{t}\left(\mathrm{x}_{t}\right),\bar{\mathrm{x}}_{\tau}\in\mathbb{R}^{n_{x}};\nonumber 
\end{align}
\noindent where 
\begin{equation}
v_{r}\left(\bar{\mathrm{x}}_{\tau},\tau\right)=-\nabla_{\bar{\mathrm{x}}}\mu_{r}\left(\bar{\mathrm{x}}_{\tau},\tau\right).\label{eq:measure-drift}
\end{equation}

Solving (\ref{eq:Kolmogorov-backward-eq-proof}) corresponds to the
Cauchy problem in reverse time $\tau\le T$, which is equivalent to
solve the stochastic system stated by (\ref{eq:lemma-state-process-backward})
and (\ref{eq:lemma-scalar-field-backward}). Therefore, because the
solution to the SDE (\ref{eq:proof-state-process-forward}) is assumed
to exist and corresponds to the solution of (\ref{eq:Kolmogorov-backward-eq-proof})
for $\tau\le T$, then there exists the equivalent reverse process
$\left\{ \bar{\mathrm{X}}_{\tau},\mathcal{V}_{\tau}\right\} _{\tau\in[t,T]}$
that solves the stochastic system (\ref{eq:lemma-state-process-backward})
and (\ref{eq:lemma-scalar-field-backward}).
\end{proof}
\smallskip{}

\begin{rem}
Despite its name, inherited from \citep{Milstein2004}, it is worth
stressing that $\left\{ \bar{\mathrm{X}}_{\tau},\mathcal{V}_{\tau}\right\} $
is the solution of a stochastic system forward in time on the interval
$[t,T]$, which may be properly understood as a smoothing process.
\end{rem}
\smallskip{}

\begin{rem}
It is clear that the solution of (\ref{eq:lemma-scalar-field-backward})
at $\tau=T$ is 
\[
\mathcal{V}_{T}\equiv\mathcal{V}_{\tau}\left(T\right)=e^{-\int_{t}^{T}v_{r}\left(\bar{\mathrm{x}}_{\tau},\tau\right)d\tau},
\]
\noindent which evokes the solution of (\ref{eq:Kolmogorov-backward-eq-proof})
by the Feynman-Kac formula 
\begin{equation}
\hat{p}\left(\mathrm{x},\tau\right)=\mathbb{E}^{\mathcal{P}}\left[e^{-\int_{\tau}^{T}v_{r}\left(\bar{\mathrm{x}}_{\tau'},\tau'\right)d\tau'}\hat{p}_{T}\left(\bar{\mathrm{x}}_{T}\right)|\bar{\mathrm{x}}_{\tau}=\mathrm{x}\right].\label{eq:lemma-Feynman-Kac-formula}
\end{equation}
A more general form of the \emph{Lemma \ref{lem:L7}} can be found
in \citep{Milstein2004}.
\end{rem}
\begin{lem}
\label{lem:L10}The reverse process $\left\{ \bar{\mathrm{X}}_{\tau},\mathcal{V}_{\tau}:t\le\tau\le T\right\} $
described by (\ref{eq:lemma-state-process-backward}) and (\ref{eq:lemma-scalar-field-backward}),
has an associated smooth probability density $\hat{p}\left(\bar{\mathrm{x}}_{\tau},\tau\right)$
that satisfies, for the initial value problem, the Kolmogorov forward
equation 
\begin{alignat}{1}
\frac{\partial}{\partial\tau}\hat{p} & =-\nabla_{\bar{\mathrm{x}}}\cdot\left(\mu_{r}\hat{p}\right)+\frac{1}{2}\nabla_{\bar{\mathrm{x}}}\cdot\left(D_{r}\nabla_{\bar{\mathrm{x}}}\hat{p}\right)-v_{r}\cdot\hat{p},\,t\le\tau\le T,\label{eq:FPE-second-lemma}\\
 & \hphantom{=-\ }\hat{p}\left(\bar{\mathrm{x}}_{\tau}(t),\tau=t\right)=\hat{p}_{t}\left(\bar{\mathrm{x}}_{t}\right),\bar{\mathrm{x}}_{\tau}\in\mathbb{R}^{n_{x}}.\nonumber 
\end{alignat}
\end{lem}
\begin{proof}
We consider a continuous function of the process $\left\{ \bar{\mathrm{X}}_{\tau},\mathcal{V}_{\tau}\right\} _{\tau\in[t,T]}$,
declared as $\varphi:\mathbb{R}^{n_{x}}\times\mathbb{R}\rightarrow\mathbb{R}$,
which is assumed to be $\varphi(\bar{\mathrm{X}}_{\tau},\mathcal{V}_{\tau})\in\mathcal{C}^{2}(\mathbb{R}^{n_{x}},\mathbb{R})$,
bounded and integrable on the product space $\mathbb{R}^{n_{x}}\times\mathbb{R}$.
Applying \^Ito's lemma to $\varphi$ and substituting (\ref{eq:lemma-state-process-backward})
and (\ref{eq:lemma-scalar-field-backward}) one obtains 
\begin{alignat}{1}
d\varphi & =(\nabla_{\bar{\mathrm{X}}_{\tau}}\varphi)^{T}d\bar{\mathrm{X}}_{\tau}+\frac{1}{2}d\bar{\mathrm{X}}_{\tau}^{T}\mathcal{H}_{\bar{\mathrm{X}}_{\tau}}\left[\varphi\right]d\bar{\mathrm{X}}_{\tau}+\partial_{\mathcal{V}_{\tau}}\varphi\,d\mathcal{V}_{\tau}\nonumber \\
 & =(\nabla_{\bar{\mathrm{X}}_{\tau}}\varphi)^{T}\cdot\left(\mu_{r}d\tau+D_{r}{}^{\nicefrac{1}{2}}d\bar{W}_{\tau}\right)+\frac{1}{2}\text{tr}\left\{ D_{r}\mathcal{H}_{\bar{\mathrm{X}}_{\tau}}\left[\varphi\right]\right\} d\tau+\partial_{\mathcal{V}_{\tau}}\varphi\cdot\left(v_{r}\mathcal{V}_{\tau}d\tau\right),\nonumber \\
 & =\left[(\nabla_{\bar{\mathrm{X}}_{\tau}}\varphi)^{T}\mu_{r}+\frac{1}{2}\text{tr}\left\{ D_{r}\mathcal{H}_{\bar{\mathrm{X}}_{\tau}}\left[\varphi\right]\right\} +\partial_{\mathcal{V}_{\tau}}\varphi\cdot v_{r}\mathcal{V}_{\tau}\right]d\tau+(\nabla_{\bar{\mathrm{X}}_{\tau}}\varphi)^{T}D_{r}{}^{\nicefrac{1}{2}}d\bar{W}_{\tau}.\label{eq:Ito-lemma-applied-lemma}
\end{alignat}

Consider the expected (average) rate of change of the projections
of $\varphi$ defined by: 
\begin{align}
\left\langle \dot{\varphi}_{\mathcal{V}}\right\rangle \left(\tau\right) & \triangleq\left\langle \frac{d}{d\tau}\int_{1}^{\mathcal{V}_{T}}\varphi(\bar{\mathrm{X}}_{\tau},\mathcal{V}_{\tau})d\mathcal{V}_{\tau}\right\rangle \nonumber \\
 & =\left\langle \int_{1}^{\mathcal{V}_{T}}\partial_{\tau}\varphi(\bar{\mathrm{X}}_{\tau},\mathcal{V}_{\tau})d\mathcal{V}_{\tau}\right\rangle \nonumber \\
 & \equiv\int_{\mathbb{R}^{n_{x}}}\int_{1}^{\upsilon_{T}}\varphi\left(\bar{\mathrm{x}}_{\tau},\upsilon_{\tau}\right)\cdot\partial_{\tau}\hat{p}\,d\upsilon_{\tau}d\bar{\mathrm{x}}_{\tau}.\label{eq:rate-of-change}
\end{align}
Substituting (\ref{eq:Ito-lemma-applied-lemma}) into (\ref{eq:rate-of-change}),
we have 
\begin{align}
\left\langle \dot{\varphi}_{\mathcal{V}}\right\rangle \left(\tau\right) & =\left\langle \int_{1}^{\mathcal{V}_{T}}\partial_{\tau}\varphi(\bar{\mathrm{X}}_{\tau},\mathcal{V}_{\tau})d\mathcal{V}_{\tau}\right\rangle \nonumber \\
 & =\left\langle \int_{1}^{\mathcal{V}_{T}}\left[(\nabla_{\bar{\mathrm{X}}_{\tau}}\varphi)^{T}\mu_{r}+\frac{1}{2}\text{tr}\left\{ D_{r}\mathcal{H}_{\bar{\mathrm{X}}_{\tau}}\left[\varphi\right]\right\} +\partial_{\mathcal{V}_{\tau}}\varphi\cdot v_{r}\mathcal{V}_{\tau}\right]d\mathcal{V}_{\tau}\right\rangle \nonumber \\
 & =\int_{1}^{\upsilon_{T}}\int_{\mathbb{R}^{n_{x}}}\left[(\nabla_{\bar{\mathrm{x}}_{\tau}}\varphi)^{T}\mu_{r}+\frac{1}{2}\text{tr}\left\{ D_{r}\mathcal{H}_{\bar{\mathrm{x}}_{\tau}}\left[\varphi\right]\right\} \right]\hat{p}\,d\bar{\mathrm{x}}_{\tau}d\upsilon_{\tau}\nonumber \\
 & +\int_{\mathbb{R}^{n_{x}}}\int_{1}^{\upsilon_{T}}\left[\partial_{\mathcal{\upsilon}_{\tau}}\varphi\,\cdot v_{r}\,\upsilon_{\tau}\right]\hat{p}\,d\upsilon_{\tau}d\bar{\mathrm{x}}_{\tau}.\label{eq:rate-of-change-2}
\end{align}
Integrating (\ref{eq:rate-of-change-2}) by parts and offsetting the
integration constant to cancel out the surface terms, we have 
\begin{align}
\left\langle \dot{\varphi}_{\mathcal{V}}\right\rangle \left(\tau\right) & =\int_{1}^{\upsilon_{T}}\int_{\mathbb{R}^{n_{x}}}\varphi\cdot\left[-\nabla_{\bar{\mathrm{x}}_{\tau}}\cdot(\mu_{r}\hat{p})+\frac{1}{2}\nabla_{\bar{\mathrm{x}}_{\tau}}\cdot\left(D_{r}\nabla_{\bar{\mathrm{x}}_{\tau}}\hat{p}\right)\right]\,d\bar{\mathrm{x}}_{\tau}d\upsilon_{\tau}\nonumber \\
 & +\int_{\mathbb{R}^{n_{x}}}\int_{1}^{\upsilon_{T}}\varphi\cdot\left[-v_{r}\partial_{\upsilon_{\tau}}(\upsilon_{\tau}\hat{p})\right]\,d\upsilon_{\tau}d\bar{\mathrm{x}}_{\tau}\nonumber \\
 & =\int_{\mathbb{R}^{n_{x}}}\int_{1}^{\upsilon_{T}}\varphi\cdot\left[-\nabla_{\bar{\mathrm{x}}_{\tau}}\cdot(\mu_{r}\hat{p})+\frac{1}{2}\nabla_{\bar{\mathrm{x}}_{\tau}}\cdot\left(D_{r}\nabla_{\bar{\mathrm{x}}_{\tau}}\hat{p}\right)\right]\,d\upsilon_{\tau}d\bar{\mathrm{x}}_{\tau}\nonumber \\
 & +\int_{\mathbb{R}^{n_{x}}}\int_{1}^{\upsilon_{T}}\varphi\cdot\left[-v_{r}\hat{p}\right]\,d\upsilon_{\tau}d\bar{\mathrm{x}}_{\tau}\nonumber \\
 & =\int_{\mathbb{R}^{n_{x}}}\int_{1}^{\upsilon_{T}}\varphi\cdot\left[-\nabla_{\bar{\mathrm{x}}_{\tau}}\cdot(\mu_{r}\hat{p})+\frac{1}{2}\nabla_{\bar{\mathrm{x}}_{\tau}}\cdot\left(D_{r}\nabla_{\bar{\mathrm{x}}_{\tau}}\hat{p}\right)-v_{r}\hat{p}\right]\,d\upsilon_{\tau}d\bar{\mathrm{x}}_{\tau}.\label{eq:rate-of-change-3}
\end{align}
The proof of the lemma is complete by comparing (\ref{eq:rate-of-change-3})
to the definition (\ref{eq:rate-of-change}) and noting that their
integrands must be equal.
\end{proof}
\begin{thm}
\label{thm:T11}Let $\left(\Omega,\mathcal{F},\mathcal{P}\right)$
to be a complete probability space and let $\left\{ \mathcal{F}_{\lambda}\right\} _{\lambda\ge0}$,
$\lambda\in[0,T]$, be an increasing family of sub $\sigma$-fields
of $\mathcal{F}$. Let $\left\{ \mathrm{X}_{\lambda}:0<\lambda\le T\right\} $
be an $\mathcal{F}_{\lambda}$-adapted process, considered to be the
signal process with state equation 
\begin{align}
d\mathrm{X}_{\lambda} & =\mu_{f}\left(\mathrm{X}_{\lambda}\right)d\lambda+D_{f}\left(\lambda\right)^{\nicefrac{1}{2}}dW_{\lambda},\label{eq:proof-state-process}\\
 & \hphantom{=\ }\mathrm{X}_{\lambda}\left(0\right)=\mathrm{x}_{0},\,0\le\lambda\le T;\nonumber 
\end{align}
\noindent for a Wiener process $\left\{ W_{\lambda}\right\} _{\lambda\in[0,T]}$
under the probability measure $\mathcal{P}$. Assume $p(\mathrm{x}_{\lambda},\lambda)$
defined on $\left(\Omega,\mathcal{F}\right)$ to be the probability
density of the measure $\mathcal{P}$, which 

(a) is the probabilistic representation of the process $\left\{ \mathrm{X}_{\lambda}\right\} _{\lambda\in[0,T]}$,$\vphantom{p(\mathrm{x}_{\lambda})\stackrel{\lambda\rightarrow T}{\longrightarrow}\pi\left(\mathrm{x}_{\lambda}\right)\propto e^{-\Phi\left(\mathrm{x}_{\lambda}\right)}}$ 

(b) is absolutely continuous with respect to the Lesbesque measure,$\vphantom{p(\mathrm{x}_{\lambda})\stackrel{\lambda\rightarrow T}{\longrightarrow}\pi\left(\mathrm{x}_{\lambda}\right)\propto e^{-\Phi\left(\mathrm{x}_{\lambda}\right)}}$ 

(c) approaches a stationary measure as ${p(\mathrm{x}_{\lambda})\stackrel{\lambda\rightarrow T}{\longrightarrow}\pi\left(\mathrm{x}_{\lambda}\right)\propto e^{-\Phi\left(\mathrm{x}_{\lambda}\right)}}$,
for a sufficiently long horizon $T$. 

\medskip{}
\noindent Let $\left\{ \bar{X}_{\tau},\mathcal{V}_{\tau}:\lambda<\tau\le T\right\} $
be the reverse process of $\left\{ \mathrm{X}_{\lambda}\right\} _{\lambda\in[0,T]}$,
as established in Lemma \ref{lem:L7} by the stochastic system (\ref{eq:lemma-state-process-backward})
and (\ref{eq:lemma-scalar-field-backward}), so that the reverse drift
and diffusion coefficients are, respectively, 
\begin{align*}
\mu_{r}\left(\bar{\mathrm{x}}_{\tau}\right) & =-\frac{1}{2}D_{r}\left(\lambda\right)\nabla_{\bar{\mathrm{x}}}\log\pi\left(\bar{\mathrm{x}}_{\tau}\right),\\
D_{r}\left(\tau\right) & =\left[-\mathcal{H}_{\bar{\mathrm{x}}}\left[\log\pi\left(\bar{\mathrm{x}}_{\tau}\right)\right]\right]^{-1}.
\end{align*}
Assume $\hat{p}(\bar{\mathrm{x}}_{\tau},\tau)$ to be the probability
density on $\left(\Omega,\mathcal{F}\right)$ describing the reverse
process $\left\{ \bar{X}_{\tau},\mathcal{V}_{\tau}\right\} _{\tau\in[\lambda,T]}$,
under the same measure $\mathcal{P}$, which must satisfy the Kolmogorov
forward equation (provided a known initial condition) according to
Lemma \ref{lem:L10}: 
\begin{alignat}{1}
\frac{\partial}{\partial\tau}\hat{p} & =-\nabla_{\bar{\mathrm{x}}}\cdot\left(\mu_{r}\hat{p}\right)+\frac{1}{2}\nabla_{\bar{\mathrm{x}}}\cdot\left(D_{r}\nabla_{\bar{\mathrm{x}}}\hat{p}\right)-v_{r}\cdot\hat{p},\,\lambda\le\tau\le T,\label{eq:proof-Kolmogorov-forward-eq}\\
 & \hphantom{=-\ }\hat{p}\left(\bar{\mathrm{x}}_{\tau}\left(\lambda\right),\tau=\lambda\right)=\hat{p}_{\lambda}\left(\bar{\mathrm{x}}_{\tau}\left(\lambda\right)\right)=\pi\left(\mathrm{x}_{T}\right),\bar{\mathrm{x}}_{\tau}\in\Omega;\nonumber 
\end{alignat}
If the stationary density is set to be 
\begin{equation}
\pi\left(\mathrm{x}\right)\coloneqq\frac{p\left(\mathrm{y}_{k}|\mathrm{x}\right)p\left(\mathrm{x}|\mathrm{y}_{1:k-1}\right)}{Z_{1}}=\frac{p\left(\mathrm{y}_{k}|\mathrm{x}\right)p_{x}\left(\mathrm{x}\right)}{Z_{1}},\label{eq:proof-stationary-density}
\end{equation}
\noindent where the prior density $p_{x}\left(\mathrm{x}\right)$
and the likelihood $p\left(\mathrm{y}_{k}|\mathrm{x}\right)$ are
integrable functions with respect to $\mathrm{x}$, $Z_{1}=p\left(\mathrm{y}_{k}|\mathrm{y}_{1:k-1}\right)$
is a normalization constant, and the discrete-time observation process
$\left\{ \mathrm{y}_{k}\in\mathbb{R}^{n_{y}}:k\in\mathbb{N}\right\} $
is described as 
\begin{equation}
\mathrm{y}_{k}=h\left(\mathrm{x}_{k}\right)+R^{\nicefrac{1}{2}}\xi_{k},\quad\xi_{k}\sim\mathcal{N}(\xi_{k};0_{n_{y}},\mathrm{I}_{n_{y}});\label{eq:proof-observation-process}
\end{equation}
\noindent then the probability density corresponding to the signal
process (\ref{eq:proof-state-process}) is equivalent to the following
filtering entity 
\begin{equation}
p(\mathrm{x},\lambda|\mathcal{F}_{\lambda})=\frac{\mathbb{E}^{P}\left[e^{h\left(\mathrm{x}_{T}\right)^{T}R^{-1}\mathrm{y}_{k}-\frac{1}{2}h\left(\mathrm{x}_{T}\right)^{T}R^{-1}h\left(\mathrm{x}_{T}\right)}|\mathrm{x}\right]p\left(\mathrm{x}|\mathrm{y}_{1:k-1}\right)}{Z}.\label{eq:proof-filtering-entity}
\end{equation}
In addition, the expression (\ref{eq:proof-filtering-entity}) can
be interpreted as the analogous to the well known result (see \citep{Bucy1965})
\begin{equation}
p(\mathrm{x},\lambda|\mathcal{F}_{\lambda})=\frac{\mathbb{E}^{P}\left[e^{\int_{0}^{T}h_{T}\left(\mathrm{x}_{\lambda}\right)^{T}R_{T}^{-1}d\mathrm{y_{\lambda}}-\frac{1}{2}\int_{0}^{T}h_{T}\left(\mathrm{x}_{\lambda}\right)^{T}R_{T}^{-1}h_{T}\left(\mathrm{x}_{\lambda}\right)d\lambda}|\mathrm{x}\right]}{Z}p\left(\mathrm{x}|\mathrm{y}_{1:k-1}\right),\label{eq:proof-filtering-entity-1}
\end{equation}
\noindent for a discrete-time observation process whose analog continuous-time
(interpolated) version has the observation function $h_{T}\left(.\right)$
and covariance matrix $R_{T}$. 
\end{thm}
\begin{proof}
By definition of the stochastic system described by (\ref{eq:lemma-state-process-backward})
and (\ref{eq:lemma-scalar-field-backward}), and \emph{Lemma \ref{lem:L10}},
the reverse process $\left\{ \bar{X}_{\tau},\mathcal{V}_{\tau}\right\} $
is known to satisfy the Kolmogorov forward equation (\ref{eq:proof-Kolmogorov-forward-eq})
in reverse time $\lambda\le\tau\le T$, for which the stationary density
$\pi$ is an initial condition (initial value problem). Using the
reverse time variable $\tau=T-\lambda$, so that $p\left(\mathrm{x}_{\lambda},\lambda\right)\equiv\hat{p}\left(\bar{\mathrm{x}}_{\tau}(T-\tau),T-\tau\right)$
and ${\partial_{\lambda}p=-\partial_{\tau}\hat{p}}$, and applying
the relations (\ref{eq:drift-correspondence}), (\ref{eq:diffusion-correspondence})
and (\ref{eq:measure-drift}) from \emph{Lemma \ref{lem:L7}}, we
rewrite the equation (\ref{eq:proof-Kolmogorov-forward-eq}) for $0\le\lambda\le T$
as 
\begin{alignat}{1}
\frac{\partial}{\partial\tau}\hat{p} & =-\nabla_{\bar{\mathrm{x}}}\cdot\left(\mu_{r}\hat{p}\right)+\frac{1}{2}\nabla_{\bar{\mathrm{x}}}\cdot\left(D_{r}\nabla_{\bar{\mathrm{x}}}\hat{p}\right)-v_{r}\cdot\hat{p},\,\lambda\le\tau\le T,\nonumber \\
\frac{\partial}{\partial\tau}\hat{p} & =-\left(\nabla_{\bar{\mathrm{x}}}\cdot\mu_{r}\right)\hat{p}-\mu_{r}^{T}\nabla_{\bar{\mathrm{x}}}\hat{p}+\frac{1}{2}\nabla_{\bar{\mathrm{x}}}\cdot\left(D_{r}\nabla_{\bar{\mathrm{x}}}\hat{p}\right)+\left(\nabla_{\bar{\mathrm{x}}}\cdot\mu_{r}\right)\hat{p},\nonumber \\
\frac{\partial}{\partial\tau}\hat{p} & =-\mu_{r}^{T}\nabla_{\bar{\mathrm{x}}}\hat{p}+\frac{1}{2}\nabla_{\bar{\mathrm{x}}}\cdot\left(D_{r}\nabla_{\bar{\mathrm{x}}}\hat{p}\right),\nonumber \\
-\frac{\partial}{\partial\lambda}p & =+\mu_{f}^{T}\nabla_{\mathrm{x}}p+\frac{1}{2}\nabla_{\mathrm{x}}\cdot\left(D_{f}\nabla_{\mathrm{x}}p\right),\,0\le\lambda\le T,\label{eq:proof-KB-eq-1}\\
 & \hphantom{=+\ }p\left(\mathrm{x}_{\lambda},T\right)=\pi\left(\mathrm{x}_{T}\right),\mathrm{x}_{\lambda}\in\Omega.\nonumber 
\end{alignat}
Now we have a Kolmogorov backward equation in $p\left(\mathrm{x}_{\lambda},\lambda\right)$
with a terminal value problem for the forward process $\left\{ \mathrm{X}_{\lambda}\right\} $.
Hence, we can apply the Feynman-Kac formula for the terminal condition
$p(\mathrm{x},T)=\pi\left(\mathrm{x}\right)$, with $\mathrm{x}_{T}=\mathrm{x}_{\lambda}(T)$,
to give 
\begin{alignat}{1}
p(\mathrm{x},\lambda) & \triangleq\mathbb{E}^{\mathcal{P}}\left[\pi\left(\mathrm{x}_{T}\right)|\mathrm{x}_{\lambda}=\mathrm{x}\right]\nonumber \\
 & =\frac{\mathbb{E}^{\mathcal{P}}\left[p\left(\mathrm{y}_{k}|\mathrm{x}_{T}\right)p\left(\mathrm{x}_{T}|\mathrm{y}_{1:k-1}\right)|\mathrm{x}_{\lambda}=\mathrm{x}\right]}{Z_{1}}\nonumber \\
 & =\frac{\mathbb{E}^{\mathcal{P}}\left[e^{-\frac{1}{2}\left(\mathrm{y}_{k}-h\left(\mathrm{x}_{T}\right)\right)^{T}R^{-1}\left(\mathrm{y}_{k}-h\left(\mathrm{x}_{T}\right)\right)}p_{x}\left(\mathrm{x}_{T}\right)|\mathrm{x}\right]}{Z_{1}\left(2\pi R\right)^{\nicefrac{n_{y}}{2}}}\nonumber \\
 & =\frac{\mathbb{E}^{\mathcal{P}}\left[e^{h\left(\mathrm{x}_{T}\right)^{T}R^{-1}\mathrm{y}_{k}-\frac{1}{2}h\left(\mathrm{x}_{T}\right)^{T}R^{-1}h\left(\mathrm{x}_{T}\right)}p_{x}\left(\mathrm{x}_{T}\right)|\mathrm{x}\right]}{Z_{1}\left(2\pi R\right)^{\nicefrac{n_{y}}{2}}e^{+\frac{1}{2}\mathrm{y}_{k}^{T}R^{-1}\mathrm{y}_{k}^{\vphantom{T}}}}\nonumber \\
 & =\frac{\mathbb{E}^{\mathcal{P}}\left[e^{h\left(\mathrm{x}_{T}\right)^{T}R^{-1}\mathrm{y}_{k}-\frac{1}{2}h\left(\mathrm{x}_{T}\right)^{T}R^{-1}h\left(\mathrm{x}_{T}\right)}p_{x}\left(\mathrm{x}_{T}\right)|\mathcal{F}_{\lambda},\mathrm{x}\right]}{Z}.\label{eq:proof-probability-density-1}
\end{alignat}

Let us reinterpret the discrete-time observation process as a continuous-time
process for which we only obtain a realization at $\lambda=T$, by
linearly interpolating it along the interval $0<\lambda\le T$ to
write 
\begin{eqnarray}
d\mathrm{y_{\lambda}} & = & \frac{1}{T}h\left(\mathrm{x}_{\lambda}\right)d\lambda+\left(\frac{R}{T}\right)^{\nicefrac{1}{2}}\xi_{k}d\lambda^{\nicefrac{1}{2}},\nonumber \\
d\mathrm{y_{\lambda}^{\mathcal{Q}}} & = & h_{T}\left(\mathrm{x}_{\lambda}\right)d\lambda+R_{T}^{\nicefrac{1}{2}}d\bar{\xi}_{\lambda}^{\mathcal{P}},\quad0<\lambda\le T.\label{eq:proof-observation-continuous}
\end{eqnarray}
\noindent where $\left\{ \bar{\xi}_{\lambda}\right\} $ is an interpolated
Wiener process, under the probability measure $\mathcal{P}$, that
produces the observation noise $\int_{0}^{T}R_{T}^{\nicefrac{1}{2}}d\bar{\xi}_{\lambda}\equiv R^{\nicefrac{1}{2}}\xi_{k}$
at $\lambda=T$. The probability measure $\mathcal{P}$ is induced
in the space of paths jointly described by the state process and observation
noise $(\{\mathrm{X}_{\lambda}\},\{\bar{\Xi}_{\lambda}\})$. The interpolation
is established such that $\mathrm{y}_{\lambda}(T)=\mathrm{y}_{k}$
is the realization of the observation process under the probability
measure $\mathcal{Q}$, which is induced in the space of paths jointly
described by the state and observation processes $(\{\mathrm{X}_{\lambda}\},\{\mathrm{Y}_{\lambda}\})$.
By applying the Girsanov theorem, we know that the Radon-Nykodym derivative
to change the measure from $\mathcal{P}$ to $\mathcal{Q}$ assumes
the form (see \citep{Zakai1969} for example) 
\begin{eqnarray}
\left.\frac{d\mathcal{Q}}{d\mathcal{P}}\right|_{\mathcal{F}_{T}} & = & e^{\int_{0}^{T}h_{T}\left(\mathrm{x}_{\lambda}\right)^{T}R_{T}^{-1}d\mathrm{y_{\lambda}}-\frac{1}{2}\int_{0}^{T}h_{T}\left(\mathrm{x}_{\lambda}\right)^{T}R_{T}^{-1}h_{T}\left(\mathrm{x}_{\lambda}\right)d\lambda}\nonumber \\
 & \propto & e^{h\left(\mathrm{x}_{T}\right)^{T}R^{-1}\mathrm{y}_{k}-\frac{1}{2}h\left(\mathrm{x}_{T}\right)^{T}R^{-1}h\left(\mathrm{x}_{T}\right)}.\label{eq:proof-Radon-Nykodyn-derivative}
\end{eqnarray}
Rewritting (\ref{eq:proof-probability-density-1}) in terms of (\ref{eq:proof-Radon-Nykodyn-derivative})
and manipulating it further, we obtain 
\begin{alignat}{1}
p(\mathrm{x},\lambda) & \propto\mathbb{E}^{\mathcal{P}}\left[e^{h\left(\mathrm{x}_{T}\right)^{T}R^{-1}\mathrm{y}_{k}-\frac{1}{2}h\left(\mathrm{x}_{T}\right)^{T}R^{-1}h\left(\mathrm{x}_{T}\right)}p_{x}\left(\mathrm{x}_{T}\right)|\mathcal{F}_{\lambda},\mathrm{x}\right]\nonumber \\
 & \equiv\mathbb{E}^{\mathcal{P}}\left[\frac{d\mathcal{Q}}{d\mathcal{P}}\left(T\right)\cdot p_{x}\left(\mathrm{x}_{T}\right)|\mathcal{F}_{\lambda},\mathrm{x}\right]\nonumber \\
 & =\mathbb{E}^{\mathcal{P}}\left[\frac{d\mathcal{Q}}{d\mathcal{P}}\left(T\right)\cdot\mathbb{E}^{\mathcal{Q}}\left[p_{x}\left(\mathrm{x}_{\lambda}\right)|\mathcal{F}_{T},\mathrm{x}\right]|\mathcal{F}_{\lambda},\mathrm{x}\right]\nonumber \\
 & =\mathbb{E}^{\mathcal{P}}\left[\frac{d\mathcal{Q}}{d\mathcal{P}}\left(T\right)|\mathcal{F}_{\lambda},\mathrm{x}\right]\cdot\mathbb{E}^{\mathcal{Q}}\left[p_{x}\left(\mathrm{x}_{\lambda}\right)|\mathcal{F}_{T},\mathrm{x}\right]\nonumber \\
 & \equiv\mathbb{E}^{\mathcal{P}}\left[e^{h\left(\mathrm{x}_{T}\right)^{T}R^{-1}\mathrm{y}_{k}-\frac{1}{2}h\left(\mathrm{x}_{T}\right)^{T}R^{-1}h\left(\mathrm{x}_{T}\right)}|\mathcal{F}_{\lambda},\mathrm{x}\right]p_{x}\left(\mathrm{x}\right),\label{eq:proof-further-work-out}
\end{alignat}
\noindent where we take into account the smoothing property for conditional
expectations as 
\begin{alignat*}{1}
p(\mathrm{x},\lambda) & \propto\mathbb{E}^{\mathcal{P}}\left[\frac{d\mathcal{Q}}{d\mathcal{P}}\left(T\right)\cdot p_{x}\left(\mathrm{x}_{T}\right)|\mathcal{F}_{\lambda},\mathrm{x}_{\lambda}=\mathrm{x}\right]\\
 & =\mathbb{E}^{\mathcal{Q}}\left[p_{x}\left(\mathrm{x}_{T}\right)|\mathcal{F}_{\lambda},\mathrm{x}\right]\\
 & =\mathbb{E}^{\mathcal{Q}}\left[\mathbb{E}^{\mathcal{Q}}\left[p_{x}\left(\mathrm{x}_{\lambda}\right)|\mathcal{F}_{T},\mathrm{x}\right]|\mathcal{F}_{\lambda},\mathrm{x}\right]\\
 & =\mathbb{E}^{\mathcal{P}}\left[\frac{d\mathcal{Q}}{d\mathcal{P}}\left(T\right)\cdot\mathbb{E}^{\mathcal{Q}}\left[p_{x}\left(\mathrm{x}_{\lambda}\right)|\mathcal{F}_{T},\mathrm{x}\right]|\mathcal{F}_{\lambda},\mathrm{x}\right],
\end{alignat*}
\noindent the $\mathcal{F}_{\lambda}$-measurability of $\mathbb{E}^{\mathcal{Q}}\left[p_{x}\left(\mathrm{x}_{\lambda}\right)|\mathcal{F}_{T},\mathrm{x}_{\lambda}=\mathrm{x}\right]$,
and 
\begin{alignat*}{1}
\mathbb{E}^{\mathcal{Q}}\left[p_{x}\left(\mathrm{x}_{\lambda}\right)|\mathcal{F}_{T},\mathrm{x}_{\lambda}=\mathrm{x}\right] & =\int p_{x}\left(\mathrm{x}\right)d\mathcal{Q}\\
 & =p_{x}\left(\mathrm{x}\right)=p\left(\mathrm{x}|\mathrm{y}_{1:k-1}\right).
\end{alignat*}

As a result, the expression (\ref{eq:proof-further-work-out}) can
be written in the normalized form (\ref{eq:proof-filtering-entity}),
proving the theorem statement. The proof is complete by inserting
the continuous-time (interpolated) version of (\ref{eq:proof-Radon-Nykodyn-derivative})
into (\ref{eq:proof-further-work-out}) to verify the analogy with
(\ref{eq:proof-filtering-entity-1}).
\end{proof}
\begin{rem}
A result more general than the one presented by \emph{Theorem \ref{thm:T11}},
in terms of McKean-Vlasov diffusions, can be found in \citep{Crisan2010}.
\end{rem}
\begin{cor}
\label{cor:C13}The signal process with state equation (\ref{eq:proof-state-process}),
under the hypotheses of Theorem \ref{thm:T11}, filters its associated
(unnormalized) probability density in accordance with the Zakai equation
\begin{equation}
dp_{u}=\mathcal{L}\left[p_{u}\right]d\lambda+p_{u}\cdot h_{T}\left(\mathrm{x}_{\lambda}\right)^{T}R_{T}^{-1}d\mathrm{y}_{\lambda},\quad0<\lambda\le T;\label{eq:Zakai-equation-corollary}
\end{equation}
\noindent where $\mathcal{L}\left[.\right]=-\nabla_{\mathrm{x}}\cdot\left(\mu\cdot\right)+\nicefrac{1}{2}\nabla_{\mathrm{x}}\cdot\left(D\nabla_{\mathrm{x}}(\cdot)\right)$
is the forward Kolmogorov operator, and $\left\{ \mathrm{y}_{\lambda}:0<\lambda\le T\right\} $
is the continuous, linearly interpolated observation process defined
by (\ref{eq:proof-observation-continuous}) for which the realization
is only taken at $\lambda=T$.
\end{cor}
\begin{proof}
Define 
\begin{align}
d\zeta_{\lambda} & =h_{T}\left(\mathrm{x}_{\lambda}\right)^{T}R_{T}^{-1}d\mathrm{y_{\lambda}}-\frac{1}{2}h_{T}\left(\mathrm{x}_{\lambda}\right)^{T}R_{T}^{-1}h_{T}\left(\mathrm{x}_{\lambda}\right)d\lambda\nonumber \\
 & =\frac{1}{2}h_{T}\left(\mathrm{x}_{\lambda}\right)^{T}R_{T}^{-1}h_{T}\left(\mathrm{x}_{\lambda}\right)d\lambda+h_{T}\left(\mathrm{x}_{\lambda}\right)^{T}R_{T}^{-\nicefrac{1}{2}}d\bar{\xi}_{\lambda},\label{eq:proof-corollary-auxiliary-variable}
\end{align}
\noindent and recognize the unnormalized probability density to be
the numerator of (\ref{eq:proof-filtering-entity-1}): 
\begin{equation}
p_{u}=\mathbb{E}^{\mathcal{P}}\left[e^{\zeta_{T}}|\mathrm{x}\right]p_{x}\left(\mathrm{x}\right).\label{eq:proof-corollary-unnormalised-density}
\end{equation}
Applying \^Ito's Lemma to $p_{u}$ we get 
\begin{align}
dp_{u} & =\partial_{\lambda}p_{u}d\lambda+\partial_{\zeta}p_{u}d\zeta_{\lambda}\nonumber \\
 & +\frac{1}{2}\left[h_{T}\left(\mathrm{x}_{\lambda}\right)^{T}R_{T}^{-\nicefrac{1}{2}}\right]\left[R_{T}^{-\nicefrac{1}{2}}h_{T}\left(\mathrm{x}_{\lambda}\right)\right]\partial_{\zeta\zeta}^{2}p_{u}d\lambda\nonumber \\
 & =\partial_{\lambda}p_{u}d\lambda+\partial_{\zeta}p_{u}d\zeta_{\lambda}\nonumber \\
 & +\frac{1}{2}h_{T}\left(\mathrm{x}_{\lambda}\right)^{T}R_{T}^{-1}h_{T}\left(\mathrm{x}_{\lambda}\right)\partial_{\zeta\zeta}^{2}p_{u}d\lambda.\label{eq:proof-corollary-Ito-lemma}
\end{align}
Because 
\begin{eqnarray*}
\partial_{\zeta}^{\vphantom{2}}p_{u} & = & \partial_{\zeta\zeta}^{2}p_{u}=p_{u},\\
\partial_{\lambda}p_{u} & = & \mathbb{E}^{\mathcal{P}}\left[e^{-\int_{0}^{T}d\zeta_{\lambda}}|\mathrm{x}\right]\partial_{\lambda}p_{x}\left(\mathrm{x}\right)=\mathcal{L}\left[p_{u}\right];
\end{eqnarray*}
\noindent the expression (\ref{eq:proof-corollary-Ito-lemma}) becomes
the Zakai equation as 
\begin{align*}
dp_{u} & =\mathcal{L}\left[p_{u}\right]d\lambda\\
 & +p_{u}\cdot\left[\frac{1}{2}h_{T}\left(\mathrm{x}_{\lambda}\right)^{T}R_{T}^{-1}h_{T}\left(\mathrm{x}_{\lambda}\right)d\lambda+h_{T}\left(\mathrm{x}_{\lambda}\right)^{T}R_{T}^{-\nicefrac{1}{2}}d\bar{\xi}_{\lambda}\right]\\
 & +p_{u}\cdot\frac{1}{2}h_{T}\left(\mathrm{x}_{\lambda}\right)^{T}R_{T}^{-1}h_{T}\left(\mathrm{x}_{\lambda}\right)d\lambda\\
 & =\mathcal{L}\left[p_{u}\right]d\lambda\\
 & +p_{u}\cdot\left[h_{T}\left(\mathrm{x}_{\lambda}\right)^{T}R_{T}^{-1}h_{T}\left(\mathrm{x}_{\lambda}\right)d\lambda+h_{T}\left(\mathrm{x}_{\lambda}\right)^{T}R_{T}^{-\nicefrac{1}{2}}d\bar{\xi}_{\lambda}\right]\\
 & =\mathcal{L}\left[p_{u}\right]d\lambda+p_{u}\cdot h_{T}\left(\mathrm{x}_{\lambda}\right)^{T}R_{T}^{-1}\left[h_{T}\left(\mathrm{x}_{\lambda}\right)d\lambda+R_{T}^{\nicefrac{1}{2}}d\bar{\xi}_{\lambda}\right]\\
 & =\mathcal{L}\left[p_{u}\right]d\lambda+p_{u}\cdot h_{T}\left(\mathrm{x}_{\lambda}\right)^{T}R_{T}^{-1}d\mathrm{y}_{\lambda}.
\end{align*}
\end{proof}

\section{Derivation of the integration rule\label{sec:Integration-rule-derivation}}

We intend to approximate the integration of the following equation
with respect to $\lambda$: 
\begin{equation}
d\mathrm{x}=\frac{1}{2}D\left(\lambda\right)\cdot\nabla_{\mathrm{x}}\,\log\pi\left(\mathrm{x}\right)d\lambda+D\left(\lambda\right)^{\nicefrac{1}{2}}d\mathrm{w}_{\lambda}.\label{eq:Ito-equation}
\end{equation}
Linearising equation (\ref{eq:Ito-equation}) w.r.t. $\mathrm{x}$
around the current state $\mathrm{x}_{n-1}$, we have 
\begin{equation}
d\mathrm{x}=A\cdot\mathrm{x}\,d\lambda+B\,d\lambda+D^{\nicefrac{1}{2}}d\mathrm{w}_{\lambda},\label{eq:linearised-Ito-equation}
\end{equation}
\noindent where 
\begin{alignat}{1}
A\left(\mathrm{x}_{n-1}\right) & =\nicefrac{1}{2}D\left(\lambda_{n-1}\right)\cdot\mathcal{H}_{\mathrm{x}}\left[\log\pi\left(\mathrm{x}\right)\right]_{\mathrm{x}_{n-1}},\nonumber \\
B\left(\mathrm{x}_{n-1}\right) & =a\left(\mathrm{x}_{n-1}\right)-A\cdot\mathrm{x}_{n-1},\label{eq:set-1}\\
a\left(\mathrm{x}_{n-1}\right) & =\nicefrac{1}{2}D\left(\lambda_{n-1}\right)\cdot\left.\nabla_{\mathrm{x}}\,\log\pi\left(\mathrm{x}\right)\right|_{\mathrm{x}_{n-1}}.\nonumber 
\end{alignat}
If we apply the definition $D\left(\lambda_{n-1}\right)=-\mathcal{H}_{\mathrm{x}}\left[\log\pi\left(\mathrm{x}\right)\right]_{\mathrm{x}_{n-1}}^{-1}$,
where $\mathcal{H}_{\mathrm{x}}\left[\cdot\right]$ is the Hessian
w.r.t. $\mathrm{x}$, we have 
\begin{equation}
A=-\frac{1}{2}\mathbb{\mathbb{I}}_{n_{x}},\label{eq:Linear-transition-matrix}
\end{equation}
\noindent where $\mathbb{\mathbb{I}}_{n_{x}}$ is the identity matrix
with dimension $n_{x}\times n_{x}$. Based on the Laplace transform,
we can obtain the solution for a homogeneous version of the equation
(\ref{eq:linearised-Ito-equation}) in discrete time by 
\begin{align}
\mathrm{x}\left(\lambda\right) & =\mathcal{L}^{-1}\left\{ \left(s\cdot\mathbb{\mathbb{I}}_{n_{x}}-A\right)^{-1}\mathrm{x}\left(\lambda_{n-1}\right)\right\} \nonumber \\
 & =\int_{\lambda_{n-1}}^{\lambda}\left(s\cdot\mathbb{\mathbb{I}}_{n_{x}}-A\right)^{-1}e^{s\cdot\mathbb{\mathbb{I}}_{n_{x}}\cdot\tau}\mathrm{x}\left(\lambda_{n-1}\right)ds\nonumber \\
 & =e^{A\cdot\left(\lambda-\lambda_{n-1}\right)}\mathrm{x}\left(\lambda_{n-1}\right),\nonumber \\
\mathrm{x}_{n} & =e^{A\cdot\Delta\lambda}\mathrm{x}_{n-1}.\label{eq:homogeneous-solution}
\end{align}

By a similar procedure, and considering the definition of a Wiener
integral for the stochastic term, we can obtain the solution of the
complete inhomogeneous equation (\ref{eq:linearised-Ito-equation})
as 
\begin{alignat}{1}
\mathrm{x}\left(\lambda_{n}\right) & =e^{A\cdot\Delta\lambda}\mathrm{x}\left(\lambda_{n-1}\right)+\int_{\lambda_{n-1}}^{\lambda_{n}}e^{-A\cdot\left(\tau-\Delta\lambda\right)}B\,d\tau\nonumber \\
 & +\sqrt{\int_{\lambda_{n-1}}^{\lambda_{n}}e^{-A\cdot\left(\tau-\Delta\lambda\right)}D\left(\lambda_{n-1}\right)e^{-A^{T}\cdot\left(\tau-\Delta\lambda\right)}d\tau}\cdot\mathrm{w}_{n},\nonumber \\
\mathrm{x}_{n} & =e^{A\cdot\Delta\lambda}\mathrm{x}_{n-1}+A^{-1}\left[e^{A\cdot\Delta\lambda}-\mathbb{\mathbb{I}}_{n_{x}}\right]\left[a-A\cdot\mathrm{x}_{n-1}\right]\nonumber \\
 & +\sqrt{\int_{0}^{\Delta\lambda}e^{A\cdot\nu}D\left(\lambda_{n-1}\right)e^{A^{T}\cdot\nu}d\nu}\cdot\mathrm{w}_{n}\nonumber \\
 & =\mathrm{x}_{n-1}+A^{-1}\left[e^{A\cdot\Delta\lambda}-\mathbb{\mathbb{I}}_{n_{x}}\right]a\left(\mathrm{x}_{n-1}\right)\nonumber \\
 & +\sqrt{\int_{0}^{\Delta\lambda}e^{A\cdot\nu}D\left(\lambda_{n-1}\right)e^{A^{T}\cdot\nu}d\nu}\cdot\mathrm{w}_{n},\label{eq:inhomogeneous-solution}
\end{alignat}
\noindent where $\mathrm{w}_{n}\sim\mathcal{N}(\mathrm{w};\,0_{n_{x}},\mathbb{\mathbb{I}}_{n_{x}})$.
Substituting (\ref{eq:Linear-transition-matrix}) into (\ref{eq:inhomogeneous-solution}),
we have 
\begin{alignat}{1}
\mathrm{x}_{n} & =\mathrm{x}_{n-1}+\left[-\frac{1}{2}\mathbb{\mathbb{I}}_{n_{x}}\right]^{-1}\left[e^{-\frac{1}{2}\mathbb{\mathbb{I}}_{n_{x}}\cdot\Delta\lambda}-\mathbb{\mathbb{I}}_{n_{x}}\right]a\left(\mathrm{x}_{n-1}\right)\nonumber \\
 & +\sqrt{\int_{0}^{\Delta\lambda}e^{-\frac{1}{2}\mathbb{\mathbb{I}}_{n_{x}}\cdot\nu}D\left(\lambda_{n-1}\right)e^{-\frac{1}{2}\mathbb{\mathbb{I}}_{n_{x}}^{T}\cdot\nu}d\nu}\cdot\mathrm{w}_{n}.\label{eq:inhomogeneous-solution-1}
\end{alignat}
By noticing that 
\begin{equation}
e^{-\frac{1}{2}\mathbb{\mathbb{I}}_{n_{x}}\Delta\lambda}=e^{-\frac{1}{2}\Delta\lambda}\mathbb{\mathbb{I}}_{n_{x}},\label{eq:exponential-matrix}
\end{equation}
\noindent the equation (\ref{eq:inhomogeneous-solution-1}) can be
simplified as 
\begin{alignat}{1}
\mathrm{x}_{n} & =\mathrm{x}_{n-1}-2\left(e^{-\frac{\Delta\lambda}{2}}-1\right)a\left(\mathrm{x}_{n-1}\right)\nonumber \\
 & +\sqrt{\int_{0}^{\Delta\lambda}e^{-\nicefrac{\nu}{2}}e^{-\nicefrac{\nu}{2}}d\nu}\cdot D\left(\lambda_{n-1}\right)^{\nicefrac{1}{2}}\cdot\mathrm{w}_{n}\nonumber \\
 & =\mathrm{x}_{n-1}+2\left(1-e^{-\frac{\Delta\lambda}{2}}\right)a\left(\mathrm{x}_{n-1}\right)\nonumber \\
 & +\sqrt{\int_{0}^{\Delta\lambda}e^{-\nu}d\nu}\cdot D\left(\lambda_{n-1}\right)^{\nicefrac{1}{2}}\cdot\mathrm{w}_{n},\nonumber \\
\mathrm{x}_{n} & =\mathrm{x}_{n-1}+\left(1-e^{-\frac{\Delta\lambda}{2}}\right)D\left(\lambda_{n-1}\right)\cdot\nabla_{\mathrm{x}}\log\pi\left(\mathrm{x}_{n-1}\right)\nonumber \\
 & +\left(1-e^{-\Delta\lambda}\right)^{\nicefrac{1}{2}}D\left(\lambda_{n-1}\right)^{\nicefrac{1}{2}}\cdot\mathrm{w}_{n}.\label{eq:final-integration-rule}
\end{alignat}

\section{Justification for the local flow linearization\label{sec:Justification-for-the-local-flow-linearization}}

When approximating the stochastic particle flow (\ref{eq:state-flow-equation})
as locally linear in the neighbourhood of a probability mass located
at $\mathrm{x}_{l}$, we expect to produce a negligible error in the
propagated moments. Given a small increment of pseudo-time $\Delta\lambda>0$,
the SDE is approximated within the region $\left\Vert \mathrm{x}-\mathrm{x}_{l}\right\Vert <\zeta$,
for a sufficiently small $\zeta\in\mathbb{R}_{+}$, as 
\begin{align}
d\mathrm{x} & =\frac{1}{2}D(\lambda)\nabla_{\mathrm{x}}\log\pi\left(\mathrm{x}\right)d\lambda+D(\lambda)^{\nicefrac{1}{2}}d\mathrm{w}_{\lambda},\quad\lambda\in(\lambda_{l},\lambda_{l}+\Delta\lambda],\,\mathrm{x}(\lambda_{l})=\mathrm{x}_{l};\label{eq:flow-local-exact}\\
d\mathrm{x} & \approx\left[C(\mathrm{x}_{l},\lambda)\cdot\mathrm{x}+c(\mathrm{x}_{l},\lambda)\right]d\lambda+D(\lambda)^{\nicefrac{1}{2}}d\mathrm{w}_{\lambda}.\label{eq:flow-local-approximation-1}
\end{align}

In this section we provide a nonrigorous argument to explain why this
local flow approximation produces admissible errors on the propagated
moments without major concern. We will look at the expected error
with respect to the intermediate marginal measures that follow from
the Langevin dynamics for $\lambda\ge\lambda_{l}$ as 
\begin{align*}
q(\mathrm{x}|\mathrm{y}_{k}) & =\int_{\mathcal{X}}p_{t}(\mathrm{x}|\mathrm{x}_{l})p(\mathrm{x}_{l}|\mathrm{y}_{k})d\mathrm{x}_{l}=\mathbb{E}_{p(l)}\left[p_{t}(\mathrm{x}|\mathrm{x}_{l})\right]\\
 & =\mathbb{E}_{p(l)}\left[\mathcal{N}\left(\mathrm{x;}\,\mathrm{x}_{l}+\int_{\lambda_{l}}^{\lambda_{l}+\Delta\lambda}\mu(\mathrm{x}_{l},\lambda)^{\vphantom{\nicefrac{1}{2}}}d\lambda,\left(\int_{\lambda_{l}}^{\lambda_{l}+\Delta\lambda}D(\lambda)^{\nicefrac{1}{2}}d\mathrm{w}_{\lambda}\right)^{2}\right)\right].
\end{align*}
\noindent where $\mu(\mathrm{x},\lambda)=\frac{1}{2}D(\lambda)\nabla_{\mathrm{x}}\log\pi\left(\mathrm{x}\right)$.

In the following analysis we will assume that the state is one-dimensional,
i.e., $\mathrm{x}\in\mathbb{R}$, just to present a short argument
that can be easily extended to the multidimensional case. If one follows
a procedure that (i) applies \^Ito's lemma to a continuous, measurable
and nicely behaved function $\mathscr{M}\left(\mathrm{x}\right)$,
(ii) substitutes in the exact stochastic differential (\ref{eq:flow-local-exact}),
(iii) takes expectation of the resulting equation, (iv) derivates
it with respect to $\lambda$, and then (v) integrates its right-hand
side by parts, one obtains the so-called moment equation: 
\begin{equation}
\frac{d}{d\lambda}\mathbb{E}_{q}\left[\mathscr{M}\left(\mathrm{x}\right)\right]=\mathbb{E}_{q}\left[\frac{\partial\vphantom{^{2}}\mathscr{M}\left(\mathrm{x}\right)}{\partial\mathrm{x}\vphantom{^{2}}}\mu\left(\mathrm{x},\lambda\right)+\frac{1}{2}D(\lambda)\frac{\partial^{2}\mathscr{M}\left(\mathrm{x}\right)}{\partial\mathrm{x}^{2}}\right].\label{eq:Moment-equation}
\end{equation}
Note that when applied to $\mathscr{M}\left(\mathrm{x}\right)=\mathrm{x}$
and $\mathscr{M}\left(\mathrm{x}\right)=(\mathrm{x}-\mu_{m})^{2}$,
for the linear approximation $\tilde{\mu}(\mathrm{x},\lambda)=C(\lambda)\cdot\mathrm{x}+c(\lambda)$,
the referred equation gives the ODEs of the approximated mean and
variance respectively 
\begin{align*}
\frac{d}{d\lambda}\tilde{\mathbb{E}}_{q}\left[\mathrm{x}\right] & =\hphantom{2}C(\lambda)\cdot\tilde{\mathbb{E}}_{q}\left[\mathrm{x}\right]+c(\lambda),\\
\frac{d}{d\lambda}\tilde{\mathbb{E}}_{q}\left[(\mathrm{x}-\mu_{m})^{2}\right] & =2C(\lambda)\cdot\tilde{\mathbb{E}}_{q}\left[(\mathrm{x}-\mu_{m})^{2}\right]+D(\lambda).
\end{align*}

Denote the deviation $\delta_{\mathbb{E}}\left[\mathrm{x}\right]=\mathbb{E}_{q}\left[\mathrm{x}\right]-\tilde{\mathbb{E}}_{q}\left[\mathrm{x}\right]$,
where $\mathbb{E}_{q}\left[\mathrm{x}\right]=\mu_{m}$ is the exact
mean propagated by the process (\ref{eq:flow-local-exact}) and $\tilde{\mathbb{E}}_{q}\left[\mathrm{x}\right]=\tilde{\mu}_{m}$
is the approximated mean propagated by the locally linearized process
(\ref{eq:flow-local-approximation-1}). First we note that the reverse
triangle inequality allows us to state 
\begin{align}
\lim_{\Delta\lambda\rightarrow0}\left|\frac{\left\Vert \delta_{\mathbb{E}}\left[\mathrm{x}(\lambda+\Delta\lambda)\right]\right\Vert -\left\Vert \delta_{\mathbb{E}}\left[\mathrm{x}(\lambda)\right]\right\Vert }{\Delta\lambda}\right| & \le\lim_{\Delta\lambda\rightarrow0}\frac{\left\Vert \delta_{\mathbb{E}}\left[\mathrm{x}(\lambda+\Delta\lambda)\right]-\delta_{\mathbb{E}}\left[\mathrm{x}(\lambda)\right]\right\Vert }{\Delta\lambda},\nonumber \\
\left|\lim_{\Delta\lambda\rightarrow0}\frac{\left\Vert \delta_{\mathbb{E}}\left[\mathrm{x}(\lambda+\Delta\lambda)\right]\right\Vert -\left\Vert \delta_{\mathbb{E}}\left[\mathrm{x}(\lambda)\right]\right\Vert }{\Delta\lambda}\right| & \le\lim_{\Delta\lambda\rightarrow0}\left\Vert \frac{\delta_{\mathbb{E}}\left[\mathrm{x}(\lambda+\Delta\lambda)\right]-\delta_{\mathbb{E}}\left[\mathrm{x}(\lambda)\right]}{\Delta\lambda}\right\Vert ,\nonumber \\
\therefore\left|\frac{d}{d\lambda}\left\Vert \delta_{\mathbb{E}}\left[\mathrm{x}(\lambda)\right]\right\Vert \right| & \le\left\Vert \frac{d}{d\lambda}\delta_{\mathbb{E}}\left[\mathrm{x}(\lambda)\right]\right\Vert ,\label{eq:Derivatives-inequality}
\end{align}
\noindent where $\left|\cdot\right|$ is the absolute value, $\left\Vert \cdot\right\Vert $
the Euclidean norm, and all limits are assumed to exist. We use inequality
(\ref{eq:Derivatives-inequality}) and the moment equation (\ref{eq:Moment-equation})
to work out 
\begin{align}
\left|\frac{d}{d\lambda}\left\Vert \mathbb{E}_{q}\left[\mathrm{x}\right]-\tilde{\mathbb{E}}_{q}\left[\mathrm{x}\right]\right\Vert \right| & \le\left\Vert \frac{d}{d\lambda}\mathbb{E}_{q}\left[\mathrm{x}\right]-\frac{d}{d\lambda}\tilde{\mathbb{E}}_{q}\left[\mathrm{x}\right]\right\Vert ,\nonumber \\
\left|\frac{d}{d\lambda}\left\Vert \mu_{m}-\tilde{\mu}_{m}\right\Vert \right| & \le\left\Vert \mathbb{E}_{q}\left[\mu(\mathrm{x},\lambda)\right]-\mathbb{E}_{q}\left[\tilde{\mu}(\mathrm{x},\lambda)\right]\right\Vert ,\quad\text{(moment equation for \ensuremath{\mathscr{M}(\mathrm{x})=\mathrm{x}})}\nonumber \\
\left|\frac{d}{d\lambda}\left\Vert \mu_{m}-\tilde{\mu}_{m}\right\Vert \right| & \le\left\Vert \mathbb{E}_{q}\left[\mu(\mathrm{x},\lambda)-\tilde{\mu}(\mathrm{x},\lambda)\right]\right\Vert \le\mathbb{E}_{q}\left[\left\Vert \mu(\mathrm{x},\lambda)-\tilde{\mu}(\mathrm{x},\lambda)\right\Vert \right],\nonumber \\
\left|\frac{d}{d\lambda}\left\Vert \mu_{m}-\tilde{\mu}_{m}\right\Vert \right| & \le\mathbb{E}_{q}\left[\left\Vert \mu(\mathrm{x},\lambda)-\left(C(\mathrm{x}_{l},\lambda)\cdot\mathrm{x}+c(\mathrm{x}_{l},\lambda)\right)\right\Vert \right]\le K_{0}\mathbb{E}_{q}\left[\left\Vert (\mathrm{x}-\mathrm{x}_{l})^{2}\right\Vert \right],\nonumber \\
\frac{d}{d\lambda}\left\Vert \mu_{m}-\tilde{\mu}_{m}\right\Vert  & \le K_{1}\zeta^{2}\nonumber \\
\left\Vert \mu_{m}-\tilde{\mu}_{m}\right\Vert  & \le K_{1}\zeta^{2}\Delta\lambda,\quad\text{for }\lambda\in(\lambda_{l},\lambda_{l}+\Delta\lambda],\,\mathrm{x}(\lambda_{l})=\mathrm{x}_{l},\,K_{0},K_{1}\in\mathbb{R}_{+},\label{eq:error-bound-mean}
\end{align}
\noindent where the modulus is dismissed because $\left\Vert \mu_{m}-\tilde{\mu}_{m}\right\Vert $
increases monotonically with $\Delta\lambda$. Similarly for the error
on the second moment 
\begin{align}
\left|\frac{d}{d\lambda}\Vert\Sigma_{m}-\tilde{\Sigma}_{m}\Vert\right| & =\left\Vert \frac{d}{d\lambda}\mathbb{E}_{q}\left[(\mathrm{x}-\mu_{m})^{2}\right]-\frac{d}{d\lambda}\tilde{\mathbb{E}}_{q}\left[(\mathrm{x}-\mu_{m})^{2}\right]\right\Vert ,\nonumber \\
\left|\frac{d}{d\lambda}\Vert\Sigma_{m}-\tilde{\Sigma}_{m}\Vert\right| & \le\left\Vert \mathbb{E}_{q}\left[2(\mathrm{x}-\mu_{m})\mu(\mathrm{x},\lambda)+D(\lambda)\right]-\mathbb{E}_{q}\left[2(\mathrm{x}-\mu_{m})\tilde{\mu}(\mathrm{x},\lambda)+D(\lambda)\right]\right\Vert ,\nonumber \\
\left|\frac{d}{d\lambda}\Vert\Sigma_{m}-\tilde{\Sigma}_{m}\Vert\right| & \le\left\Vert 2\mathbb{E}_{q}\left[(\mathrm{x}-\mu_{m})(\mu(\mathrm{x},\lambda)-\tilde{\mu}(\mathrm{x},\lambda))\right]\right\Vert \le2\mathbb{E}_{q}\left[\left\Vert (\mathrm{x}-\mu_{m})(\mu(\mathrm{x},\lambda)-\tilde{\mu}(\mathrm{x},\lambda))\right\Vert \right],\nonumber \\
\left|\frac{d}{d\lambda}\Vert\Sigma_{m}-\tilde{\Sigma}_{m}\Vert\right| & \leq2K_{0}^{\prime}\mathbb{E}_{q}\left[\left\Vert (\mathrm{x}-\mu_{m})(\mathrm{x}-\mathrm{x}_{l})^{2}\right\Vert \right]=2K_{0}^{\prime}\mathbb{E}_{q}\left[\left\Vert (\mathrm{x}_{l}-\mu_{m})(\mathrm{x}-\mathrm{x}_{l})^{2}+(\mathrm{x}-\mathrm{x}_{l})^{3}\right\Vert \right],\nonumber \\
\left|\frac{d}{d\lambda}\Vert\Sigma_{m}-\tilde{\Sigma}_{m}\Vert\right| & \le2K_{0}^{\prime}\mathbb{E}_{q}\left[\left(\left\Vert \mathrm{x}_{l}-\mu_{m}\right\Vert \cdot\left\Vert (\mathrm{x}-\mathrm{x}_{l})^{2}\right\Vert +\left\Vert (\mathrm{x}-\mathrm{x}_{l})^{3}\right\Vert \right)\right],\quad\text{(triangle, Cauchy-Schwarz)}\nonumber \\
\left|\frac{d}{d\lambda}\Vert\Sigma_{m}-\tilde{\Sigma}_{m}\Vert\right| & \le K_{1a}^{\prime}\zeta^{2}+K_{1b}^{\prime}\zeta^{3},\nonumber \\
\frac{d}{d\lambda}\Vert\Sigma_{m}-\tilde{\Sigma}_{m}\Vert & \le K_{1}^{\prime}\zeta^{2},\nonumber \\
\Vert\Sigma_{m}-\tilde{\Sigma}_{m}\Vert & \le K_{1}^{\prime}\zeta^{2}\Delta\lambda,\quad\text{for }\lambda\in(\lambda_{l},\lambda_{l}+\Delta\lambda],\,\mathrm{x}(\lambda_{l})=\mathrm{x}_{l},\,,\,K_{0}^{\prime},K_{1a}^{\prime},K_{1b}^{\prime},K_{1}^{\prime}\in\mathbb{R}^{+},\label{eq:error-bound-variance}
\end{align}
\noindent where the modulus is suppressed because $\Vert\Sigma_{m}-\tilde{\Sigma}_{m}\Vert$
increases monotonically with $\Delta\lambda$.

It is very important to mention that the collection of factors $K_{0},\,K_{1},\,K_{0}^{\prime},\,K_{1a}^{\prime},\,K_{1b}^{\prime},\,K_{1}^{\prime},$
can be different for each possible interval $(\lambda_{l},\lambda_{l}+\Delta\lambda]$.
Rigorously speaking those coefficients can depend on pseudo-time $\lambda$
because 
\[
K_{0}^{\vphantom{\prime}},\,K_{0}^{\prime}\propto\frac{1}{2}\left\Vert \frac{\partial^{2}}{\partial\mathrm{x}^{2}}\mu(\mathrm{x},\lambda)\right\Vert _{\mathrm{x}=\mathrm{x}_{l}}.
\]
However, in accordance with the methodology of the stochastic particle
flow, we select these factors so that inequalities (\ref{eq:error-bound-mean})
and (\ref{eq:error-bound-variance}) hold for a specific interval
within which the diffusion coefficient is kept fixed as $D(\lambda_{l})\equiv D(\mathrm{x}(\lambda_{l}))=D(\mathrm{x}_{l})$.
Given that the diffusion coefficient is piecewise constant in $\lambda$,
the referred factors are also piecewise constant in $\lambda$ and
directly dependent on the diffusion coefficient. If the target density
is Gaussian, then $K_{0}^{\vphantom{\prime}},\,K_{0}^{\prime}=0$
and the error commited due to the local linearization is null. For
a fixed diffusion coefficient, $K_{0}^{\vphantom{\prime}},\,K_{0}^{\prime}>0$
if and only if the target log-density has third or higher-order non-zero
derivatives at $\mathrm{x}_{l}$.

Notice that integrating (\ref{eq:flow-local-exact}) by the Euler-Maruyama
scheme would produce $\mathbb{E}_{q}\left[\left\Vert \tilde{\mathrm{x}}_{l+1}-\mathrm{x}_{l+1}\right\Vert \right]\le K\Delta\lambda^{\nicefrac{1}{2}}$
and, therefore, it would be reasonable to expect $\Vert\mu_{m}-\tilde{\mu}_{m}\Vert\le K_{2}^{\vphantom{\prime}}\Delta\lambda^{2}$
and $\Vert\Sigma_{m}-\tilde{\Sigma}_{m}\Vert\le K_{2}^{\prime}\Delta\lambda^{2}$
for some $K_{2}^{\vphantom{\prime}},K_{2}^{\prime}\in\mathbb{R}_{+}$.
In the multivariate case, by applying the method used in this section,
the curious reader should learn that the bounds are multiplied by
the dimension $n_{x}$, to give $\Vert\mu_{m}-\tilde{\mu}_{m}\Vert_{2}\le K_{1}\zeta^{2}\Delta\lambda\cdot n_{x}$
and $\Vert\Sigma_{m}-\tilde{\Sigma}_{m}\Vert_{2}\le K_{1}^{\prime}\zeta^{2}\Delta\lambda\cdot n_{x}$.

\section{Discrete-time stochastic IDM\label{sec:Discrete-time-stochastic-IDM}}

This section of the appendix presents the resulting discrete-time
approximation of the stochastic Intelligent Driver model. Define the
state equation for the discrete-time IDM to be 
\begin{equation}
\mathrm{x}_{k}=A\cdot\mathrm{x}_{k-1}+B+\mathrm{w}_{k},\label{eq:state-equation}
\end{equation}
\noindent where the state vector is represented for $\alpha$ vehicles
by 
\begin{equation}
\mathrm{x}_{k}=\left[\begin{array}{c}
p_{1\hphantom{-\alpha}}\\
p_{2\hphantom{-\alpha}}\\
\vdots\\
p_{\alpha-1}\\
p_{\alpha\hphantom{-1}}\\
v_{1\hphantom{-\alpha}}\\
v_{2\hphantom{-\alpha}}\\
\vdots\\
v_{\alpha-1}\\
v_{\alpha\hphantom{-1}}
\end{array}\right]_{k}.\label{eq:state-vector}
\end{equation}

The variables $p_{i}$ and $v_{i}$ are the position and velocity
of the $i$th vehicle respectively. The state-transition matrix can
be written as 
\begin{equation}
A=\left[\begin{array}{cccc}
A_{1,1} & A_{1,2} & \dots & A_{1,2\alpha}\\
A_{2,1} & A_{2,2} &  & \vdots\\
\vdots &  & \ddots\\
A_{2\alpha,1} & \dots &  & A_{2\alpha,2\alpha}
\end{array}\right]_{k-1}.\label{eq:state-transition-matrix-1}
\end{equation}
For $i,j\in\mathbb{N}$, the diagonal elements of the state-transition
matrix are given by 
\begin{equation}
A_{i,i}=\begin{cases}
1, & i\in[1,\alpha];\\
1+\frac{\partial\dot{v}_{i-\alpha}}{\partial v_{i-\alpha}}\cdot\Delta t, & i\in(\alpha+1,2\alpha];
\end{cases}\label{eq:A-elements}
\end{equation}
\noindent and the off-diagonal elements given by 
\begin{equation}
A_{i,j}=\begin{cases}
\Delta t, & i\in[1,\alpha],\,j=i+\alpha;\\
\frac{\partial\dot{v}_{i-\alpha}}{\partial p_{\alpha\hphantom{-i}}}\hphantom{_{-1}}\cdot\Delta t, & i=\alpha+1,\,j=\alpha;\\
\frac{\partial\dot{v}_{i-\alpha\hphantom{-1.}}}{\partial p_{i-\alpha-1}}\cdot\Delta t, & i\in(\alpha+1,2\alpha],\,j=i-1-\alpha;\\
\frac{\partial\dot{v}_{i-\alpha}}{\partial v_{\alpha\hphantom{-i}}}\hphantom{_{-1}}\cdot\Delta t, & i=\alpha+1,\,j=2\alpha;\\
\frac{\partial\dot{v}_{i-\alpha\hphantom{-1.}}}{\partial v_{i-\alpha-1}}\cdot\Delta t, & i\in(\alpha+1,2\alpha],\,j=i-1;\\
0, & \text{otherwise;}
\end{cases}\label{eq:A-elements-2}
\end{equation}
\noindent where 
\begin{alignat}{1}
\frac{\partial\dot{v}_{n\hphantom{-1}}}{\partial p_{n-1}} & =+2a\left(\frac{s\left(v_{n},\Delta v_{n}\right)^{2}}{s_{n}^{3}}\right),\label{eq:A-derivative-1}\\
\frac{\partial\dot{v}_{n}}{\partial p_{n}}\hphantom{_{-1}} & =-2a\left(\frac{s\left(v_{n},\Delta v_{n}\right)^{2}}{s_{n}^{3}}\right),\label{eq:A-derivative-2}\\
\frac{\partial\dot{v}_{n\hphantom{-1}}}{\partial v_{n-1}} & =+2a\left(\frac{s\left(v_{n},\Delta v_{n}\right)}{s_{n}^{2}}\right)\left(\frac{v_{n}}{2\sqrt{a\cdot b}}\right),\label{eq:A-derivative-3}\\
\frac{\partial\dot{v}_{n}}{\partial v_{n}}\hphantom{_{-1}} & =-a\left(\frac{\delta}{v_{0}}\right)\left(\frac{v_{n}}{v_{0}}\right)^{\delta-1}\nonumber \\
 & -2a\left(\frac{s\left(v_{n},\Delta v_{n}\right)}{s_{n}^{2}}\right)\left(T_{h}+\frac{2v_{n}-v_{n-1}}{2\sqrt{a\cdot b}}\right).\label{eq:A-derivative-4}
\end{alignat}

The model takes into account the fact that, on a ring road, the last
vehicle in the convoy can be regarded the one potentially in front
of the vehicle leading the queue, assuming that the first vehicle
can complete the circuit faster and approach the last one from behind.
This is represented by the terms $\partial_{p_{\alpha}}\dot{v}_{1}\cdot\Delta t$
and $\partial_{v_{\alpha}}\dot{v}_{1}\cdot\Delta t$ that appear in
(\ref{eq:A-elements-2}) when $i=\alpha+1$, which shall be calculated
respectively according to expressions analogous to (\ref{eq:A-derivative-1})
and (\ref{eq:A-derivative-3}). The constant term is defined as 
\begin{equation}
B=\left[\begin{array}{c}
B_{1\hphantom{\alpha}}\\
B_{2\hphantom{\alpha}}\\
\vdots\\
B_{2\alpha}
\end{array}\right]_{k},\label{eq:IDM-constant-term}
\end{equation}
\noindent where 
\begin{alignat}{1}
B_{i} & =\begin{cases}
0, & i\in[1,\alpha];\\
\left\langle \dot{v}_{i}\right\rangle -\frac{\partial\dot{v}_{i\hphantom{-1}}}{\partial p_{i-1}}\cdot p_{i-1}-\frac{\partial\dot{v}_{i}}{\partial p_{i}}\cdot p_{i}\\
\hphantom{\left\langle \dot{v}_{i}\right\rangle }-\frac{\partial\dot{v}_{i\hphantom{-1}}}{\partial v_{i-1}}\cdot v_{i-1}-\frac{\partial\dot{v}_{i}}{\partial v_{i}}\cdot v_{i}, & i\in[\alpha+1,2\alpha];
\end{cases}\label{eq:B-elements}
\end{alignat}
\noindent and 
\begin{equation}
\left\langle \dot{v}_{i}\right\rangle =a\left[1-\left(\frac{v_{i}}{v_{0}}\right)^{\delta}-\left(\frac{s\left(v_{i},\Delta v_{i}\right)}{s_{i}}\right)^{2}\right].\label{eq:IDM-average-acceleration}
\end{equation}

The covariance matrix $Q_{k}=\mathbb{E}\left[\mathrm{w}_{k}\mathrm{w}_{k}^{T}\right]$
is defined as 
\begin{equation}
Q_{k}=\left[\begin{array}{cccc}
Q_{1,1} & Q_{1,2} & \dots & Q_{1,2\alpha}\\
Q_{2,1} & Q_{2,2} &  & \vdots\\
\vdots &  & \ddots\\
Q_{2\alpha,1} & \dots &  & Q_{2\alpha,2\alpha}
\end{array}\right]_{k},\label{eq:IDM-process-noise-covariance-matrix}
\end{equation}
\noindent with diagonal elements 
\begin{equation}
Q_{i,i}=\sigma_{q}^{2}\times\begin{cases}
Q_{i,i}^{(1)}, & i\in[1,\alpha];\\
Q_{i,i}^{(2)}, & i=\alpha+1;\\
Q_{i,i}^{(3)}, & i\in(\alpha+1,2\alpha];
\end{cases}\label{eq:Q-elements}
\end{equation}
\noindent where 
\begin{alignat}{1}
Q_{i,i}^{(1)} & =\frac{\Delta t^{3}}{3}+\Delta t,\nonumber \\
Q_{i,i}^{(2)} & =\left(\frac{\partial\dot{v}_{i-\alpha}}{\partial p_{\alpha\hphantom{-i}}}+\frac{\partial\dot{v}_{i-\alpha}}{\partial p_{i-\alpha}}+\frac{\partial\dot{v}_{i-\alpha}}{\partial v_{\alpha\hphantom{-i}}}+\frac{\partial\dot{v}_{i-\alpha}}{\partial v_{i-\alpha}}\right)\cdot\frac{\Delta t^{3}}{3}\nonumber \\
 & +\hphantom{(}\frac{\partial\dot{v}_{i-\alpha}}{\partial v_{i-\alpha}}\cdot\Delta t^{2}+\Delta t,\nonumber \\
Q_{i,i}^{(3)} & =\left(\frac{\partial\dot{v}_{i-\alpha\hphantom{-1}}}{\partial p_{i-\alpha-1}}+\frac{\partial\dot{v}_{i-\alpha}}{\partial p_{i-\alpha}}+\frac{\partial\dot{v}_{i-\alpha\hphantom{-1}}}{\partial v_{i-\alpha-1}}+\frac{\partial\dot{v}_{i-\alpha}}{\partial v_{i-\alpha}}\right)\cdot\frac{\Delta t^{3}}{3}\nonumber \\
 & +\hphantom{(}\frac{\partial\dot{v}_{i-\alpha}}{\partial v_{i-\alpha}}\cdot\Delta t^{2}+\Delta t;\label{eq:Q-diagonal-elements}
\end{alignat}
\noindent and off-diagonal elements 
\begin{equation}
Q_{j,i}=Q_{i,j}=\sigma_{q}^{2}\times\begin{cases}
Q_{i,j}^{(4)}, & i\in[1,\alpha],\,j=i+\alpha;\\
Q_{i,j}^{(5)}, & i=\alpha+1,\,j=\alpha;\\
Q_{i,j}^{(6)}, & i\in(\alpha+1,2\alpha],\,j=i-1-\alpha;\\
Q_{i,j}^{(7)}, & i=\alpha+1,\,j=2\alpha;\\
Q_{i,j}^{(8)}, & i\in(\alpha+1,2\alpha],\,j=i-1;\\
0, & \text{otherwise;}
\end{cases}\label{eq:Q-elements-2}
\end{equation}
\noindent where 
\begin{alignat}{1}
Q_{i,j}^{(4)} & =\frac{\partial\dot{v}_{i}}{\partial v_{i}}\cdot\frac{\Delta t^{3}}{3}+\left(\frac{\partial\dot{v}_{i}}{\partial p_{i}}+1\right)\cdot\frac{\Delta t^{2}}{2},\nonumber \\
Q_{i,j}^{(5)} & =\frac{\partial\dot{v}_{i-\alpha}}{\partial v_{\alpha\hphantom{-i}}}\cdot\frac{\Delta t^{3}}{3}+\frac{\partial\dot{v}_{i-\alpha}}{\partial p_{\alpha\hphantom{-i}}}\cdot\frac{\Delta t^{2}}{2},\nonumber \\
Q_{i,j}^{(6)} & =\frac{\partial\dot{v}_{i-\alpha\hphantom{-1}}}{\partial v_{i-\alpha-1}}\cdot\frac{\Delta t^{3}}{3}+\frac{\partial\dot{v}_{i-\alpha\hphantom{-1}}}{\partial p_{i-\alpha-1}}\cdot\frac{\Delta t^{2}}{2},\nonumber \\
Q_{i,j}^{(7)} & =\frac{\partial\dot{v}_{i-\alpha}}{\partial v_{\alpha\hphantom{-i}}}\cdot\frac{\Delta t^{2}}{3},\nonumber \\
Q_{i,j}^{(8)} & =\frac{\partial\dot{v}_{i-\alpha\hphantom{-1}}}{\partial v_{i-\alpha-1}}\cdot\frac{\Delta t^{2}}{3}.\label{eq:Q-off-diagonal-elements}
\end{alignat}

\bibliographystyle{ieeetr}
\bibliography{references}

\end{document}